%% file: mainVerily.tex
\documentclass[rmp,letterpaper,superscriptaddress,aps,twocolumn, floatfix]{revtex4}
\usepackage{amsmath,amssymb}
\usepackage{bm}
\usepackage{graphicx,color}
\usepackage{placeins}
\usepackage[dvipsnames]{xcolor}
\usepackage{soul}       
\usepackage[colorlinks,linkcolor=darkstrawberry,citecolor=darkstrawberry]{hyperref}

\newcommand{\beq}{\begin{equation}}
\newcommand{\eeq}{\end{equation}}
\newcommand{\beqa}{\begin{eqnarray}}
\newcommand{\eeqa}{\end{eqnarray}}

\renewcommand{\vec}[1]{\bm{#1}}     
\newcommand{\header}[1]{\emph{#1.} --}

\newcommand{\kB}{k_\text{B}}        
\newcommand{\kT}{k_\text{B}T}       
\newcommand{\tauB}{\tau_\text{B}}   
\newcommand{\pecl}{\operatorname{P\kern-.08em e}}
\newcommand{\Pe}{\pecl}

\definecolor{myblue}{rgb}{0.003921,0.07058,0.47450}
\definecolor{teal}{rgb}{0.0,0.5664,0.5742}
\definecolor{strawberry}{rgb}{1.0,0.0,0.5}
\definecolor{darkstrawberry}{rgb}{0.875,0.0,0.4375}
\definecolor{royallBlue}{rgb}{0.0039,0.0705,0.4745}
\definecolor{blueberry}{rgb}{0.015686275,0.2,1}
\definecolor{dukeblue}{RGB}{1,33,105}
\definecolor{darkgreen}{RGB}{0,150,0}

\newcommand{\expcolor}[1]{{\color{darkstrawberry} #1}}

\definecolor{orange}{rgb}{1.0, 0.65, 0.39}

\begin{document}

\title{Colloidal Hard Spheres: Triumphs, Challenges and Mysteries}

\author{C. Patrick Royall}
\email{paddy.royall@espci.psl.eu}
\affiliation{Gulliver UMR CNRS 7083, ESPCI Paris, Universit\' e PSL, 75005 Paris, France.}
\affiliation{School of Physics, HH Wills Physics Laboratory, University of Bristol, Tyndall Avenue, Bristol, BS8 1TL, UK.}
\affiliation{School of Chemistry, University of Bristol, Cantock's Close, Bristol, BS8 1TS, UK.}

\author{Patrick Charbonneau}
\affiliation{Department of Chemistry, Duke University, Durham, North Carolina 27708, USA.}
\affiliation{Department of Physics, Duke University, Durham, North Carolina 27708, USA.}

\author{Marjolein Dijkstra}
\affiliation{Soft Condensed Matter, Debye Institute for Nanomaterials Science, Utrecht University, Princetonplein 1, 3584  CC, Utrecht, The Netherlands.}

\author{John Russo}
\affiliation{Department of Physics, Sapienza University of Rome, Piazzale Aldo Moro 5, 00185 Rome, Italy.}

\author{Frank Smallenburg}
\affiliation{Universit\'{e} Paris--Saclay, CNRS, Laboratoire de Physique des Solides, 91405 Orsay, France.}

\author{Thomas Speck}   
\affiliation{Institut f\"{u}r Theoretische Physik IV, Universit\"{a}t Stuttgart, Heisenbergstr. 3, 70569 Stuttgart, Germany.}

\author{Chantal Valeriani}
\affiliation{Departamento de Estructura de la Materia, Fisica Termica y Electronica, Facultad de Ciencias Fisicas, Universidad Complutense de Madrid, 28040 Madrid, Spain.}

\begin{abstract}
The simplicity of hard spheres as a model system is deceptive. Although the particles interact solely through volume exclusion, that nevertheless suffices for a wealth of static and dynamical phenomena to emerge, making the model an important target for achieving a comprehensive understanding of matter. In addition, while real colloidal suspensions are typically governed by complex interactions, Pusey and Van Megen [\emph{Nature} \textbf{320} 340--342 (1986)] demonstrated that suitably tuned suspensions result in hard-sphere like behavior, thus bringing a valuable experimental complement to the celebrated theoretical model. Colloidal hard spheres are thus both a material in their own right and a platform upon which phenomena exhibited by simple materials can be explored in great detail. These various purposes enable a particular synergy between experiment, theory and computer simulation. Here we review the extensive body of work on hard spheres, ranging from their equilibrium properties such as phase behavior, interfaces and confinement to some of the non--equilibrium phenomena they exhibit such as sedimentation, glass formation and nucleation.
\end{abstract}

\maketitle

\tableofcontents

\input{Introduction/introduction}
\input{HistoricalBackground/historicalBackground}
\input{Realizing/realizing}
\input{Simulations/simulations}
\input{Theory/theory}
\input{PhaseBehavior/phaseBehavior}

\input{BinaryHardSphereMixtures/binaryHardSphereMixtures}
\input{Confinement/confinement}
\input{FarFromEquilibriumPhenomena/farFromEquilibriumPhenomena}
\input{GlassTransition/glassTransition}
\input{NucleationCrystallisation/nucleationCrystallisation}
\input{SummaryOutlook/summaryOutlook}

\section*{List of symbols}

\begin{tabular}{cl}
$\sigma$ & Hard-sphere diameter \\
$\sigma_\mathrm{eff}$ & Effective diameter\\
$\phi$ & Volume (Packing) fraction\\
$\phi_\text{cp}$ & Close packing fraction\\
$\rho$ & Number density \\
$P$ & (Osmotic) pressure \\
$Z$ & Compressibility factor $\beta P/\rho$\\
$\eta$ & Solvent viscosity\\
$\tau_\text{B}$ & Brownian time\\
$\tau_\alpha$ & $\alpha$-relaxation time\\
$\tau_\beta$ & $\beta$-relaxation time\\
$\beta$ & $1/\kB T$\\
$u(r)$ & Pair interaction potential\\
$\kappa$ & Debye length \\
$\lambda_B$ & Bjerrum length \\
$D_S$ & Short-time diffusion\\
$D_L$ & Long-time diffusion\\
$D_0$ & Free diffusion\\
$J$ & Nucleation rate\\
$\xi_g$ & Gravitational length \\
$v$ & Velocity\\
$q$ & Size ratio \\
$k$ & Wavevector\\
$S(k)$ & Structure factor \\
$g(r)$ & Radial distribution function\\
$g(\sigma^{+})$ & Contact value of $g(r)$\\
$\gamma$ & Interfacial free energy \\
$\tau$ & Shear stress\\
$\gamma_s$, $\dot\gamma_s$ & Strain (rate) \\
\textit{s} & polydispersity \\
\end{tabular}

\begin{acknowledgments}
The authors would like to thank
Malcolm Faers,
Thomas Palberg,
Peter Pusey
and 
Brian Vincent
for helpful comments and discussions.
Peter Pusey, Alice Thorneywork and Eric Weeks are gratefully thanked for kindly helping with graphics,
Josh Robinson for digging up morphometric data, Lars K\"{u}rten for providing experimental data, and Ian Williams for remembering what some of the authors had long forgotten.
Bob Evans, Daan Frenkel, Peter Pusey and Brian Vincent are gratefully acknowledged for their helpful and insightful comments on the manuscript.
CPR and FS gratefully acknowledge support from the ANR via the grant DiViNew.
PC acknowledges support from the Simons Foundation (Grant No. 454937).
MD acknowledges financial support from the European Research Council (ERC Advanced Grant No. ERC-2019-ADV-H2020 884902,
SoftML). JR acknowledges support from the European Research Council Grant DLV-759187, and from ICSC – Centro Nazionale di Ricerca in High Performance Computing, Big Data and Quantum Computing, funded by European Union – NextGenerationEU.
CV acknowledges financial support from the MINECO (EUR2021-122001 and IHRC22/00002).
Data relevant to this work have been archived and can be accessed at the Duke Digital Repository~\cite{data}.
\end{acknowledgments}


\end{document}

%% file: Introduction/introduction.tex
\section{Introduction}
\label{sectionIntroduction}

\begin{figure}[t]
\centering
\includegraphics[width=40 mm]{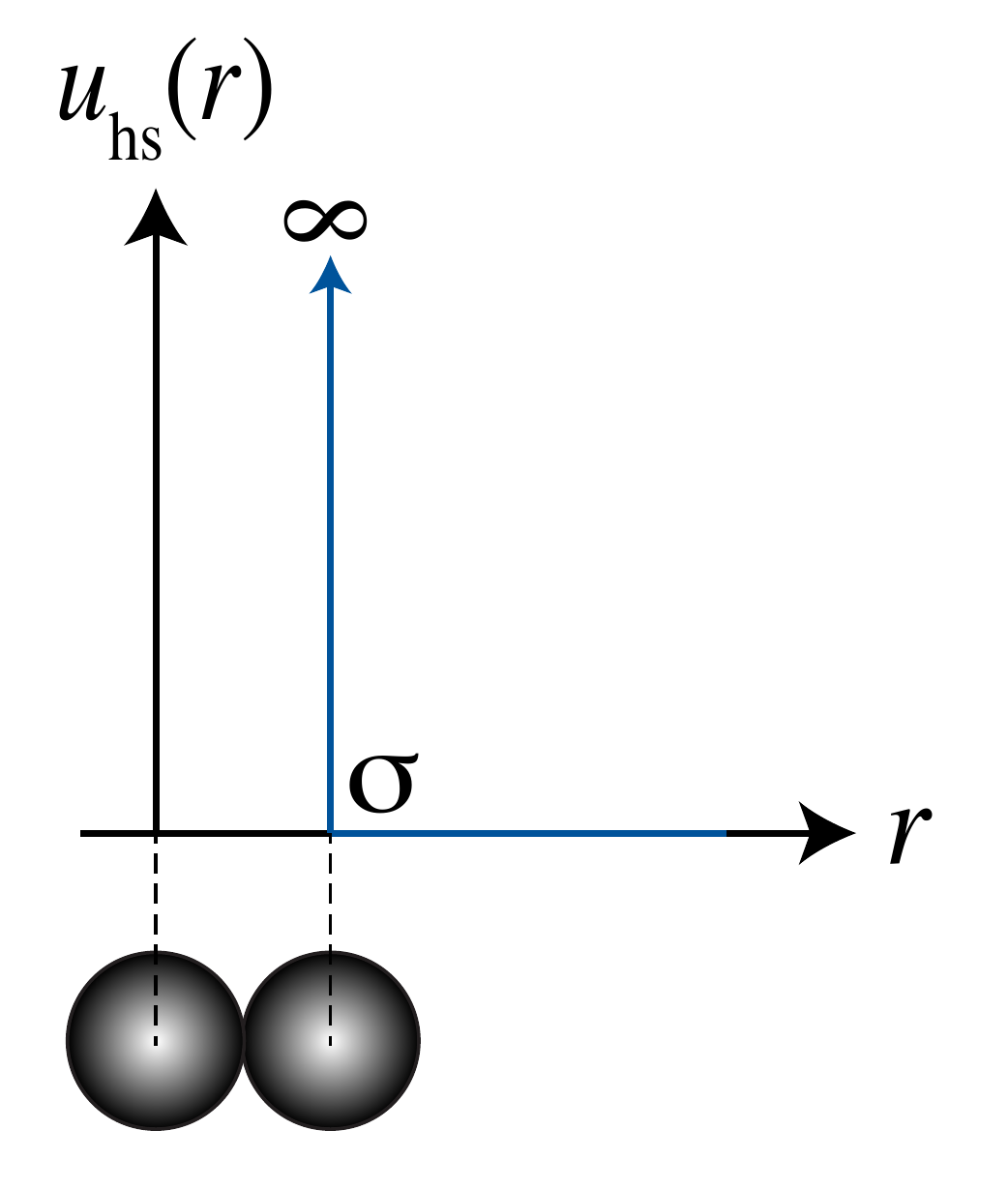}
\caption{Hard-sphere pair interaction potential $u_\mathrm{hs}(r)$ as a function of the center-to-center distance $r$ between particles of diameter $\sigma$.
\label{figUHS}}
\end{figure}

Polymers, liquid crystals, surfactants, and colloidal dispersions are key material pillars of soft matter. While physicists have long focused on the study of the first three, colloids did not receive comparable attention until relatively recently. Only thanks to significant advancements in the synthesis of colloidal particles in the second half of the 20th century -- motivated by both industrial interest~\cite{vincent2012,tadros,vanderhoff1956} and scientific curiosity~\cite{fijnaut1978} -- did the field come to light. The potential of hard--sphere--like colloidal dispersions as physical models of ``simple liquids'' exhibiting a fluid-crystal phase transition
~\cite{poon1996,frenkel2002,evans2019} 
was subsequently developed by Hachisu and co--workers~\cite{hachisu1974, kose1974}
and by \citet{vrij1978}. Hard-sphere particles, which are impenetrable at distances less than their diameter $\sigma$ and otherwise exhibit no interaction (see Fig.~\ref{figUHS}), had long been the object of theoretical enquiry, but had until then largely remained but an experimental fantasy (see Sec.~\ref{sectionHistorical}).

\begin{figure*}[t]
\centering
\includegraphics[width=175mm]{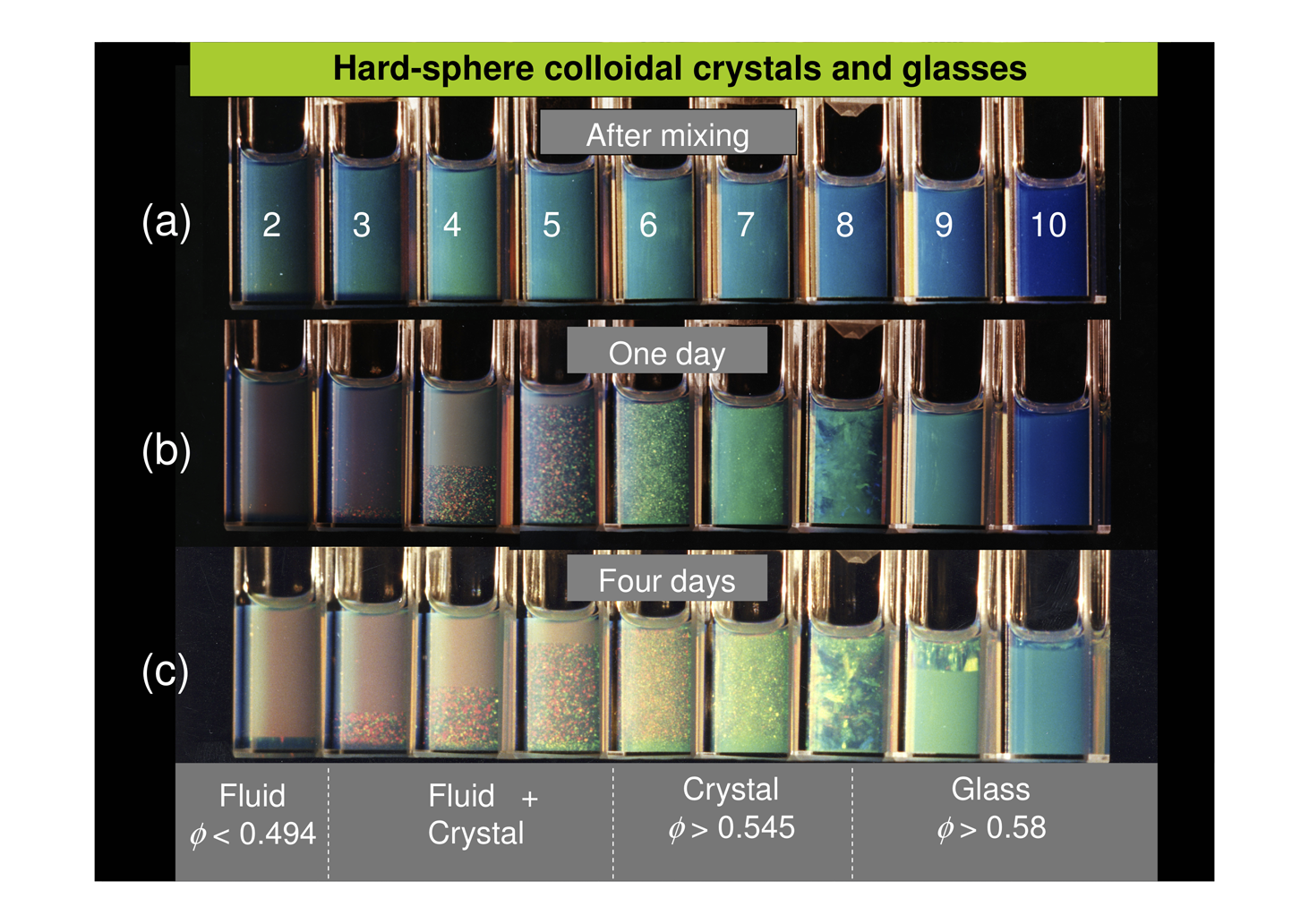} \caption{The colloidal hard spheres observed by Pusey and Van Megen are depicted (a) immediately, (b) one, and (c) four days after mixing. After four days, the system is presumed to have completed its phase separation. Volume fraction increases from left to right, with effective values (determined by reference to the phase behavior) $\phi = 0.491, 0.517, 0.525, 0.542, 0.568, 0.593, 0.611, 0.637$  and $0.654$ for samples 2--10 respectively. The samples range from a fluid phase to a fluid coexisting with an iridescent crystal; at slightly larger  volume fractions the whole sample is crystalline while at yet  higher volume fractions glassy amorphous states are encountered. These states initially ``coexist'' with the crystal until the entire sample ultimately becomes glassy. Reproduced from \citet{pusey2009} with permission.}
\label{figPuseyVanMegen} 
\end{figure*}

\begin{figure}[h!]
\centering
\includegraphics[width=85mm,keepaspectratio]{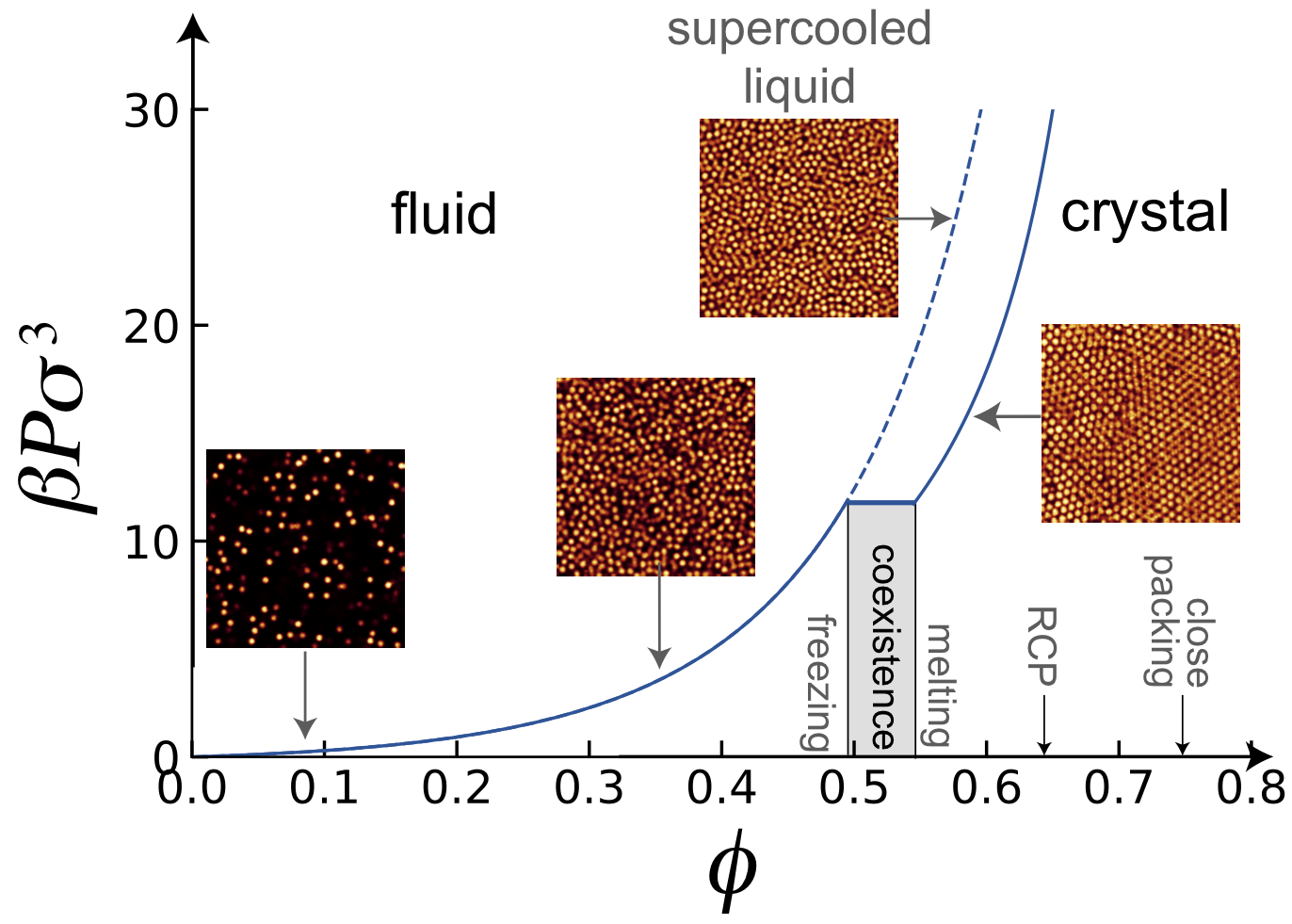} 
\caption{Equation of state and phase diagram of hard spheres. The pressure $\beta P \sigma^3$ as a function of volume fraction $\phi$ (solid blue) is (approximately) given by the Carnahan-Starling expression~\cite{carnahan1969} for the fluid and by that of Hall~\cite{hall1972} for the crystal. The metastable extension of the fluid branch (dashed blue) is where slow dynamics and the hard--sphere supercooled liquids and glass transition can be found, see Sec~\ref{sectionGlass}. 
Indicated on the $x$--axis are the fluid and crystal phases, along with freezing volume fraction of about $\phi_f=0.492$ and melting volume fraction $\phi_m=0.543$ and phase coexistence at pressure $\beta P \sigma^3=11.6$ (see Sec.~\ref{sectionEquilibriumPhaseBehavior}). Also indicated are random close packing $\phi_\mathrm{rcp}\approx0.64$ and crystal close packing $\phi_\mathrm{cp}=\pi/(3\sqrt{2})=0.740\ldots$. Approximate state points of the confocal microscopy images are denoted by grey arrows.
}
\label{figHSPhaseDiagram} 
\end{figure}

During the 1960s, '70s and '80s, advancements in colloid synthesis in industry~\cite{barrett1973,walbridge1966} and by pioneering colloid scientist Ron Ottewill~\cite{cairns1976,antl1986}, combined with the application of light scattering by soft matter physicists Peter Pusey and Bill van Megen, led to the development of a well-controlled colloidal model system (Fig.~\ref{figPuseyVanMegen})
~\cite{vanmegen1985,pusey1986,vanmegen1987}. Pusey and Van Megen then not only convincingly demonstrated hard-sphere freezing (Fig.~\ref{figHSPhaseDiagram}), thus experimentally validating theoretical~\cite{kirkwood1940} and 
computer simulation~\cite{alder1957,wood1957} predictions from a generation prior, but also---and perhaps more importantly---elegantly demonstrated the epistemic potential of colloidal suspensions.

Underlying the newfound physical interest is the fact that colloidal particles---like atoms and molecules---exhibit thermal motion which allows them to explore configurational space, and self-assemble into different phases, such as \emph{colloidal} crystals, liquids, and  gases (Fig.~\ref{figHSPhaseDiagram}). Of the plethora of possible colloidal systems, hard spheres have naturally emerged as the benchmark. Despite the challenge of synthesizing perfectly hard colloids, (micron-sized) hard spheres have four key strengths:
(\emph{i}) they have a single control parameter, the volume fraction $\phi$;
(\emph{ii}) they are \emph{classical} in nature, thus enabling accurate comparison with a broad array of theoretical predictions and large--scale computer simulation;
(\emph{iii}) they diffuse their own radius on the order of seconds, thus making their dynamics readily accessible in experiment;
(\emph{iv}) their size makes them 
amenable to optical techniques such as light scattering and confocal microscopy, thus enabling accurate measurement of spatial and dynamical correlations and even particle positions, without the need for large facilities such a synchotrons. Colloidal hard spheres are therefore prized as model systems. Quantitative tests and validations with theory nevertheless remain challenging. 
In this review, we specifically take a critical look at what has been achieved and what challenges remain to be faced in using colloids to verify theory. We also emphasize the experimental observations that have yet to be given an accurate theoretical description.

In order to manage the scope of this review, we largely restrain our consideration to experiments with colloids. Only where exceptionally relevant do we mention work with nanoparticles (see \citet{boles2016} for a dedicated review) or with granular matter (for which reviews are available in~\cite{liu2010,vanhecke2010,charbonneau2017,torquato2010rmp,arceri2022,forterre2008,bi2015}). Where possible, we have referenced relevant review papers from these other fields, but we humbly beg for grace from the reader regarding important material that we may have missed. We nevertheless hope that our review conveys that similar behavior can be observed in very different systems, and that the transfer of ideas between these fields can be a fruitful source of future research objectives.

Even after taking these thematic restrictions into account, the table of contents hints at the wide reach of the remaining scope. Because at first sight going through this whole review 
may seem daunting, we note that Secs.~\ref{sectionHistorical}-\ref{sectionTheory} deal with the historical perspective and methodology, with the results from hard sphere colloids being from Sec.~\ref{sectionEquilibriumPhaseBehavior} onwards. We also emphasize a few highlights:
\begin{itemize}
    \item phase behavior---colloidal experiments confirm the entropy-driven fluid-crystal transition in hard-sphere systems (see Sec.~\ref{sectionBulk});
    
    \item fluid structure---experimental measures validate pair structure estimates and track the development of even higher--order structures (see Sec.~\ref{sectionBulk});

    \item interfaces---the interfacial free energy underlies the barrier to nucleation, and grain boundaries in hard-sphere systems form a fundamental model for a key failure mechanism in crystalline materials 
    (see Sec.~\ref{sectionInterfaces});

    \item binary mixtures---colloidal suspensions have fueled the exploration of the rich phase behavior of multi--component systems (see Sec.~\ref{sectionBinary});

    \item confinement---the stabilization of a wealth of different structural arrangements of hard spheres has been achieved in both experiments and simulations (see Sec.~\ref{sectionConfinement});

    \item far-from-equilibrium behavior---hard spheres have been instrumental in disentangling hydrodynamic interactions from distortions of the local structure to, \emph{inter alia}, shed light onto the phenomenon of shear thickening (see Sec.~\ref{sectionFarFromEq});
    
    \item glass transition---colloidal hard spheres have markedly advanced the understanding of glasses, thanks to their simplicity and the possibility to image them in real space, thus providing access to a host of properties that are hard to access in molecular systems (see Sec.~\ref{sectionGlass});

    \item nucleation---the ability to directly image hard spheres enabled the first direct observation of a critical nucleus, which makes it possible to test more stringently the approximations that underlie classical nucleation theory (see Sec.~\ref{sectionNucleation}).
\end{itemize}
We trust that this selection will motivate at least a few of the more hesitant readers to continue through.

Before embarking on the bulk of this review, a few essential quantities must be defined. We have already encountered the first one: the hard-sphere diameter $\sigma$ (see Fig.~\ref{figUHS}). Although in theory it is a well defined quantity, in practice it might refer to the effective diameter, the hydrodynamic diameter, or the mean diameter in a suspension of slightly size polydisperse colloidal particles, see Sec.~\ref{sectionMapping}. The second has also been alluded to above: the number density of $N$ colloids within a volume $V$, i.e., $\rho=N/V$. More often, we will employ the dimensionless volume (or packing) fraction $\phi=Nv_1/V=\pi \sigma^3 \rho/6$ with $v_1=\pi\sigma^3/6$ the volume of a single three-dimensional (3d) sphere. The conjugate variable to density is the pressure $P$, our third quantity. It is often beneficial to consider the dimensionless \emph{reduced pressure} $Z=\beta P/\rho$ (also known as the compressibility factor), which combines pressure with density and the inverse temperature $\beta=(\kB T)^{-1}$ where $\kB$ is Boltzmann's constant and $T$ is temperature. Note that i the case of colloidal hard spheres, the relevant quantity is the \emph{osmotic} pressure. The final quantity is the (Brownian) time taken by a single sphere to diffuse by its own radius 
\begin{equation}
\tauB = \frac{(\sigma/2)^2}{D_0} = \frac{3\pi\eta\sigma^3}{4 \kB T}
\label{eqTauB}
\end{equation}
where $D_0$ is the (bare) diffusion coefficient and $\eta$ is the solvent viscosity.

As Fig.~\ref{figUHS} makes clear, the potential energy in a hard sphere system is zero. This means that only entropy contributes to the phase behavior. Furthermore, provided the temperature is non-zero, ie the system is thermal, temperature plays no role in the behavior of the system beyond scaling the timescale as shown in Eq.~\ref{eqTauB}. It is this independence of temperature that leads to density or volume fraction or pressure being the sole control parameter for hard sphere.

%% file: HistoricalBackground/historicalBackground.tex
\section{Historical background}
\label{sectionHistorical}

Given the extended and intricate history of hard spheres before they became an  object of study experimentally, a \emph{longue-durée} overview of the topic is in order. This section summarizes the distinct histories of hard spheres as a model of matter and as  colloidal particles that mimick that model.

\subsection{History of hard-sphere models}

Describing atoms as hard elastic spheres -- akin to billiard balls -- finds its origin in the early days of the kinetic theory of gases~\cite{brush2003}. The laws of impact derived in the 17th century underlie the first such theory, which was formulated by Bernoulli in the early 18th century. That approach, however, was deemed disputable at the time and did not have much immediate scientific impact. Further formalization by Maxwell in the mid-1800s, which built explicitly on ``systems of particles acting on each other only by impact", was more kindly received, and gave rise to a large body of work on kinetic theories~\cite{brush2003}. In the words of science historian Stephen Brush, this \emph{exercise in mechanics} ``not only helped establish the theory, but laid the foundations for modern statistical mechanics''.

It is therefore unsurprising that Van der Waals' attempt at a microscopic model of condensation, a couple of decades later, used a model effectively consisting of \emph{hard spheres} with weak, fairly short-range attractive forces~\cite{nairn1972}~\cite[p.~407]{brush1976}\footnote{We now know the Van der Waals equation of state to be exact only for a one-dimensional system with infinitely long range and weak attraction~\cite{kac1962,hansen,niss2018}.}. 
Interestingly, Van der Waals' description gave the correct second virial coefficient, $B_2$, for the reference model~\cite{brush1976}. Virial coefficients, which characterize how the equation of state for the pressure $P$ as a function of density  $\rho$ deviates from the ideal gas reference, i.e.,
\begin{equation}
Z=\frac{\beta P}{\rho}=1+\sum_{i=1} B_{i+1} \rho^i,
\label{virial}
\end{equation}
quickly became an important tool to characterize models of intermolecular interactions. (An example is shown in Fig.~\ref{figHSPhaseDiagram}. See Secs.~\ref{sectionIntegral} and \ref{sectionEquationOfState} for further discussion.) For hard spheres, $B_3$ and $B_4$, were hence determined by Boltzmann before the end of the 19th century~\cite{nairn1972}. Beyond that point, however, the algebra is intractable. It therefore took more than fifty years and the advent of electronic computers for estimates of higher-order coefficients to be obtained. The numerical determination of $B_5$ by \citet{rosenbluth1954} renewed the quest for accurate  estimates, which currently extend up to $B_{12}$~\cite{clisby2006,wheatley2013}. 

Hard spheres also formed the basis for studying liquid \emph{structure}. Early experiments with lead shot in the 1920s~\cite{smith1929} and with hard gelatin balls in the 1930s~\cite{morrell1936} were followed after the war by Bernal's extensive work with ball bearings~\cite{bernal1960,finney2013} and by more controlled theoretical descriptions, as further discussed in Sec.~\ref{sectionJamming}. Hard spheres also contributed to establishing the scaled particle theory of liquids~\cite{reiss1959}, and were considered by Wertheim as a solution of the Percus-Yevick approximation to the Ornstein–Zernike equation that formally describes the liquid structure~\cite{wertheim1963}. Hard spheres were subsequently a natural object for other integral equation approaches~\cite{hansen,rowlinson2005} (see also Sec. \ref{sectionIntegral}).

A parallel research effort into hard spheres stemmed from Kirkwood's speculation~\cite{kirkwood1939}, which he quickly substantiated with his postdoc Elizabeth Monroe~\cite{kirkwood1940,kirkwood1941,kirkwood1951}, that a hard-sphere fluid becomes unstable to crystallization at high density~\cite{charbonneau2022,hoddeson1992} (see Fig.~\ref{figHSPhaseDiagram}). The somewhat surprising proposal that the ordered phase could be entropically more favorable than the dense (and disordered) liquid was independently formulated by \citet{fisher1955} a decade later, and supported by early simulations using molecular dynamics by \citet{alder1957} and  Monte Carlo sampling by \citet{wood1957}\footnote{It is important to note that the contributions of coders for many of these works were not acknowledged through authorship~\cite{battimelli}. The author list of these works therefore does not fully reflect the full scope of intellectual contributions that went into their realization.}. The ensuing debate about the physicality (and eventual validation) of such a phenomenon led to various computational advances, as was recently reviewed by \citet{battimelli}.

The study of the crystal state of hard spheres also has a long history. (That of disordered jammed packings of hard spheres is somewhat more recent; see Sec.~\ref{sectionJamming}.) The Kepler conjecture for their densest packing -- their infinite-pressure state with a packing fraction $\phi_\text{cp}=\pi/(3\sqrt{2})$ -- dates back to the 17th century and was proven by Hales only relatively recently~\cite{aste2008,hales2017}. Finite-pressure descriptions of hard-sphere crystals are less ancient, but have reasonably rich antecedents as well. The familiar description in terms of free volume, known as cell theory (Sec.~\ref{sectionCellTheory}) takes root in various attempts to characterize the \emph{liquid state} in the 1930s~\cite{rowlinson2005}. The free-volume description only became a useful reference for hard-sphere crystals in the 1950s~\cite{wood1952,salsburg1962} and persists to this day in the pedagogical literature (see, e.g., \citet{barrat2003,kamien2007soft}). However, because the associated corrections are less well controlled and are computationally harder to evaluate than for the virial series (see \citet{charbonneau2021three} and references therein), cell theory has not been as quantitatively influential.

\subsection{The foretelling: How PMMA colloids came to be}
\label{sectionForetelling}

Despite the extended theoretical importance of hard spheres, the system long remained but a distant abstraction as far as experiments were concerned. It is only through a gradual increase in the control over colloidal experiments that the state of affairs has changed. In this section, we focus  specifically on the developments that have led to the sterically stablized poly-methyl methacrylate (PMMA) system. Although other systems have been utilized to model hard spheres~\cite{russel,hunter,royall2013myth} (see Sec.~\ref{sectionSynthesizing}), PMMA-based colloids have predominated since their introduction. During the first 10-15 years of its use,~\footnote{Specifically before the density matching of larger PMMA particles for confocal microscopy , see Sec.~\ref{sectionInteractions} around the millenium.} and as the field burgeoned,any particle softness is attributed to the steric stabilization layer, which is much shorter than the particle diameter. Because PMMA particles have a bulk modulus that is orders of magnitude larger, the tiny stablization layer only slightly perturbs the system away from true hard sphere behavior (see Sec.~\ref{sectionInteractions}).

The origins of modern colloid science may be traced back to the latter part of the nineteenth century with the emergence of physical chemistry as a distinct discipline~\cite{vincent2012,vincent2018}, and early discoveries, such as the first synthetic suspension of (optically active) gold nanoparticles by~\citet{faraday} and the use of colloidal sedimentation profiles to determine Boltzmann's constant by Perrin~\cite{perrin}. A major challenge to the use and application of colloidal suspensions is that they typically aggregate irreversibly when concentrated. Understanding of this phenomenon took a great leap forward with the independent development---during the Second World War---by Derjaguin and Landau~\cite{derjaguin1941} in the Soviet Union and by Verwey and Overbeek~\cite{verwey1948} in the Netherlands of the DLVO theory for the interactions between charged colloids. DLVO theory notably provided a first-principles explanation for the long-noted \emph{salting out} phenomenon for aqueous suspensions of both colloids and proteins, in which salt addition leads to aggregation and precipitation.

Colloid synthesis further progressed in the postwar years~\cite{vanderhoff1956,ihler} not least as certain industrial products, especially paints and coatings, called for novel stabilization schemes~\cite{barrett1973}. It had long been known that addition of (colloidal) spheres to a solution increases viscosity~\cite{batchelor}, hence when in the 1960s control was sought over the rheology of paints and coatings, adding colloids seemed like a natural solution. However, these largely oil--based products ---this being long before legislation promoted the use of water--based coatings which release fewer volatile organic compounds as they dry~\cite{derksen1995}---called for stabilizing colloids in non--aqueous solvents. In these systems, electrostatic charging was assumed to be negligible and therefore incapable of stabilizing colloidal dispersions. (It was not until the turn of the third millennium that electrostatic charging in non--aqueous solvents would be investigated thoroughly (see Sec.~\ref{sectionInteractions})~\cite{yethiraj2003,roberts2007,royall2013myth,royall2021jpcm}.) Another method for stabilization was therefore sought out.

Inspiration came from the burgeoning developments in polymer physics, notably from the work of Flory~\cite{degennes}. The loss of configurational entropy of polymers grafted to surfaces was proposed as a colloidal stablization mechanism. Early work included the stablization of carbon black through phy\-si\-sorption of aromatic hydrocarbons~\cite{vanderwaarden1950,overbeek1966}. In the 1960s, steric stabilization was first combined with a (relatively) monodisperse nonaqueous colloidal system at Imperial Chemical Industries (ICI), a major paint and coatings manufacturer of the time, by coating polymethyl methacrylate (PMMA) particles with polyhydroxy steric acid (PHSA)~\cite{walbridge1966,barrett1973}. In the UK, at that time leading academic research in colloids took place in Cambridge University, University College London and the University of Bristol \cite{vincent2012}. In Bristol, in particular, Ottewill had strong links with ICI from where the sterially stablized PMMA was initially obtained. Once synthesized in Bristol~\cite{cairns1976}, these particles played a central role in the hard-sphere story. They not only closely approximated the model of interest, but they were amenable to solvents which match their refractive index and mass density. As we will see, the former is essential for light scattering (Sec.~\ref{sectionLightScattering}), and both are necessary for confocal microscopy (Sec.~\ref{sectionOptical}).

In 1974, Ottewill met Pusey who was then based at Malvern, close to Bristol, and the two became interested in exploring light scattering as a means to study strongly interacting colloidal systems~\cite{pusey1974,brown1975} as a means to study the PMMA system~\cite{brown1975}. With the arrival in Bristol of Van Megen from the Royal Melbourne Institute of Technology and through a series of extended back-and-forth visits, Ottewill's ability to synthesize the hard-sphere-like PMMA and Pusey's interest in these system naturally led to the series of seminal papers that gave rise to Fig.~\ref{figPuseyVanMegen}, see e.g.,~\cite{pusey1986,pusey1987prl,vanmegen1985,vanmegen1987}.

%% file: Realizing/realizing.tex
\section{Realizing hard-sphere systems}
\label{sectionRealizing}

In this section, we briefly discuss the main experimental methods that have been brought to bear in the study of colloidal hard spheres more specifically. For a more general treatment of colloidal synthesis, we  refer the reader to appropriate reviews~\cite{russel,hunter,ihler}. It is worth emphasizing that, in experiments, truly hard spheres do not exist~\cite{poon2012,royall2013myth}.  In order to understand how far real experimental systems deviate from hard spheres, different mappings to hard spheres are discussed in Sec.~\ref{sectionMapping}.

\subsection{Experimental methods to prepare hard-sphere-like systems}
\label{sectionExperimental}

Experimental techniques for colloidal hard spheres into three main categories: synthesis, characterization and observation. The first has a chemical flavor and is typically carried out by chemists, or by specially-trained physicists. The second and third, by contrast, are more typically practised by physicists (and are discussed in Sec.~\ref{sectionMeasuring}). These tasks can be conducted independently, as it is common to buy colloidal particles (especially silica and polystyrene) from commercial suppliers. However, it is worth noting that there is no commercial supplier for the (canonical) sterically stabilized PMMA suspensions discussed in Sec.~\ref{sectionForetelling}~\footnote{Worse, the PHSA sythesis has proven to be challenging to challenging to reproduce, leading to very polydisperse, or colloidally unstable, PMMA particles. At present, the global stocks of PHSA are less than 10kg.}. (A short review of PMMA synthesis is also given in~\citet{dullens2006sm}.) Here, we mainly focus  on the characterization and observation of colloidal hard-sphere suspensions.

Aside from identifying the ``hardness'' of the particles (see Sec.~\ref{sectionInteractions}), a key property of these suspensions is their polydispersity $s$. This is typically expressed as the standard deviation of the particle diameter normalized by the mean diameter expressed as a percentage. For PMMA synthesis, a ``good'' level of polydispersity is considered to be $s \lesssim 5\%$ such that the system crystallizes as readily as in the monodisperse case. (In Sec.~\ref{sectionPolyPhase} we will report a detailed  discussion of the impact of polydispersity on the phase behavior.) Colloids produced from other materials, such as polystyrene, can exhibit significantly lower polydispersity, down to $s\approx1\%$~\cite{russel,vanderhoff1956}.

It is worth emphasizing that polydispersity distributions in experimental systems cannot be prescribed.  While in numerical simulations one may easily change from Gaussian, to top hat or  lognormal distributions, in experiment the distribution is at the mercy of the particular synthesis. Even bimodal distributions can sometimes arise (due to secondary nucleation events, in which a second wave of colloids form during the reaction)~\cite{kawaguchi2005}.

\subsection{Synthesizing hard-sphere-like colloidal particles}
\label{sectionSynthesizing}

The method adopted for  PMMA synthesis was \emph{dispersion polymerisation}~\cite{kawaguchi2005,barrett1973}. In this approach, the methyl-methacrylate monomers and methacrylic-acid are dispersed in a suitable solvent (typically an alkane). Under the action of an \emph{initiator}, polymerisation begins at a somewhat elevated temperature (around 80$^\circ$C) and in principle proceeds until the monomer is exhausted. Relatively early in the polymerization process, the initially transparent solution acquires a blueish tint, due to the growing polymers falling out of solution and condensing into colloids that scatter predominately blue light. Polyhydroxy-stearic acid (PHSA) is added to the synthesis and is typically physisorbed to provide steric stabilization. In this way, the strerically stabilized polymethyl methacrylate (PMMA) system was born~\cite{walbridge1966,barrett1973,cairns1976,antl1986}. The size of the particles can be controlled with the monomer concentration, enabling for example binary systems with a given size ratio to be synthesized, or a system whose size optimized density matching (easier with small particles) against higher-quality imaging (better with large particles)~\cite{bosma2002,poon2012}.

Developments in synthesis include (\emph{i}) partial control over  the particle size by the monomer concentration~\cite{poon2012,bosma2002},  (\emph{ii}) locking the stabilizer, such that it is covalently bonded to the PMMA polymer backbone, and (\emph{iii}) crosslinking the PMMA chain. The latter prevents  particles swelling and dissolving in a good solvent. Using such crosslinked particles as a core is important for confocal microscopy because it enables a PMMA \emph{shell}  to be grown around the particle~\cite{dullens2003,dullens2006sm}\footnote{The particles used by~\citet{kose1974} had little to do with modern PMMA particles, except their chemical nature. In particular, they were not sterically stabilized.}: 
if the initial step (the \emph{core}) is labeled with a fluorescent dye and the shell is unlabeled, then tracking particle locations with confocal microscopy  is much easier, thanks to the clean separation between different fluorescent cores  (see Sec.~\ref{sectionOptical})~\cite{vanblaaderen1992langmuir,ivlev}.

While many key early hard-sphere experiments were carried out with sterically stabilized PMMA, and indeed the majority of the work we shall consider used PMMA colloids, it is far from being the only material of interest. Silica may also be stabilized in a non--aqueous solvent~\cite{vanhelden1980}. Aqueous systems of polystyrene and silica colloids form reasonable approximations to hard spheres, as do microgel particles~\cite{royall2013myth} and emulsion droplets~\cite{dong2022}. Because the diameter of microgel particles can be tuned \emph{in-situ}, the effective volume fraction to be changed at will, and therefore a single sample can access multiple state points~\cite{yunker2014}.

Two promising systems have been developed more recently. One involves the use of an aqueous copolymer of fluorinated methacrylate and methacrylate with a tunable density and refractive index, as demonstrated by~\citet{kodger2015}. The other involves the use of 3-(trimethoxysilyl)propyl methacrylate (TPM) particles~\cite{liu2016,liu2019}. Impressive results also have been obtained with nanoparticles, whose small size enables their faster diffusion than for the larger colloids and their self--assembly into larger structures~\cite{boles2016} (see Sec.~\ref{sectionSize}). 
Early work in the same size range used microemulsions, which are thermodynamically stable droplets of \emph{nanometer} dimensions (despite their name).

\subsection{Interactions in real hard-sphere-like systems}
\label{sectionInteractions}

Central to identifying the state point of a hard-sphere-like system is determining  the (effective) volume fraction $\phi_\mathrm{eff}$, as reviewed in~\cite{poon2012,royall2013myth}. The notion of  effective hard-sphere diameter will be properly discussed in Sec.~\ref{sectionMapping}, but we here first discuss interactions commonly used in experiments to mimic hard spheres. We consider two sizes, $\sigma=200$ nm and $\sigma=2000$ nm, which roughly reflects the range of colloid sizes used in the work discussed here. We note the impact of the colloid size upon the system dynamics (Sec.~\ref{sectionSize}) and their applicability for particular experimental techniques (see Sec.~\ref{sectionMeasuring}).

\begin{figure}[h!]
\centering
\includegraphics[width=85mm,keepaspectratio]{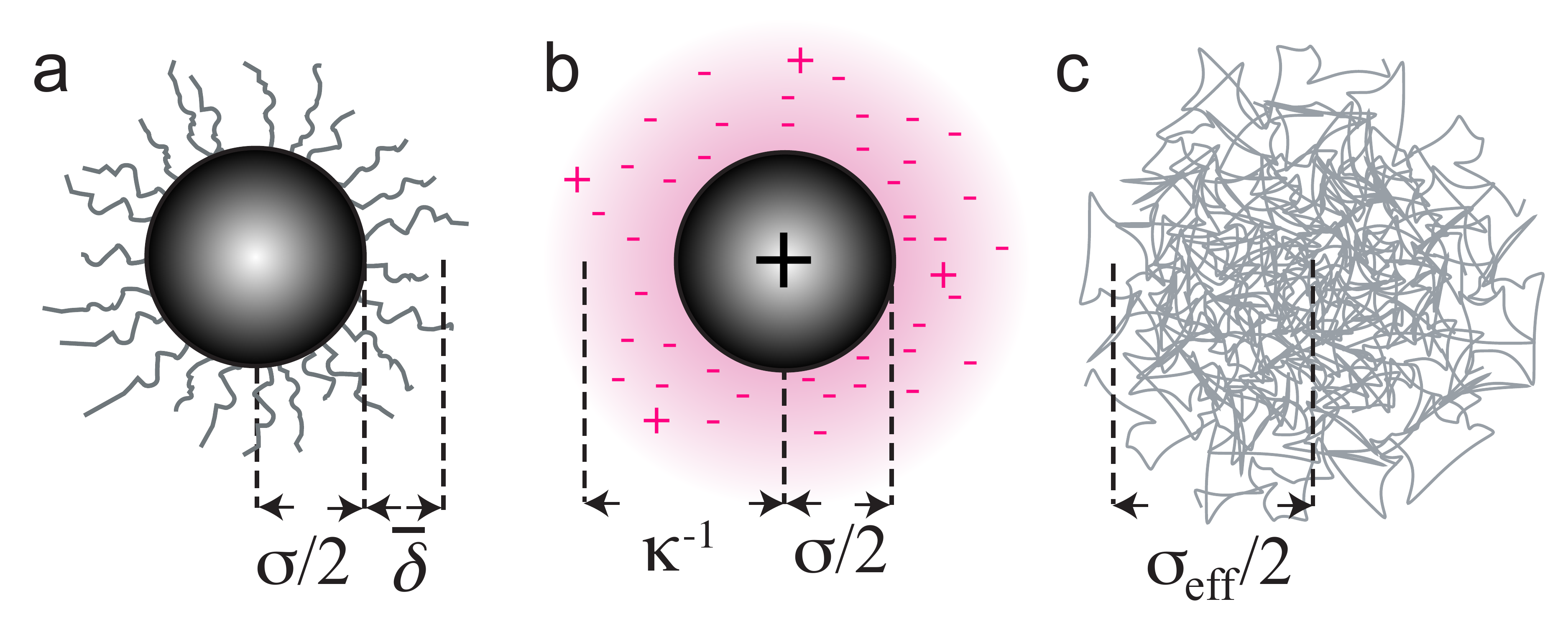} 
\caption{Schematic representation of various models for hard-sphere colloids.  
(a) A sterically-stabilized particle has surface ``hairs'' (not to scale) of average
thickness $\bar{\delta}$, resulting in core-stabilizer diameter $\sigma_\mathrm{cs}=\sigma_{c}+2\bar{\delta}$. 
(b)  A charged colloid has an electrical double layer (shaded) that gives rise to an effective diameter $\sigma_\mathrm{eff}$. 
(c) A microgel particle is a heavily cross-linked polymer. Figure reproduced from~\cite{royall2013myth}.
}  
\label{figRealHard} 
\end{figure}

While we have discussed the origins of some hard-sphere experimental systems, we should assess how close to hard spheres these truly are~\cite{royall2013myth,poon2012}. Here, we provide a summary for the purposes of this review. Practical hard-sphere--like systems fall into three broad categories, as illustrated in Fig.~\ref{figRealHard}. Colloidal stabilisation can be achieved either via (a) steric stabilization or (b)  charge stabilization. Because (c) microgel particles consist of densely crosslinked polymers in a swelling solvent, they do not explicitly need stabilization~\cite{schneider2017,lyon2012}. The hardness of microgel colloids depends on the quality of the solvent and the degree of crosslinking and can be density dependent~\cite{royall2013myth}. Emulsion droplets are an intermediate case. They use a molecularly thin layer of surfactant for stabilization. Although these particles are liquid, even a \emph{microscopic} ($\lesssim1$nm$^2$) change in surface area leads to an interfacial free energy cost of order $\kB T$. The softness of mesoscopic-scale droplets is, therefore, very limited~\cite{dong2022,morse1993,lacasse1996}.

\begin{figure}[h!]
\centering
\includegraphics[width=85mm,keepaspectratio]{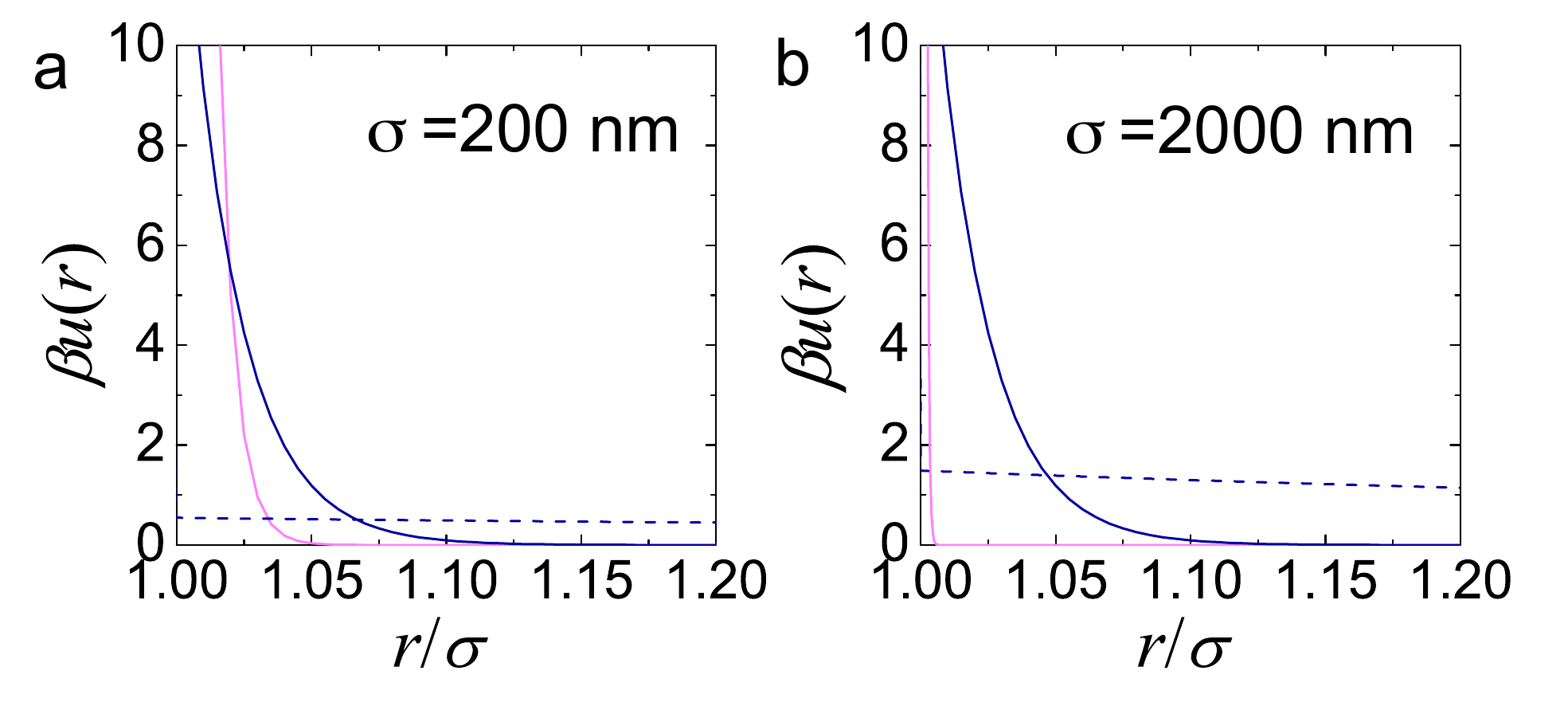} 
\caption{Estimation of effective colloid-colloid interactions for various hard-sphere-like scenarios for particle diameters (a) $\sigma=200$ nm and (b) $\sigma=2000$ nm. Shown is steric stabilization (pink line), strong electrostatic interaction (solid blue line) and weak electrostatic interaction (dashed blue line). See text for additional details. Figure reproduced from~\cite{royall2013myth}.} 
\label{figU} 
\end{figure}

\header{Softness due to stabilization} A sterically stabilized particle is shown schematically in Fig.~\ref{figRealHard}(a). One of us~\cite{royall2013myth} has previously reviewed ways to obtain an effective hard-sphere diameter for these particles via mapping various hard-sphere properties (see Sec.~\ref{sectionMapping}). A mapping is also possible via direct measurement of colloidal interactions. This procedure has been applied to poly-hydroxy-steric acid-stabilized PMMA particles, whose interactions were measured with the surface force apparatus and were found to be quite well described by an inverse power law potential with energy scale $\epsilon_\mathrm{ipl}$~\cite{bryant2002}\footnote{While an inverse power law of course is long-ranged, unlike the steric stablization, here we follow the work of~\citet{bryant2002}.}: 
\begin{equation}
u_{s}(r) \approx \epsilon_\mathrm{ipl} \left(\frac{\sigma}{r}\right)^n, 
\label{eqUs}
\end{equation}
where $\sigma$ denotes the particle diameter and $r$ the interparticle distance. The relative range of $u_{s}$ depends on the particle size. For example, $n=170$ was determined for particles with a diameter of $\sigma=200$ nm. Likewise, the strength of the interactions also depends on the particle size, with $u_{s}(\sigma)=146$ $\kB T$ was reported for  $\sigma=200$ nm, as shown by the pink line  in Fig.~\ref{figU}(a). For larger particles ($\sigma=2000$ nm), $n=1800$ and $u_{s}(\sigma)\approx1800$ $\kB T$, were determined~\cite{bryant2002} [Fig.~\ref{figU}(b), pink line]. The results of Bryant \emph{et al.}~\cite{bryant2002} quantify what is intuitively obvious, namely, that for a fixed length of stabilizing ``hairs'', larger particles are relatively harder. Similar conclusions were obtained in the rheological study of~\citet{mewis1989}, which varied the thickness of the stabilizing layer.

\header{Electrostatics}
It is now generally accepted that immersion of a colloid in a liquid medium systematically gives rise to some degree of charging~\cite{yethiraj2003,leunissenthesis,roberts2007,royall2013myth}. This charge is a potential source of softness that should always be considered in experiments. In determining the degree of softness due to electrostatic interactions, the linearized Poisson-Boltzmann (PB) theory is incorporated into the Derjaguin-Landau-Verwey-Overbeek (DLVO) theory~\cite{verwey1948} to describe the interactions between charged colloids. The original DLVO potential consists of Van der Waals (VdW) and electrostatic components. That contribution often becomes very small due to refractive index matching between colloids and solvent as in the case of light scattering or (3d) microscopy. In the case of quasi-2d systems, or techniques which do not require index matching such as rheology, electrostatic stabilization ensures that particles do not come sufficiently close together for VdW interactions to become important. Therefore, we shall henceforth neglect VdW interactions except where explicitly stated.

We are often interested in systems in which sterically-stabilized particles become charged. In this case, we can consider an inter-particle potential consisting of a steric repulsion, $u_s(r)$, and an electrostatic interaction, which in linearized Poisson--Boltzmann theory has a Yukawa form, $u_\mathrm{yuk}(r)$:
\begin{eqnarray}
u(r)&=&u_{s}(r)+ u_\mathrm{vdW}(r) + u_\mathrm{yuk}(r),
\label{eqU}\\
u_\mathrm{yuk}(r)& =& \epsilon_\mathrm{yuk}\frac{\exp \left[-\kappa ( r-\sigma) \right] }{r/\sigma},\label{eqYuk}
\end{eqnarray}
where $\kappa$=$\sqrt{4\pi\lambda_{B}\rho_\mathrm{ion}}$ is the inverse Debye screening length with $\rho_\mathrm{ion}$ the number density of monovalent ions, and $\epsilon_\mathrm{yuk}$ denotes the contact value of the Yukawa interaction given by
\begin{equation}
\beta\epsilon_\mathrm{yuk}=\frac{Z_e^{2}}{(1+\kappa\sigma/2)^{2}}\frac{\lambda_{B}}{\sigma},
\label{eqEpsilonYuk}
\end{equation}
\noindent
with $Z_e$ the number of electronic charges on the colloid. The Bjerrum length 
\begin{equation}
\lambda_{B}=\frac{e^{2}}{4\pi\epsilon_{0}\epsilon_{r}\kB T}, 
\label{eqBjerrum}
\end{equation}
is the distance at which the interaction energy between two elementary charges equals  $\kB T$,  $e$ is the electronic elementary charge, $\epsilon_{0}$ the permittivity of the vacuum, and $\epsilon_{r}$ the relative dielectric constant.

We first consider the case for aqueous systems in which the charging is rather strong. In 3d, most work with aqueous systems has used rather small colloids. We provide  an estimation of the electrostatic interaction potential by plotting the solid blue line in Fig.~\ref{figU}(a),  where we evaluate the Yukawa interaction   Eq.~\eqref{eqYuk} as follows. We assume that salt is added such that the Debye length is 4 nm. It has been suggested that an upper bound for the effective charge in many systems~\cite{alexander1984} can be described by the rule of thumb $Z_e \lambda_B / \sigma \approx 6$. For higher values of the colloid charge, the electric field at distances beyond the Stern layer reduces to the electric field corresponding to the effective charge  due to ion condensation~\cite{alexander1984,russel}. With water as the solvent, this leads to an (effective) colloid charge of $Z_e=1500$ for the colloid diameter of interest, $\sigma=200$ nm. The resulting electrostatic interaction (represented by the solid blue line in Fig.~\ref{figU}(a)) demonstrates a noticeable degree of softness.

Sterically stabilized PMMA is used in non-aqueous solvents. For example, the work of Pusey and Van Megen used a mixture of cis-decalin and carbon disulphide  with dielectric constant $\epsilon_r = 2.64$~\cite{pusey1986},
which leads to a Bjerrum length $\lambda_{B}\approx 21$ nm. 
With ionic sizes in the range of $\lesssim 1$ nm, one expects strong coupling between oppositely charged ions and consequently very little dissociation of surface groups. It was therefore a long-held assumption that electrostatics could be safely neglected. However, more recent work has demonstrated that some electrostatic charging is \emph{always} present~\cite{yethiraj2003,roberts2007,klix2013,leunissenthesis}.

To evaluate electrostatic interactions in solvents with such low dielectric constants, we  again use the $Z_e \lambda_B / \sigma \approx 6$  criterion with a Debye length of $\kappa^{-1} = 5\mu$m, which is consistent with measurements of the very small ionic strength found in such solvents~\cite{klix2013}. We can then evaluate Eq.~\eqref{eqYuk} for particles with diameters of  $\sigma=200$ nm  and  $\sigma=2000$ nm, which are represented by blue dashed lines in Fig.~\ref{figU}(a) and Fig.~\ref{figU}(b), respectively.  We see that the electrostatic repulsion changes very little over the chosen range. It is worth noting that for the smaller particles ($\sigma=200$ nm), these parameters correspond to an electrostatic charge of just $Z_e=2$. Although small, comparable values have been measured in experiments~\cite{klix2013}. For this case, it is highly probable that the spherically symmetric DLVO approach would not be accurate. However, the effects seem small enough that one can reasonably neglect electrostatic charging for this combination of  solvent dielectric constant and particle size. Larger colloids, however, require density matching to suppress sedimentation. Since two of the denser solvents used in experiment, namely tetrachloroethylene (TCE) and carbon tetrachloride, can be readily absorbed by PMMA, both the  density and the refractive index of the particles change, resulting in a turbid system characterized by substantial Van der Waals interparticle attractions. The hard-sphere-like behavior of the particles is then lost~\cite{ohtsuka2008,royall2013myth}. Very recent work has nevertheless found excellent hard-sphere behavior using a solvent mixture with TCE~\cite{kale2023}.

Finally, we consider stronger electrostatic charging in the PMMA system often applied in studies using confocal microscopy for larger colloids with $\sigma=2000$ nm. The solvents used for the smaller PMMA particles used for light scattering have a relative dielectric constant $\epsilon_r\approx2$. Relative to these, the density matching mixture of cyclo-hexyl bromide (CHB) and \emph{cis}-decalin has a dielectric constant of $\epsilon_r=5.4$, which is due to the CHB component with $\epsilon_r=7.9$. While non-aqueous, the dielectric constant is nevertheless much higher than that mentioned in the preceding paragraph and tetrabutyl ammonium bromide (or a similar salt) is typically used to screen the electrostatic interactions~\cite{yethiraj2003}. CHB advantageously appears to be less aggressive towards the PMMA particles than does TCE for example, with rather less absorption and swelling. If we again use the same criterion $Z_e \lambda_B / \sigma \approx 6$ we arrive at $Z_e\approx500$. The Debye length corresponding to a saturated solution of this salt is around $\kappa^{-1}\approx 100$ nm~\cite{leunissenthesis,royall2006,royall2003}, Evaluating Eq.~\eqref{eqYuk} then gives the solid blue line in Fig.~\ref{figU}(b), which exhibits a very considerable degree of softness. Therefore for the larger PMMA particles, which require density matching, one can either accept some softness and add salt, or risk attractions due to solvent absorption (in the case of TCE for example).

Imaging quasi-2d or strongly confined systems -- obtained, for example, by sedimenting particles onto a substrate~\cite{ivlev} -- is somewhat relieved from such drawbacks. For example, polystyrene colloids in an aqueous solvent with a suitable amount of salt added are rather hard as the Debye length of a few nm is very much less than the micron scale particle diameter~\cite{royall2013myth}. Particles of size 3-5 $\mu$m are readily imaged and their gravitational length~\footnote{The gravitational length is a measure of vertical movement due to thermal energy, see Sec.~\ref{sectionNonEquilibriumSedimentation}.} can be $\approx 0.01 \sigma$ so that out--of--plane motion is rather small~\cite{williams2013,williams2015jcp}.

\subsection{Mapping soft spheres to hard spheres}
\label{sectionMapping}

As discussed above, real colloids inevitably display some degree of softness. It is therefore important to be able to map their behavior to that of perfect hard spheres for the purpose of comparison with theory and simulation results.  Mapping entails finding an effective hard-sphere diameter $\sigma_{\rm eff}$, such that one may translate the experimentally-controllable particle number density $\rho$ to an effective hard-sphere volume fraction $\phi_{\rm eff}$ using $\phi_{\rm eff}=\pi\sigma_{\rm eff}^{3}\rho/6$.  Two distinct approaches can be used to determine $\sigma_{\rm eff}$: directly from the interaction potential and indirectly from the observed behavior of the system. The latter method rests on the assumption that the weight fraction (which may be accurately determined) at freezing may be taken to correspond to the freezing volume fraction. This provides a calibration of the volume fraction through an accurately known quantity. Developments of this method to address the effects of graviational settling are discussed in Sec.~\ref{sectionBulk}.

\begin{figure}[h!]
\centering
\includegraphics[width=80mm,keepaspectratio]{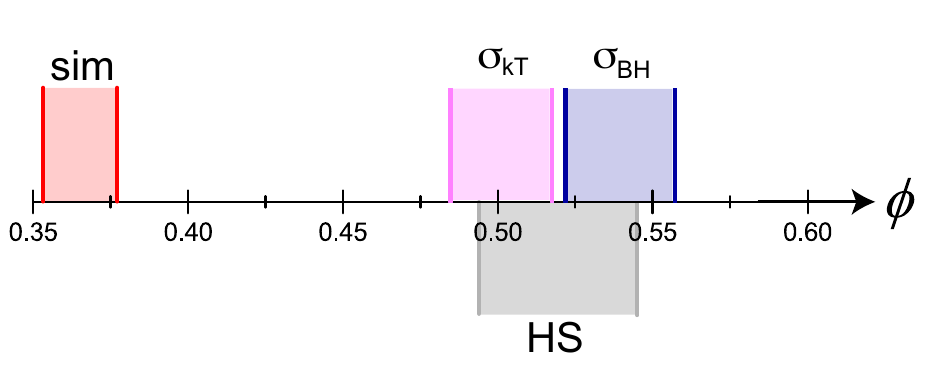} 
\caption{Mapping the phase behavior of weakly charged hard-core colloids to pure hard spheres. The hypothetical particles have an electrostatic  charge $Z = 500$, and the solvent has $\sigma = 2\mu$m with $\kappa^{-1} = 100$nm, which is typical for  non--aqueous solvents~\cite{royall2005}. The fluid-solid coexistence gap of the hypothetical charged-colloid system as a function of packing fraction $\phi$ is taken from simulations~\cite{hynninen2003} (red, ``sim''), and is mapped to pure hard spheres assuming that the effective hard-sphere diameter corresponds to the interaction potential $\beta u(\sigma_\mathrm{kT})=1$ (lilac, $\sigma_\mathrm{kT}$), and  using the mapping of Eq.~\eqref{eqBH} (blue, $\sigma_\mathrm{BH}$). The phase diagram of pure hard spheres (grey, HS) is shown for comparison.}
\label{figMapping} 
\end{figure}

If the interaction potential is known, several analytical routes to estimate an effective hard-sphere diameter have been proposed. Arguably the simplest one is to set an effective hard-sphere diameter $\sigma_\mathrm{kT}$ such that the inter-particle repulsive energy at this centre-to-centre separation between two particles is equal to the thermal energy $\kB T$. A more sophisticated approach entails taking into account the functional behavior of the pair interaction as proposed by Barker and Henderson (BH)~\cite{barker1967}, 
\begin{equation}
\sigma_\mathrm{BH}=\intop_0^\infty dr\left[1-\exp\left(-\beta u(r)\right)\right].
\label{eqBH}
\end{equation}
\noindent
An alternative mapping due to Andersen, Weeks, and Chandler~\cite{andersen1971} takes into account the structure of a hard-sphere reference fluid at the same density, making the effective diameter density-dependent. This approximation has shown to be effective for mapping the structural relaxation time of soft spheres near the glass transition to the hard-sphere model~\cite{schmiedeberg2011}. For the commonly used Weeks-Chandler-Andersen potential~\cite{weeks1971}, both the Barker-Henderson and Andersen-Weeks-Chandler mapping have been shown to provide excellent phase behavior predictions at sufficiently low temperatures~\cite{attia2022,dasgupta2020}. However, for higher temperatures -- or equivalently, particles that are less hard-sphere-like -- significant deviations emerge. As a result, care is needed when interpreting the results of this type of mapping in colloidal systems~\cite{royall2013myth}. The situation is illustrated for a model system of charged colloids in Fig.~\ref{figMapping}, where the  fluid-crystal phase boundaries from simulations~\cite{hynninen2003} are mapped to hard spheres using $\sigma_\mathrm{kT}$ and $\sigma_\mathrm{BH}$. Both approaches show significant deviations from the true hard spheres. Importantly, the fluid-crystal coexistence gap is also narrowed. Therefore, as corroborated by simulation work showing that
nucleation barriers continue to be sensitive to the softness 
even at screening lengths on the order of a few percent of the particle diameter~\cite{dejager2022}, simply obtaining $\sigma_\mathrm{eff}$ may not suffice to confidently reproduce hard-sphere behavior.

The second option is to measure quantities that depend on  volume fraction. Structural quantities such as the radial distribution function may be compared with simulation or theory (see Sec.~\ref{sectionStructureDynamics})~\cite{royall2007jcp,royall2018jcp,thorneywork2014,kale2023}. However, for dense hard spheres, these pair correlations vary rather slowly with $\phi$. Alternatively, one can choose a quantity which strongly depends on $\phi$, such as the relaxation time at high volume fraction (see Fig.~\ref{figLucaCipeletti}, Sec.~\ref{sectionGlass}) or higher--order structural observables (see Fig.~\ref{figTCCrho}, Sec.~\ref{sectionStructureDynamics})~\cite{royall2018jcp,pinchaipat2017}. However, it is important to note that there is absolutely no guarantee that different methods give the same mapping. For example, higher-order structure measurements are highly sensitive to missing -- or ``ghost'' particles (see Sec.~\ref{sectionOptical}), while measuring dynamics in real space is more sensitive to errors in particle positions.

In short, nearly-hard spheres, whether realized in experiments or in simulations, can closely approach the behavior of true hard spheres, but subtle differences remain. Choosing an effective diameter is an essential step in assigning an effective packing fraction to an experimental system, and hence in comparing its behavior to hard-sphere simulations or theory. However, no unambiguous definition of such an effective packing fraction exists, and different approaches to obtain one result in different estimates.

\subsection{Accuracy in hard-sphere experiments}
\label{sectionAccuracy}

The above discussion makes clear that, with care, an effective hard sphere volume fraction may be arrived at for an experimental system. However, a degree of uncertainty will persist, caused by  
(\emph{i})  softness, due to electrostatics, stabilization or other sources;
(\emph{ii}) attraction, due to Van der Waals interactions or other sources; and
(\emph{iii}) polydispersity, which is typically represented via the second moment of the distribution. (An unbiased way to characterize the distribution remains to be found. Even the laborious method of sizing every particle by electron microscopy likely samples from a biased set of particles, due to inhomogeneities during evaporation in preparing the sample.) These qualities pertain to the particles, but, especially for non-aqueous systems, the parameters which might effect the state point include the solvent choice, its age and purity as well as the ambient temperature and humidity (which are often not controlled in colloid laboratories).

Even if these uncertainties could somehow be mitigated, there remains a fundamental aspect of colloidal systems that sets them apart from most materials in that \emph{model colloids are a synthetic material}. As noted above, PMMA, for instance, cannot be purchased commercially; it needs to be synthesized, typically in small batches of 10-1000g. Therefore, even if the quantities above were somehow characterized for a given synthesis (which would entail a very large amount of work), once that had been exhausted, the operation would need to be carried out again for the next synthesis.

As a consequence, \emph{colloidal systems are not as well--characterized as atomic and molecular systems}. A container of argon of sufficient purity has identical characteristics to another. However, like home-made cakes, \emph{colloidal hard spheres from two different syntheses will never have identical characteristics}.
That is to say, the concept of an accurate measurement of a physical property fundamentally differs. For example, the triple point temperature of argon is 83.8058 K, i.e. known to six significant figures, drawing on many successive and increasingly accurate measurements of the same material. Such an accuracy is essentially impossible to achieve in colloids. This brings us to the question of just how much accuracy we can expect in colloids.~\citet{poon2012} concluded that a \emph{relative error} in volume fraction around the freezing transition of 6\% might be reasonable. Our opinion is that when careful comparisons are made, as described in Sec.~\ref{sectionMapping}, an absolute error of 1\% in volume fraction seems achievable, albeit not always realized. In the remainder of this review, we shall revisit the impact of colloidal--level accuracy where appropriate.

A more far-reaching interpretation of accuracy in the context of this review is to consider what is meant by \emph{hard} spheres. This matter is not as straightforward as it might first seem, as whether a system is sufficiently similar to hard spheres or not depends on the physical property considered. For example, it has long been known that even quite soft repulsive particles have a fluid structure similar to that of hard spheres~\cite{weeks1971}. By contrast, phase boundaries are highly sensitive to the interaction details. For the purposes of this review, we can clearly state that systems that are \emph{not} hard-sphere--like are (\emph{i}) so soft that they form a bcc crystal, i.e., their phase behavior differs qualitatively from that of hard spheres; (\emph{ii}) attract each other significantly, by which we mean a well depth of 0.5 $\kB T$ or above; (\emph{iii}) use temperature rather than volume fraction as control parameter. Beyond these three criteria, the degree of hardness required for our purposes is therefore context--dependent. The interactions of the particles in the work that we discuss in this review then is sufficiently hard for appropriate hard sphere like behavior to be observed in the context in question. We discuss certain instances where the degree of hardness is particularly relevant.

\section{Measuring hard spheres \emph{in vitro}}
\label{sectionMeasuring}
We now turn to techniques for analyzing  colloidal hard spheres in experiments. The configurations adopted by the particles and their dynamics are the main quantities of interest. The experimental work that we review  primarily used direct observation through microscopy or scattering methods.  (Rheological approaches have been carefully reviewed elsewhere~\cite{wagner2022,larson,jacob2019}.) While some scattering studies of hard spheres have used neutron~\cite{cebula1981,dekruif1988} and small-angle X-ray scattering~\cite{petukhov2002} and more exotic X-ray scattering methods~\cite{wochner2009}, most have used light scattering. This last method is therefore the focus of our discussion.

\subsection{Light scattering}
\label{sectionLightScattering}

\begin{figure}[t]
\centering
\includegraphics[width=85mm,keepaspectratio]{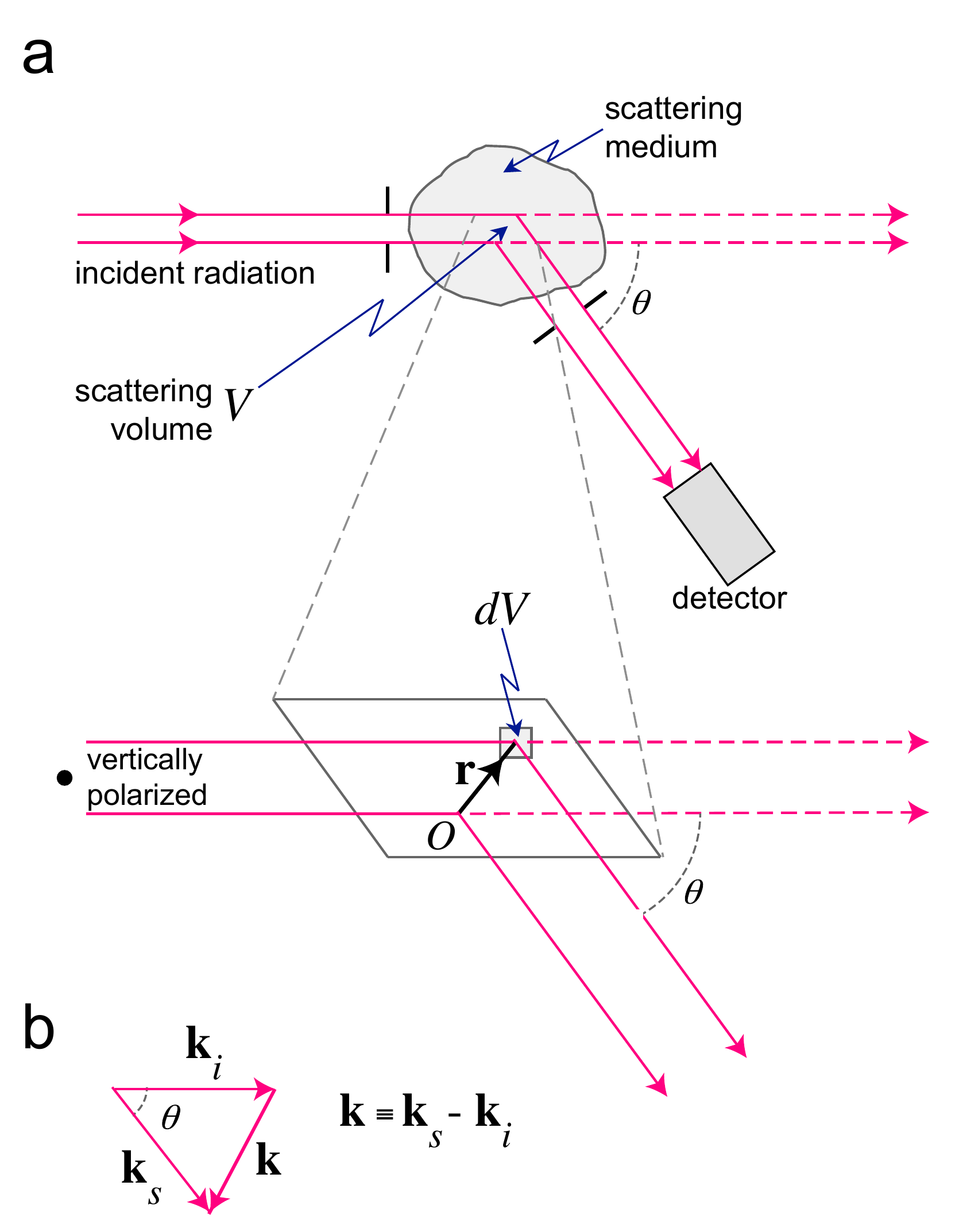} 
\caption{Schematic of a light-scattering set-up. (a) Global view of the set-up. In the expanded view, scattering though an angle $\theta$ is indicated in a small volume $dV$. (b) Close up of the scattering volume with the scattering wavevector $\mathbf{k}$ defined in terms of the wavevector of the incident $\mathbf{k}_i$ and scattered $\mathbf{k}_s$ light, and also the scattering angle $\theta$. Inspired by~\citet{pusey2002ch1}.
} 
\label{figPeterLight} 
\end{figure}

Characterization of materials via the scattering of electromagnetic radiation dates back almost to the discovery of X-rays, whose wavelength is comparable to length scales relevant to atomic and molecular systems. The larger length scales of colloids correspond to the wavelength of visible light, which is scattered by spatial and dynamical fluctuations in the refractive index of the material. A typical light scattering set--up is illustrated in Fig.~\ref{figPeterLight}, by a global view and by a close up of the scattering volume with the scattering wavevector $\mathbf{k}$ defined in terms of the wavevector of the incident $\mathbf{k}_i$ and scattered $\mathbf{k}_s$ light, together with  the scattering angle $\theta$.

The key observables in light scattering are the intensity of the scattered light $I(\mathbf{k},t)$ and its time correlation. Here $\mathbf{k}$ is the wavevector through which the light is scattered, $t$ is the time and $\mathbf{r}$ is the particle position (Fig.~\ref{figPeterLight}). The scattered intensity is then~\cite{pusey2002ch1}
\begin{eqnarray}
\nonumber I(\mathbf{k},t)&&= \frac{E_0^2}{r^2} \sum_{j=1}^N \sum_{k=1}^N b_j(\mathbf{k},t) b_k^*(\mathbf{k},t) \cdot \\ 
 && \exp \left[-i\mathbf{k}\cdot\left(\mathbf{r}_j(t)-\mathbf{r}_k(t) \right) \right], 
\label{eqLightScattering}
\end{eqnarray}
where $E_0$ is the amplitude of the incident light field, $b_i$ is the scattering length of the $i$th particle and is closely related to the form factor.

\header{Static light scattering}
In the case of static light scattering (SLS) temporal quantities are averaged over time. Then  
from Eq.~\eqref{eqLightScattering} for a given $b(\mathbf{k},t)$ (which can be readily calculated for spheres made of a material with a fixed refractive index) we can obtain the structure factor~\cite{pusey2002ch1} 
\begin{equation}
S(\mathbf{k})=\frac{1}{N}\sum_{j=1}^N\sum_{k=1}^N \langle \exp \left[-i\mathbf{k}\cdot\left(\mathbf{r}_j-\mathbf{r}_k \right) \right] \rangle,
\label{eqLightScattering_static}
\end{equation}
and therefore its Fourier transform, the radial distribution function $g(r)$. This quantity also provides a link to the thermodynamics of the system, since in the long-wavelength limit it is related to the isothermal (osmotic) compressibility $\chi_T$
\begin{equation}
    \lim_{k\to 0} S(k) = \rho \kB T \chi_T = \kB T \left(\frac{\partial \rho}{\partial P}  \right)_T. \label{eqCompressibility}
\end{equation}

Light scattering has been applied to soft-matter systems since the 1930s~\cite{doty1951}. For example,~\citet{doty1952} studied solutions of Bovine Serum Albumin (BSA) protein, and soon were followed by famed crystallographer Rosalind Franklin and coworkers~\cite{klug1959} who studied colloidal-type crystals of charged viruses. However, the approach only really took off after 1962, when Spectra Physics \emph{lasers} became commercially available. The intense light produced by these sources revolutionized the technique by dramatically increasing the signal-to-noise ratio, making it much more practical and widely available. For typical scattering angles $\theta$ and a HeNe laser wavelength of  633 nm, sub-micron sized colloids fall within the ideal size range for light scattering. By the mid-1970s, the measurement technology was ready for the emergence of experimental hard spheres from the point of view of two-point spatial correlation functions.

Pusey was an early champion of light scattering, moving from the  study of critical phenomena in molecular liquids to the study of viruses as colloidal particles~\cite{pusey1972}. Because synthetic colloids held the promise of being more robust than viruses, he arranged to receive polystyrene samples from the Bristol colloid group via Ottewill and Vincent (see Sec.~\ref{sectionForetelling}). This enabled his pioneering work on determining static structure factors from the mid 1970s onward, starting with charged polystyrene colloids~\cite{brown1975}, which complemented earlier work on colloidal crystals~\cite{hiltner1969,williams1974}. When using light scattering for denser suspensions (by which we mean denser than the dilute limit, $\phi\gtrsim 0.01$), precise refractive index matching between colloids and solvent is essential as noted in Secs.~\ref{sectionForetelling} and \ref{sectionExperimental}. For polystyrene, however, this is challenging due to its high refractive index (relative to the solvent). Early work, which explicitly matched the refractive index of the colloids and solvent, was carried out by~\citet{vrij1983}.

Analysis of light scattering data in dense suspensions can be complicated by the occurrence of multiple scattering, whereby incident photons are scattered more than once before exiting the sample. For both static and dynamic light scattering, however, this issue can be circumvented by using scattering set-ups that isolate the signal from singly scattered light by cross-correlating the intensity of two different wave lengths of light~\cite{segre1995pre, schatzel1991, moussaid1999pre}.

\header{Dynamic light scattering} While the discussion above outlines early efforts to extract structural information such as the static structure factor from SLS, as Eq.~\eqref{eqLightScattering} makes clear, determining time correlations in the fluctuations in refractive index -- in the scattered form of the so--called \emph{speckle pattern} -- can reveal dynamical behavior, i.e., \emph{dynamic light scattering} (DLS). In the 1960s, the early pioneers in the field, Benedek and coworkers~\cite{clark1970} and~\citet{cummins1970}, used a spectral technique to analyze the speckle pattern, which was experimentally tedious. The now dominant technique, time correlation of detected photons (frequently called “photon correlation spectroscopy”), was pioneered by Pike and coworkers~\cite{foord1970}.

The key quantity measured in DLS is the normalized time correlation function 
\begin{equation}
g^{(2)}(k,\tau)\equiv \frac{\langle I(\mathbf{k},\tau) I(\mathbf{k},0) \rangle}{\langle I(\mathbf{k}) \rangle^2}.
\label{eqTimeCorrelation}
\end{equation}
Importantly, $g^{(2)}(k,\tau)$ is intimately related to a key dynamical property, the intermediate scattering function $F(k,\tau)$, via the Siegert relation 
\begin{equation}
F(k,\tau) = \sqrt{g^{(2)}(k,\tau)-1}, 
\label{eqISFLight}
\end{equation}
which constituted the mainstay of dynamical measurements of hard-sphere systems until dynamical analysis using confocal microscopy became available around the turn of the millennium.

A significant advantage of DLS over confocal microscopy is the possibility to  access  a broader range of time scales. Microscopy is limited to a relatively small time window. Fast time scales ($ < 10^{-2}$ s) and very long ones  ($> 10^4$ s) are inaccessible, due to limitations in sample stability. By contrast, DLS can probe much shorter time scales. (See also the discussion concerning the consequences of particle size in Sec.~\ref{sectionSize}.)

\header{More exotic light scattering: small angles, multiple length scales and multiple color} While the essential methods of light scattering were established by the mid--1970s, the technique has since enjoyed considerable development. As Fig.~\ref{figPeterLight} makes clear, going to very small angles (small-angle light scattering) probes smaller wave vectors and larger real--space length scales~\cite{cipelletti1999}. Mounting multiple detectors, such as one corresponding to a ``standard'' wave vector around $2\pi/\sigma$ and one at a smaller angle $\ll2\pi/\sigma$, then enables multiple length scales to be probed simultaneously. This approach has since been implemented to great effect~\cite{franke2014,tamborini2012}. Other developments include two-color techniques which enable rather more turbid samples ot be studied~\cite{segre1995jmo}. More recent advancements such as multi-speckle correlation spectroscopy enable simultaneous measurement of dynamical and spatial heterogeneities~\cite{golde2013,golde2016}.

\subsection{Optical microscopy and particle-resolved studies}
\label{sectionOptical}

\begin{figure}[h!]
\centering
\includegraphics[width=65mm,keepaspectratio]{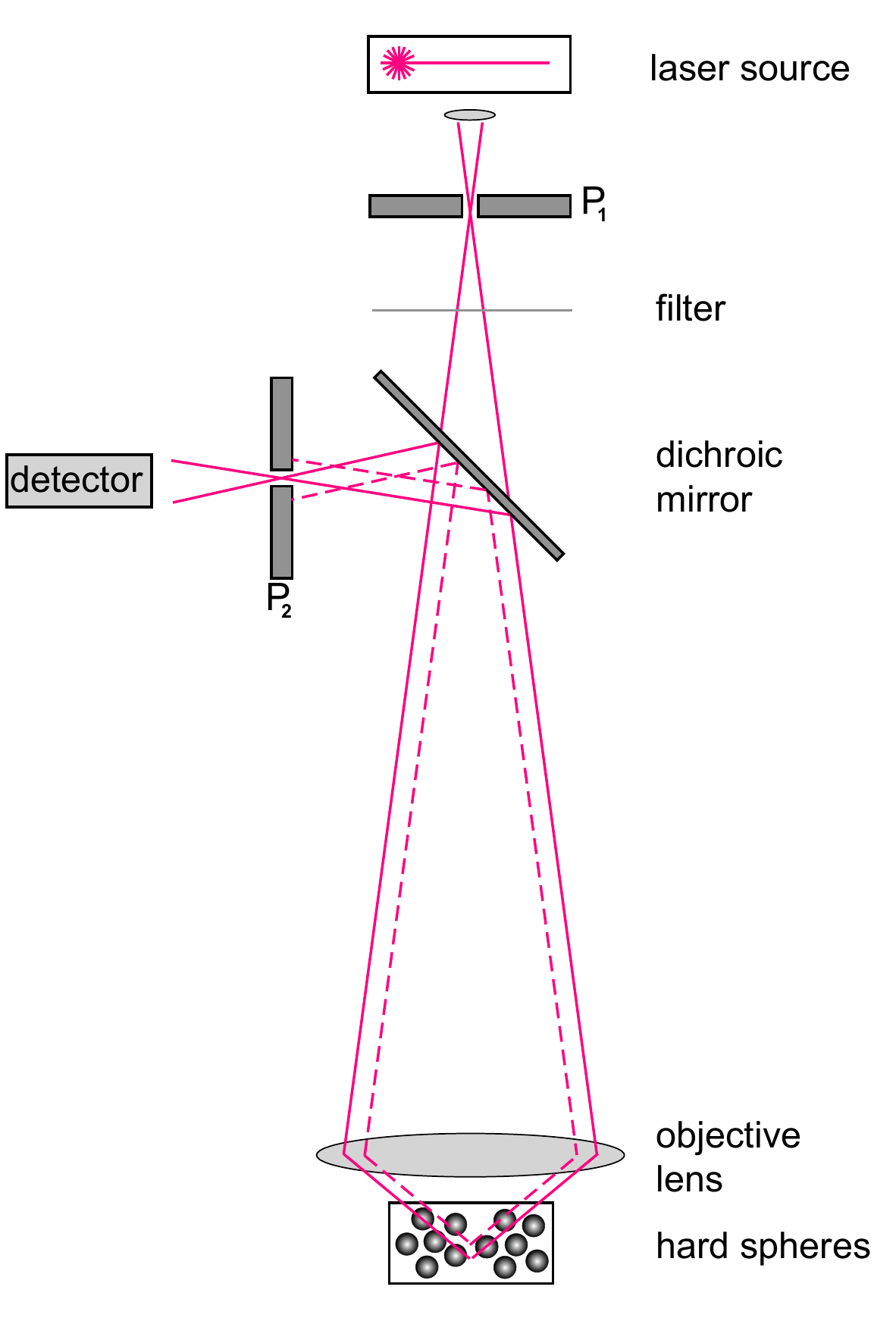} 
\caption{The principle of confocal microscopy in the conventional \emph{Epi--}illumination mode. In \emph{Epi--}illumination, the objective and condenser lenses are the same, and the dichroic mirror allows the illuminating beam to pass through, while reflecting the returning beam to the detector. Light is focused into a point in the sample by the condenser lens. The confocal pinhole rejects all light except that from the point in focus (dotted lines show reflection of light from out-of-focus regions). In this way, only one point in a 3d sample is in focus and the incident beam is scanned relative to the sample to generate a 3d image. \emph{Epi--}illumination records fluorescent light emitted by the sample, so the distribution of fluorescent dye is imaged.}
\label{figConfocal} 
\end{figure}

Unlike atomistic systems, colloids are amenable to \emph{in-situ} observation via light microscopy techniques. Analysis of such real-space images of colloidal hard-sphere systems dates back at least to the early 1970s, with the work of Hachisu and coworkers~\cite{kose1973ii,kose1976,okamoto1977}, but antecedents can be found in the Bragg scattering study of ``colloidal'' crystals of viruses of Franklin and coworkers~\cite{klug1959}. Microscopy work  may be broadly classified into two categories--one  in which  individual particles are resolved (so--called particle--resolved studies~\cite{ivlev}) and the other in which particles are too small to be resolved.

In the case of particle-resolved studies of 3d systems, some kind of 3d imaging is clearly important. Here, the development of \emph{confocal} microscopy would prove to be transformative. A schematic of a confocal microscope is shown in Fig.~\ref{figConfocal}. Confocal microscopy was proposed in the mid-20th century with Minsky developing an operational device in 1955~\cite{minsky1957}. Like light scattering, confocal microscopy was transformed by the advent of lasers. By the mid-1980s, stage controlled systems, which scan the stage relative to the beam, were commercially available from Oxford Optoelectronics, later acquired by Biorad, the company which became synonymous with early confocal microscopes.

Since the signal recorded in (\emph{epi--}illumination) confocal microscopy is the distribution of fluorescent dye, some form of labeling of colloidal particles (or solvent) is necessary. Novel fluorescent particles suitable for confocal microscopy had been developed by Van Blaaderen \emph{et al.}~\cite{vanblaaderen1992langmuir}. The additional synthetic step of growing an undyed shell around the fluorescently labeled core contributed greatly to the ability to distinguish particles (see Sec.~\ref{sectionSynthesizing}). To our knowledge, the first 3d imaging of hard--sphere-like systems with confocal microscopy was that of Van Blaaderen \emph{et al.}~\cite{vanblaaderen1992langmuir}, who observed colloidal crystals and glasses. The field, which has since grown markedly, has also been extensively reviewed~\cite{ivlev,hunter2012}. Confocal microscopy has since been developed to incorporate a shear stage as discussed in Sec.~\ref{sectionFarFromEq} and \ref{sectionFarGlass}~\cite{besseling2009,lin2014rsi} and optical tweezers (see below).

\header{Particle tracking} In a seminal paper,~\citet{crocker1996} developed an algorithm for tracking the coordinates of colloidal particles  from microscopy data. This approach interpreted bright pixels as potential particle centres. In the Crocker and Grier method, the trial centers are subsequently refined through a  series of criteria, such as being the centre of an object weighted by pixel intensity and being far enough apart from other objects, so that other particles do not overlap. It is worth noting that the majority of work on 2d hard-sphere systems has  been carried out using microscopy, as such systems are less amenable to scattering techniques.

While the algorithms introduced by Crocker and Grier for quasi-2d systems~\cite{crocker1996} have undoubtedly formed the backbone of most work since, it is important to note that more sophisticated approaches are available. Among the pioneers of the field,~\citet{vanblaaderen1995} sought to improve the tracking accuracy in the axial direction by fitting a Gaussian to the integrated intensities in each plane that constituted a given colloid. The early work examined hard-sphere glasses and crystals, i.e., colloidal solids in which diffusion could be neglected~\cite{vanblaaderen1995}. Even though this method made imaging easier, the slow scan rates needed to acquire images with low noise left  dynamical information beyond reach. For a quasi--2d system, time--resolved tracking was carried out by the mid-1990s~\cite{marcus1996}. In 3d systems, two breakthrough papers, published simultaneously in the year 2000, obtained dynamical information by tracking particle coordinates over time~\cite{weeks2000,kegel2000science}. This local information led to a breakthrough experimental measurement of dynamical heterogeneity, a particularly important characteristic of glass physics (see Sec.~\ref{sectionGlass}).

More recently,~\citet{jenkins2008} pushed the limits of the technique as they sought to identify contacts between colloids in a dense sediment through ultra-high precision coordinate location. Such precision was achieved by first determining an empirical image of a colloid, which could then be compared to the original image. Further systematic improvements were made by~\citet{gao2009}. Other, more specialized, developments include that of Lu \emph{et al.}~\cite{lu2006}, whose algorithmic advances for image processing made it possible to follow experiments in real time, rather than via {\it post facto} analysis. This distinction is particularly significant. The spatial region of interest can often drift out of the field of view, which is usually quite small given the large file sizes generated by time-resolved 3d imaging, but real-time tracking allowed feedback to be applied to the stage, such that the region of interest could be followed irrespective of drift.

Ever-present polydispersity is typically neglected, except in the simple case of binary systems wherein two-color imaging enables the two species to be differentiated~\cite{royall2007jcp}. More recently, algorithms which identify the size of \emph{each} particle have been developed~\cite{kurita2012,leocmach2013sm}. Ideas to extract further information than possible using conventional techniques have also been implemented, such as comparing simulated trial coordinates and experimental data~\cite{statt2016} as well as maximal precision methods~\cite{bierbaum2017}. One might imagine that machine-learning techniques could be applied to the vexing problem of obtaining particle coordinates from confocal microscopy images, as proposed by~\citet{bailey2022} for a dilute  suspension of  out-of-equilibrium (active) colloids. However, this is a challenging problem in the case of concentrated colloidal dispersions in 3d, not least because of the paucity of the ``ground truth''  (other than synthetic data) against which a neural network can be trained~\cite{kawafi2023}.

When comparing experimental particle-resolved data for hard spheres with numerical simulations, the effects of polydispersity and particle tracking errors should never be omitted. \emph{High-quality} data nevertheless implies an error of $0.05\sigma$ in particle position and a few percent missing or additional ``ghost'' particles~\cite{ivlev,royall2007jcp}. This problem may be mitigated to some extent by mimicking such features into computer simulation data~\cite{royall2007jcp,pinchaipat2017}, and through other techniques, as extensively reviewed~\cite{ivlev,lu2013,hunter2012}.

\header{Optical tweezers} The development of lasers had yet another important consequence for colloid experiments. A tightly focused laser beam enables trapping of colloidal particles and their manipulation~\cite{ashkin1970}. These optical tweezers have great potential to enable \emph{in-situ} observation of phenomena which depend upon exquisite control of constituent particles~\cite{grier1997,grier2003}. Like microscopy, 3d implementation of optical tweezers carries challenges, mainly related to the requirement of refractive index matching colloids and solvent for imaging while simultaneously \emph{mismatching} refractive indexes for tweezing. The first fully 3d system for colloids was introduced by~\citet{vossen2004} and single-lens 3d tweezer methods have since been developed~\cite{curran2014}. It is important to note, however, that the maximum number of particles which can be tweezed is about 40 and that tweezing leads to a relatively weak confinement, of the order of a few $\kB T$ at most~\cite{williams2013}.

\subsection{Size \emph{does} matter: colloid dynamics in different experiments}
\label{sectionSize}

Light scattering and microscopy techniques have typically been used for rather different sized colloidal particles. While of course the fundamental behavior of the system remains unaltered, there are significant consequences concerning the ability to access phenomena of interest. In particular, light scattering can operate with particles in the 100s of nanometer size range, while microscopy typically uses particles larger than a micron.

As Eq.~\eqref{eqTauB} shows, the Brownian time scales as the volume of the colloidal particle, $\tau_B\sim\sigma^3$. In other words, a factor of ten drop in particle size means that the Brownian  time falls by a factor of 1000, with major experimental  implications. For instance, a phenomenon observed on the time scale of a day in a light scattering set-up with 200 nm diameter colloids might take a year to observe with confocal microscopy with 2~$\mu$m diameter particles. In practice, that would make it  unobservable. This distinction is particularly important for rare events, such as crystal nucleation for which the ``discrepancy regime'' is inaccessible to confocal microscopy (see Sec.~\ref{sectionNucleation}), and for phenomena with long time scales, such as the glass transition (see Sec.~\ref{sectionGlass}).

%% file: Simulations/simulations.tex
\section{Hard-sphere systems \emph{in silico}} 
\label{sectionSimulation}

From a numerical simulation perspective, hard spheres might seem to be as simple as it gets.
Their interaction is pairwise and short-ranged, and is also trivial to calculate. These features should result in efficient simulations, as is indeed the case for Monte Carlo (MC) \cite{frenkel} and event-driven molecular dynamics (EDMD) \cite{rapaport} schemes. Both of these methods perfectly sample the structure and phase behavior of a pure hard-sphere model, and EDMD simulations faithfully reproduce the dynamics of perfect hard spheres in a vacuum. In practice, however, experiments using colloidal spheres inevitably also include a solvent, which 
has profound effects on dynamics especially in out-of-equilibrium systems  (see Secs.~\ref{sectionStructureDynamics} and \ref{sectionFarFromEq}). These hydrodynamic effects can be included at different levels of detail, with Brownian dynamics (BD) simulations being  the most straightforward implementation that reproduces the diffusive short-time motion of colloidal particles. Simulating hard-sphere colloidal systems away from thermal equilibrium, however, requires more complex methods that explicitly model a coarse-grained version of the background solvent. It is nevertheless important to note that for dense fluids close to equilibrium, the difference between different simulation methods in terms of the dynamics is rather small \cite{berthier2007jpcm, montero2017}. Detailed comparison of EDMD with experimental data, for instance, found no meaningful difference  \cite{royall2018jcp}. 

This section provides an overview of the simulation techniques most commonly used for hard-sphere models. Because not all of these simulation methods are easily applicable to perfectly hard spheres, we additionally briefly discuss commonly used models of \emph{nearly} hard spheres.

\subsection{Monte Carlo methods}
\label{sectionSimulationMC}

Since MC simulations were first used to extract the radial distribution functions and pressure of hard-sphere fluids~\cite{rosenbluth1954}, the scheme has remained instrumental for evaluating much of the physical behavior of the model. A key strength of the approach is its adaptability. Different thermodynamic ensembles, boundary conditions, or biased sampling can be considered~\cite{frenkel}. Although standard implementations of these methods are often straightforward,  DL\_MONTE~\cite{brukhno2021}, a general purpose code with advanced sampling methods, and the HOOMD-blue package \cite{anderson2020}, which includes (CPU- or GPU-based) parallelization for fast simulation of large systems~\cite{anderson2013}, should nevertheless be noted. 

The simplest local Metropolis algorithm for hard spheres consists of single-particle attempted displacements, which are accepted only if no overlap is created. (Interestingly, in the limit of small step sizes, this algorithm approximates BD \cite{cichocki1990, sanz2010}.) In many scenarios, hard-sphere MC simulations can be made more efficient by incorporating clever collective displacements, such as cluster moves \cite{dress1995, almarza2009, ashton2011,dijkstra1997,buhot1998,liu2004}, or by employing non-local sampling schemes. One particularly effective example is to exchange the diameter of particles in dense mixtures of hard spheres of different sizes~\cite{berthier2016prl,ninarello2017}. 
In dense mixtures of not-too-dissimilar hard spheres, this \emph{swap} MC approach leads to a major efficiency gain in configurational sampling. In glass forming hard-sphere mixtures, in particular, swap MC can probe the equilibrium fluid structure at packing fractions far beyond the point at which conventional sampling would be essentially arrested (see Sec.~\ref{sectionDeeper})~\cite{berthier2016prl, ninarello2017}. 

Another effective strategy for speeding up configurational sampling of hard spheres is the rejection-free event-chain MC moves introduced in \citet{bernard2009}. In event-chain MC, particle displacements involve a chain of neighboring particles. A particle is displaced until it collides with a neighbor, which then becomes the next particle to be displaced. These rejection-free moves can accelerate configurational sampling by over an order of magnitude compared to single-particle MC in dense systems of hard spheres in 2d or 3d \cite{bernard2009, klement2019,li2022}. A variant of this approach called Newtonian event-chain MC, which incorporates aspects of Newtonian dynamics into the event-chain moves, was shown to lead to a speed-up of yet another factor of about five for dense hard-sphere systems \cite{klement2019}.

\subsection{Event-driven simulations}
\label{sectionSimEventDriven}

Historically, MC simulations were soon followed by EDMD simulations, which evolve the 
positions of elastic hard spheres in vacuum \cite{alder1957}. In other words, each particle interacts purely via instantaneous elastic collisions, and moves at constant velocity between such collisions \cite{rapaport}.  The efficiency of these simulations hinges strongly on the efficiency of the algorithms used to predict and schedule future collisions \cite{lubachevsky1991, marin1993, paul2007, donev2005}, and can rival event-chain MC \cite{klement2019, smallenburg2022} for equilibrating dense systems. As a result, they are vastly more efficient than MC methods based on single-particle moves. From a technical standpoint, EDMD simulations are more challenging to implement and adapt, but publicly available codes have greatly helped spread the approach~\cite{donev2005, bannerman2011, smallenburg2022}. EDMD simulations have also been adapted to, e.g., reproduce Brownian motion \cite{scala2007}, to include swap moves \cite{bommineni2019}, and to incorporate isotropic compression \cite{donev2005}.

\subsection{Brownian dynamics simulations}
\label{sectionSimBD}

Suspensions of colloidal hard spheres embedded in a solvent effectively undergo Brownian motion. The stream of collisions with the much smaller solvent molecules results in the spheres experiencing random forces. The ensuing equation of motion is
\begin{equation}
    \frac{d\mathbf{r}}{dt} = \frac{D_S}{\kB T} \mathbf{f}(t) + \sqrt{2 D_S}\mathbf{R}(t), \label{eqBD}
\end{equation}
where $\mathbf{r}(t)$ is the colloid position at time $t$, $D_S$ is the short-time self-diffusion coefficient, and $\mathbf{R}(t)$ is a Gaussian random process with zero mean and unit variance. The total force $\mathbf{f}(t)$ on the particle includes its interactions with neighbors as well as any external forces present in the system. 

Simulating Brownian dynamics entails numerically integrating Eq.~\eqref{eqBD}. In practice, this integration is often achieved using a fixed small time step. For hard spheres, however, this approach does not work, because any fixed time step inevitably leads to overlaps. Since the event-driven approach mentioned above is not particularly efficient~\cite{scala2007} -- frequent stochastic updates require re-predicting collisions for all particles after each velocity change --  BD simulations are most commonly performed on model systems that interact via a continuous approximation of the hard-sphere potential. (Variants are presented in Sec.~\ref{sectionHSApproxModels}.) The implementation approach for BD simulations closely follows standard molecular dynamics (MD) schemes. Each time step consists of calculating the forces on all particles, followed by updating particle positions. As a result, MD codes can be readily adapted to perform BD simulations, and MD packages such as LAMMPS \cite{thompson2022}, HOOMD-Blue \cite{anderson2020}, and Espresso \cite{weik2019} provide efficient implementations of BD, including options for parallelization or GPU computation.

\subsection{Simulations including hydrodynamics}
\label{sectionSimHydro}

All simulation methods described thus far neglect hydrodynamic coupling, which is (necessarily) present for colloids moving in a solvent. While hydrodynamic interactions do not modify static \textit{equilibrium} properties, they can play a significant role in the dynamics of colloidal hard spheres (see Sec.~\ref{sectionStructureDynamics}), especially when out-of-equilibrium phenomena are considered (an issue to which we return in Sec.~\ref{sectionFarFromEq}). A number of different numerical methods have therefore been developed to account for hydrodynamic effects in hard-sphere simulations.

\textit{Stokesian dynamics}~\cite{brady1988} assumes an incompressible fluid and a low Reynolds number, both of which are typically valid for colloidal hard spheres. Hydrodynamic interactions between colloids are then taken into account via a many-particle mobility matrix that models the solvent flow as a superposition of Stokeslets centered on each colloid. Due to the slow algebraic decay of this effect, the matrix effectively couples the motion of all particles in the system, making the method computationally expensive, particularly when Brownian forces are included. Significantly better performance can be obtained using the so-called accelerated Stokesian dynamics (ASD) \cite{sierou2001, banchio2003}. Publicly available implementations can be found in the simulation package Espresso \cite{weik2019}, as a plugin \cite{fiore2019} for HOOMD-Blue \cite{anderson2020}, as well as in specialized codes \cite{yan2020}.

\textit{Lattice Boltzmann} (LB) methods~\cite{ladd2001, cates2004jpcmsimulating, dunweg2009} divide the solvent into lattice sites, each carrying a set of local solvent densities associated with a discrete set of permitted solvent velocities. These methods reproduce the dynamics of a Newtonian liquid with a given shear viscosity, and recover the relevant hydrodynamic variables as moments of the one-particle velocity distribution functions \cite{succi2001}.
The effect of thermal fluctuations can also be included directly into the discrete Boltzmann equation~\cite{adhikari2005,gross2011}.
The fluid dynamics then emerges from the evolution of the one-particle distribution function. During each simulation time step, these densities are updated by considering both the streaming of the solvent to neighboring lattice sites and a collision step, which relaxes the distribution of velocities at each lattice site towards the equilibrium distribution. The lattice-based solvent is then coupled to the colloidal particles by imposing boundary conditions at their surface, thus allowing momentum exchange between the solvent and the particles.  
While explicit treatment of the solvent inherently increases the computational cost, LB methods scale well with particle size and can be readily parallelized for large systems. Simulation codes are also publicly available~\cite{desplat2001, weik2019, bonaccorso2020}.

\textit{Multi-particle collision dynamics} (MPCD) simulations \cite{malevanets1999,gompper2009,howard2019,kapral2008} are similar to LB methods in that they explicitly incorporate the solvent as a simplified fluid evolved via a streaming and a collision update. In MPCD approaches, however, the solvent is modeled as discrete \emph{effective particles} moving in continuous space. At each time step, all particles first move forward according to their instantaneous velocities (streaming), and then exchange momentum with other nearby solvent particles during a collision step. In the most commonly used collision algorithm, the system is divided into cells, and the velocities of all particles within a cell are mixed via a collective rotation that conserves their total momentum. Due to this rotation, MPCD methods using this update are also known as \textit{stochastic rotational dynamics} (SRD) \cite{malevanets1999, ihle2001}. Coupling to the colloids can be done either by including the colloidal particles as collision partners in the collision step, or via direct interactions between solvent particles and colloids during the streaming step. It should be noted that MPCD methods model compressible solvents. In terms of implementation, MPCD methods are closely connected to standard MD, and public (parallellized) implementations are available in, e.g., LAMMPS \cite{thompson2022} and HOOMD-Blue \cite{anderson2020, howard2018}.

\textit{Dissipative particle dynamics}~\cite{hoogerbrugge1992, espanol1995, groot1997, espanol2017} (DPD) is similar to MPCD in spirit but introduces soft repulsive interactions that include both a dissipative and a stochastic term to replace the collision step. These interactions are idealized in such a way that the dynamics of the solvent particles can be resolved with time steps much larger than those used for the (steeper) colloid interactions, thus significantly lowering the computational cost compared to full-solvent simulations. Coupling to the colloids is achieved by associating each colloid particle with one or more solvent particles that overlap with it, and interact with the remaining solvent. Similar to the MPCD approach, DPD can be seen as a variation on classical MD, and efficient implementations can be found in several  simulation packages, including LAMMPS \cite{thompson2022}, Espresso \cite{weik2019}, and HOOMD-Blue \cite{anderson2020}.

\textit{Fluid particle dynamics}~\cite{tanaka2000} (FPD) is based on a direct numerical simulation of the Navier-Stokes equations, resolving the problem of coupling the fluid and the colloidal particles by approximating each colloidal particle as highly viscous particle with a smooth viscosity profile at its interface with the fluid. The biggest advantage of the method is that it avoids the difficulties associated with moving solid-fluid boundary conditions and that it allows the fluid to couple to additional fields. The fluid itself is also incompressible. Thermal fluctuations can additionally be considered by integrating the fluctuating hydrodynamics equations~\cite{furukawa2018}. A method akin to FPD is the \emph{smooth profile method} (SPM)~\cite{nakayama2005}, which avoids introducing a large viscosity in the particle domain  by separating the calculation of the hydrodynamic forces from that of the boundary conditions. The approach has found application to a large variety of passive and active soft-matter systems~\cite{yamamoto2021}. 

\subsection{Advanced sampling and free-energy methods}
\label{sectionAdvancedSimulation}

For the purpose of studying more detailed aspects of hard-sphere behavior, such as phase boundaries, defects, interfacial properties and nucleation, a wide variety of advanced simulation techniques have been combined with the general approaches outlined above.

Arguably the most immediate way of obtaining phase boundaries is to simulate the {\it direct coexistence}  between the two phases of interest, e.g., a hard-sphere fluid and a hard-sphere crystal. In practice, however, this approach suffers from large finite-size effects, in addition to being sensitive to  statistical error as well as to details in the set-up of the initial configuration \cite{espinosa2013}. Moreover, this approach is usually unsuitable for determining crystal-crystal coexistence, due to its slow equilibration dynamics. As a result, studies of equilibrium phase behavior in hard spheres often make use of {\it free-energy calculations}. Although free energies can typically not be directly measured in simulations, thermodynamic integration can be used to determine free-energy differences with respect to some known reference state \cite{frenkel}. The series of simulations that define the integration path between the system of interest and a reference system can be chosen in a variety of ways. Common choices are the ideal gas for the fluid phase \cite{frenkel} and an Einstein crystal for an ordered phase \cite{frenkel1984, vega2007}, but a variety of other reference states have been employed for hard-sphere systems, including cell models \cite{hoover1968, nayhouse2011} and pinning schemes \cite{schilling2009, moir2021}. Note that the hard-sphere system itself has also often been used as a reference system for more complex systems (e.g. \citet{dijkstra1999jpcm, bolhuis1997jpcm, patey1973, smallenburg2013}).
After obtaining the free energy of all relevant phases for a range of state points, equilibrium phase boundaries can be determined by identifying which phase (or coexistence of phases) has the lowest free energy at any given state point. As an alternative approach, accurate phase coexistence conditions can  be extracted by  mapping out the free-energy landscape connecting the fluid and crystal phases \cite{martin2011, fernandez2012}. Note that all of these methods require initial knowledge of the potential phases of interest, which may require an extensive search for possible crystal structures \cite{filion2009, filion2009pre, kummerfeld2008, otoole2011, hudson2011, hopkins2012}, especially in multicomponent mixtures.

Free-energy integration methods also provide routes to calculate the effect of interfaces and defects on the free energy of hard-sphere systems. In particular, the equilibrium concentration of point defects in hard-sphere crystals has been determined by setting up a thermodynamic integration path between systems with and without a defect \cite{pronk2001, vandermeer2017}. Similar approaches can also be used to determine the fluid-solid interfacial free energy of hard spheres (Sec.~\ref{sectionInterfaces})---by constructing an integration path between a direct fluid-solid coexistence and a state of two separate bulk phases \cite{davidchack2000, davidchack2010, schmitz2015}---or between a pure fluid and a fluid in coexistence with a slab of crystal \cite{espinosa2014,sanchez2021}. Alternatively, interfacial free energies can be determined from interfacial fluctuations \cite{davidchack2006}, or via  thermodynamic integration \cite{bultmann2020}. The  free-energy difference between face-centered-cubic (fcc) and hexagonal-close-packed (hcp) crystals and the hcp-fcc interfacial free energy can be computed using the lattice-switch method \cite{bruce1997,pronk1999,wilding2000}.

For studying rare events, such as {\it nucleation}, it is common to use sampling schemes that purposefully bias the simulation towards sampling rare configurations. In particular, umbrella sampling \cite{torrie1977, kastner2011, filion2010} has been used extensively to determine the barrier to nucleating hard-sphere crystals by calculating the free-energy cost of creating crystalline nuclei of different sizes. This approach, however, assumes that the cluster is in local equilibrium at each cluster size, and therefore ignores the possibility of any non-equilibrium dynamics. Hence, to explore nucleation trajectories that take into account the dynamics of the system, more specialized methods such as forward-flux sampling \cite{allen2005, allen2006simulating, allen2006forward, allen2009, hussain2020}, transition path sampling \cite{bolhuis2002}, or transition interface sampling \cite{moroni2004} are required.
Biasing methods have also found uses in other areas of hard-sphere simulations, such as with potentials that suppress crystallization to study hard-sphere glasses \cite{valeriani2011, taffs2016}.

\subsection{Simulation models}
\label{sectionHSApproxModels}

The general strategy to approximate the hard-sphere interaction is to consider a harshly-repulsive continuous interaction potential. Perhaps the simplest of such forms is the inverse power-law potential of exponent $n$, as given in Eq.~\eqref{eqUs}. In the limit $n\rightarrow\infty$, this potential converges to the true hard-sphere potential with diameter $\sigma$. Similar to hard spheres, the thermodynamic behavior of the inverse power-law model with a given exponent $n$ is effectively controlled by a single parameter, since a change in the interaction strength $\epsilon$ and the particle size $\sigma$ both have the same trivial effect of scaling the total energy of the system. Experimental measurements of the interactions between sterically stablized colloids
have further been shown to agree well with power-law potentials \cite{bryant2002} (see Sec.~\ref{sectionExperimental}). However, the inverse power law does not naturally capture the short-range nature of the soft interaction that results from steric stabilization.

Another commonly used continuous approximation is the Weeks-Chandler-Andersen (WCA) potential. Originally introduced to consider the separate roles of attraction and repulsion on the structure of simple liquids~\cite{weeks1971}, the WCA potential corresponds to a purely repulsive variant of the Lennard-Jones potential  
\begin{equation}
    u(r) = \begin{cases} 
      4\epsilon\left(
       \left(\frac{\sigma}{r}\right)^{12} -
       \left(\frac{\sigma}{r}\right)^{6} +
       \frac{1}{4}\right) & r \leq 2^{1/6}\sigma \\
      0 &  r > 2^{1/6}\sigma
   \end{cases},\label{eqWCA}
\end{equation}
with energy parameter $\epsilon$. This interaction form smoothly approaches zero at the cutoff distance $r_c = 2^{1/6} \sigma$. At sufficiently low temperatures (a common choice is $\epsilon /\kB T = 40$), the WCA potential has been shown to map well to hard spheres in terms of, e.g., the equation of state and nucleation rates \cite{dasgupta2020,richard2018jcpkinetics, filion2011jcp}. 
Variations of Eq.~\eqref{eqWCA}, with different (typically higher) exponents have been designed to match the hard-sphere behavior even more closely. A notable example of this is the so-called pseudo-hard-sphere potential \cite{jover2012}, for which the freezing transition lies very close to that of true hard spheres \cite{espinosa2013}. This class of potentials has a long history as a substitute for hard spheres, and has the advantage of being purely short-ranged and being continuous, thus allowing for an easy BD implementation. However, the thermodynamic behavior of this class of models depends on both the chosen temperature and density.

Note that regardless of the exact functional form chosen, a trade off must generally be made between simulation efficiency and the accuracy of the approximation of the true hard-sphere potential. Generally, although a sharper interaction potential provides a better approximation of hard spheres, it also results in a more rapid variation of the interaction forces as two particles approach. Since MD and BD simulations numerically integrate the force experienced by each particle over time, these faster variations then necessitate a smaller integration time step to maintain accurate results, thus slowing simulations down.{}

%% file: Theory/theory.tex
\section{Hard-sphere systems in theory}
\label{sectionTheory}

As noted in Secs.~\ref{sectionIntroduction} and~\ref{sectionHistorical}, hard spheres came into existence as a minimal model system for exploring condensed matter, and have been extensively reviewed  \cite{hansen,santos2020,roth2010,tarazona1984}. Here we only cover a couple of key theories which are mentioned in later sections, and which we believe had a significant impact in the success of the hard-sphere system.

\subsection{Integral equation theory}
\label{sectionIntegral}

Integral equation theory lies at the heart of descriptions of the liquid state \cite{hansen,barker1976}. In the context of hard-sphere colloids as model atoms and molecules, it provides an elegant and simple 
theoretical framework for obtaining predictions that 
may be directly probed in experiments. 
For our purposes, integral equation theory boils down to finding approximate solutions (or closures) for the Ornstein--Zernike (OZ) relation 
\begin{equation}
h(r)=c(r)+\rho \int c(r) h(r) d\mathbf{r},
\label{eqOZ}    
\end{equation}
\noindent
which relates the total correlation function $h(r)=g(r)-1$ (with $g(r)$ the radial distribution function) to the direct correlation function $c(r)$.

In the framework of classical density functional theory (see Sec.~\ref{sectionCDFT}), the direct correlation function is related to the second functional derivative of the excess part of the free-energy functional with respect to the one-particle density of the system \cite{hansen}. A homogeneous and isotropic system at density $\rho$ and temperature $T$ interacting with a pairwise additive potential $u(r)$ is therefore \emph{uniquely} defined by its  $h(r)$ and thus by its $g(r)$, which is measurable in experiments (see Sec.~\ref{sectionRealizing}). Measuring $g(r)$ -- or its Fourier transform $S(k)$ -- is therefore an important means of characterizing the structure and thermodynamics of an experimental system.

The OZ relation can also be solved using a second (closure) relation. A relatively simple  relation, first proposed by Percus and Yevick (PY) \cite{percus1958}, is given by 
\begin{equation}
c(r)=(\exp[-\beta u(r)]-1)(h(r)-c(r)+1).
\label{eqPY}
\end{equation}
This relationship is highly accurate for strongly repulsive and short-range interactions, such as for hard spheres. In 1963, Wertheim and Thiele independently showed that the PY closure to the OZ equation for a fluid  of hard spheres with diameter $\sigma$ and  volume fraction $\phi = \pi \sigma^3\rho/6$ yields  for the direct correlation function \cite{wertheim1963,thiele1963}
\begin{equation}
  \hspace{-0.2cm}   c(r) = \nonumber \\
 \hspace{0.0cm}    \begin{cases}
   \frac{-(1+2 \phi)^2+ 6 \phi(1+ \frac{1}{2}\phi)^2 \frac{r}{\sigma}- \frac{1}{2}\phi(1+ 2\phi)^2 (\frac{r}{\sigma})^3}{(1-\phi)^4} & r \leq \sigma \\
       0 &  r > \sigma.
   \end{cases}
\end{equation}
$c(r)$ can be analytically Fourier transformed, from which the structure factor $S(k)$ follows using the Fourier transform of the Ornstein-Zernike equation. 
The pair correlation function $g(r)$ can subsequently be obtained by a numerical Fourier transform of $S(k)$. The theoretical predictions agree  well with computer simulation results for volume fractions $0\leq \phi \lesssim  0.5$, i.e., over the entire stable fluid regime. Examples of  experimental comparisons against the PY approximation to the OZ equation are shown in Fig.~\ref{figPeterSq} in Sec.~\ref{sectionBulk}. 

Given $g(r)$, three independent routes can then be followed to extract thermodynamic quantities:
\begin{eqnarray}
P&=&\rho \kB T - \frac{\rho^2}{6}\int d{\bf r} r u'(r)g(r), 
\label{pvir}\\
\frac{E}{V}&=&\frac{3}{2}\rho \kB T + \frac{\rho^2}{2}\int d{\bf r} u(r)
g(r), \\
\kB T\left(\frac{\partial\rho}{\partial P}\right)_T&=&1+\rho
\int d{\bf r} \big(g(r)-1\big),\label{compr}
\end{eqnarray}
which are the virial, caloric, and compressibility routes, respectively.
An important exact sum rule for hard spheres can be obtained straightforwardly from Eq.~\eqref{pvir}
\begin{equation}
    \frac{\beta P}{\rho}= 1 + \frac{2 \pi \rho \sigma^3 g(\sigma^{+})}{3},
\end{equation}
which relates the pressure of the hard-sphere fluid to the contact value of $g(r)$. 
Using $g(r)$ from the PY approximation to the OZ equation in Eq.~\eqref{compr}  yields  the compressibility pressure $P_c$  
\begin{equation}
    \frac{\beta P_c}{\rho} = \frac{1 + \phi + \phi^2}{(1-\phi)^3}, 
\end{equation}
while the virial pressure $P_v$ from Eq.~\eqref{pvir} gives 
\begin{equation}
    \frac{\beta P_v}{\rho} = \frac{1 + 2\phi + 3\phi^2}{(1-\phi)^2}. 
\end{equation}
The compressibility equation of state overestimates  the pressure obtained from simulations, whereas the virial equation of state underestimates the simulation results \cite{hansen}.
It turns out that the Carnahan and Starling (CS) linear combination $P_\mathrm{CS} = (2P_c + P_v)/3$ \cite{carnahan1969},   
\begin{equation}
   \frac{\beta P_\mathrm{CS}}{\rho} = \frac{1 + \phi + \phi^2-\phi^3}{(1-\phi)^3}. 
   \label{eqCS}
\end{equation}
is essentially indistinguishable from  simulations up to $\phi\simeq 0.5$ \cite{hansen}. It should be noted, however, that Carnahan and Starling did not use simulation results but rather the pressure from the virial expansion for the equation of state Eq.~\eqref{virial} to motivate their expression \cite{carnahan1969}. Using the analytical expressions for the virial coefficients $B_2, B_3$ and $B_4$, and the numerical expressions for $B_5$ and higher, the equation of state for hard spheres becomes
\begin{eqnarray}
    \frac{\beta P}{\rho} &=& 1 + 4 \phi + 10 \phi^2 + 18.365 \phi^3 + 28.225 \phi^4 + 39.74 \phi^5 \nonumber \\
    & & + 53.5 \phi^6 + 70.8\phi^7 + \cdots.
\end{eqnarray}
Approximating coefficients as $B_{i+1}= (\pi \sigma^3/6)^i(i^2+3 i)$ then gives Eq.~\eqref{eqCS} \cite{carnahan1969}.

While integral equation theory is successful in predicting the structure and thermodynamics of equilibrium hard-sphere fluids, the closures are somewhat uncontrolled and not systematically improvable. A clear manifestation of this effect is that predictions for binary mixtures of hard spheres turn out to be extremely sensitive to the details of the approximation scheme. A more robust approach in this sense is density functional theory, and especially fundamental measure theory, described in Refs.~\cite{tarazona2008,roth2010} and Sec.~\ref{sectionCDFT}.

\subsection{Cell theory}
\label{sectionCellTheory}

The crystal branch of the equation of state of hard spheres is less amenable to integral equation theory, due to the broken translational and rotational symmetry in the system. One early mean-field theory that can be used as an approximate analytical prediction was originally introduced by \citet{lennardjones1937} to estimate fluid free energies, and is now commonly known as cell theory or free volume theory when applied to solids and especially crystals \cite{wood1952,salsburg1962}. In this approach, it is assumed that the Helmholtz free energy $F(N, V, T)$ of the crystal phase can be subdivided into contributions from each particle, and that each can be computed by assuming that all other particles are fixed exactly at their lattice site. In that approximation, the free energy $f_1$ of a single particle in an FCC lattice is given by
\begin{equation}
    \beta f_1(\rho) = -\log \frac{V_\mathrm{free}(\rho)}{\Lambda^3},
\end{equation}
where $V_\mathrm{free}$ is the free volume available to the particle confined in the cage of its neighbors, and $\Lambda$ is the thermal De Broglie wavelength. This free volume is then typically approximated as a sphere with diameter $a - \sigma$, with $a$ the nearest-neighbor distance, which is given by $a = (\rho_\mathrm{max} / \rho)^{1/3}$. Here, the  close-packed density of crystal lattices of monodisperse hard spheres is $\rho_\mathrm{max}\sigma^3 = \sqrt{2}$. Hence, cell theory gives the approximate total free energy
\begin{equation}
    \frac{\beta F}{N} \simeq \beta f_1 \simeq -\log\left[\frac{4\pi \sigma^3}{3\Lambda^3}\left(\left(\frac{\rho_\mathrm{max}}{\rho}\right)^{1/3}  - 1\right)^3\right].
\end{equation}
Taking the derivative with respect to the volume gives the pressure
\begin{equation}
    \frac{\beta P}{\rho} = \left(1 - \left(\rho/\rho_\mathrm{max}\right)^{1/3}\right)^{-1}.
\end{equation}
Although this expression significantly underestimates the pressure (on the order of $\sim 1 \kB T / \sigma^3$), cell theory provides a quick and physically intuitive estimate for the crystal free energy and equation of state. As a result, it is commonly used for pedagogical purposes \cite{barrat2003,kamien2007soft}, as well as for estimating phase transitions. Cell theory has been extended in a variety of ways (see, e.g.,~\cite{rudd1968, koch2005, charbonneau2021three}), including for binary mixtures of hard spheres \cite{cottin1993, cottin1995, vandermeer2020} and glassy systems. It should be noted that when cell theory is applied to disordered systems, such as glasses, it cannot be expected to take into account the configurational entropy of the system  (see Sec.~\ref{sectionGlass}). Furthermore, in the fluid regime, particles are no longer trapped in cells in any meaningful way, and hence the description should be expected to break down there as well.

\subsection{Classical density functional theory}
\label{sectionCDFT}

Classical density functional theory (DFT) provides an exact theoretical framework for describing the thermodynamic and structural properties of interacting many-body systems, starting from a microscopic description of the interparticle interactions. The approach is based on the observation that the grand potential of a specified inhomogeneous fluid is a functional $\Omega\lbrack \rho(\mathbf{r}) \rbrack$ of the variational  one-body density profile $\rho(\mathbf{r})$, with the properties that (\emph{i}) the equilibrium density profile $\rho_0(\mathbf{r})$ minimizes the functional $\Omega\lbrack \rho(\mathbf{r})\rbrack$, and (\emph{ii}) this minimum equals the equilibrium grand potential $\Omega_0\lbrack \rho_0(\mathbf{r}) \rbrack$. From $\Omega_0$,  all thermodynamic properties then follow. For instance, one can obtain the homogeneous system pressure 
\begin{equation}
P=-\frac{\Omega_0}{V}
\end{equation}

\noindent
Functional derivatives of the Helmholtz free-energy functional additionally provide correlation functions. In particular, the OZ two-body direct correlation
function is related to the second functional derivative of the functional from which the two-body structure of the system follows. However, because the exact Helmholtz free-energy functional for a given interaction potential is unknown,  DFT depends on approximate free-energy density functionals. The efficacy of a given DFT approximation is therefore commonly tested by comparing  DFT predictions with computer simulations.

Exploiting the wealth of available information on the thermodynamics and structure of hard-sphere systems, high-accuracy approximations of that functional have
been constructed for hard-sphere mixtures starting with Rosenfeld's Fundamental Measure Theory (FMT) in 1989 \cite{rosenfeld1989}. In FMT, the Helmholtz free-energy functional is based on  weighted densities that are convolutions of the density profiles with weight functions depending on the geometrical properties of the spheres.  Various subsequent extensions and modifications have been proposed to improve the description of various inhomogeneous systems of hard spheres. Further details on this approach can be found in the comprehensive and excellent reviews by Tarazona \cite{tarazona2008} and Roth \cite{roth2010}. 

%% file: PhaseBehavior/phaseBehavior.tex
\section{Bulk equilibrium hard spheres}
\label{sectionBulk}

Given the wide array of experimental, numerical and theoretical methods described in the previous sections, we can now consider the equilibrium behavior of hard-sphere systems. This behavior not only serves as a reference point in nearly all subsequent sections of this review, but is also an important scientific matter in its own right.

Naturally, the first hard-sphere systems studied by computer simulation were monodisperse~\cite{alder1957,wood1957}, but only \emph{quasi}--monodisperse in colloidal experiments. However, because--as we will see later in this section--a  mean size polydispersity $s\lesssim5\%$ barely changes the equilibrium phase diagram~\cite{wilding2010}, both experimental and numerical work nevertheless focused on equivalent systems. In this context, a natural starting topic is their bulk phase behavior. Other topics considered include the equation of state as well as fluid structure and dynamics for both 3d and (quasi--)2d hard spheres.

\subsection{Equilibrium phase behavior of monodisperse hard spheres}
\label{sectionEquilibriumPhaseBehavior}

Since hard spheres do not interact beyond contact, their (Helmholtz) free energy only comprises an entropic contribution. Temperature plays but a trivial role, and hence the hard-sphere equilibrium phase diagram depends solely on the volume fraction $\phi$ (density) of the system. At low $\phi$ the system is naturally in a fluid phase, whereas at high $\phi$ the system is in a crystal phase. (In 3d, the two are separated by a first-order phase transition.) Put differently, at sufficiently high $\phi$, the crystal entropy is higher than that of the fluid; the regular arrangement of spheres on the crystal lattice provides each particle with more local free volume to move around than it would have in a fluid at the same density.

As mentioned in Sec.~\ref{sectionHistorical}, the first numerical validations of this theoretical prediction were obtained in the 1950s by studying the melting of a face-centered cubic (fcc) crystal~\cite{alder1957,wood1957}. (The crystal phase with the fcc structure is narrowly thermodynamically favored over competing close packed structures, as discussed later in this section.) Since then, considerable efforts have been devoted to pinning down the details of this transition, including characterizing finite-size effects~\cite{polson2000}. Table \ref{tableCoex} provides an overview of numerical results for the coexistence packing fractions, pressures, and chemical potentials. Apart from a few outliers, often associated with small system sizes, results for these quantities are in good agreement, and have established that the freezing and melting volume fractions of monodisperse hard spheres are $\phi_f \simeq 0.492$ and $\phi_m \simeq 0.543$, respectively. 
Similar predictions have also been obtained from various theoretical treatments, notably fundamental measure theory~\cite{roth2010}. Experimental measurements on hard-sphere colloids, such as those shown in Fig.~\ref{figPuseyVanMegen}~\cite{pusey1986}, agree reasonably well with this result, provided a suitable $\sigma_\mathrm{eff}$ is chosen to account for any residual softness (see Sec.~\ref{sectionInteractions}). 

Note that while these studies do not typically take into account the presence of vacancies in the equilibrium crystal phase, its effect on the coexistence pressure is expected to be smaller than the typical error bars considered here ($\beta \Delta P \sigma^3 \simeq -0.0026$)~\cite{pronk2001}. Simulation results have revealed that the equilibrium fcc crystal is indeed nearly flawless. The equilibrium fraction of defective sites at melting has been computed to be approximately $10^{-4}$ for vacancies and $10^{-8}$ for interstitials~\cite{bennett1971, pronk2001, pronk2004}. However, comparing these predictions to experiments is inherently difficult beyond the usual concerns about coexistence determination, given (\emph{i}) the low concentrations involved, (\emph{ii}) their sensitivity to the crystal packing fraction, and (\emph{iii}) the possibility that defects become kinetically trapped during crystal formation (Sec.~\ref{sectionInteractions})~\cite{poon2012,royall2013myth}.

Interestingly, various properties of the bulk crystal phase, including its elastic constants~\cite{runge1987, pronk2003, sushko2005} and defect diffusivity~\cite{vandermeer2017}, have been studied in detail in simulation but remain to be systematically investigated in experiments.

\newcommand{\markconsensus}{$^\dagger$}
\begin{table*}
\caption{\label{tableCoex}
Table summarizing prior reports of hard sphere fluid-fcc phase coexistence using computer simulations (black) and DFT (\textcolor{darkgreen}{green}). For each result, we list the main method used to determine the free energy or stability of the solid phase (see also Sec.~\ref{sectionSimulation}), the freezing and melting packing fractions, the coexistence pressure and chemical potential, and the system size considered. A system size of $\infty$ indicates that results were extrapolated to the thermodynamic limit $N\rightarrow\infty$, but the system sizes considered and the extrapolation scheme vary between studies.  The final consensus averages results from five studies with high reported accuracy (marked by \markconsensus{}), namely \citet{frenkel, fortini2006, fernandez2012, pierprzyk2019, moir2021}
}
\newcommand{\theo}{\textcolor{darkgreen}}
\begin{ruledtabular}
\begin{tabular}{lcccccc}
\textbf{Source} & Method & $\phi_f$ & $\phi_m$ & $\beta P_\mathrm{coex}\sigma^3$ & $\beta\mu_\mathrm{coex}$ & $N$\\
\hline
  \citet{hoover1968}     & Single-occupancy cell & 0.494     & 0.545     & 11.70     &           & $\infty$ \\
  \citet{speedy1997}     & Unspecified           & 0.491(1)  & 0.543(1)  & 11.55(11) &           & \\
  \citet{davidchack1998} & Direct coexistence    & 0.491     & 0.543     & 11.55(5)  &           & 10752 \\
  \citet{wilding2000}    & Phase switch          &           &           & 11.49(9)  &           & $\infty$  \\
  \markconsensus\citet{frenkel}        & Ladd-Frenkel          & 0.4917    & 0.5433    & 11.567    & 16.071    & $\infty$\\
  \citet{roth2002}       & \theo{FMT (White Bear I)}    & \theo{0.489}     & \theo{0.536}     &           &           & \\
  \citet{errington2004}  & Phase switch          &           &           & 11.43(2)  &           & $\infty$ \\
  \markconsensus\citet{fortini2006}    & Ladd-Frenkel          & 0.4915(5) & 0.5428(5) &  11.57(10)& 16.08(10) & \\
  \citet{vega2007}       & Einstein molecule     & 0.492     & 0.543     & 11.54(4)  & 16.04     & $\infty$ \\
  \citet{noya2008}       & Direct coexistence    & 0.491(1)  & 0.543(2)  & 11.54(4)  &           & 5184   \\
  \citet{odriozola2009}  & Parallel tempering    & 0.492(4)  & 0.545(4)  & 11.43(17) &           & $\infty$ \\ 
  \citet{zykova2010}     & Direct coexistence    & 0.492     & 0.545     & 11.576(6) &           & $160000$\\    
  \citet{nayhouse2011}   & Single-occupancy cell & 0.4912(6) & 0.5425(5) & 11.502(19)&           & $\infty$ \\
  \citet{oettel2010}     & \theo{FMT (White Bear II)}   & \theo{0.495}     & \theo{0.545}     & \theo{11.89}     & \theo{16.40}     & \\
  \citet{demiguel2008}   & Ladd-Frenkel          & 0.489(1)  & 0.541(2)  & 11.43(11) & 15.92(11) & 864 \\
  \markconsensus\citet{fernandez2012}  & Free-energy landscape & 0.49188(2)& 0.54312(5)& 11.5727(10)&          & $\infty$ \\
  \citet{ustinov2017}    & Kinetic MC            & 0.491     & 0.543     & 11.534    & 16.054    & 3000 \\
  \markconsensus\citet{pierprzyk2019}  & Literature values     & 0.4917(5) & 0.5433(5) & 11.5712(10)&16.0758(20)& $\infty$ \\
  \markconsensus\citet{moir2021}       & Tethered particle     & 0.49161(3)& 0.54305(4)& 11.550(4) & 16.053(4) & $\infty$ \\
  \hline
  Consensus               &                       & 0.4917(2) & 0.5431(2)  & 11.56(2)  & 16.06(2)          & $\infty$
\end{tabular}
\end{ruledtabular}
\end{table*}

\header{Phase behavior in experiment}
As noted in Sec.~\ref{sectionHistorical}, interest in hard spheres received a tremendous boost from the experiments of Pusey and Van Megen (see Fig.~\ref{figPuseyVanMegen})~\cite{pusey1986}, but earlier work by Hachisu and coworkers~\cite{kose1974,hachisu1974} had already demonstrated that hard sphere-like colloids crystallize. Additionally, results for the strongly screened charged stabilized aqueous suspension of the Japanese group were consistent with theoretical and simulation predictions~\cite{hachisu1974}. When it came to non-aqueous systems (for which electrostatic effects were expected to be minimal), \citet{kose1974} used  what would now be called microgel colloids of crosslinked PMMA dispersed in benzene~\footnote{It is worth noting that benzene is now typically avoided, given its toxicity.}.
Given the difficulty of determining effective volume fractions for such particles (see Sec.~\ref{sectionInteractions})~\cite{poon2012,royall2013myth}, however, a quantitative comparison was challenging for this system~\cite{kose1974}. Later work indeed found that the softness of microgel particles (which can be tuned via the degree of crosslinking used~\cite{lyon2012}) indeed can have a profound effect on the phase coexistence ~\cite{paulin1996}, as may be inferred from the discussion of Fig.~\ref{figMapping}. Experimental efforts to determine the osmotic pressure at freezing were pioneered by Takano and Hachisu~\cite{takano1977,hachisu1982}, who, after drying and measuring the total mass of particles from the fluid phase in a sedimented system, found reasonable agreement with  simulation results.

\begin{figure}
\centering
\includegraphics[width=85mm]{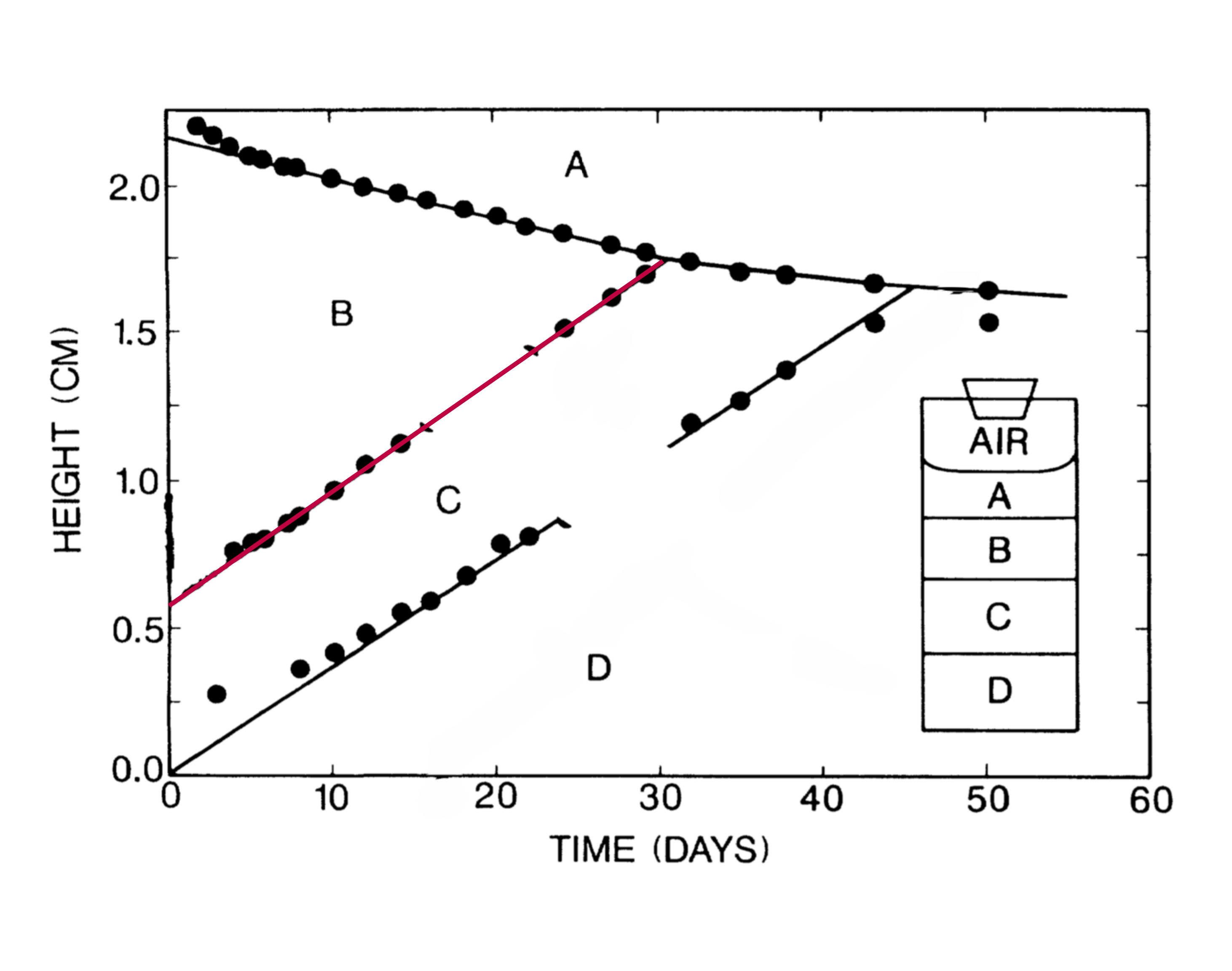}
\caption{Determining the equilibrium coexistence volume fraction from the time--evolution of a  sedimenting suspension. The height of a system in the fluid-crystal coexistence region undergoing sedimentation is recorded as a function of time. After a short waiting period of two days, sharp interfaces form between the clear supernatant (A), the fluid phase (B), the crystal phase (C), and a dense polycrystalline sediment (D).  Extrapolating the fluid-crystal (B-C) interface (red line) back to time $t=0$ provides an estimate of the phase boundary prior sedimentation settling in, i.e., in the absence of gravity. Reproduced with permission from~\cite{paulin1990}.}
\label{figPaulin}
\end{figure}

The subsequent development of sterically stabilized PMMA (see Secs.~\ref{sectionForetelling} and \ref{sectionSynthesizing}) brought better controlled hard-sphere-like experimental systems to the field. A key hurdle to achieving quantitative accuracy, however, has been to properly determine the (effective) $\phi$ in experiment (see Sec.~\ref{sectionInteractions})~\cite{poon2012,royall2013myth}. One might think that even if particles have some degree of softness (as is always the case), phase boundaries should nevertheless lead to volume fractions that can be mapped onto the predicted values. In fact, as discussed in Sec.~\ref{sectionInteractions} and Fig.\ref{figMapping}, it is not that simple, as softness effects the relative coexistence gap between fluid and crystal. Furthermore, as noted by Pusey and Van Megen, sedimentation complicates matters~\cite{pusey1986}.

One clever method to address the effects of sedimentation was proposed by~\citet{paulin1990}. As shown in Fig.~\ref{figPaulin}, the fluid-crystal phase boundary (in vials such as those shown in Fig.~\ref{figPuseyVanMegen}) tends to move upwards as the fluid phase slowly crystallizes. While the system starts as a metastable fluid, with an effective volume fraction that falls within the coexistence regime (see Fig.~\ref{figHSPhaseDiagram}), the final sedimentation-diffusion equilibrium state is a crystal with only a thin fluid phase above. (The thickness of the fluid phase is of the order of the gravitational length defined in Sec.~\ref{sectionFarFromEq}, which in this case is very much less than the container size.) In Fig.~\ref{figPaulin}, the top of the sediment is characterized by the supernatant (A)--colloidal fluid (B) interface. Sedimentation proceeds over the course of the experiment--some two months--but after approximately 27 days, the colloidal fluid (B) is no longer visible to the naked eye. The fluid ``disappearance'' reflects the approach to sedimentation--diffusion equilibrium. The colloidal fluid--crystal (C) phase separation, however, took only about one day. (An additional, distinct layer in the form of a polycrystalline sediment (D) also formed, akin to that observed in Fig.~\ref{figPuseyVanMegen}(c), but plays no role in the subsequent analysis.)

Phase diagram determination from these observations requires a key assumption: that the boundary in the sedimenting system  extrapolated back to $t=0$ reflects the phase-separated system prior to any effects from sedimentation. In other words, $t=0$ should correspond to a phase-separated system in the \emph{absence} of gravity. Using different samples of varying initial $\phi$, application of the lever rule then gives the equilibrium fluid-crystal coexistence region. This method, therefore requires a separation of timescales between that of sedimentation (1 month) and that of fluid-crystal phase separation (1 day). The resulting phase behavior was later found to be in reasonable agreement with the experimental equation of state~\cite{rutgers1996,phan1996}. (Section~\ref{sectionEquationOfState} discusses what \emph{reasonable} agreement means in the context of colloidal experiments.)

\header{Polymorphism} At sufficiently high $\phi$, one expects the stable crystal to be one of the close-packed structures, known as Barlow packings, considered in the Kepler conjecture \cite{hales2017}, which consist of stacked layers of spheres arranged at the vertices of a hexagonal (or triangular) lattice. The non-uniqueness of the stacking follows from the fact that each additional layer has two different positions in which it can be placed with respect to the previous one. Computer simulations combined with free-energy calculations have demonstrated that the fcc structure (with an  ABCABC stacking order) is slightly entropically favored over the hexagonal  close-packed (hcp) structure (with an ABAB stacking order)~\cite{frenkel1984,bolhuis1997,mau1999,polson2000}. Given that   the free energy difference is only about $10^{-3} \kB T$ per particle, however, spontaneously formed hard-sphere crystals often contain defects in the fcc stacking sequence, resulting in a random-hexagonal-close-packing (rhcp) structure. This polytype\footnote{Polytypes are polymorphs whose symmetry only differs in one direction.} is unlikely to anneal out to the equilibrium fcc crystal phase over any realistic time scale in 
simulation~\cite{pronk1999}. However, annealing of rhcp into fcc crystal structures has been observed in   X-ray crystallographic experiments of suspensions of quite small particles (600nm)~\cite{kegel2000jcp,martelozzo2002}. Small external biases can also play a significant role. For instance, crystallization resulting from sedimentation under gravity seems to more strongly favor the formation of fcc~\cite{hoogenboom2002, marechal2011}, whereas experiments in microgravity result in rhcp crystals~\cite{zhu1997,cheng2001ApOpt}. A possible explanation is that mechanical instability  occurring during the packing process of the former selects the fcc structure~\cite{Heitkam2012}.

\subsection{Impact of polydispersity on the equilibrium phase diagram}
\label{sectionPolyPhase}

As mentioned in Sec.~\ref{sectionRealizing}, preparing a  sample of colloidal spheres necessarily leads to a size distribution $p(\sigma)$ of diameters instead of the idealized identical-sphere model. The standard deviation of $p(\sigma)$ relative to the mean diameter defines the polydispersity $s$, which typically quantifies this effect.

The phase behavior of systems with small polydispersity, $s\leqslant 5$\%, is essentially indistinguishable ($\lesssim 2\%$ difference in coexistence $\phi$) from that of monodisperse hard spheres.  Going to larger polydispersities, however, strongly suppresses crystallization. The existence of a ``terminal polydispersity'' ($s_\text{t}\approx5-12\%$ depending on the higher moments of the distribution), beyond which the crystal cannot be stabilized was therefore initially suggested~\cite{barrat1986jphysfrance,pusey1987jphysfrance,bolhuis1996,phan1998,bartlett1999,ackerson1995}.

\begin{figure}[h!]
\centering
\includegraphics[width=85mm]{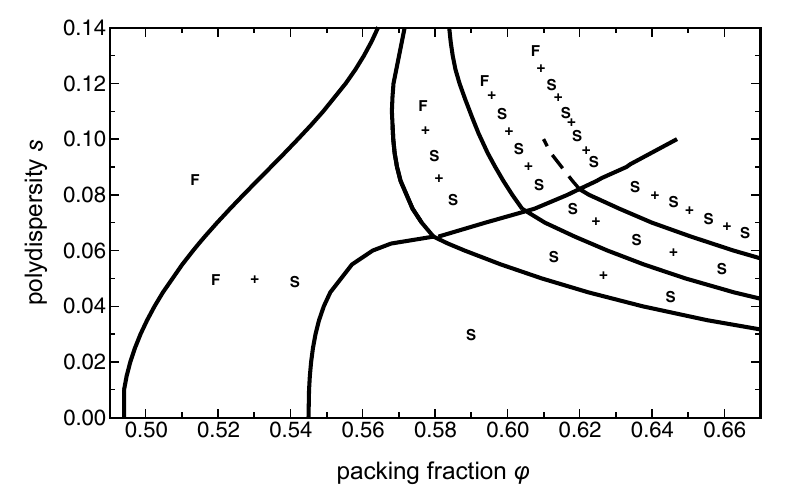}
\caption{Phase diagram of polydisperse hard spheres. Data are from Sollich and Wilding's free-energy calculations for a top hat distribution~\cite{fasolo2004}. Beyond 6\%, the single solid phase fractionates into coexisting solids of different size distributions with smaller polydispersity. The larger the polydispersity and packing fraction, the larger the number of coexisting phases needed to satisfy the polydispersity constraint on individual crystal phases\cite{sollich2010}. Reproduced with permission from~\citet{fasolo2004}.}
\label{figPoly}
\end{figure}

Follow-up simulations, however, revealed that polydisperse suspensions circumvent this barrier through \emph{fractionation} (see Fig.~\ref{figPoly}). The solid phase then splits into coexisting solid phases with smaller polydispersity (each being limited to $s_\ast\approx6\%$)~\cite{kofke1999}. In other words, instead of a single crystal with widely different particle sizes (implying large local strains), the system favors the formation of separate crystalline domains of small and large particles. This proposal has since been put onto solid theoretical ground through (approximate) free-energy calculations~\cite{fasolo2003,fasolo2004}.

Support for fractionation can be obtained from computer simulations, albeit with particular methodological care. In standard fixed-volume simulations, accessing equilibrated fractionated solids is obfuscated by the extremely slow nucleation dynamics (see Sec.~\ref{sectionNucleation}) and the slow dynamics of domain boundaries. \citet{sollich2010} have side-stepped these difficulties by employing constant pressure Monte-Carlo simulations that prescribe a size-dependent chemical potential $\Delta\mu(\sigma)$ (instead of fixing the parent distribution $p(\sigma)$). This demonstration of qualitative agreement between simulations and theoretical work (see Fig.~\ref{figPoly}~\cite{wilding2010,sollich2011}), is supported by some evidence for fractionation in experiments using light scattering~\cite{martin2003}.

Intriguingly, recent swap-assisted EDMD simulations at fixed volume (see Sec.~\ref{sectionSimulationMC}) report ordering of polydisperse hard spheres into complex crystals with a large unit cell, such as Laves and Frank-Kasper phases, instead of fractionating into fcc crystals~\cite{lindquist2018,bommineni2019}. These large unit cells, which incorporate small and large particles, are reminiscent of what is observed in binary mixtures (see Sec.~\ref{sectionBinary}).  Similarly, polydisperse mixtures with a non-Gaussian size distribution have also been observed to partially crystallize into an AlB$_2$ structure~\cite{coslovich2018}. The spontaneous formation of these phases suggests that their formation might be thermodynamically preferred over fractionation into crystal (see Fig. \ref{figPoly}), but this proposal has yet to be investigated using free-energy calculations. Although this phenomenon has not yet been observed in experimental colloidal hard spheres either, polydisperse charged silica nanospheres do form comparably complex crystal phases~\cite{cabane2016}.  The question of the true equilibrium phase diagram of strongly polydisperse hard sphere mixtures therefore remains open.

\subsection{Equation of state}
\label{sectionEquationOfState}

Equations of state (EoS) are some the most fundamental descriptions of equilibrium systems. On the theoretical and numerical side, high-accuracy EoS are now available for both liquid and crystal phases of hard spheres~\cite{pierprzyk2019}. Such a work caps decades of systematic improvements (see e.g.,~\citet{alder1968,speedy1997,speedy1998,almarza2009, santos2020} and Secs.~\ref{sectionTheory} and \ref{sectionHistorical}). Despite the quantitative success of these efforts, physical insight largely emerges from approximation schemes, which also suffice as reference and calibration for colloid experiments. The expression obtained by \citet{carnahan1969} for the fluid phase (Sec.~\ref{sectionTheory}, Eq.~\eqref{eqCS}), in particular, has proven particularly useful. For the crystal, fitted data to computer simulation results, such as Hall's fit~\cite{hall1972} and the Speedy equation of state~\cite{speedy1998jpcm}, play a similar role.

\begin{figure}[h!]
\centering
\includegraphics[width=6cm]{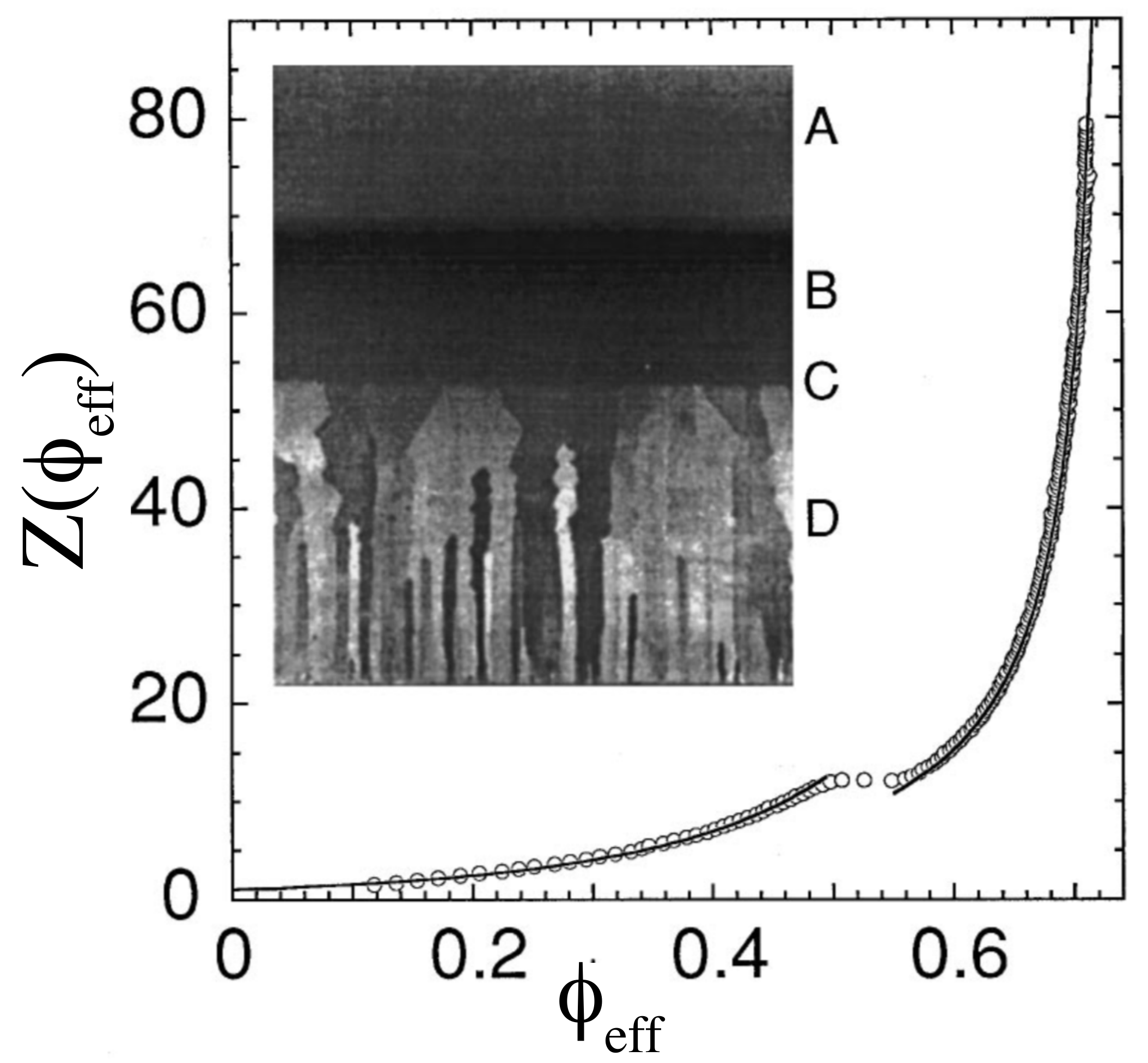}
\caption{Experimental determination of the equation of state, i.e., $Z(\phi_{\textrm{eff}})=\beta P(\phi_{\textrm{eff}})/\rho_{\textrm{eff}}$ versus $\phi_{\textrm{eff}}$, for $\sigma=0.720 \mu\mathrm{m}$ PMMA particles from the observed sedimentation-diffusion equilibrium: 
(a) supernatant, $\phi_{\textrm{eff}}\approx 0$; 
(b) fluid phase, $0<\phi_{\textrm{eff}} \le 0.492$; 
(c) sharp interface; 
(d) crystalline phase, $0.543 \le \phi_{\textrm{eff}} \le  0.74$. Lines denote the fluid and solid EoS for true hard spheres. The inset shows a photograph of 1 cm of the sample. Adapted from \citet{rutgers1996}.
}
\label{figPaulChaikinEoS}
\end{figure}

Early experimental attempts at determining the fluid EoS were carried out by \citet{vrij1983}, and marked methodological improvements were later achieved by~\citet{piazza1993}. Although both approaches obtained the EoS by integrating the equilibrium density profile in a single sedimentation experiment, the later work used a charge--stabilized system of poly-tetrafluoro ethylene (a polymer colloid similar in some respects to polystyrene but with a rather lower refractive index) with a Debye length much smaller than the particle diameter ($\kappa^{-1}=2$ nm vs.~$\sigma=146$ nm), thus yielding a good hard-sphere approximation  (see Sec.~\ref{sectionMapping}). These later measurements compared well with the Carnahan--Starling expression for the fluid and reasonably well to Hall's fit to simulation data for the crystal~\cite{hall1972}. As shown in Fig.~\ref{figPaulChaikinEoS}, subsequent work by Rutgers \emph{et al.} using X-ray densiometry~\cite{rutgers1996} also found reasonable agreement with theory for the fluid branch as well as for the solid branch for $\phi$ approaching close packing. 

What is \emph{reasonable} agreement in this context? As discussed in Sec.~\ref{sectionAccuracy}, although colloidal systems are excellent model systems to demonstrate a range of phenomena of condensed matter, 
highly accurate standard measures, they are not, at least compared to atomic and molecular systems. While the accuracy limit has not been systematically characterized, it seems reasonable that an error in volume fraction of about $\delta \phi \sim 0.01$ should result from measurements that base their phase boundaries on the method of \citet{paulin1990} discussed in Sec.~\ref{sectionEquilibriumPhaseBehavior}.

At low $\phi$, however, the results of~\citet{piazza1993} exhibited a systematic drift far in excess of the expected error. The measured volume fraction dropped much slower than anticipated as a function of height. In other words, the sedimentation profile was \emph{extended}. The physical mechanism that underlies such extension was later identified by \citet{vanroij2003}. In these systems, colloids and small ions decouple under a gravitational field---the latter being only  negligibly affected on the experimental length scale. The resulting macroscopic electric field then partially counters gravity, thus extending the sedimentation profile. Subsequent experiments confirmed that interpretation~\cite{rasa2004,royall2005}.

Particle-resolved studies have enabled new developments in EoS measurements. For example, \citet{dullens2006pnas} performed numerical Widom particle insertion on experimental particle configurations to determine the chemical potential of colloidal hard spheres under the assumption of a hard-sphere pair potential. The resulting thermodynamic properties, including the equation of state are quite good, but  become numerically inaccessible at higher volume fraction ($\phi \gtrsim 0.43$). The relation between free volume measurements and free energy has separately been explored in the supercooled fluid regime ($\phi\geq0.54$) with reasonable quantitative agreement~\cite{zargar2013,dang2022} despite the crudeness of the assumptions underlying cell theory in this regime (Sec.~\ref{sectionCellTheory}). Furthermore, as elegant as this approach can seem, the effects of polydispersity and tracking errors can hamper such analyses based on particle coordinates. An analysis based on cavity averages which compared with simulation data found errors in the pressure to be 100\% and up to 10k$_B$T in the chemical potential \cite{schindler2015}.

\subsection{Structure and dynamics of the hard-sphere fluid}
\label{sectionStructureDynamics}

\begin{figure*}
\centering
\includegraphics[width=18cm,keepaspectratio]{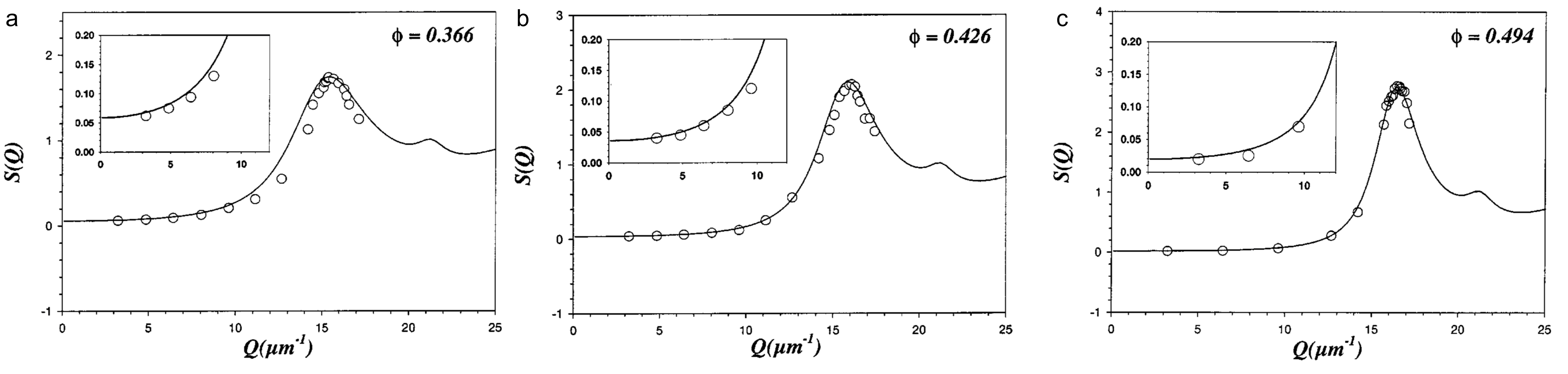} 
\caption{Reciprocal-space pair structure of hard sphere fluids determined using static light scattering~\cite{moussaid1999pre}. 
Experimental data is shown as points, and compared with Percus-Yevick predictions (solid line) at the specified volume fractions.
}
\label{figPeterSq} 
\end{figure*}

\header{Structure}
As discussed in Sec.~\ref{sectionTheory}, the radial distribution function $g(r)$ and structure factor $S(k)$ of the hard-sphere fluid can be obtained from the Percus-Yevick closure to the Ornstein-Zernike equation. The resulting theoretical predictions have been extensively tested against simulation data (e.g., \citet{ree1966, frenkel1986, hansen}). Experimental studies of colloidal fluids using static light scattering (SLS, see Sec.~\ref{sectionLightScattering}), also provide a natural access to the pair structure. Early SLS experiments~\cite{vrij1983, dekruif1985} found that $S(k)$ for hard-sphere--like colloidal suspensions was consistent with the Percus-Yevick expression in the small wave vector limit $k\to 0$. Because $S(k\to 0)$ is directly linked to the isothermal compressibility of the fluid under consideration, as described by Eq.~\eqref{eqCompressibility}, this correspondence  also experimentally validated the hard-sphere EoS. Subsequent small-angle neutron scattering~\cite{dekruif1988} and light scattering experiments~\cite{moussaid1999pre} extended this correspondence over a range of  $k$ that included the first peak of $S(k)$ (see Fig.~\ref{figPeterSq}).

\begin{figure}[h!]
\centering
\includegraphics[width=7cm,keepaspectratio]{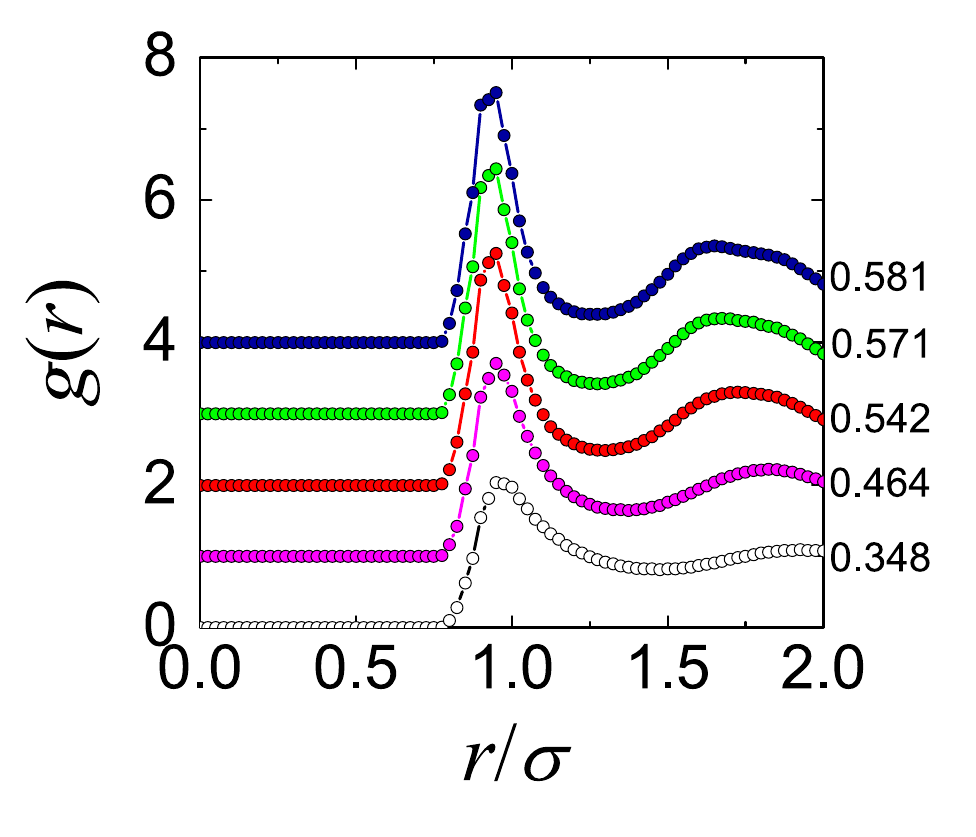} 
\caption{Radial distribution function  of hard-sphere fluids in real space at various (effective) volume fractions. Points are experimental data, lines are simulations of hard sphere fluids with the experimental polydispersity. Adapted from \citet{royall2018jcp}, with the permission
of AIP Publishing.  }
\label{figG} 
\end{figure}

Access to the real-space structure of bulk colloidal fluids, made possible with the advent of confocal microscopy, led to the first experimental measurements of $g(r)$ of hard spheres~\cite{vanblaaderen1995}. The example in Fig.~\ref{figG} compares experimentally measured $g(r)$ with EDMD simulation results. Note, however, that the measurements, especially the first peak of $g(r)$, are sensitive both to polydispersity and to tracking errors~\cite{ivlev,royall2007jcp, mohanty2014}. Both effects can be explicitly taken into account in simulations, in order to improve the match with {\it in silico}  predictions of the experimental $g(r)$. The resulting pair correlation functions match to a high degree of accuracy. Very recently, a range of solvents has been employed to obtain a behavior indistinguishable from hard spheres when real--space $g(r)$ comparisons were carried out. That is to say, the effects of softness (see Sec.~\ref{sectionInteractions}) were found to be negligible~\cite{kale2023}.

\begin{figure}[h!]
\centering
\includegraphics[width=85mm,keepaspectratio]{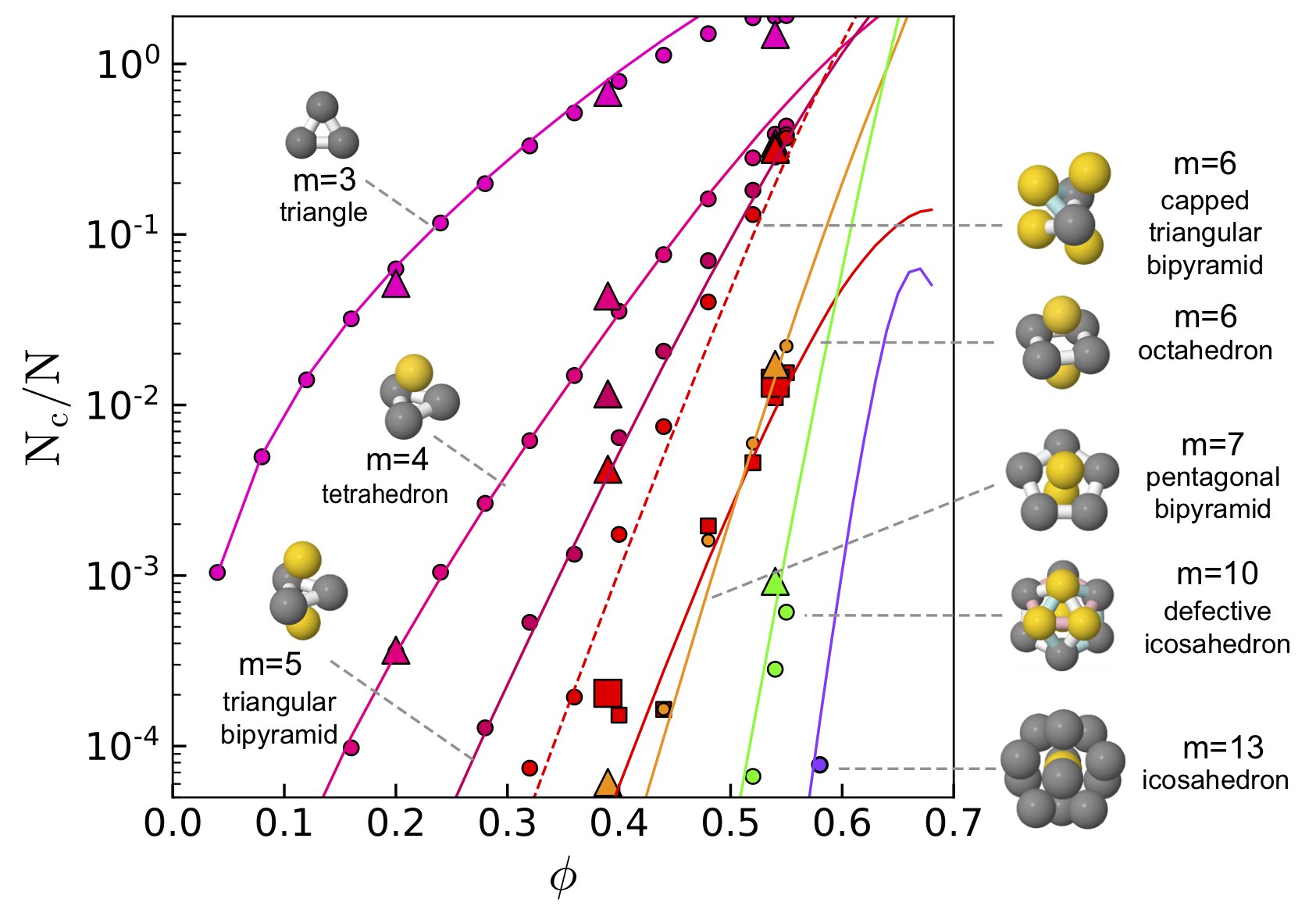} 
\caption{Higher--order structure in the bulk hard-sphere fluid. Populations of higher-order structures $N_c/N$ containing $m=3$ to $m=13$ particles, where $N_c$ is the total number of clusters of a certain topology. Lines are from morphometric theory ~\cite{robinson2019,robinson2020}. Small data points are from monodisperse Monte-Carlo simulations, except for the purple line for $m=13$ (with $s=8$\% polydispersity). 
Large data points are from confocal microscopy experiment using particles with $\sigma=2.0 \mu$m. Simulations and experiments are analyzed with the topological cluster classification using a simple ``bond length'' (or pair distance) of $1.2\sigma$~\cite{malins2013tcc}. }
\label{figTCCrho} 
\end{figure}

Many-body correlation functions and structural features are also easier to extract from real-space than from reciprocal-space information~\cite{royall2015physrep}. In particular, it has  been possible to compare experiments ~\cite{taffs2013}, simulations and theoretical predictions~\cite{robinson2019,robinson2020} for the occurrence of clusters that are known to minimise packing constraints~\cite{manoharan2003,robinson2019}. As demonstrated in Fig.~\ref{figTCCrho}, theoretical predictions of the population of such clusters show close agreement between the various approaches.

\begin{figure*}
\centering
\hspace{0.5cm}\includegraphics[width=155 mm]{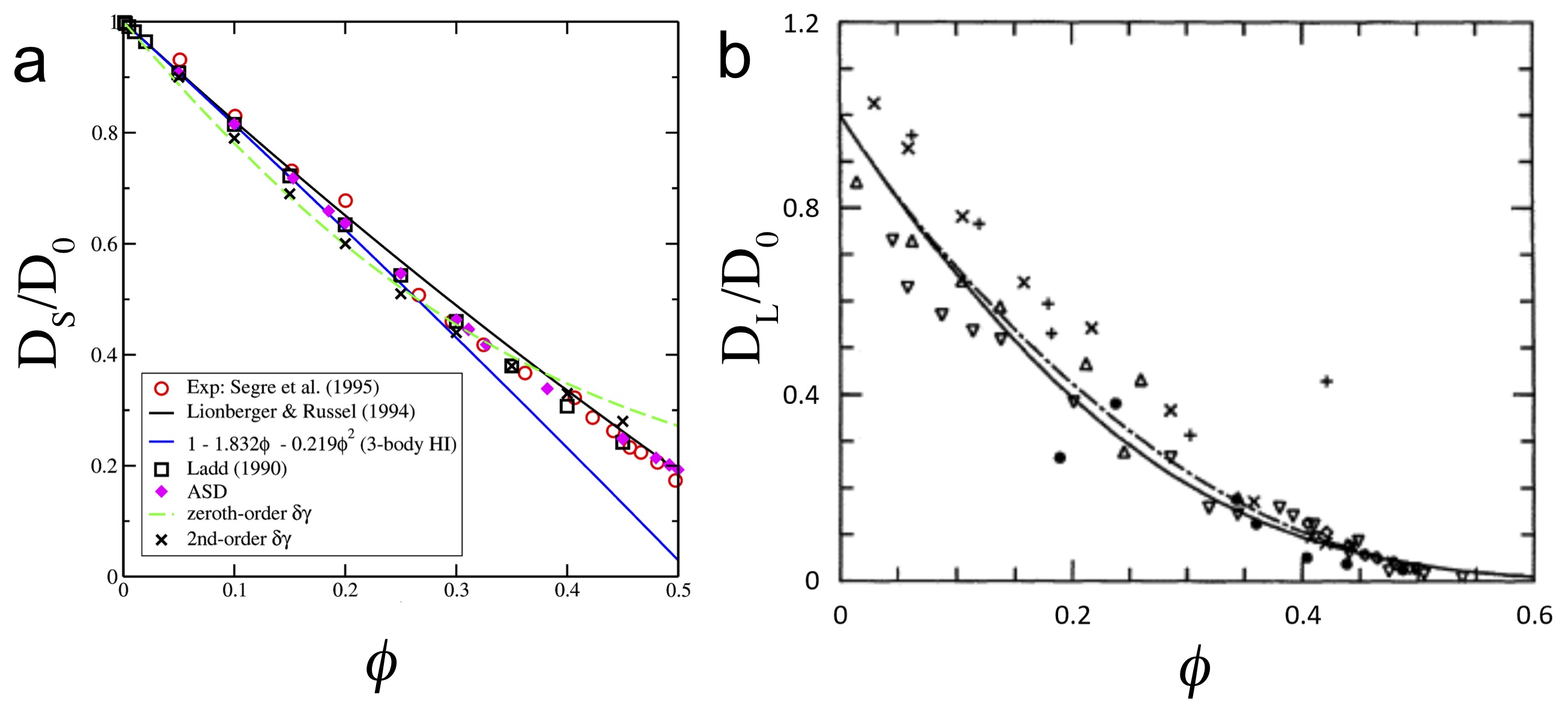}\\
\caption{Self-diffusion coefficient of colloidal hard spheres.
(a) Short-time diffusion coefficient obtained from various sources: experiments~\cite{segre1995pre}, Eq.~\eqref{eqLionberger} by \citet{lionberger1994}, theoretical prediction based on up to three-body hydrodynamic interactions~\cite{cichocki1999}, force multipole simulations by~\cite{ladd1990}, accelerated Stokesian dynamics (ASD) simulations~\cite{banchio2008}, and theoretical predictions based on the theory of Beenakker and Mazur~\cite{beenakker1983, banchio2008}. Figure taken from ~\cite{banchio2008}.   
(b) Long-time diffusion coefficient from various sources: experiments with hard-sphere like suspensions by \citet{vanmegen1987} ($\triangledown$), \citet{segre1995prl} ($\diamond$), \citet{vanblaaderen1992jcp} ($+, \triangle$), \citet{imhof1995pre} ($\times$), ~\cite{kops1982} ($\bullet$) as well as the theory of \citet{cohen1998} (solid line) and of \citet{medina1988} (dash-dotted line). Adapted from~\cite{cohen1998}.}
\label{figDiffusion}
\end{figure*}

\header{Dynamics}
At very short times (prior to colliding with a significant number of solvent molecules), colloidal motion  is ballistic~\cite{franosch2011,hammond2017}. At slightly longer times, however, hard-sphere colloids behave  diffusively~\cite{tough1986}; colloidal dynamics then stems from Brownian motion. Due to the frequent collisions of colloids with solvent molecules, inertia can typically be neglected, thus making their motion overdamped, as in Eq.~\eqref{eqBD}. 

Before particles have had the opportunity to interact with their neighbors, their diffusivity can be described by a short-time diffusion coefficient $D_S$. In the dilute limit $\phi \to 0$, neglecting  hydrodynamic interactions between colloids, the short-time diffusion coefficient coincides with  the free diffusion coefficient,  $D_S=D_0$, which for a single colloidal sphere in a solvent with viscosity $\eta$ is accurately described by the Stokes-Einstein relation
\begin{equation}
D_0 = \frac{\kB T}{3 \pi \eta \sigma}. 
\label{eqStokesEinstein}
\end{equation}
At higher volume fractions, the short-time diffusion coefficient is reduced by hydrodynamic interactions.  This slowdown has been demonstrated in a variety of experimental hard-sphere realizations~\cite{ottewill1987, vanmegen1987, qiu1990, vanmegen1990, zhu1992, segre1995pre}, and reproduced in  simulations that specifically account for such interactions~\cite{ladd1990, segre1995pre, banchio2008}. 
Various theoretical and empirical descriptions of the relation between $D_0$ and $D_S$ have also been proposed~\cite{beenakker1983, beenakker1984, pusey1983fara, dhont1984, cohen1998, cichocki1999}. A common semi-empirical  expression was introduced by \citet{lionberger1994}
\begin{equation}
\frac{D_S}{D_0} \simeq (1-1.56 \phi)(1 - 0.27 \phi). \label{eqLionberger}
\end{equation}
Figure \ref{figDiffusion}(a) compares experimental results of $\frac{D_S}{D_0}$ with Eq.~\eqref{eqLionberger}, as well as with several other theoretical and simulation approaches. Clearly, different simulation methods (force multipole simulations by~\citet{ladd1990} and accelerated Stokesian dynamics by~\citet{banchio2008}) accurately describe the experimental data, while the theoretical predictions based on~\citet{beenakker1983} agree well up to reasonably high $\phi$. Good agreement with the same theory up to $\phi \simeq 0.4$ was also obtained by the experimental work of~\citet{orsi2012}. The wave vector dependence of $D_S$ has also been investigated and theoretical predictions verified~\cite{vanmegen1985,segre1995prl}.

At still longer times, interactions with neighboring particles hinder colloid mobility. This crowding leads to a subdiffusive regime (as shown by  the slope of the mean squared displacement  at intermediate times) before motion ultimately becomes diffusive once more. This long-time diffusive behavior is now described by a different diffusion coefficient  $D_L < D_S$~\cite{tough1986}.  As discussed in Sec.~\ref{sectionGlass},  $D_L$ is strongly suppressed at very high $\phi$, in the regime of glassy dynamics. Here, we focus on the dynamics in the equilibrium fluid phase alone.

The long-time diffusion coefficient and other transport properties of pure hard spheres (i.e. in the absence of hydrodynamics) have been extensively explored by means of  theory and simulations. For instance, early simulations considered the agreement between low-density fluid results with predictions from Enskog's kinetic theory of gasses~\cite{chapman,alder1967, alder1970, easteal1983, speedy1987}. When appropriately rescaled, these results are essentially independent of the choice of microscopic dynamics (i.e. Monte Carlo, Brownian, or Newtonian dynamics)~\cite{scala2007, sanz2010}. As a result, for pure hard spheres the choice of simulation methods typically does not significantly impact dynamical behavior.

In experimental colloidal hard-sphere fluids, the long-time dynamics is influenced by collisions with neighbors as well as by hydrodynamics. While hydrodynamic interactions typically only lead to an overall (density-dependent) scaling of the dynamical time scales for systems close to equilibrium, they can significantly influence non-equilibrium behavior (see Sec.~\ref{sectionFarFromEq} and \ref{sectionDiscrepancy}). Experimental measurements of the diffusion coefficient in a variety of systems~\cite{vanmegen1987, segre1995prl, vanblaaderen1992jcp, imhof1995pre, kops1982} have shown broadly consistent results for $D_L / D_0$ as a function of $\phi$ (see Fig.~\ref{figDiffusion}b). Theoretical approaches based on the generalized Langevin equation~\cite{medina1988}, on the expected scaling of diffusion near random close packing \cite{brady1994}, or on a mean-field description~\cite{cohen1998} provide a reasonably good description of this behavior. It should be noted that although the experimental data in Fig.~\ref{figDiffusion}(b) represent a variety of systems with different polydispersities and electrostatic charge, these effects are largely hidden by the experimental noise. Since both charge and polydispersity shift the effective hard sphere volume fraction, it is natural to expect a corresponding shift in the dynamics. This effect can be more clearly observed when considering viscosity measurements~\cite{segre1995prl, vanderwerff1989, dekruif1985jcp, jones1991jcis, papir1970, imhof1994}, and is closely matched  by the theoretical treatment of \citet{cohen1998}. Charged colloids also exhibit the same scaling and it has been suggested that the collapse may be more coincidence than something of fundamental significance.

The reciprocal-space dynamics of hard sphere has also been extensively studied. Intriguingly, experiments by Segr\'e {\em et al.}~on sterically stabilized PMMA particles found an unexpected collapse of the dynamical behavior for large wave vectors: both the wave vector-dependent diffusion coefficient $D(k, t)$ and the intermediate scattering function $F(k,t)$ collapsed onto a single curve for $k\sigma \gtrsim 5$~\cite{segre1996, segre1997}. 
Later work on charge-stabilized polystyrene spheres, however, could not reproduce that collapse~\cite{lurio2000}, thus suggesting that residual charges may have played a role in that study even if their effect can be hardly distinguished in the static structure factor. This interpretation is further supported by the fact that both mode-coupling theory calculations on pure hard spheres~\cite{fuchs1999} and additional experiments on sterically stabilized PMMA spheres~\cite{orsi2012} reproduce the approximate collapse of $F(k,t)$ observed by Segr\'e and coworkers.

While not strictly pertaining to \emph{colloidal} hard spheres, because the work was carried out using MD simulations of hard spheres in a vacuum, we note \emph{en passant} that some attention has been given to the tail of the velocity autocorrelation function. Around the freezing volume fraction, this observable turns from positive to negative a back to positive~\cite{williams2006,martinez2014}, as a result of collisional back-scattering in high-density fluids. Unlike some authors, however, we conclude based on the nature of the phase transition and its equivalence in systems with other microscopic dynamics that the emergence of this dynamical phenomenon is merely coincidental with the thermodynamic freezing point and not reflective of a deeper physical relationship.

To summarize, we have a clear and robust theoretical understanding and experimental validation of the diffusion of colloidal hard-sphere fluids at low and intermediate densities.

\subsection{Bulk hard spheres in two dimensions}
\label{sectionEquilibrium2d}

\begin{figure}
\includegraphics[width=85 mm]{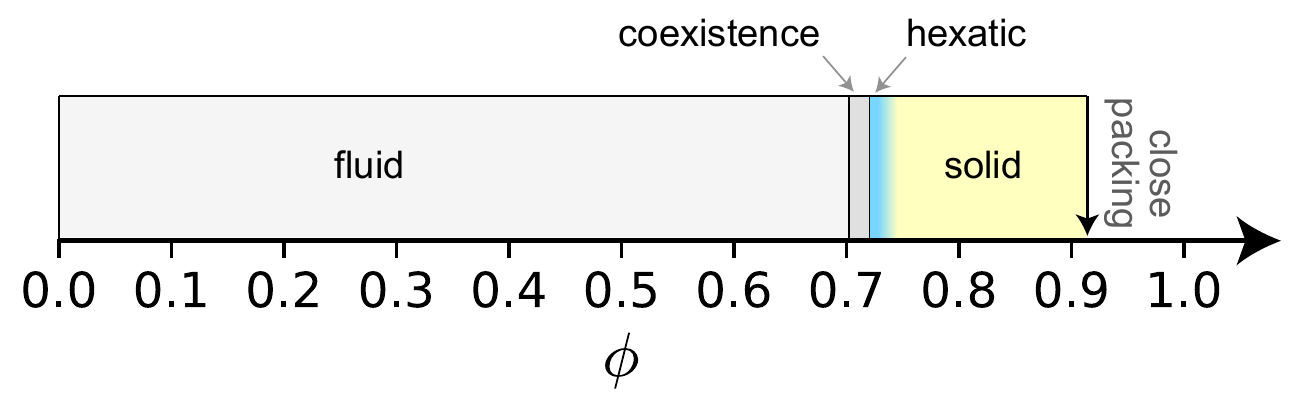}
\caption{Hard-sphere phase behavior in 2d as a function of area fraction. 
Close packing is indicated at $\phi_\mathrm{cp}=0.907\ldots$. The first-order fluid-hexatic transition is marked by a narrow coexistence region of $\phi \approx 0.68--0.70$. The hexatic-solid transition is second order at $\phi \approx 0.73$ as indicated by the shading. Phase boundaries are taken from~\citet{bernard2011}.}
\label{figHDPD}
\end{figure}

\header{Equilibrium phase behavior of hard disks}
When monodisperse hard spheres are confined to  a strictly two-dimensional set-up, they behave as hard disks, a  model system that has itself long been the focus of theoretical and computational study. Interestingly, the phase behavior of hard disks is profoundly different to that of hard spheres. As shown in Fig.~\ref{figHDPD}, a qualitatively new phase -- the hexatic phase -- then appears between the fluid and the solid (which, strictly speaking, is not a crystal as it lacks periodic positional order due to Goldstone modes fluctuations at  small wavevector $k$).

In the perfect crystal, each particle has six neighbors and at close packing $\phi_\text{cp}=\pi/2\sqrt{3}=0.907\ldots$ of the area is covered. The hexagonal nature is captured through the extent of spatial  correlations of the bond-orientation order (BOO) parameter $\psi_6$, which takes a value of unity for perfect hexagonal order. The hexatic phase is characterized by an algebraic decay of orientational correlations, in contrast to the exponential decay and incomplete decay (i.e. long-range order) of these same correlations in the fluid and the solid, respectively. There is an intimate link between the XY model in 2d, for which the  Kosterlitz-Thouless-Halperin-Nelson-Young (KTHNY) theory predicts two continuous phase transitions~\cite{kosterlitz1973, halperin1978, young1979} as first dislocation pairs appear and in a second step these dislocation pairs unbind and proliferate, thus destroying orientational order in the fluid.

How closely hard disks follow this scenario was debated for decades. Simulations finally settled the question about a dozen years ago~\cite{bernard2011,engel2013}. The key difficulty is that finite-size effects cut off the algebraic decay of correlations. Very large system sizes are therefore required to assess their relevance. This requirement is compounded by the slow decay of structural correlations in the regime of interest. This dual hurdle was first overcome thanks to event-chain MC simulations (Sec.~\ref{sectionSimulationMC}). The results clearly revealed that the solid-hexatic transition is continuous, but that the hexatic-fluid transition is weakly first order with a narrow coexistence region $0.700<\phi<0.716$, wherein domains of long-range orientational correlations coexist with short-range orientational correlations, see Table~\ref{tableCoex2d}. In order to relate these predictions to experimental systems, the effect of out-of-plane fluctuations was investigated using event-driven molecular dynamics simulations~\cite{qi2014}. These simulations showed that the two-stage melting scenario of  hard disks persists even for monolayers of hard spheres with out-of-plane buckling as high as half a particle diameter. These simulations have further shown  that the solid-hexatic transition is of the Kosterlitz-Thouless type and occurs via dissociation of bound dislocation pairs, whereas the hexatic-liquid transition  is  driven by a spontaneous proliferation of grain boundaries.   

\begin{table*}
\caption{\label{tableCoex2d}
Table summarizing prior reports on the phase behavior of  hard disks in 2D   using computer simulations. For each result, we list  the area fractions  of the coexisting fluid $\phi_\mathrm{f}$ and hexatic $\phi_\mathrm{h}$ phases,  the area fraction at the hexatic-solid transition $\phi_\mathrm{hs}$,  the coexistence pressure at the fluid-hexatic phase transition $\beta P^\mathrm{fh}_\mathrm{coex} \sigma^2$ as well as the system size considered. Note that the area fraction is defined as $\phi=\pi \sigma^2 N/4A$  with $A$ the system area. Note also that the accuracy is to the last reported digit.
}
\newcommand{\theo}{\textcolor{darkgreen}}
\begin{ruledtabular}
\begin{tabular}{lcccccc}
\textbf{Source} &  $\phi_\mathrm{f}$ & $\phi_\mathrm{h}$ & $\phi_\mathrm{hs}$ & $\beta P^\mathrm{fh}_\mathrm{coex}\sigma^2$ &  $N$\\
\hline
  \citet{bernard2011}     &  0.700     & 0.716     & 0.720     &   9.185         & $1024^2$ \\
  \citet{engel2013}     &  0.700  & 0.716  & 0.720 &     9.185     & $1024^2$\\
  \citet{qi2014}   & 0.699     & 0.717     & 0.725  &   9.183        & $1024^2$ 
\end{tabular}
\end{ruledtabular}
\end{table*}

Early experimental work aiming at elucidating the nature of the melting transition of colloidal monolayers of PMMA spheres suggested a first-order phase transition between the fluid and hexatic phases as well as a first-order phase transition between the hexatic and solid phase~\cite{marcus1997}. However, the authors argued that this observation could be linked to (non-hard) attractive or repulsive interactions between  particles, consistent with an earlier simulation study ~\cite{bladon1995}.

\begin{figure*}
\includegraphics[width=160mm]{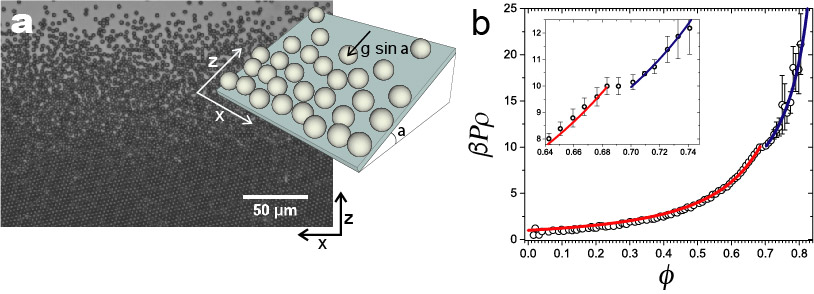}
\caption{Experimental determination of the hard-sphere phase behavior in 2d.
(a) Experimental image of quasi-2d hard spheres in sedimentation-diffusion equilibrium under a slight tilt.
(b) The equation of state $P/\rho \kB T$. 
The inset shows an expanded view of EoS around the discontinuity. The solid red line is a prediction of scaled particle theory for the fluid regime  (Eq.~\ref{eqSPT2D}).
The solid blue line shows is a semiempirical fit $P/\rho \kB T=a/(\phi_\mathrm{cp}-\phi)$. Reproduced from \cite{thorneywork2017prl}}
\label{figAlice}
\end{figure*}

A definitive experimental elucidation of the nature of the hard-disk phase diagram was later provided by experiments on  a tilted monolayer of colloidal hard spheres in sedimentation-diffusion equilibrium (Fig.~\ref{figAlice})~\cite{thorneywork2017prl}. By measuring the density profile at different tilts of the plane, the full equation of state was extracted via the barometric law, unveiling a discontinuity at $\phi=0.68$. By calculating the height-resolved BOO correlation function in time and a modified Lindemann parameter, this discontinuity was identified as a first-order fluid-hexatic phase transition, and a study of the thermal fluctuations of the interface identified the width of the fluid-hexatic coexistence region as $\Delta\phi\simeq 0.02$. In addition, no finite discontinuities were found between the hexatic-solid transition. In short, a reasonably good agreement with the simulation results was observed (with essentially the same level of accuracy as that noted above in Sec.~\ref{sectionEquationOfState}).  Later direct comparisons of correlations in translational and bond-orientational orders in the various phases similarly demonstrated excellent agreement between this experimental model system and simulations~\cite{thorneywork2018jpcm}.

Interestingly, whether the hexatic phase is present or not depends strongly on the details of the system. For instance, the fluid-hexatic phase transition becomes metastable with respect to a first-order fluid-solid transition for binary mixtures of large (L) and small (S) disks with diameter ratio $q=\sigma_L/\sigma_S=1.4$, for molar fractions of small disks as small as $1\%$~\cite{russo2017,russo2018}. The two-stage melting scenario of a continuous solid-hexatic and a first-order fluid-hexatic transition also becomes metastable with respect to a first-order fluid-solid transition for hard disks with less than $1\%$ of the particles pinned to a triangular lattice~\cite{qi2015softmatter}. These findings are corroborated with a renormalization group analysis based on the KTHNY theory, which shows that the renormalized Young's modulus of the crystal increases in the presence of pinned particles~\cite{qi2015softmatter}. The melting transition was shown to be qualitatively changed to the KTHNY scenario for polydisperse disks~\cite{sampedro2019}, for which a significantly enlarged stability range for the hexatic phase and re-entrant melting have been observed.

\header{Equation of state}
The EoS of hard disks has been explored numerically since the very introduction of the Metropolis algorithm in 1953 ~\cite{metropolis1953}, and has been studied extensively ever since (see e.g. ~\cite{li2022}). Unlike 3d hard spheres, however, the Percus-Yevick closure to the Ornstein-Zernike equation cannot be solved analytically in 2d, thus necessitating the use of numerical integration. As a result, a large number of different approximations for the equation of state 
has been proposed (see e.g. ~\cite{helfand1961, henderson1975,tejero1993, santos1995, mulero2009, boublik2011}). Scaled particle theory offers a particularly simple expression~\cite{helfand1961},
\begin{equation}
\frac{\beta P}{\rho} = \frac{1}{(1-\phi)^2}, 
\label{eqSPT2D}
\end{equation}
which agrees reasonably well with experimental results (see Fig.~\ref{figAlice}(b)). When more accurate results are required, a common approach is to use the polynomial fits of simulation data provided by \citet{kolafa2006}. 

The experimental equation of state has been measured  by~\citet{brunner2003} in the fluid regime and by ~\citet{thorneywork2017prl} for the whole $\phi$ regime (see Fig.~\ref{figAlice}).  Comparison using
test-particle insertion agrees remarkably well with simulations~\cite{stones2018}. A very good agreement between theory and simulations has also been reported for $g(r)$ at various $\phi$~\cite{marcus1997,santanasolano2001,brunner2003,thorneywork2014}.

\header{Dynamics of the hard-disk fluid}
The dynamics of 2d confined hard-sphere colloids has been explored experimentally, with early work investigating the decay of the short-time diffusion to its long-time limit~\cite{marcus1996}. This study was followed  by explorations of hydrodynamic effects on the short--time dynamics and related these to the 3d system~\cite{carbajaltinoco1997,santanasolano2001,bonillacapilla2004}.

Another way to form a 2d colloidal system is for particles to adhere to an interface between two liquids. The dynamics in such a system provides a good estimate of the drag coefficient of the particles~\cite{peng2009}. Other work has emphasised the importance of hydrodynamic interactions at short times~\cite{thorneywork2015}. Surprisingly, while higher $\phi$ lead to a lower short-time diffusion coefficient $D_S$, as expected, the behavior of the long-time diffusion coefficient $D_L$ agrees well with simulation results of a pure hard-disk system. Hence, although hydrodynamics slows down the individual motion of particles at short times, its effect is largely compensated at longer time scales. Recently, the degree of confinement to a plane has also been investigated using simulations incorporating hydrodynamic interactions. The long-time diffusion was then found to be enhanced in weaker confinement~\cite{tian2022}.

\section{Interfaces in hard spheres}
\label{sectionInterfaces}

Hard-sphere systems spontaneously form interfaces as a result of fluid-crystal phase separation as well as grain boundaries in the crystal phase. Interfacial effects can also arise by bringing hard spheres in contact with a boundary. This section will delve into these various phenomena, starting with the conceptually important case of a single hard wall.

\subsection{Fluid in contact with a wall}
\label{sectionSinglePlanarWalls}

\header{Planar wall}
A bulk fluid of hard spheres next to a single hard wall can be considered as  the simplest inhomogeneous system. Intuitively, the presence of a flat wall induces some local structuring of the fluid, which in turn promotes the formation of the crystal phase. Early simulations by \citet{courtemanche1992,courtemanche1993} indeed demonstrated that crystalline layers  could  spontaneously form near a planar hard wall, even at pressures below bulk freezing. However, it is important to note that in simulations, because of periodic boundary conditions, a single wall is essentially a confined system with two walls at a plate separation determined by the system size. Therefore, complete wetting of a hard-sphere crystal at a single wall, also termed pre-freezing, is challenging to distinguish from capillary freezing~\cite{kegel2001,groot1987,dijkstra2004}. 
Careful simulation studies have now demonstrated that (\emph{i}) a (111) hard-sphere crystal completely wets a hard wall-fluid interface at about  $98.3\%$ of the bulk freezing  density, via interfacial free-energy calculations~\cite{davidchack2007}, and (\emph{ii}) a crystalline film at a single wall, independent of the plate separation, grows logarithmically, and is clearly distinguishable from capillary freezing~\cite{dijkstra2004}.

\begin{figure}
\includegraphics[width=0.8 \linewidth]{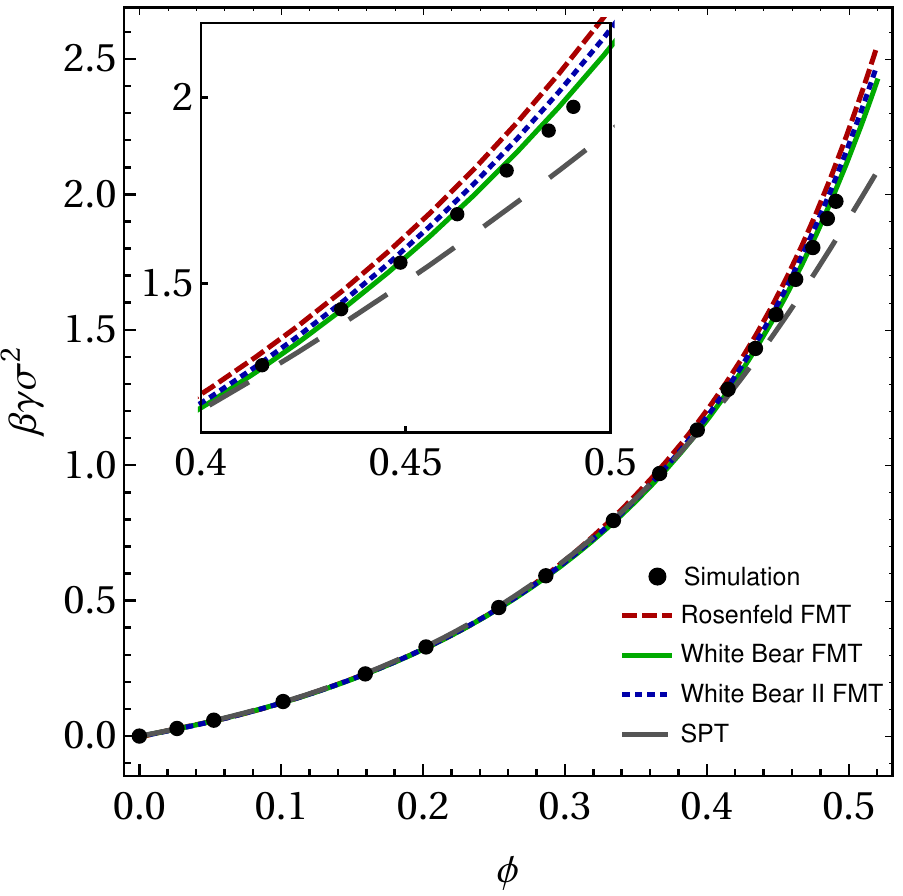}
\caption{Interfacial free energy of a hard-sphere fluid in contact with a flat hard wall, as a function $\phi$. The lines are theoretical predictions based on fundamental measure theory (FMT) and scaled particle theory (SPT). The points are simulation data. Simulation and FMT data taken from \citet{davidchack2016}; the SPT expression is based on \citet{heni1999}.}
\label{figWallGamma}
\end{figure}

The interface between a hard-sphere fluid and a hard wall also gives rise to an
interfacial free-energy $\gamma$, which depends on the fluid volume fraction $\phi$. Although a virial-like expansion for this quantity is possible, too few coefficients are known for it to be of much practical use~\cite{charbonneau2010}. \citet{davidchack2016}  numerically studied a hard-sphere fluid close to  a planar wall, and computed the density profiles, $\gamma$ and the excess adsorption  by means of computer simulations and density functional theory (DFT) (see Sec.~\ref{sectionCDFT}). They detected systematic deviations  between numerical and DFT results for $\phi > 0.3$ (see Fig.~\ref{figWallGamma}). 
Similar deviations were previously reported in a study of confined hard-sphere fluids~\cite{deb2011}, which suggested that nontrivial correlations between  fluid particles beyond nearest-neighbor shell might develop at higher $\phi$. These correlations could not, however, be accurately captured by DFT.

In experiments~\cite{hoogenboom2003jcp}, the presence of a bottom wall was found to initially induce layering in a sedimenting colloidal fluid; as sediment thickness increased, crystallization occurred within these fluid layers. Crystallization in the first layer appeared to proceed via a first-order transition consistent with predictions from BD simulation and theory~\cite{biben1994}. Using grand-canonical Monte Carlo simulations including a gravitational field, it was later shown that the fluid crystallized via a first-order freezing transition at which several fluid layers close to the bottom of the sample froze at the same chemical potential \cite{marechal2007}. The number of such layers simultaneously freezing decreases for higher gravitational field strengths.

The presence of a wall can also affect colloid dynamics. 
Evanescent dynamic light scattering experiments have studied sterically stabilized PMMA particles for various volume fractions over a range of scattering wave vectors. In the dilute regime, \citet{michailidou2009} observed that the near wall short-time diffusion was  slowed down due to particle-wall hydrodynamic interactions. However, for a concentrated suspension, the wall effect progressively diminished at all vectors $k$ and  many-body hydrodynamic interactions became less relevant~\cite{liu2015smnearwall,michailidou2009}.

\header{Patterned wall}
The wetting behavior of the hard-sphere crystal has also been studied for patterned walls~\cite{heni2000}. In the case of a (111) structured surface,  complete wetting of the hard-sphere crystal sets in already at $29\%$ of the bulk freezing density. Even crystal structures that are unstable in  bulk can be promoted by the wall surface pattern. As an example, hcp (rather than fcc) has been epitaxially grown on a structured template~\cite{hoogenboom2003prl}. By contrast, surface wetting by the crystal below freezing can  be completely suppressed by wall patterns that are incommensurate with the equilibrium crystal structure~\cite{espinosa2019}.

\subsection{Fluid-solid interface}
\label{sectionXtalFluidInterfaces}
At coexistence, the fluid phase is separated from the ordered solid by a (thermally fluctuating) fluid-solid interface. Understanding the properties of this interface on a microscopic scale is fundamental to assess crystal nucleation (see Sec.~\ref{sectionNucleation}). For a planar fluid-crystal interface at coexistence, one can define the interfacial (or surface) free energy $\gamma$ as the (reversible) work needed to form a unit area of a flat interface. The quantity $\gamma$  generally depends on the orientation of the interface normal ($\widehat n$) relative to the crystalline axes. When considering the fluctuations of that same interface, the quantity of interest is instead the interfacial stiffness $\tilde{\gamma}$. We here discuss efforts to quantify both $\gamma$ and $\tilde{\gamma}$.

\header{Interfacial free energy} 
In a pioneering numerical simulation, \citet{davidchack1998} characterized the face-centered cubic (100) and (111) fluid-solid
interfaces, thus demonstrating that  the transition from crystal to fluid occurred over few crystal planes made of domains of  coexisting crystal and fluid phases. They later extended the method developed by Broughton and Gilmer~\cite{Broughton1986}, based on a thermodynamic integration along a reversible path defined by cleaving, to determine $\gamma$ for hard-sphere fluids, whose value depended on the crystalline lattice structure~\cite{davidchack2000}.  Their results were then sufficiently precise to assess the anisotropy of the fluid-solid interfacial free energy.
This work was followed by more simulation-based efforts to determine $\gamma$, using a wide variety of approaches including the analysis of interfacial fluctuations using capillary wave theory \cite{mu2005,davidchack2006, hartel2012}, non-equilibrium work methods \cite{davidchack2010}, tethered Monte Carlo \cite{fernandez2012}, classical nucleation theory \cite{cacciuto2003}, mold integration \cite{espinosa2014} and thermodynamic integration \cite{bultmann2020,benjamin2015} (see also Sec.~\ref{sectionSimulation}). \citet{hartel2012} additionally used fundamental measure theory (FMT, see Sec.~\ref{sectionTheory}), although this approach resulted in significantly higher values than in simulations. An overview of results for $\gamma$ is provided in Table \ref{tableGamma}, listing results for both individual interfaces and an orientationally averaged $\bar{\gamma}$, which is a key quantity in crystal nucleation (see Sec.~\ref{sectionNucleation}). Clearly, significant variations exist between different methods, especially when compared to the relatively small error bars reported.  \citet{Schmitz2014,schmitz2015} suggested that systematic errors  (related to finite-size effects) could   be a possible explanation for this disagreement. \citet{benjamin2015}  proposed a theoretical tool to obtain reliable estimates for $\gamma$ in the thermodynamic limit.

\begin{table*}
\caption{\label{tableGamma}
Interfacial free-energy $\gamma$ and  stiffness $\tilde{\gamma}$, where ($hjk$) are the Miller indices of the interfacial crystal plane. For the stiffness  we might also indicate the symmetry of the short (in plane) direction. $\bar{\gamma}$ is averaged over the (100), (110) and (111) (of an fcc crystal). Various approached have been employed for the determination: cleaving [CL], capillary wave [CW], non-equilibrium work [NEW], mold integration [MI], tethered Monte Carlo [TET], CNT analysis of the nucleation rate [CNT], thermodynamic integration [TI], ensemble switch [SW], fundamental measure DFT \textcolor{darkgreen}{[FMT] (theory)},  and \expcolor{experiments $[]_\mathrm{exp}$}.}
\begin{ruledtabular}
\begin{tabular}{ccc}
\textbf{Interfacial free energy} ($\gamma$) & \textbf{Interfacial stiffness} ($\tilde{\gamma}$)  &\\
\hline
  $\gamma$(111)$[\kB T/\sigma^2]$                &$\tilde{\gamma}$(111)$[\kB T/\sigma^2]$   &     \\
  0.58  $\pm$ 0.01 ~\cite{davidchack2000}  [CL]   &     0.80 [$\bar{1}10$] ~\cite{mu2005} [CW]   &    \\
   0.61 $\pm$ 0.02 ~\cite{mu2005}  [CW]      &    0.78  $\pm$ 0.04   [$\bar{1}10$]  ~\cite{davidchack2010} [NEW]    &     \\
   0.546 $\pm$ 0.016 ~\cite{davidchack2006}  [CW] &  \textcolor{darkgreen}{ 1.025 $\pm$ 0.079 }   ~\cite{hartel2012} \textcolor{darkgreen}{ [FMT] }   &     \\
      0.5416 $\pm$ 0.0031 ~\cite{davidchack2010}  [NEW]         & 0.810 $\pm$ 0.005 ~\cite{hartel2012} [CW]   &     \\
 \textcolor{darkgreen}{  0.636 $\pm$ 0.001} ~\cite{hartel2012} \textcolor{darkgreen}{ [FMT] }        &    \expcolor{ 0.41}  ~\cite{vanloenen2019} \expcolor{[CW]$_\mathrm{exp}$} &     \\  
  0.600$\pm$0.011 ~\cite{hartel2012}  [CW]         &    &     \\ 
  0.556$\pm$ 0.003 ~\cite{benjamin2015}  [TI]         &    &     \\ 
 0.544$\pm$ 0.008 ~\cite{schmitz2015}  [SW]         &    &     \\  
    0.554$\pm$0.006 ~\cite{sanchez2021}  [MI]         &    &     \\  
  \hline
     $\gamma$(100)$[\kB T/\sigma^2]$               &  $\tilde{\gamma}$(100)$[\kB T/\sigma^2]$&    \\
  0.62 $\pm$ 0.01  ~\cite{davidchack2000} [CL] &  0.55 [$001$]   \cite{mu2005} [CW] &   \\
  0.64 $\pm$ 0.02  ~\cite{mu2005} [CW]          &  0.44 $\pm$ 0.03   [$001$] \cite{davidchack2006} [CW]   &    \\
  0.574$\pm$ 0.017 ~\cite{davidchack2006}  [CW]   &  0.40 $\pm$ 0.02  [$001$]   \cite{davidchack2010} [NEW]    &     \\
0.582 $\pm$ 0.002  ~\cite{davidchack2010} [NEW]    &   0.49 $\pm$ 0.02   [$001$] \cite{zykova2010} [CW]  &    \\
      0.636 $\pm$ 0.011  ~\cite{fernandez2012} [TET]     & \textcolor{darkgreen}{ 0.53$\pm$  14  [$001$] }\cite{hartel2012} \textcolor{darkgreen}{ [FMT] }&    \\
\textcolor{darkgreen}{ 0.687 $\pm$ 0.001}  ~\cite{hartel2012} \textcolor{darkgreen}{ [FMT] }   & 0.419$\pm$  0.005  [$001$] \cite{hartel2012} [CW]  &    \\     
0.639 $\pm$ 0.0011  ~\cite{hartel2012} [CW]    &   &    \\     
 0.586 $\pm$ 0.008  ~\cite{espinosa2014} [MI]          &   &    \\
  0.596$\pm$ 0.006  ~\cite{benjamin2015} [TI]          & \expcolor{1.2} \cite{hernandez2009} \expcolor{[CW]$_\mathrm{exp}$} &     \\             
  0.581$\pm$ 0.003  ~\cite{schmitz2015} [SW]          &  \expcolor{1.3 $\pm$ 0.3 }\cite{ramsteiner2010} \expcolor{[CW]$_\mathrm{exp}$} &    \\ 
     0.589 $\pm$ 0.001  ~\cite{bultmann2020} [TI]          &  \expcolor{ 0.47}  ~\cite{vanloenen2019} \expcolor{[CW]$_\mathrm{exp}$}  &    \\
 0.586 $\pm$ 0.006  ~\cite{sanchez2021} [MI]          &   &    \\     
\hline
     $\gamma$(110)$[\kB T/\sigma^2]$  &     $\tilde{\gamma}$(110)$[\kB T/\sigma^2]$ &   \\
 0.64  $\pm$ 0.01 ~\cite{davidchack2000} [CL] &    0.71 [$001$]   \cite{mu2005} [CW]      &     \\
   0.62 $\pm$ 0.02  ~\cite{mu2005}  [CW]&         0.49 [$\bar{1}10$]   \cite{mu2005} [CW]           &    \\
   0.557$\pm$ 0.017 ~\cite{davidchack2006}  [CW]         &  0.42   $\pm$ 0.03  [$\bar{1}10$]   \cite{davidchack2006}  [CW]    &     \\      
     0.559 $\pm$ 0.002  ~\cite{davidchack2010}  [NEW]&   0.70  $\pm$ 0.03  [$001$]   \cite{davidchack2006}  [CW]               &    \\
\textcolor{darkgreen}{ 0.616 $\pm$ 0.011 } ~\cite{hartel2012} \textcolor{darkgreen}{ [FMT] }   &   0.34  $\pm$ 0.05  [$\bar{1}10$]   \cite{davidchack2010}  [NEW]    &    \\     
 0.600$\pm$0.011 ~\cite{hartel2012}  [CW]         &  0.74  $\pm$ 0.02  \ [$001$]   \cite{davidchack2010} [NEW]   &     \\  
  0.577 $\pm$ 0.004  ~\cite{benjamin2015}  [TI]&  \textcolor{darkgreen}{ 0.283 $\pm$ 0.035  [$\bar{1}00$]}~\cite{hartel2012}\textcolor{darkgreen}{  [FMT] }        &    \\   
  0.559 $\pm$ 0.001 ~\cite{schmitz2015}  [SW]&  0.401   $\pm$ 0.005  [$\bar{1}00$]~\cite{hartel2012} [CW]         &    \\    
0.572 $\pm$ 0.007  ~\cite{sanchez2021}  [MI] &   \textcolor{darkgreen}{  0.86  $\pm$ 0.14  [01$\bar{1}$]}~\cite{hartel2012}\textcolor{darkgreen}{ [FMT]  }      &    \\ 
  & 0.769 $\pm$ 0.005  [01$\bar{1}$]~\cite{hartel2012} [CW] &    \\ 
& \expcolor{ 1.0 $\pm$ 0.2} \cite{ramsteiner2010} \expcolor{ [CW]$_\mathrm{exp}$} &    \\ 
  & \expcolor{0.53}  ~\cite{vanloenen2019} \expcolor{ [CW]$_\mathrm{exp}$} &    \\ 
   \hline
     $\bar{\gamma}$ $[\kB T/\sigma^2]$  &   & \\
 0.617  $\pm$ 0.006 ~\cite{davidchack2000} [CL] & &     \\
 \expcolor{ 0.27}  ~\cite{gasser2001} \expcolor{[CNT]$_\mathrm{exp}$} & &     \\
 0.616  $\pm$ 0.003 ~\cite{cacciuto2003} [CNT] & &     \\
   0.62 $\pm$ 0.02  ~\cite{mu2005}  [CW] &  &    \\
    0.56  $\pm$ 0.01 ~\cite{davidchack2010} [NEW] & &     \\
\textcolor{darkgreen}{ 0.66  $\pm$ 0.01 }~\cite{hartel2012}\textcolor{darkgreen}{ [FMT]} & &     \\
0.5916 \cite{sanchez2021} [CNT] & &     \\
\end{tabular}
\end{ruledtabular}
\end{table*}

\begin{figure}[h!]
\includegraphics[width=1.0 \linewidth]{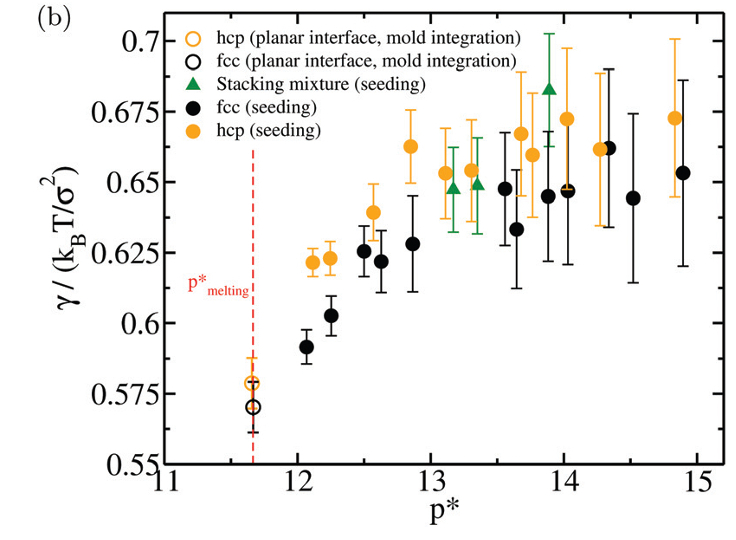}
\caption{Interfacial free energy $\gamma$ for different  polytypes at coexistence. For planar interfaces (at the bulk melting pressure), an average $\bar\gamma$  was obtained by averaging over different crystal orientations for both polymorphs. The values at higher pressures correspond to $\bar\gamma$ estimated for spherical nuclei (from the seeding technique), using CNT~\cite{sanchez2021}.}
\label{figGamma}
\end{figure}

To establish whether the fcc or hcp crystal is favored due to a difference in interfacial free energy,  \citet{sanchez2021} computed $\gamma$ by means of the mold integration method considering both ordered phases. The authors concluded that  the $(11\bar{2}0)$ plane of the hcp crystal had a slightly higher $\gamma$ than any of the faces of the fcc polytype. They also found a higher orientationally averaged $\bar{\gamma}$ for the hcp phase when considering spherical nuclei using a seeding approach (see Fig.~\ref{figGamma}).

Only more recently have experiments based on colloidal suspensions studied the phenomenon via confocal microscopy~\cite{gasser2001,dullens2006prl,hernandez2009}.
Experimentally, $\gamma$ can be   obtained  from indirect measurements of crystal nucleation rates using classical nucleation theory (see Sec.~\ref{sectionNucleation}). 
Using this method on PMMA colloids, \citet{gasser2001} obtained a surprisingly low  $\bar{\gamma}$ (see Table \ref{tableGamma}),
possibly due to the colloids  being softened  by electrostatic interactions (see Sec.~\ref{sectionInteractions}).

\header{Interfacial stiffness} 
If the fluid-crystal interface is  rough, capillary wave theory (CWT) can be applied to define the interfacial stiffness $\tilde{\gamma}$, that  controls the resulting capillary waves. The tensorial expression for $\tilde{\gamma}$ is then
\begin{equation}
\tilde{\gamma}_{\alpha \beta}(\widehat n) = \gamma(\widehat n) + \frac{\partial^2 \gamma(\widehat n) }{\partial \widehat n_{\alpha} \partial \widehat n_{\alpha}},
\label{eqInterfacialStiffness}
\end{equation}
with $\widehat n_{\alpha}$ and $\widehat n_{\beta}$ two directions orthogonal to $\widehat n$~\cite{fisher1983}. 
While the first contribution ($\gamma(\widehat n)$)  describes the free-energy cost of increasing the interfacial area, the second contribution accounts for the free energy required to locally change the crystal orientation. Note that the applicability of CWT requires  the interface to be rough. This condition is only valid above the (system-dependent) roughening transition temperature for thermal systems \cite{zykova2010,schmid1992}, but is always so for hard spheres.

If the anisotropy of the interfacial free energy is known, $\tilde{\gamma}$ can be calculated directly from Eq.~\eqref{eqInterfacialStiffness}, as in \citet{hartel2012}. Alternatively, it can be obtained from interfacial fluctuations measured in simulations using capillary wave theory~\citet{davidchack2006,mu2005,hartel2012,zykova2009,zykova2010}. Values from various different approaches are given in Table \ref{tableGamma}. Overall,  $\tilde{\gamma}$ shows significantly more anisotropy than $\gamma$, but there are also significant discrepancies between different measurements.

\citet{hernandez2009} used confocal microscopy to study the interfacial fluctuations of a solid-fluid interface and determined the interfacial stiffness using capillary wave theory. They obtained a value about twice as high as that obtained from computer simulations (see Table \ref{tableGamma}). This discrepancy might be caused by slightly charged particles, but that remains to be verified.

\begin{figure}[h!]
\includegraphics[width=0.8 \linewidth]{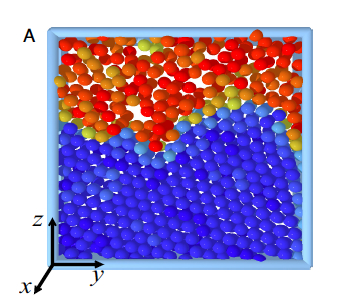}
\caption{Rendering of particle coordinates determined from confocal microscopy showing the fluid-crystal interface.
Shown is a slice that is two crystalline layers thick.
Particles are colored according to the number of ordered neighbors they have~\cite{hernandez2009}.}
\label{figCapillary}
\end{figure}

Similar values of the interfacial stiffness were obtained by \citet{ramsteiner2010}. After sedimenting hard-sphere silica colloids  onto (100)  and  (110) oriented templates, they located the interface  by confocal microscopy,  and again used capillary wave theory to determine $\tilde{\gamma}$. For all three main crystal orientations, they noted that the Fourier amplitudes are  independent of the in-plane direction of the associated wave vectors in the long wavelength limit. This result directly contradicts simulation results for the (110) interface, from which a difference consistent  with the stiffness tensor of a cubic crystal is expected. \citet{ramsteiner2010} suggested that the relatively thin fluid layer and the small gravitational length could explain this discrepancy.

Recently, \citet{vanloenen2019}  studied the interfaces between hard-sphere colloidal fluids and fcc crystals sedimented onto differently oriented templates.  The $\tilde{\gamma}$ they obtained from the capillary waves were not only much closer to the simulation results than earlier estimates by \citet{hernandez2009}, but their order also was more consistent with simulation predictions (see Table \ref{tableGamma}).

As these different results show, experimental measurements of the stiffness  of nominally the same system have yielded different results. A very recent study by ~\citet{macdowell2023} suggests that the surface stiffness of colloidal hard spheres, as measured by interface  fluctuations, might include a gravity-dependent contribution which could account for this discrepancy. The external field dependence of the interfacial stiffness could then be explained by an improved interface Hamiltonian that provides corrections to the capillarity theory equation, thus reconciling experimental and theoretical results.

\subsection{Grain boundaries}

\begin{figure}
    \centering
    \includegraphics[width=0.95\linewidth]{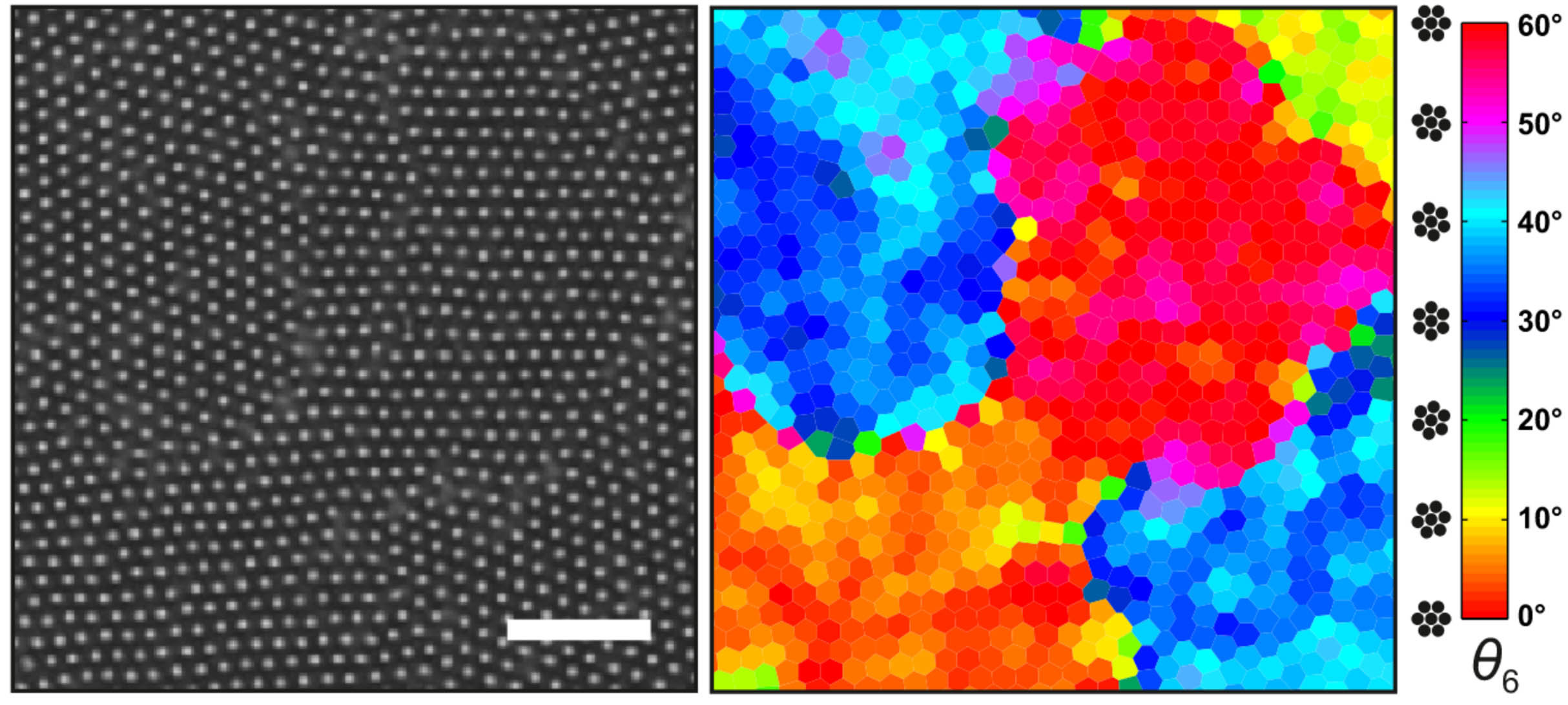}
    \caption{Grain boundaries in a polycrystalline 2d solid of colloidal hard spheres. The left panel shows the particles (scale bar is 20$\mu$m). The right panel shows the Voronoi cells of the particles, colored by the local orientation of the BOO parameter, as encoded in the color bar, which allows for easy distinction between the different domains. Figure adapted from \citet{lavergne2017}.}
    \label{figGrainBoundaries}
\end{figure}

Grain boundaries form spontaneously in crystalline materials at finite temperature or, in the case of hard spheres, for volume fractions below close packing, i.e., $\phi<\phi_\mathrm{cp}$. In a  polycrystalline material, crystalline grains with differing orientations are separated by  an  interface made of amorphous grain boundaries. Most of the work on grain boundaries has been performed in 2d ~\cite{zhang2009,gray2015,lavergne2015}. Grain boundaries and other defects  reduce  long-range order of a single crystal~\cite{zhang2009}, deforming the underlying hexagonal order (in 2d)  due to a preference for five- and seven-fold coordinated particles to be adjacent to impurities~\cite{gray2015}.

Considering a polycrystalline monolayer of colloidal hard spheres (see Fig.~\ref{figGrainBoundaries}), \citet{lavergne2017} followed the grain growth process and detected an anomalous slow growth of the BOO correlation length. The authors invoked the curvature-driven coarsening of the large-angle grain boundaries at a rate dependent on the grain boundary length to explain the effect. When dealing with a polycrystalline monolayer of hard spheres with embedded impurities, \citet{lavergne2016} demonstrated that the size of the impurities  determined whether they behaved as interstitial or substitutional impurities in the bulk crystal. Once formed, grain boundaries can also shrink. Spontaneous shrinkage of circular grain boundaries has been studied in 2d colloidal crystals by \citet{lavergne2019}, who demonstrated that the shrinkage can be driven by three mechanisms:  curvature-driven migration, coupled grain boundary migration, and grain boundary sliding.

Grain boundary dynamics are driven by their local curvature and have been thoroughly studied by means of point sampled surface analysis techniques~\cite{lavergne2015}. While the structure on length scales larger than the grain boundaries distances  strongly depends on the defects concentration, local structural distortions close to a grain boundary occur only over short distances compared to the grain boundary size, and are independent of the defect concentration~\cite{gray2015}. The kinetics of  grain boundaries is closely related to the topological constraints imposed on their  dislocation structure. As an example, \citet{lavergne2018} showed that a local rotational deformation of a 2d colloidal crystal with an optical vortex might originate a grain boundary loop, thus underlining the relevance of  defects in the kinetics of grain boundaries.

In 3d, studies of hard-sphere grain boundaries are much more sparse. Recently, \citet{orr2021} introduced a new methodology to detect and characterize grain boundaries in experimental data on colloidal grain boundaries, revealing detailed misorientation distributions and grain boundary structures, but much remains to be done.

%% file: BinaryHardSphereMixtures/binaryHardSphereMixtures.tex
\section{Binary hard-sphere mixtures}
\label{sectionBinary}

\begin{figure*}
\includegraphics[width=185mm]{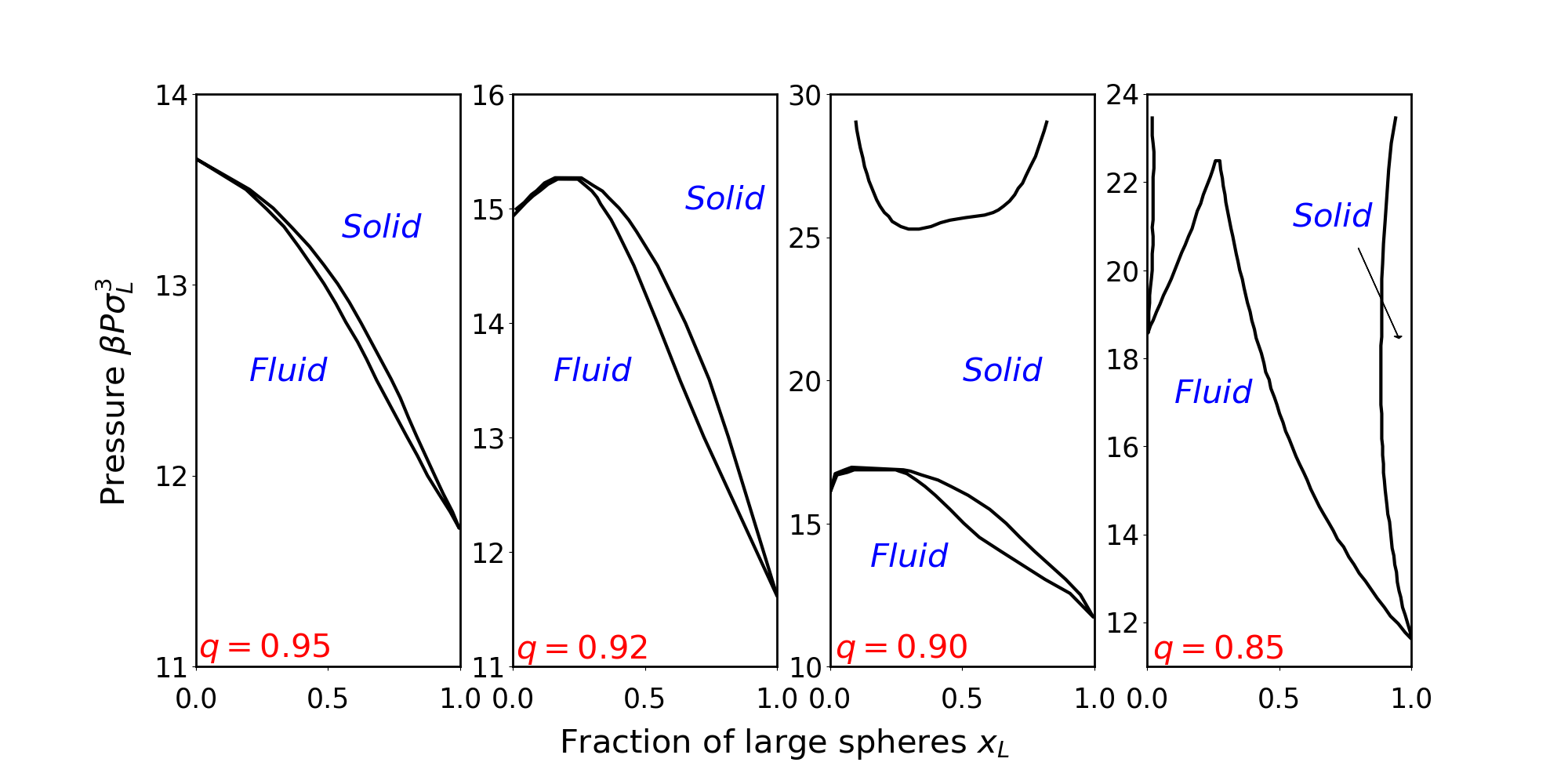}
\caption{Phase diagrams of binary hard-sphere mixtures  displaying a spindle-like phase behavior at size ratio $q=0.95$, an azeotropic phase diagram at $q=0.92$, an azeotropic phase behavior with a solid-solid demixing transition at $q=0.90$, and an eutectic phase diagram at $q=0.85$ as obtained from simulations. Data taken from ~\cite{kranendonk1991}.}
\label{figBinaryPhD}
\end{figure*}

The phase behavior of binary mixtures of large (L) and small (S) hard spheres with diameters $\sigma_\mathrm{L}$ and $\sigma_\mathrm{S}$, respectively, is remarkably rich. Even at close packing, complex binary crystals emerge as the size ratio $q=\sigma_\mathrm{S}/\sigma_\mathrm{L}$ changes~\cite{dijkstra2014, hopkins2012,hudson2011}. In this section, we review some of the salient thermodynamic features of these systems, including the diversity of  binary crystal structures they can form, the structural crossover in binary fluids, fluid-fluid demixing and the quest for its critical point as well as the behavior of sedimented monolayers of binary hard-sphere mixtures.

\subsection{Crystal regime} 
\label{sectionBinaryXtal}

In binary mixtures of hard spheres of two very similar sizes, the most stable state at infinite pressure is phase separated into two separate fcc crystals: one of large spheres and one of small spheres. Density functional theory, computer simulations, and scaled particle theory have revealed further enrichment of the  crystal regime at finite pressure, away from close packing. In the limit $q=\sigma_\mathrm{S}/\sigma_\mathrm{L}\rightarrow 1$, the system reduces to a one-component hard-sphere system for which the phase behavior is discussed in Sec.~\ref{sectionEquilibriumPhaseBehavior}. As the two components become more dissimilar in size, i.e., $q\lesssim 1$,  the freezing transition changes from a spindle-like via an azeotrope to an eutectic phase diagram ~\cite{barrat1986prl,barrat1987,zeng1990,denton1990,kranendonk1991,kofke1991,cottin1997}, as shown in Fig.~\ref{figBinaryPhD}.

In the case of spindle-like phase behavior, the coexistence between a fluid and a substitutionally disordered fcc crystal phase is narrow, in the sense that only a small composition difference between the two phases develops. When spheres become more dissimilar in size, the fluid-solid region broadens and an azeotropic point appears at around $q=
0.94$~\cite{kranendonk1991}. At higher packing fractions a coexistence region between two substitutionally disordered fcc solids appears when the spheres become sufficiently dissimilar, which shifts to lower densities as the size ratio is further increased. When this miscibility gap in the solid phase interferes with the fluid-crystal coexistence, the phase diagram becomes eutectic at $q\simeq0.875$ ~\cite{kranendonk1991}.  Note that only substitutionally disordered fcc phases appear for the size ratio range $q\in(0.85,1)$, in contrast to what happens at smaller $q$.

\begin{figure}
\centering
\includegraphics[width=85mm]{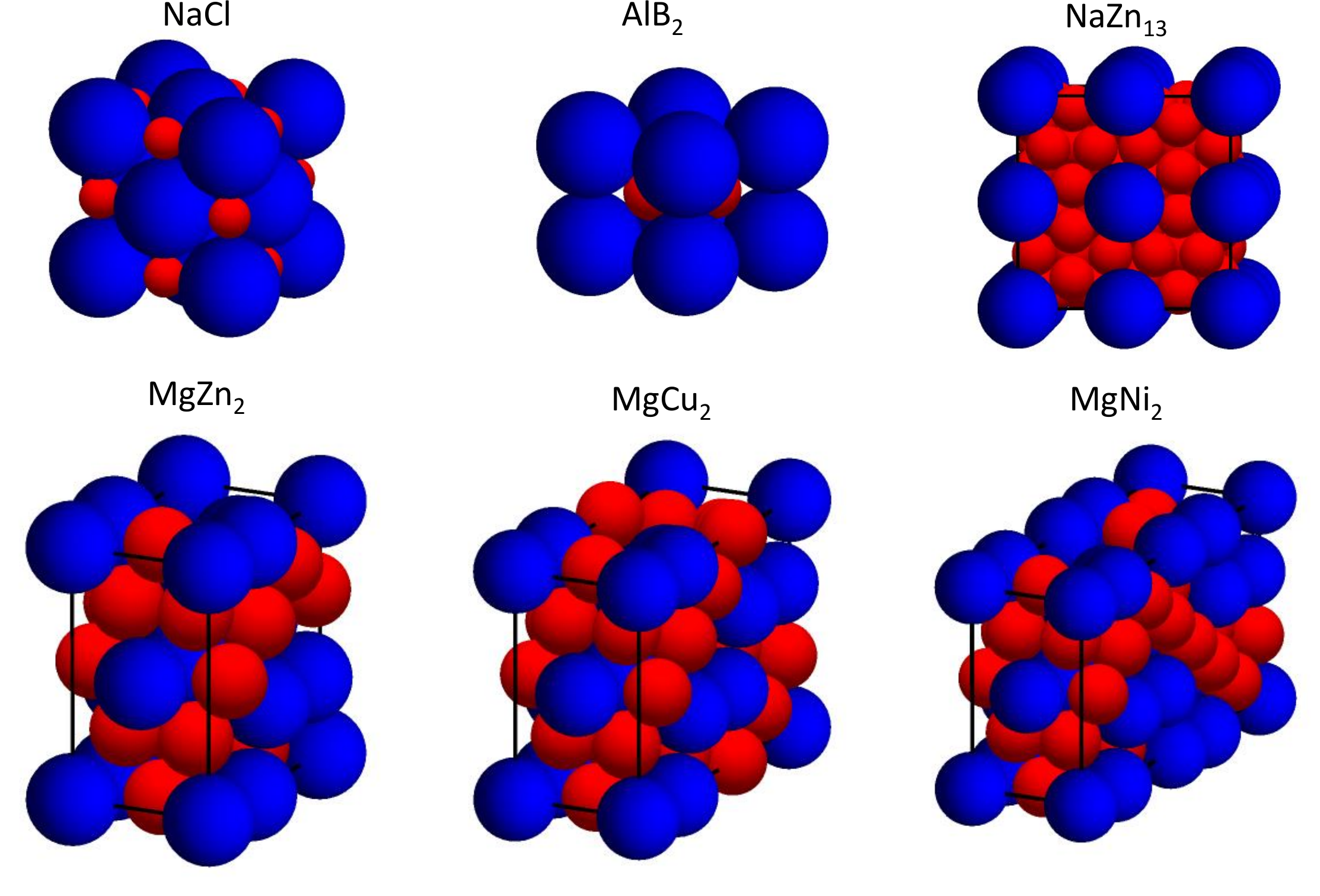}
\caption{Binary crystal phases of binary hard spheres. Schematic representations of binary crystal structures known to be (nearly) stable in binary hard-sphere mixtures.}
\label{figBinaryCrystals}
\end{figure}

The phase behavior of more asymmetric mixtures was studied in experiments using light scattering ~\cite{bartlett1990,bartlett1992, pusey1994}. For size ratios $q=0.58$ and $0.62$,  complex binary LS$_2$ and LS$_{13}$ superlattice structures were found for sterically stabilized PMMA spheres ~\cite{bartlett1990,bartlett1992}.  The stability of these superlattice structures, which are analogous to their atomic counterparts, AlB$_2$ and NaZn$_{13}$ as shown in Fig.~\ref{figBinaryCrystals}, was subsequently confirmed by computer simulations for size ratios $0.42<q<0.625$ ~\cite{eldridge1993nature,eldridge1993molphysab13,eldridge1993molphysab2,eldridge1995,dijkstra2014} and by density functional approaches~\cite{xu1992}.

For size ratios $0.2<q<0.42$, a superlattice structure isostructural  to the rocksalt NaCl was  predicted by computer simulations ~\cite{trizac1997}, and  verified experimentally ~\cite{hunt2000,vermolen2009} (see Fig.~\ref{figBinaryCrystals}).  Surprisingly, the experimental observations of NaCl always showed a crystal phase with many vacancies in the sublattice of the small spheres in such a way that the L:S stoichiometry of the crystal is not 1:1 but rather 1:$x$ with  $x\leq 1$~\cite{hunt2000,vermolen2009}. Free-energy calculations from Monte Carlo simulations subsequently showed that an interstitial solid solution constructed by filling the octahedral holes of an fcc crystal of large spheres with small spheres is indeed the stable phase ~\cite{filion2011prl}. Upon increasing pressure, the fraction of octahedral holes filled with a small sphere can be completely tuned from zero---corresponding to a fcc phase of pure large spheres---to one---corresponding to a NaCl phase. For less assymetric size ratios, non-equilibrium interstitial solid solutions are found, which are long--lived on the experimental timescale~\cite{riosdeanda2017}.

For larger size ratios, another set of LS$_2$ phases, known as Laves phases, were found in computer simulations ~\cite{hynninen2007,hynninen2009, bommineni2020}. Three Laves phase structures exist: hexagonal MgZn$_2$, cubic MgCu$_2$, and hexagonal MgNi$_2$. Each is characterized by the stacking of  large-sphere dimers in the crystal structure,  as shown in Fig.~\ref{figBinaryCrystals}. The MgCu$_2$ phase  is particularly interesting because it consists of a diamond lattice of large (Mg) spheres and a pyrochlore lattice of small (Cu) spheres. By selectively removing one of the species, one can readily obtain either the diamond or pyrochlore phase. Because both the diamond and pyrochlore phases have a photonic band gap ~\cite{hynninen2007}, which makes them potential targets for various applications in optics, Laves phases are strongly sought after. However, these phases are  notoriously difficult to self-assemble. At the high packing fractions at which they become thermodynamically stable, the fluid phase is dynamically sluggish ~\cite{dasgupta2020} and may show instabilities under sedimentation ~\cite{milinkovic2011}. Additionally, the Laves phases are predicted to contain a significant concentration of substitutional defects ~\cite{vandermeer2020}, which can disrupt crystal growth and hence hinder self-assembly ~\cite{dasgupta2018}.

"Laves phases have nevertheless been observed experimentally for binary nanoparticle
suspensions, which diffuse much faster than micron-scale  colloids ~\cite{shevchenko2006,evers2010,cabane2016} (see Sec.~\ref{sectionExperimental}), as well as sub-micron-sized spheres that interact with soft repulsive potentials ~\cite{nakagaki1983,hasaka1984,ma1994,gauthier2004}. The observation of Laves phases in softer spheres is consistent with recent simulation work showing that interparticle softness pushes the glass transition in the binary fluid phase to higher densities and hence enhances crystallization of Laves phases in nearly hard spheres ~\cite{dasgupta2020}. However, \citet{schaertl2018} have recently also demonstrated self-assembly of Laves phases in a slightly off-stoichiometric mixture of (nearly) hard microgel particles, using static light scattering. "

Simulation and theoretical studies using a variety of methodologies have explored other dense possible crystal structures of binary hard sphere mixtures ~\cite{filion2009, filion2009pre, kummerfeld2008, otoole2011, hudson2011, hopkins2012}. These surveys have revealed a rich variety of dense packings at different size ratios and compositions that are expected to be stable in the limit of infinite pressures ~\cite{hopkins2012}. Interestingly, many of them have yet to  be observed in simulations or experiments at finite pressure.

\subsection{Binary fluids: Structural crossover, demixing and dynamics}
\label{sectionBinaryFluid}

\begin{figure*}
\centering
\includegraphics[width=160mm]{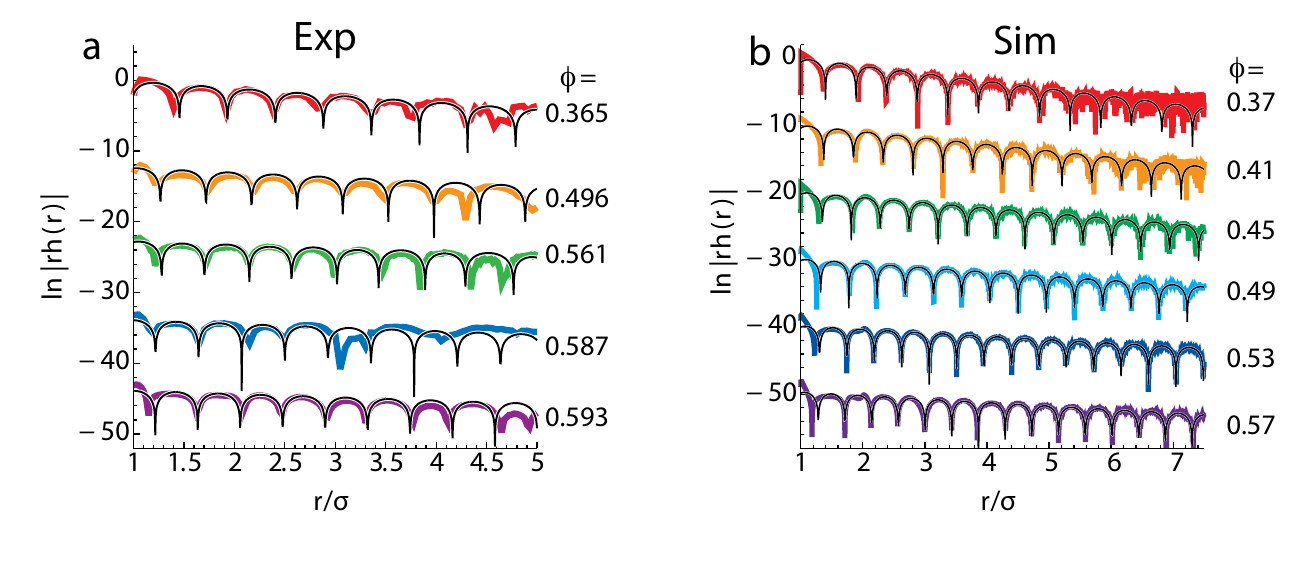}
\caption{Structural crossover in binary hard-sphere fluids.
Plots of  $\ln|r h_{ij}(r)|$ obtained by real-space experiment (top) and simulation (bottom) for L-L, L-S, and S-S total correlation functions. The packing fraction $\phi_\mathrm{S}$ (marked) is increasing from top (blue) to bottom (red) for each case. Curves are shifted vertically for clarity. The black lines in (a) and (b) are fits from which parameters controlling the crossover are obtained~\cite{grodon2005}. Simulations are for a binary hard-sphere mixture with a size ratio $q=0.648$.
Reproduced from ~\citet{statt2016}.}
\label{figAntonia} 
\end{figure*}

\header{Structure: crossover and demixing}
DFT and computer simulations have identified  a \emph{structural crossover} line in the phase diagram of  binary hard-sphere fluids. This line marks a rather abrupt change in the wavelength that dominates the asymptotic decay of the radial distribution function ~\cite{grodon2004,grodon2005}. Experimental evidence for this effect was also found in sedimented quasi-2d hard-sphere glasses ~\cite{baumgartl2007} and subsequently in 3d hard-sphere fluids where quantitative agreement with theory and simulation was found (Fig.~\ref{figAntonia}) ~\cite{statt2016}.

For smaller size ratios, it has long been debated whether a stable fluid-fluid demixing transition exists or not. Such a spinodal instability in the fluid mixture can be ascribed to the \emph{depletion} mechanism, which is known to drive phase separation in colloid-polymer suspensions~\cite{asakura1954,long1973,vrij1977,lekkerkerker1992}. The depletion effect, first described by Asakura and Oosawa in 1954 ~\cite{asakura1954,oosawa2021}, induces an attractive interaction between large  colloids due to an unbalanced osmotic pressure of the small (polymeric) spheres. Alternatively, this effect can be explained by the increase in free volume available to the small spheres upon clustering of the large spheres. The resulting entropy gain for the small spheres would then drive phase separation of the colloids.

The effective depletion interaction between the larger spheres, due to the smaller ones, has been calculated theoretically~\cite{attard1989spherically,attard1990hypernetted,dickman1997,mao1995,gotzelmann1998,gotzelmann1999,roth2000} and extracted from computer simulation ~\cite{dickman1997,biben1996,gotzelmann1999,ashton2011}. The results exhibit a short-range attractive well close to the surface of large hard spheres, followed by several oscillations around zero at larger distances.

\begin{figure}
\centering
\includegraphics[width=85mm]{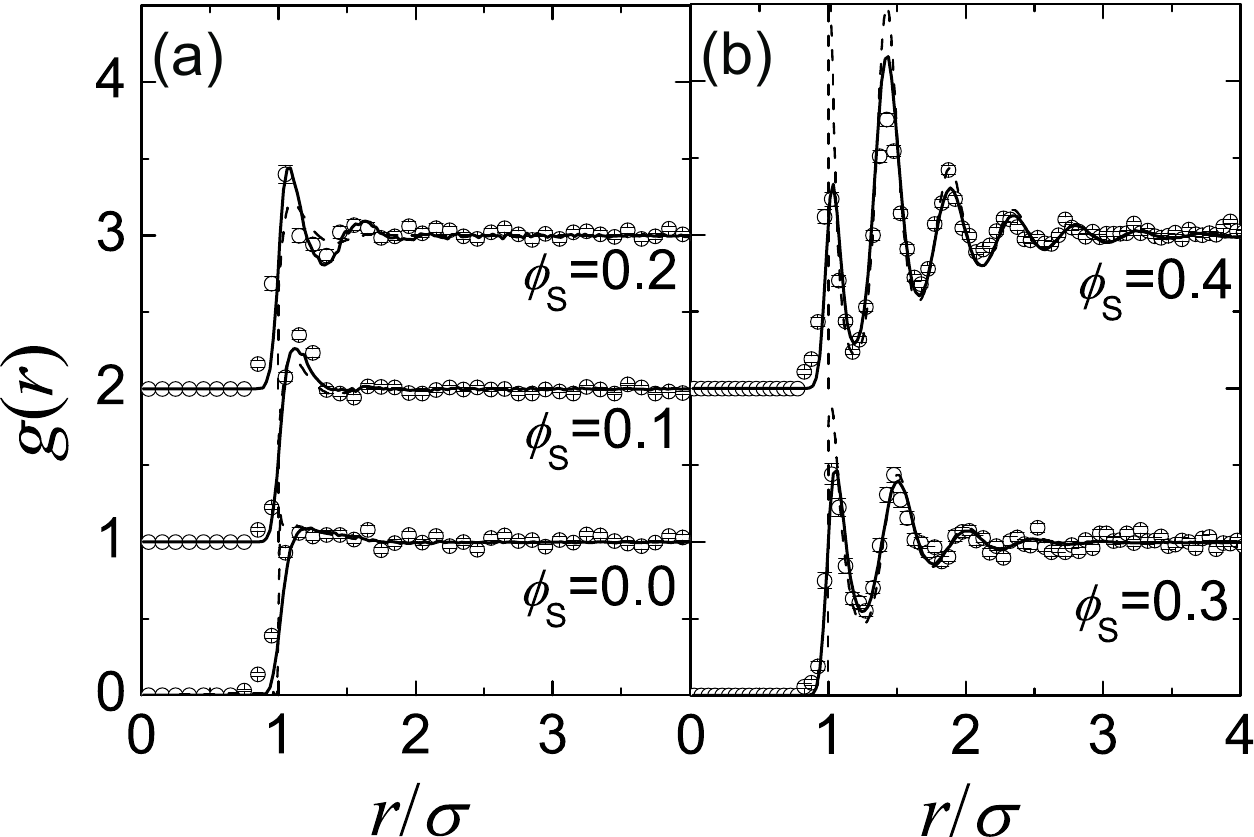}
\caption{Radial distribution function of large colloids in binary hard spheres at various concentrations of the smaller species for:  (a) low $\phi_\mathrm{S} \le 0.25$; (b) high $\phi_\mathrm{S} > 0.25$. Monte Carlo simulation at small colloid volume fraction $\phi_\mathrm{S}$ are denoted by solid lines, and  compared to the experimental data (circles). Dashed lines correspond to the relation $g(r)\approx \exp(-\beta u(r))$ where $u(r)$ is the effective interaction between the large colloids consisting of the depletion interaction induced by the smaller and a residual electrostatic contribution. 
In addition to experimental resolution and polydispersity, MC simulations account for residual colloid charge. Reproduced from ~\cite{royall2007jcp}.}
\label{figGBHS}
\end{figure}

Experimental investigations of the depletion interaction include tracking the Brownian trajectory of a large sphere  near a wall in a suspension of small spheres with video microscopy ~\cite{kaplan1994prldirect} and using optical tweezers to study the interaction between two large colloids~\cite{crocker1999}. This latter work showed a discrepancy with subsequent theory~\cite{roth2000}. Later work which determined radial distribution functions from which the interaction may be inferred showed better agreement with simulation and theory (Fig.~\ref{figGBHS})~\cite{royall2007jcp,roth2000}.

Other experimental investigations probing fluid-fluid demixing have suggested that it might be strongly coupled to freezing. For instance, Sanyal {\em et al.}, using mixtures of polystyrene spheres with $q=0.2$, observed segregation in regions rich in large spheres and rich in small spheres  in the sediment at the bottom of their samples~\cite{sanyal1992}. However, when they suspended the mixture in a density-matched solvent, neither sedimentation nor demixing was seen. ~\citet{vanduijneveldt1993} observed a phase instability in a fairly narrow concentration range of small and large sterically stabilized silica particles with $q=0.1667$, but sedimentation obscured whether the transition corresponded to fluid-fluid demixing or to freezing.  Experiments by Kaplan {\em et al.}~on mixtures of polystyrene particles with $0.069<q<0.294$ revealed the existence of a single homogeneous disordered phase, a coexistence between two disordered phases, and a coexistence between one or two disordered phases and a surface crystal on the sample wall~\cite{kaplan1994prlentropically}. However, bulk crystallization was not observed. A possible reason why no surface crystallization was found in the experiments of van Duijneveldt {\em et al.}~is that silica spheres settle quickly compared to polystyrene particles. In the experiments of Dinsmore {\em et al.}~on mixtures of polystyrene particles with $0.083<q<0.149$ ~\cite{dinsmore1995}, a phase separation into a disordered fluid phase consisting primarily of small spheres and a crystalline solid of large spheres permeated by a disordered fluid of small spheres was observed at sufficiently high volume fractions. They additionally showed that the crystallites on the surface of the sample cell had the same structure as the bulk crystals, which they attributed to wetting of the bulk phase. Finally, Imhof and Dhont observed a fluid-solid type of phase separation but no fluid-fluid demixing in experiments on a binary mixture of silica spheres with $q=0.1075$ ~\cite{imhof1995prl,imhof1995pre}.

Conflicting results have also been reported from various theoretical approaches. Integral equation theory approaches predict either that the homogeneous fluid phase of a binary mixture of large and small hard spheres is stable with respect to demixing ~\cite{lebowitz1964,mansoori1971}, that a spinodal instability occurs in the fluid phase~\cite{biben1990,biben1991prl,biben1991jpcm,rosenfeld1994}, or that the fluid-fluid demixing transition is metastable with respect to a broad fluid-solid transition by using a simple freezing criterion ~\cite{caccamo1997}. Free volume theory approaches have further predicted that the fluid-fluid phase separation is metastable with respect to freezing ~\cite{poon1994,dinsmore1997}. Furthermore, it has been shown  that large spheres could crystallize at the wall well below those corresponding to bulk phase separation due to  the presence of small spheres, in agreement with experimental observations. A broad fluid-solid coexistence was also found in DFT~\cite{xu1995freezing}, while theoretical  approaches based on virial coefficients predicted either a fluid-fluid phase separation in the limit of highly asymmetric sizes~\cite{saija1996,coussaert1997,coussaert1998erratum}, a demixing transition metastable with respect to either a broad or a narrow fluid-solid phase transition ~\cite{coussaert1998jcp}, a simple narrow fluid-solid phase transition in the limit $q\rightarrow 0$ ~\cite{vega1998}, or a complete absence of a fluid-fluid demixing transition ~\cite{deharo2013}. What this broad range of predictions makes clear is that the phase behavior is too sensitive to the details of the specific approximations used in the integral equation theories and virial expansion approaches to reach a clear conclusion.

Dijkstra {\em et al.}~followed a different approach by mapping the binary hard-sphere mixture onto an effective one-component system by formally integrating out the degrees of freedom of the small spheres in the partition function ~\cite{dijkstra1998prl,dijkstra1999jpcm}. By using the two-body (depletion potential) contribution to the effective Hamiltonian in simulations, this effort revealed that the fluid-solid phase coexistence region significantly broadens as $q$ becomes very small. In addition, it found a stable isostructural solid-solid demixing transition for $q<0.05$ and fluid-fluid demixing that remains metastable with respect to the fluid-solid transition for $q<0.10$. These predictions were later validated by simulations of the true binary hard-sphere mixture~\cite{dijkstra1999prl,dijkstra1999pre}. However, only recently has the \emph{critical point} for the fluid-fluid demixing transition been reported ~\cite{kobayashi2021}. Using a two-level simulation approach based on a coarse-grained description with effective two- and three-body interactions and the full (fine-grained) binary mixture, ~\citet{kobayashi2021} managed that feat by matching the probability distribution for the number of large particles to the 3d Ising universality class scaling form. However, a non-trivial dependence on $q$ was also found.

In summary, it is now well-settled that a fluid-fluid demixing transition exists in a binary mixture of hard spheres for sufficiently large size asymmetries, but this transition is metastable with respect to a broad fluid-solid transition.

\header{Dynamics in binary hard spheres}
Dynamics in binary fluids, as probed with multiple light scattering, has  shown reasonable agreement with hydrodynamic theory~\cite{kaplan1992}. Later, ~\citet{imhof1995pre} measured the diffusion in strongly asymmetric mixtures by photobleaching part of the sample and examining the recovery of the fluoresence as the system rearranged. They found good agreement with previous theory by ~\citet{batchelor1983} for the diffusion of the small spheres in the dilute regime, but observed significant deviations for the large spheres, which they attributed to the large size disparity in their system. Significant experimental and simulation  work has also focused on the dense fluid regime ~\cite{foffi2003, marin2020, williams2001motions, gotze2003}, thus revealing a sensitive dependence of the diffusivity behavior of both species on the size ratio and composition of the chosen mixtures. This is of particular importance when studying glassy systems (see Sec.~\ref{sectionGlass}), in which, depending on $q$, one might find a single glass (where only the large particles are dynamically arrested) or a double glass (where both species are arrested) ~\cite{voigtmann2011, lazaro2019, laurati2018}.

\subsection{Sedimented monolayers of binary hard spheres}

Analogous to the monodisperse case, binary mixtures of hard spheres can be confined to a quasi-2d setup by, e.g., allowing them to sediment into a monolayer on a substrate. In this case, the system can be mapped onto a binary hard-disk model, which typically has to be non-additive in order to account for  the centers of the small spheres being positioned below those of the large spheres ~\cite{thorneywork2014}. Thorneywork {\em et al.}~performed a detailed characterization of the radial distribution function ~\cite{thorneywork2014}, structure factor ~\cite{thorneywork2018molphys}, and self-diffusion coefficient ~\cite{thorneywork2017pre} of these systems, and found their experimental results to be in good agreement with simulations and theory of model mixtures. An intriguing consequence of the depletion interaction in sedimented binary hard spheres is that the large particles are in fact \emph{repelled} from the edge of a raised surface~\cite{dinsmore1996}.

As in the bulk 3d case, the introduction of a second size of particles vastly increases the phase diagram complexity. In 2d hard disks, fluid-fluid demixing is generally expected only for positively non-additive hard disks ~\cite{sillren2010}, which would be difficult to achieve in colloidal experiments with hard spheres. In the crystal regime, however, binary hard disks can stabilize extremely diverse behavior. Considering only structures in the limit of infinite pressure, Likos and Henley identified a wide range of distinct periodic crystals as well as the possibility of a quasicrystal with 12-fold symmetry ~\cite{likos1993}. For extreme size ratios, the densest possible crystal phases are expected to consist of a hexagonal packing of large disks with an increasing number of small particles inserted into the triangular voids ~\cite{uche2004}. For the specific case of hard spheres sedimented onto a substrate, simulations interestingly predict the spontaneous self-assembly of both 12-fold and 8-fold quasicrystals ~\cite{fayen2023}.

It should be noted that while monolayers of hard-sphere mixtures closely approximate true 2d systems, simulations incorporating hydrodynamics on binary mixtures confined between two plates suggest that even small amounts of vertical freedom can significantly speed up dynamics of quasi-2d binary mixtures ~\cite{tian2022}.

%% file: Confinement/confinement.tex
\section{Confinement}
\label{sectionConfinement}

As detailed in Sec.~\ref{sectionBulk}, the entropy-driven equilibrium phase diagram of \emph{bulk} hard-sphere systems is fairly well understood. The insertion of one or more confining walls in the fluid, however, decreases the number of possible configurations and  generally alters the system structure and  phase behavior ~\cite{bechinger2002,araujo2023}. In this section, we specifically consider the interfacial behavior of hard-sphere fluids between two hard walls, under quasi-1d and quasi-2d conditions as well as within cylindrical, spherical, and flexible immurement. Note that the case of a single planar wall, which results in an interface albeit not in confinement proper is presented with in Sec.~\ref{sectionSinglePlanarWalls}.

\subsection{Quasi-2d confinement}
\label{sectionQuasi2dConfinement}

The effect of confinement between two planar hard walls was first investigated in ~\citet{pieranski1983}using colloidal polystyrene spheres.  They observed a sequence of layered solid structures with triangular ($\triangle$) and square ($\square$) 
symmetry, $1\triangle \rightarrow 2\square \rightarrow 2 \triangle \rightarrow 3\square \rightarrow 3\triangle \ldots \rightarrow n\square$ upon increasing the   plate separation, where $n$ denotes the number of layers. Using MC simulations and cell theory (Sec.~\ref{sectionCellTheory}), Schmidt and L\"owen later mapped out the phase diagram for plate separations   ranging from one to two particle diameters, and found additional buckled and rhombic phases ~\cite{schmidt1996, schmidt1997}. The sequence of high-density
structures was determined more accurately in subsequent experiments ~\cite{neser1997,fontecha2005}, further identifying prism phases with both square and triangular symmetry.

\begin{figure}
\includegraphics[width=0.95\linewidth]{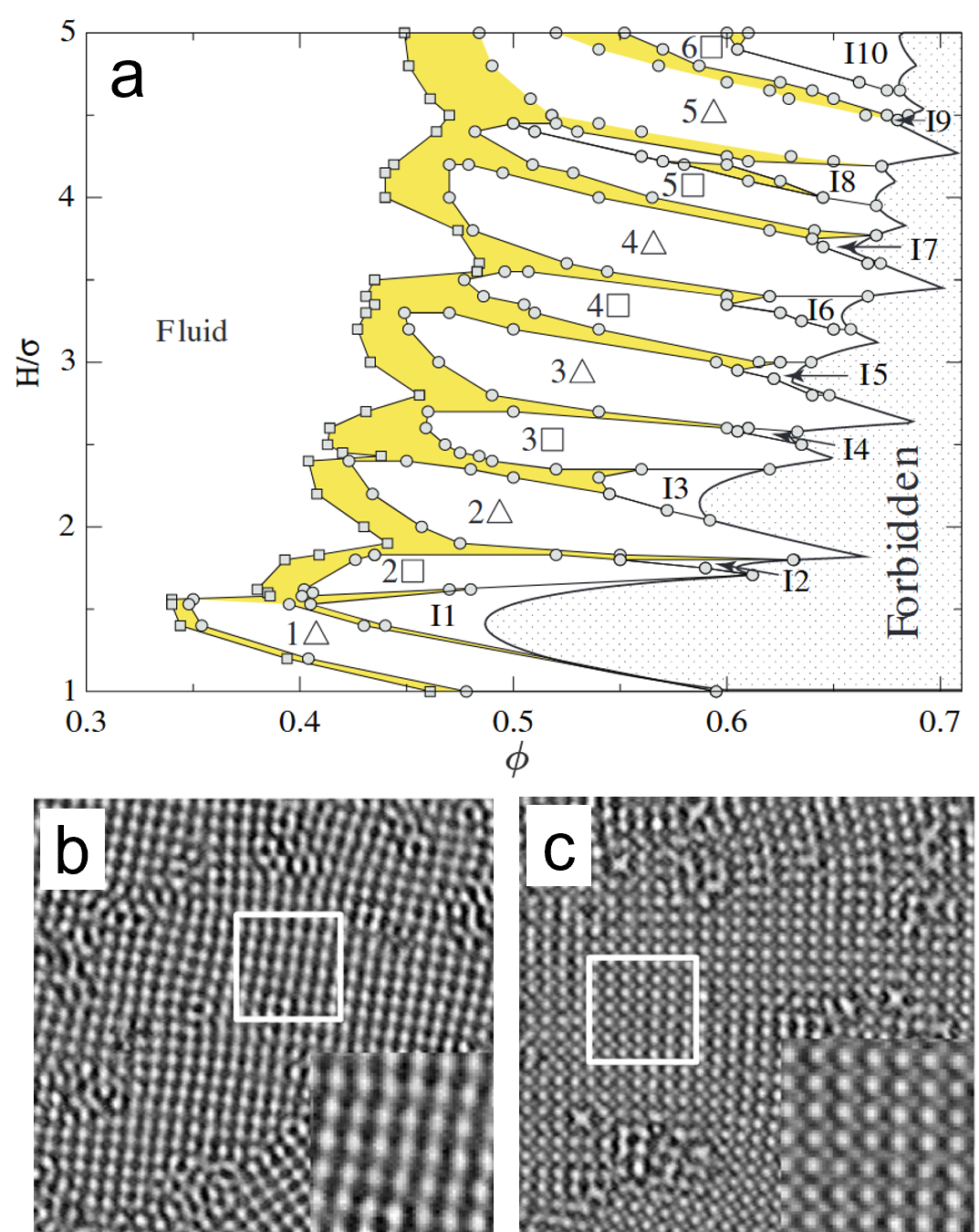}
\caption{a) Equilibrium phase diagram of hard spheres confined between two parallel hard walls with plate separation $H$ versus packing fraction  representation. The white, shaded (yellow) and dotted regions indicate the stable one-phase region, the two-phase coexistence region, and the forbidden region, respectively. 
(b,c) Optical microscopy image of charged spheres in an aqueous system confined in a wedge geometry. Shown are both a 2$\Delta$ phase (b) and a $2\square$ phase (c). Reproduced with permission from (a) ~\cite{fortini2006} and (b,c) ~\cite{fontecha2005}. Copyright IOP Publishing. All rights reserved. }
\label{figParallelPlates}
\end{figure}

Using extensive free-energy calculations in MC simulations, the full phase diagram for  plate separations from one to five hard-sphere diameters was mapped out as a function of packing fraction  ~\cite{fortini2006}, see Fig.~\ref{figParallelPlates}. These results identify a first-order fluid–solid transition, corresponding to either capillary freezing
or capillary melting depending on the plate separation, with the coexisting solid phases consisting of crystalline layers with either triangular or square symmetry.
 At high densities, prism, buckled, and rhombic phases are found to be thermodynamically stable in agreement with experiments ~\cite{neser1997,fontecha2005,ouguz2012}.

~\citet{curk2012} used simulations to investigate hard spheres in soft quasi-2d confinement with a parabolic potential along one dimension. Like in hard confinement, a sequence of confined hexagonal and square-symmetric packings was found, but none of  the intervening ordered phases were observed, the system undergoing phase separation instead.

\subsection{Cylindrical confinement}
\label{sectionCylindricalConfinemnt}

Hard spheres perfectly confined to a 1d line form a \emph{Tonks gas}. This model, which has a rich theoretical history~\cite{lieb1966}, captures reasonably well the behavior of various condensed-matter systems, such as linear chains of mercury in Hg$_{3-\delta}$AsF$_6$ ~\cite{spal1980,chaikin}\footnote{Note that the analysis of Refs.~\onlinecite{spal1980,chaikin} did not take advantage of the exact 1d solution. This analysis is left as an exercise to the reader.}, and chains  of cobalt atoms confined by platinum steps ~\cite{gambardella2002}. (Its quantum mechanical relative, the Tonks-Girardeau model, has also been experimentally realized~\cite{bloch2008}.)

The Tonks gas can be solved by a transfer matrix (TM) scheme. A rich set of structural and thermodynamic observables can thus be obtained at minimal computational cost. With appropriate discretization schemes, the TM scheme can also be extended to quasi-1d systems~\cite{barker1962}. Both hard disks between hard lines~\cite{kofke1993,gurin2013,godfrey2014,robinson2016,hu2018,hicks2018,zhang2020} and hard spheres within a hard cylinder~\cite{kamenetskiy2004,hu2018} (as well as more exotic hard shapes~\cite{kantor2009,gurin2015,gurin2016,gurin2017pre,gurin2017jcp}) have thus been considered using this approach.

\begin{figure*}
\includegraphics[width=160mm]{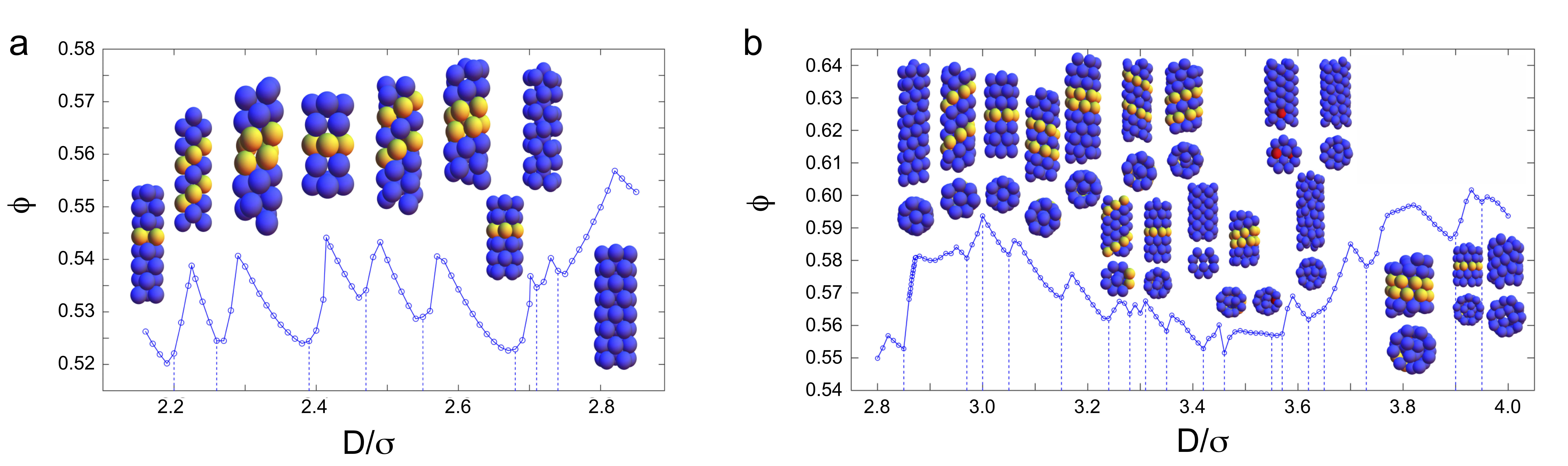}
\caption{Close packing fraction of hard spheres confined in cylinders of diameter $D$ along with some of the structures. These packings (a) are phyllotactic-like for $D/\sigma\lesssim2.8$, and (b) exhibit a core--shell structure beyond that point. Reproduced from ~\cite{fu2016} with permission from the Royal Society of Chemistry.}
\label{figQuasi1D}
\end{figure*}

The TM size, however, jumps markedly as the number of interacting particles increases, which limits the maximal cylinder width that can be computationally resolved. At present, it is nevertheless possible to study systems with up to next-nearest-neighbor interactions. Spheres in a cylinder with a diameter of $2\sigma$ (and for disks a line spacing of up to $5\sigma/2$) can thus also be solved this way~\cite{hu2018}. Despite the limitations of this size regime, it suffices for zig-zag and helical order to emerge at high density.

The analysis of these systems provides key insight into quasi-1d ordering. For instance, although 1d systems with finite-range pairwise interactions have long been known to be unable to exhibit a phase transition~\cite{vanHove1950} (see also ~\cite[Thm.~5.6.7]{ruelle1999}), various numerical simulations of cylindrically confined hard spheres have identified somewhat abrupt structural changes as density increases~\cite{gordillo2006,varga2011,koga2006,duran-olivencia2009}. Similarly, it has been proposed that confined disks could exhibit a Kosterlitz-Thouless--type phase transition  ~\cite{huerta2020,hu2021}. The TM scheme neatly resolves these apparent paradoxes. First, their consideration of infinite and fully equilibrated systems removes all hints of a thermodynamic singularity (as given by the largest TM eigenvalue). Second, the marked structural changes find a microscopic explanation. Because the correlation length that describes the spatial decay of structural order is given by the ratio of the two largest (magnitude-wise) TM eigenvalues, $\xi\sim\ln(\lambda_1/\lambda_2)$, crossing of sub-dominant TM eigenvalues can give rise to marked structural changes without any thermodynamic singularity ~\cite{hu2018}.

In order to observe richer order types, larger cylinder diameters need to be considered. The diameter dependence indeed does not result in continuous structural changes, as is most obvious at close packing. (The effect is also notable at finite pressure~\cite{mon2000}.) The associated morphological richness was first studied by~\citet{pickett2000}, who noted that chiral order spontaneously develops for certain diameter ratios. Mughal {\em et al.}~\cite{mughal2011,mughal2012} later noted that for cylinders of diameter up to $2\geq D/\sigma=1+1/\sin(\pi/5)\approx2.70$ close-packing is described using a phyllotactic construction because all spheres then coat the cylinder wall. Beyond this diameter regime, not all spheres touch the cylinder wall, which eventually results in a separation between core and shell particles~\cite{fu2016}. Exotic arrangements, complex helices and limit periodic structures then follow. More or less systematic numerical exploration of these structures ends around $D/\sigma=4$, but it is conceivable that larger diameters might accommodate even more exotic structures.

Considerable interest has also been paid to the transport dynamics of these systems. In particular, for the hard-sphere case, Mon and Percus identified a crossover from single-file to Fickian diffusion for sufficiently wide cylinders, i.e., $D\geq2\sigma$~\cite{mon2002,mon2003}. A transition state theory description of the hopping mechanism that enables sphere passing was later proposed~\cite{wanasundara2012}, and the impact of microscopic dynamics~\cite{sane2010,flomenbom2010} as well as size dispersity~\cite{wanasundara2012} on these has been considered.

From the experimental standpoint, hard spheres in quasi-1d confinement have notably been used to rationalize the packing behavior of C$_{60}$ in nanotubes~\cite{mickelson2003,khlobystov2004,khlobystov2005,troche2005}, of confined nanoparticles~\cite{tymczenko2008,sanwaria2014}, and of vacuolated cells in an embryonic structure~\cite{norman2018}. A few colloid-based realizations have also been achieved~\cite{lohr2010,jiang2013,liu2015smreconfigurable,fu2017}, as have granular-scale ones~\cite{bogomolov1990}. In all cases, a good correspondence between theory and experiments is obtained, although the assembly pathway may need to be taken into account to rationalize the observed packings~\cite{mughal2012,fu2017}. Nevertheless, relatively few of the above numerical and theoretical predictions have been tested directly in experiments.

\subsection{Quasi-1d confinement}
\label{sectionQuasi1dConfinement}

A lower-dimensional generalization of cylindrical confinement considers hard disks between parallel lines a distance $D$ apart. These models have largely been studied for the insight they offer into the physics of disordered systems. Given that the geometry of these systems is a lot simpler than that of spheres in cylindrical confinement, only fairly unremarkable structures form the densest close packing as $D$ increases. Subtleties in these packings nevertheless lead to the emergence of nontrivial local features~\cite{ashwin2009,zhang2020}. Such irregularities have allowed these models to find particular use in clarifying the physics of low-dimensional disordered systems.

Given the robustness of the jamming phenomenology down to 2d systems, such as the algebraic scaling of weak forces and small gaps (see Sec.~\ref{sectionJamming}), an interesting target has been to consider the structural criticality of quasi-1d systems. For $D<1+\sqrt{3}\sigma/2$,  jammed states are then isostatic and have a nonzero complexity~\cite{ashwin2009,bowles2011,ashwin2013,godfrey2018}, but do not exhibit a critical structure. For  $1+\sqrt{3}/2<D<2$, however, it is possible to consider states that are critical. Then, depending on the system details, different critical exponents have been reported~\cite{ikeda2020,zhang2020}. This sensitivity of the structural criticality likely follows from these systems being below the lower critical dimension for jamming.

These models have also been used to study dense fluids~\cite{godfrey2014,robinson2016,hicks2018}, so as to assess how some of the mean-field theory predictions then fare. Unsurprisingly for such low-dimensional models, a qualitatively different physics is observed. The findings should nevertheless caution against possible confounding physical factors in the study of higher-dimensional fluids, such as the importance of local structure and the possibility of crossovers at fairly large system sizes.

Given the rich physics of such simple models, a seemingly open area of research is their experimental study. Microfluidic channels with colloids~\cite{mark2010}, for instance, might be a promising approach to consider.

\subsection{Spherical confinement} 
\label{sectionSphericalConfinement}

Arguably the most natural way of confining colloidal spheres in three dimensions is inside a larger sphere. As a natural toy model system for small many-body systems, even extremely small systems of only a few hard spheres in spherical confinement have attracted significant theoretical attention exploring, e.g., the effect of the thermodynamic ensemble on the observed structure ~\cite{gonzalez1998, gonzalez1997}, and the thermodynamic properties of the confined fluid ~\cite{urrutia2011jcp, urrutia2014jcp, urrutia2012jcp}.

As one would expect, hard-sphere fluids confined to larger spherical cavities are known to show structuring ~\cite{macpherson1987, chui1991, zhou1989, calleja1991} near the cavity walls. The dynamics inside the cavity then depends strongly on  whether the cavity walls are rough or smooth ~\cite{nemeth1999}, with rough walls more strongly inducing dynamical arrest. In particular, when a glassy fluid of hard spheres is confined in a spherical cavity with rough walls consisting of pinned particles, the dynamics approaches that of a bulk hard-sphere glass for large cavities, providing a route to probe dynamical correlations in these systems ~\cite{nemeth1999, zhang2016}.

Experimentally, self-assembly of hard-sphere colloids in spherical confinement can be achieved by confining the colloids inside emulsion droplets ~\cite{yi2002, manoharan2003,manoharan2006}. When the droplets shrink (due to either evaporation or Ostwald ripening ~\cite{schmitt2016}), the slowly increasing density of the colloids can eventually lead to self-assembly. The resulting structure depends sensitively on the number of colloidal particles within the cluster. Clusters containing only a handful of colloids form small polyhedral clusters whose geometry can often be understood as the densest configuration possible in the circumstances. However, in these small clusters, capillary forces during the final stages of solvent evaporation also play a significant role, and in many cases drive the cluster to minimize the second moment of its mass distribution ~\cite{manoharan2003, cho2005chemmat, yi2004, lauga2004}. For hard spheres, this set of minimal clusters has been studied extensively by ~\citet{sloane1995}.

\begin{figure}[t]
    \includegraphics[width=0.95\linewidth]{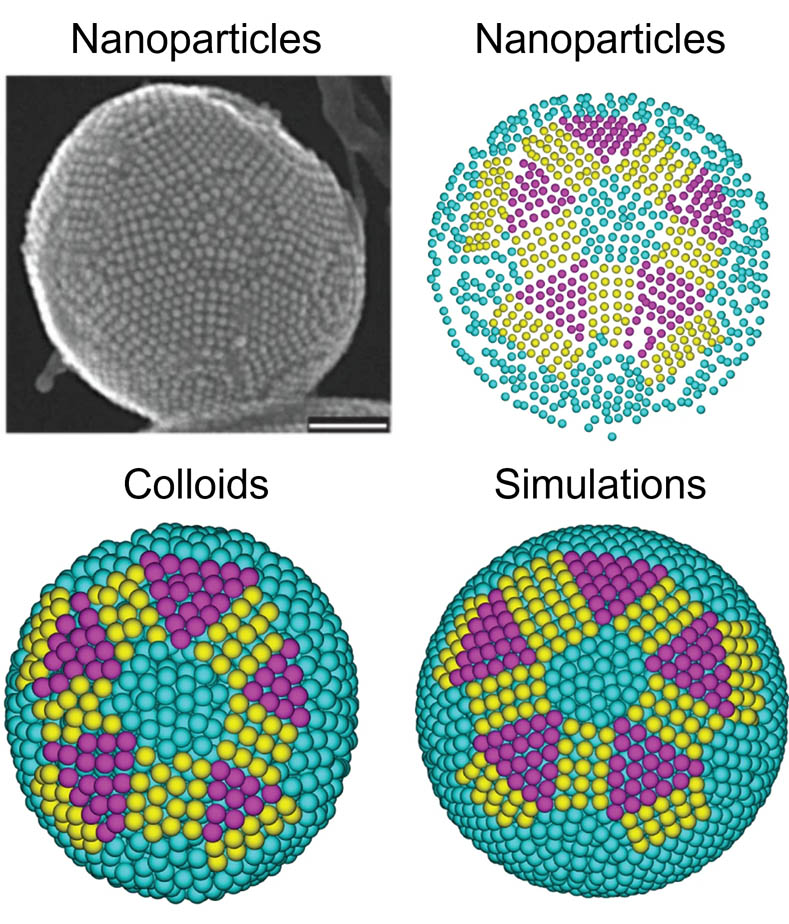}
    \caption{Icosahedral clusters of spheres self-assembled in spherical confinement. The top row shows both an electron microscopy image of a cluster of nanoparticles and a reconstruction of its surface layer. The bottom row shows (left) a cluster obtained in experiments on micron-sized colloids and (right) one obtained in simulations of hard spheres. Figure adapted with permission from Ref.~\cite{denijs2015}.}
    \label{figIcosahedralclusters}
\end{figure}

For larger clusters of monodisperse spheres in spherical confinement, the natural tendency of hard spheres to crystallize into an fcc crystal  competes with the curvature of the surface. Although sufficiently large clusters ($N \gtrsim 10^5$) form a simple fcc structure, intermediate cluster sizes ($100\lesssim N \lesssim 10^5$) spontaneously  form  clusters with icosahedral symmetry ~\cite{denijs2015}. The core of these icosahedral clusters consists of tetrahedral-shaped domains of distorted fcc crystal, with one hexagonal plane from each domain forming the faces of an icosahedron, often called Mackay clusters ~\cite{mackay1962}. For sufficiently large clusters, these domains can be capped by additional surface layers, resulting in a family of ``anti-Mackay'' sphere packings with varying surface reconstructions  ~\cite{denijs2015, wang2018} (see Fig.~\ref{figIcosahedralclusters}). Free-energy calculations based on computer simulations have shown that these clusters are indeed thermodynamically stable configurations for pure hard spheres in spherical confinement ~\cite{denijs2015}, with clusters containing certain ``magic numbers'' of spheres corresponding to defect-free clusters being particularly stable ~\cite{wang2018, wang2019}.

Naturally, this already complex behavior can be further tuned by considering, e.g., binary mixtures of spheres in spherical confinement. For small clusters, complex anisotropic colloidal supraparticles can result ~\cite{cho2005jacs}. For larger clusters, Wang {\it et al.} demonstrated the formation of clusters consisting of the thermodynamically stable MgZn$_2$ Laves phase when there is an excess of small spheres ~\cite{wang2021ncomm}, but the clusters turn  into an icosahedral cluster consisting of tetrahedral domains of  the less stable MgCu$_2$ Laves phase when the fluid composition is iso-stoichiometric with Laves phase  ~\cite{wang2021nphys}.

As the self-assembly behavior of colloidal spheres under spherical confinement is robust on both the nanometer and micrometer scale ~\cite{denijs2015}, and for a variety of materials ~\cite{yi2004}, it provides a versatile route for tuning material structure across various length scales. Supraparticles created via this technique have been proposed as building blocks for additional self-assembly steps, e.g., for creating materials with structural order on two different length scales ~\cite{bai2007}. Additionally, the icosahedral nature of intermediate-sized clusters of colloidal spheres may lead to intriguing optical effects, including (iridescent) structural color patterns ~\cite{wang2020} that can be used to track the self-assembly dynamics inside the clusters in real time.

\subsection{Flexible confinement}
\label{sectionFlexibleConfinement}

Thus far in this section, we have seen that a large variety of packing geometries can be obtained by confining spheres in a container of fixed shape. A  different question concerns the most efficient packing of spheres in a natural bin, which is the smallest convex hull (volume-wise) that can enclose a certain number of spheres. Counterintuitively, the result is not always a compact cluster of spheres. For a  packing of spheres with up to $n=55$, along with $n=57, 58, 63, 64$, the linear conformation in which the spheres lie on a straight line, also called a sausage, is denser than a cluster-like or plate-like configuration~\cite{gandini1992}. The optimal packing for $n=56$ is not fcc like, and the exact configuration remains unknown. (By contrast, in 4d, this sudden transition from a sausage packing to a cluster shape is conjectured to happen at $n=375769$ spheres, and is therefore referred as the ``sausage catastrophe''~\cite{henk2021}.) Hard spheres in a flexible container can be used to model colloids in a fluctuating vesicle, which was studied theoretically and in simulations by ~\citet{maibaum2001}. Recently,  such a system of non-close-packed colloids in a flaccid lipid vesicle has been experimentally realized~\cite{marin2023}. Through a combined experimental and simulation study, the authors obtained a state diagram that includes linear, planar, and cluster conformations of spheres, as well as bistable states, which alternate between cluster-plate and plate-linear conformations due to membrane fluctuations. Additionally, ~\citet{marin2023} have identified truncated polyhedral packings of  $56\leq N \leq 70$ spheres (excluding $N=57$ and 63) that pack more efficiently than linear arrangements.

In experiment, flexible confinement may be induced by the use of optical tweezers. To the best of our knowledge, in the context of hard spheres, using lasers to induce structuring may be traced to the work of colloid pioneers Bruce Ackerson and Noel Clark and coworkers in their use of lasers to induce freezing in quasi--2d systems~\cite{chowdhury1985}. A more recent approach was to confine a system inside a ring of particles held in optical traps. Such traps have well--defined potentials, which means that the osmotic pressure and therefore the EoS could be measured. At low density, the results were found to agree with the bulk EoS. At higher density, by contrast, the fluid takes on a layered structure, thus reflecting the  confinement. Interestingly, at still higher density, a bi-stable state between the layered fluid and a hexagonal structure was observed, reminiscent of the bulk hexatic phase~\cite{williams2013}.

%% file: FarFromEquilibriumPhenomena/farFromEquilibriumPhenomena.tex
\section{Out-of-equilibrium phenomena}
\label{sectionFarFromEq}

In this section, we discuss how hard spheres have shed some light on material phenomena away from thermal equilibrium. When studying out-of-equilibrium suspensions it is useful to distinguish linear response around thermal equilibrium, which can be treated perturbatively within the framework of statistical mechanics, from phenomena that occur due to strong driving beyond the linear regime. Non-equilibrium effects often appear as fluxes of macroscopic observables and transport phenomena. Because particle motion remains governed by thermal fluctuations, the relative magnitude of this transport compared to diffusion, as captured by the various dimensionless P\'{e}clet numbers (Table~\ref{tablePeclet}), e.g. for gravitation, shear flow~\cite{russel}, or other fields such as optical tweezers~\cite{williams2016}, is a crucial measure how strongly a system is driven.

\begin{table*}

  \centering
    \begin{tabular}{c|c|c|c|c|c}
  \hline\hline
    & general force & gravitation & drift speed & shear flow & glassy flow  \\
    drive & $F$ & $g$ & $v$ & $\dot\gamma$ & $\dot\gamma$  \\
    \hline
    $\tau_x$ & $\frac{\sigma}{2\mu F}$ & $\frac{\sigma}{2\mu m_b g}$ & $\frac{\sigma}{2v}$ & $\dot\gamma^{-1}$ & $\dot\gamma^{-1}$  \\
    \hline
    $\pecl_x$ & $\frac{F\sigma}{2\kT}$ & $\frac{\pi\sigma^4\delta\rho g}{12\kT}$ & $\frac{2\sigma v}{D}$ & $\dot\gamma\tauB$ & $\dot\gamma\tau_\alpha$  \\
  \hline\hline
  \end{tabular}
  \caption{P\'eclet numbers $\pecl_x=\tauB/\tau_x$ relate the diffusive time $\tau_\text{B}$, as defined in Eq.~\eqref{eqTauB} to the time scale $\tau_x$ of a directed transport process using the particle radius $\sigma/2$ as reference length. $\mu=D_0/\kT$ is the particle mobility. Sedimentation (second column) is due to the gravitational force $m^\ast g$ with the gravitational  $g$ and buoyant mass $m_b=(\pi\sigma^3/6)\delta\rho$, where $\delta\rho$ is the density difference between particle and solvent. In this case the P\'eclet number can also be written $\pecl_g=\sigma/(2\xi)$ with gravitational length $\xi$. In the case of shear flow, the timescale of directed transport is $\dot\gamma^{-1}$. For normal fluids, the Brownian time $\tauB$ is appropriate, but for glassy systems the Weissenberg number $\dot\gamma\tau_\alpha$ is often used (see Sec.~\ref{sectionFarGlass}).
  }
  \label{tablePeclet}
\end{table*}

Thus far in this review, the solvent has largely been considered as a structureless quiescent medium. Out of equilibrium, however, this simplification is often no longer appropriate. Solvent dynamics has to be taken into account. In this section, we therefore shift our focus away from hard spheres as a model system and consider instead the colloidal material as a complex fluid. In that context, the full dynamics of a particle-laden Newtonian solvent is described by the Navier-Stokes equation together with a suitable (typically no-slip) boundary condition on the particle surfaces. By eliminating the solvent, forces between suspended particles effectively couple through the mobility tensor which can, in principle, be derived from the Oseen tensor~\cite{dhont}. In bulk systems,  hydrodynamic coupling is long-ranged due to momentum conservation of the solvent. Taking these hydrodynamic interactions into account in simulations is feasible but computationally costly (see Sec.~\ref{sectionSimHydro} for available methods). A recurring theme is the importance of colloidal forces over hydrodynamic coupling. If the latter can be  neglected, one can resort to computationally cheaper BD simulations (Sec.~\ref{sectionSimBD}).

\subsection{Non-equilibrium sedimentation in hard-sphere colloids} 
\label{sectionNonEquilibriumSedimentation}

\begin{figure}
\centering
\includegraphics[width=\linewidth]{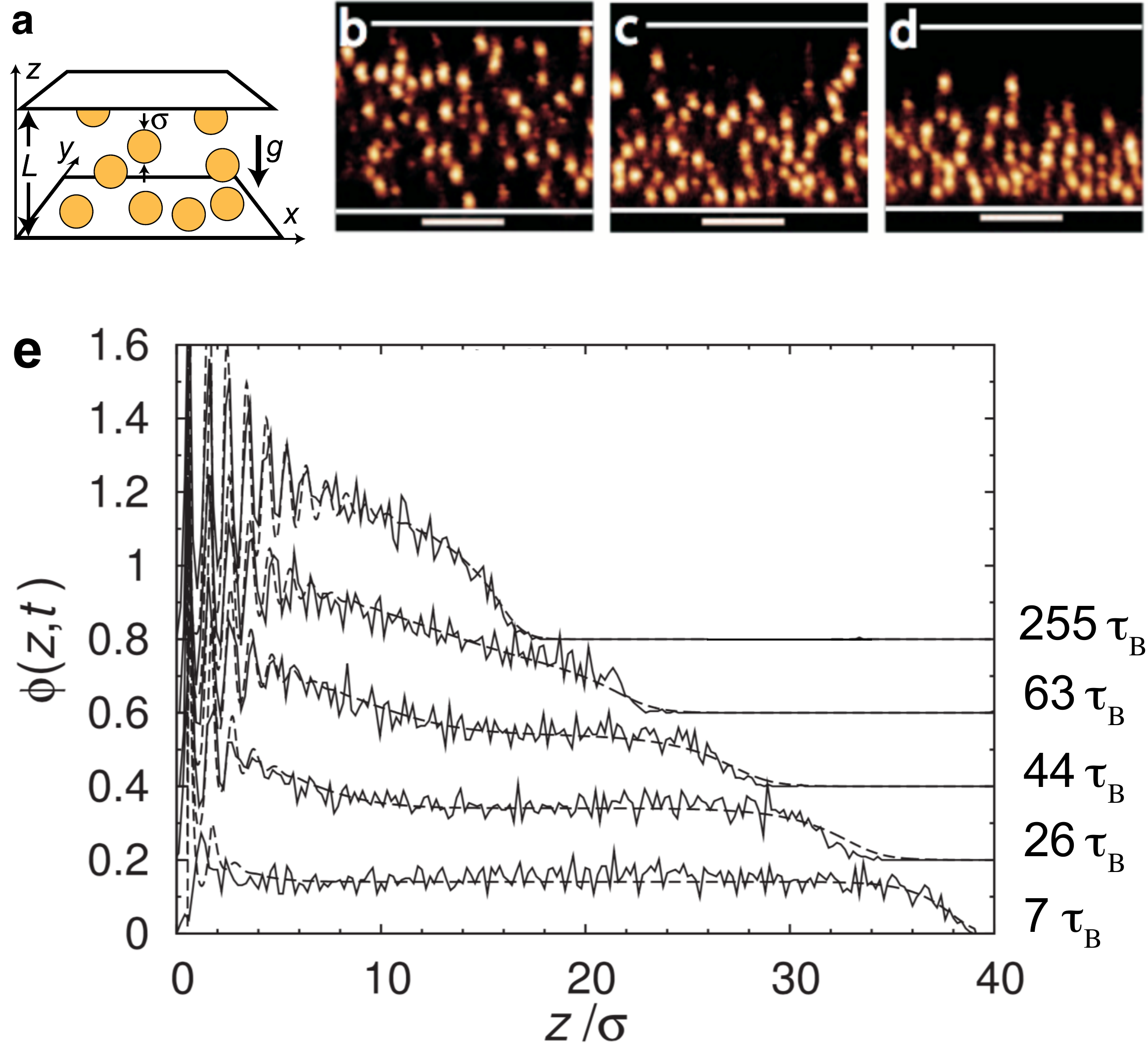}
\caption{Non-equilibrium sedimentation of hard spheres on the particle scale. (a) Illustration of a system of sterically stabilized PMMA particles under the influence of gravity $g$ and vertically confined between two walls separated a distance $L$. 
(b-d) Time series of confocal micrographs taken in the (vertical) $xz$ plane at times (b) $t=3$, (c) $26$, and (d) $200 \tauB$. The scale bars denote 20 $\mu$m; the horizontal lines indicate the position of the walls. For (b-d) the sedimentation P\'{e}clet number is $\Pe_g=0.625$. (e) Time evolution of the sedimentation profile $\phi(z,t)$ for a system with $\Pe_g=1.11$. Solid lines are  experimental data from particle-resolved studies,  dashed lines are from dynamical DFT. No fit parameters are used. Reprinted with permission from Royall \emph{et al.}, Phys. Rev. Lett. \textbf{98}, 188304 (2007) \cite{royall2007prl}. Copyright (2007) by the American Physical Society.}
\label{figMatthias}
\end{figure}

\begin{figure*}
\centering
\includegraphics[width=140mm]{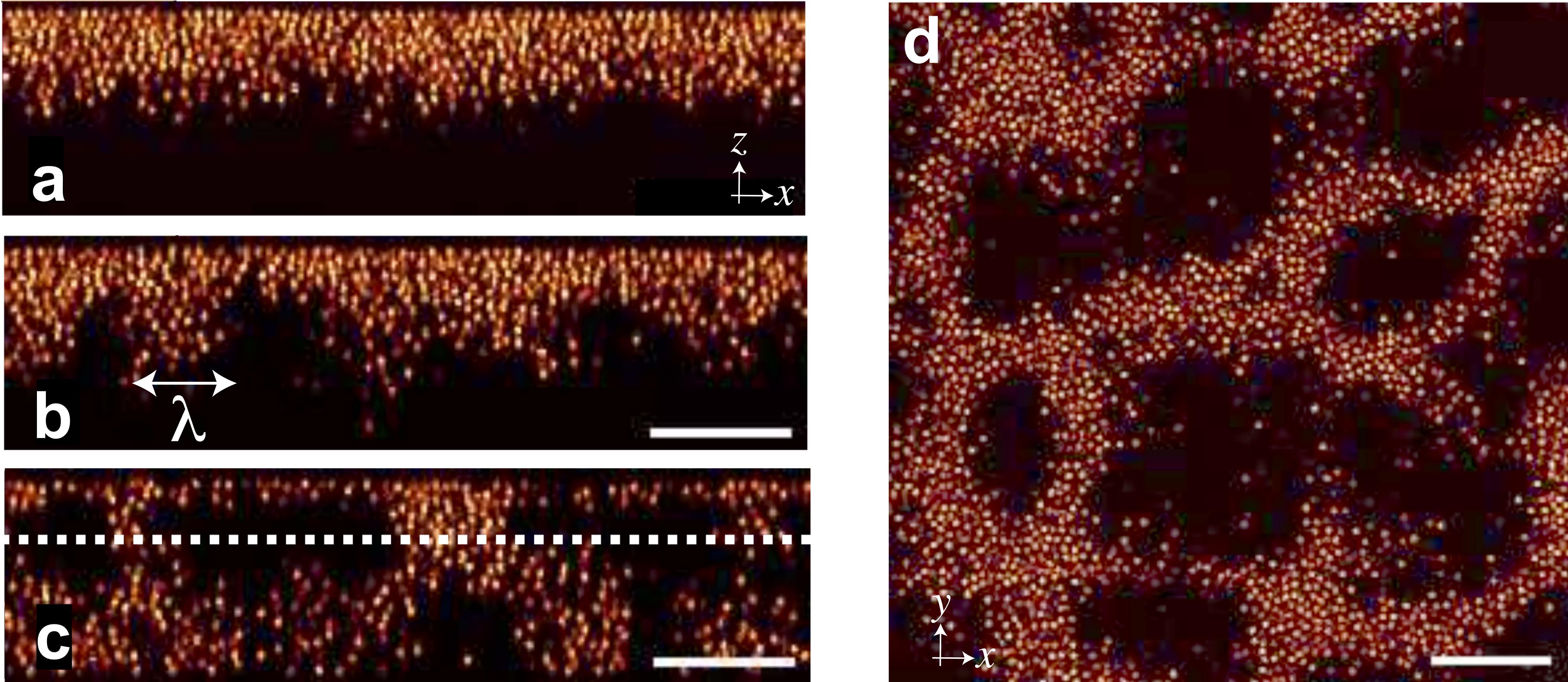}
\caption{Rayleigh-Taylor--like instability in hard-sphere colloids. 
(a-c) Time series of images taken with a confocal microscope, for $1.43$, $5.5$ and $11.2$ $\tauB$. $\lambda$ denotes the characteristic growing wavelength of the Rayleigh-Taylor instability.
(d) Image in $xy$ plane at the height of the dashed line in (c). Reproduced from \citet{wysocki2009} with permission from the Royal Society of Chemistry.}
\label{figAdamWysocki}
\end{figure*}

Let us first consider what is arguably the simplest non-equilibrium situation, namely the gravitational settling of a suspension. As discussed in Sec.~\ref{sectionEquationOfState},
waiting for the suspension to relax yields a density profile from which the equation of state may be inferred. While such equilibrium profiles are straightforward to calculate, the approach to equilibrium, from an  out-of-equilibrium starting point is a challenging problem to address.

As alluded to above, a useful way to distinguish different regimes is through the \emph{gravitational} (or sedimentation) P\'{e}clet number
\begin{equation}
  \Pe_g = \frac{\tauB}{\tau_g} = \frac{\sigma/2}{\xi_g} = \frac{\pi \sigma^4 \delta \rho g}{12k_\mathrm{B}T}
\label{eqSedPecl}
\end{equation}
defined as the ratio of the Brownian diffusive time $\tauB$ [Eq.~\eqref{eqTauB}] to the time $\tau_g$ for a sphere to  sediment its own radius, where $\delta\rho$ is the mass density difference between colloidal particles and the solvent. Alternatively it can be expressed through the gravitational length $\xi_g=\kB T/m^\ast g$ with $m^\ast$ the buoyant mass. Note that without loss of generality, sedimentation is assumed to occur in the $z$ direction and has no dependence on the $x,y$ coordinates. Equation~\ref{eqSedPecl} describes the sedimentation of a single colloid, and so holds in the dilute limit.

For more concentrated systems, when the density varies on lengthscales sufficient for local packing effects to be neglected, and for which gravitational settling is slow, i.e., $\Pe_g\ll1$, the time evolution of the sedimentation profile $\phi(z,t)$ largely follows a \emph{batch settling} process \cite{russel}. Under these conditions, the volume fraction at any height and at any time may be captured by a set of relatively simple coupled equations. (For more strongly driven systems, in the granular regime, strong swirls due to hydrodynamic coupling can be observed \cite{segre2001nature,segre1997prl}.)

For smaller systems, for which packing effects can be more noticeable (see Sec.~\ref{sectionConfinement}), classical DFT (see Sec.~\ref{sectionCDFT}) is needed. Using dynamical DFT, it is indeed possible to propagate these density profiles forward in time, as shown in Fig.~\ref{figMatthias}(d,e). The time-dependent density profiles predicted by dynamical DFT agree remarkably well with particle-resolved results from BD simulations. With the inclusion of a simple treatment for the volume fraction-dependent slowdown in dynamics due to hydrodynamic interactions (treated with the Hayakawa-Ichiki method), the time evolution of the density profiles of sedimenting hard spheres in experiment could be accurately described (see Fig.~\ref{figMatthias}(a-e))~\cite{royall2007prl,schmidt2008}.

This description of sedimentation presumes translational invariance in the $(x,y)$ plane, which holds (empirically) for a starting configuration that is approximately homogeneous \cite{royall2007prl,schmidt2008}. An inhomogeneous configuration such as that shown in Fig.~\ref{figAdamWysocki}(a) presents quite a different--and at first sight surprising--proposition. Density fluctuations with a characteristic length scale (Fig.~\ref{figAdamWysocki}(b)) then develop, leading to a behavior reminiscent of the catastrophic Rayleigh-Taylor instability when two immiscible liquids are prepared with the denser liquid above. What is particularly surprising here is that the hard-sphere fluid, which of course is a single phase, then behaves like a phase-separated system. These complex time-dependent patterns can further be accurately captured by multi-particle collision dynamics simulations \cite{wysocki2009,wysocki2010}. 
Therefore, even  phenomena quite far-from-equilibrium  can be accurately captured by theory and simulation.

Adding a second colloidal species, whose gravitational P\'{e}clet number can be tuned independently, provides a further means of controlling the structural properties of the swirls. Interestingly, these swirls depend mainly  on the
relative magnitudes of the P\'{e}clet numbers of the two species, and much less on the composition of the mixture which would be the case closer to equilibrium (see Sec.~\ref{sectionBinary}) \cite{milinkovic2011}.

\subsection{Rheology, flow, and shear-induced order}
\label{sectionRheology}

The most prevalent way to probe mechanical properties is to subject a sample to external forces and measure its deformation (rate)~\cite{wagner2022,larson,chen2010annurev}. In addition to the packing fraction $\phi$, the strain $\gamma$ (strain rate $\dot\gamma$, not to be confused with the interfacial free energy in Sec.~\ref{sectionXtalFluidInterfaces}) with conjugate stress $\tau$ are then needed to characterize a hard-sphere system. The absence of cohesive forces between hard spheres implies that their solids deform easily. The elastic moduli, which encode how a solid linearly deforms in response to an applied force, nevertheless diverge as $\propto \kB T/[\sigma^3(\phi_\text{cp}/\phi-1)^2]$ upon approaching close packing~\cite{stillinger1967,farago2000}.

\begin{figure}
    \centering
    \includegraphics[width=\linewidth]{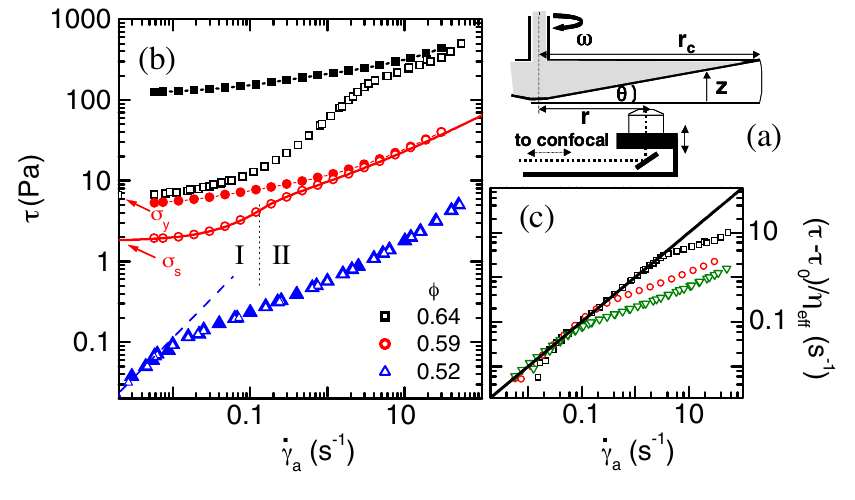}
    \caption{
    (a)~Sketch of a cone-plate rheometer together with a confocal microscope. 
    (b)~Measured shear stress $\tau$ as a function of applied strain rate $\dot\gamma_\text{a}$ for several packing fractions $\phi$ below and above the glass transition. Open symbols indicate uncoated smooth plates while full symbols show data for coated plates with no-slip boundary conditions. 
    (c)~The excess stress $\tau-\tau_0$ is linear with slope $\eta_\text{eff}$ defining an effective viscosity. Reprinted with permission from Ballesta \emph{et al.}, Phys. Rev. Lett. \textbf{101}, 258301 (2008) \cite{ballesta2008}. Copyright (2008) by the American Physical Society.}
    \label{figRheology}
\end{figure}

Similarly, a hard-sphere fluid starts flowing in response to shear forces, thus entering a dissipative non-equilibrium state. In certain geometries (e.g., Couette or cone-plate shear cells, cf. Fig.~\ref{figRheology}(a)), a uniform strain rate $\dot\gamma$ can be achieved, thus defining viscosity $\eta=\tau/\dot\gamma$. How strongly a suspension is sheared is described through the dimensionless shear P\'{e}clet number $\Pe_{\dot\gamma}=\dot\gamma\tauB$, again using the Brownian time $\tauB$ [Eq.~\eqref{eqTauB}] (Table~\ref{tablePeclet}), characterizing the importance of advection over diffusion. Figure~\ref{figRheology}(b) shows typical flow curves $\tau(\dot\gamma;\phi)$ for dense hard spheres. For the fluid ($\phi=0.52$) one observes shear thinning with a stress that is smaller than the initial linear increase. The denser ($\phi\geqslant0.59$) no-slip samples behave as Herschel–Bulkley fluids following the empirical flow curve $\tau-\tau_0\propto\dot\gamma^n$ with some exponent $n$. They approach a finite yield stress $\tau_0$ in the limit $\dot\gamma\to0$, thus indicating that the sample behaves as a (disordered) solid. Rheology therefore provides a mechanical route to probe the glass transition, to which we return in Sec.~\ref{sectionFarGlass}.

Traditional rheological studies probe macroscopic volumes through small-amplitude oscillatory shear, so as to only weakly perturb the material from thermal equilibrium. In this linear response regime with $\Pe_{\dot\gamma}\ll1$, the material settles in a periodic state with stress
\begin{equation}
\tau(t) = \int_{-\infty}^t ds\; G(t-s)\dot\gamma(s)
\end{equation}
leading to the storage modulus $G'(\omega)$ and loss modulus $G''(\omega)$, the real and imaginary part of the complex shear modulus $G(\omega)$, as a function of external frequency $\omega$.

For a dilute hard-sphere suspension, Einstein showed that (to linear order in $\phi$) $\eta/\eta_\text{sol}=1+\tfrac{5}{2}\phi$~\cite{einstein1906,mewis}. In other words, viscosity increases with respect to the solvent viscosity $\eta_\text{sol}$ due to volume excluded by the suspended particles. The quadratic order correction in $\phi$ arises from the hydrodynamic coupling between spheres. The exact value of the coefficient, however, depends on the analytical approach (\citet{batchelor1977}, for instance, found $6.2\phi^2$).

At higher strain rates beyond linear response (for $\Pe_{\dot\gamma}\approx 1$), shear thinning sets in. Viscosity then decreases upon approaching a plateau $\eta_\infty$~\cite{vanderwerff1989}, as has been observed in BD simulations without hydrodynamics~\cite{strating1999}. The phenomenon can therefore be linked to a change in the local arrangement of particles~\cite{xu2013} through the excess shear stress
\begin{equation}
    \Delta\tau = -\frac{1}{2}\kT\rho^2\int d^3\vec r\; \frac{xy}{r}g(\vec r)\delta(r-\sigma).
    \label{eqShearStress}
\end{equation}
Importantly, the pair distribution function $g(\vec r)$ is then no longer isotropic~\cite{lin2013,lin2014sm}, because an external force or flow defines a preferred direction. In particular, its contact value, $g^+(\sigma,\theta)$, using polar coordinates within the $xy$-plane through the particle center, varies with orientation and derived quantities like pressure become anisotropic with the off-diagonal component determining the shear stress, cf.~Eq.~\eqref{eqShearStress}. In linear response, \citet{brady1993} determined the Brownian stress as the equilibrium $g(\sigma^{+})$ at contact divided by the short‐time self‐diffusivity.

\begin{figure}
    \centering
    \includegraphics{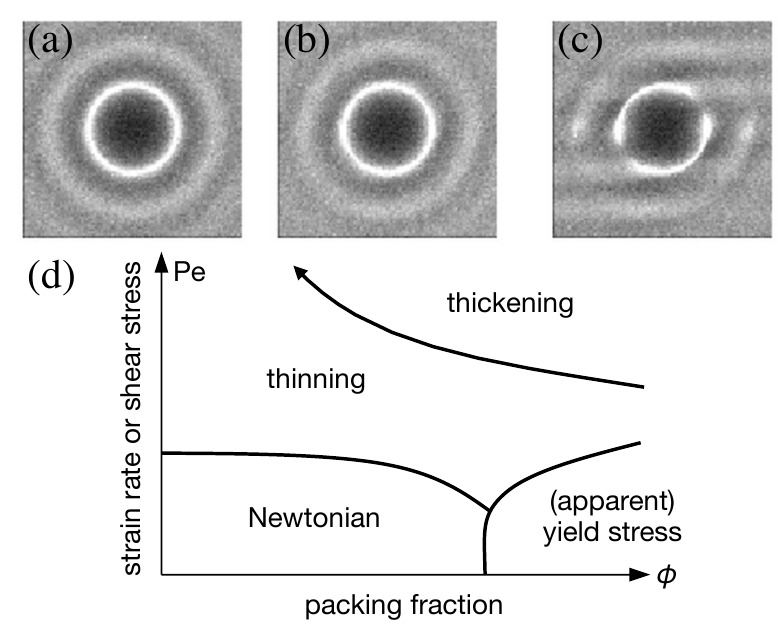}
    \caption{The microstructure $g(\vec r)$ deforms in shear flow at (a)~$\Pe_{\dot\gamma}=0.1$, (b)~$\Pe_{\dot\gamma}=1$, and (c)~$\Pe_{\dot\gamma}=10$. Reprinted with permission from Foss and Brady, J. Rheol. \textbf{44}, 629 (2000) \cite{foss2000}. Copyright 2000, The Society of Rheology. (d)~Sketch of the different flow regimes of hard sphere suspensions. Lines indicate crossovers, with shear thinning being predominately due to colloidal forces and shear thickening having a hydrodynamic origin. Shear thickening has also been observed in dilute suspensions at high strain rates~\cite{bergenholtz2002}. At high packing fractions and high strain rates shear-induced ordering is possible.}
    \label{figFlow}
\end{figure}

The deformation at intermediate P\'eclet numbers is shown in Fig.~\ref{figFlow}(a-c). The relationship between macroscopic material properties and microscopic structure is at the heart of theoretical approaches. Much effort has been devoted to predict the deformation of the pair distribution in response to external forces and flows~\cite{squires2005}, allowing to calculate flow curves $\tau(\dot\gamma)$ from first principles. Coupling of confocal microscopy (of sufficient frame rate) and rheology provides a powerful means to access the local structure in hard-sphere fluids far from equilibrium~\cite{besseling2009,lin2014rsi}.

At even higher strain rates (or shear stresses) and moderate to high packing fractions ($\phi>0.3$), some experiments report an abrupt (sometimes discontinuous) viscosity increase as the stress is increased, i.e., shear thickening~\cite{d1993,bender1996,meeker1997,cheng2002pre}. A comprehensive understanding of this somewhat counter-intuitive behavior (for instance, it is absent in atomistic and molecular liquids) has been an open challenge~\cite{brown2014}. For Brownian hard spheres it is now broadly accepted to be a hydrodynamic phenomenon caused by the (reversible) formation of ``hydroclusters'' held together through lubrication forces~\cite{wagner2009}. Confocal microscopy evidence~\cite{cheng2011} has also been used to investigate the hydrodynamic and contact force contributions to shear thickening~\cite{lin2015}. Shear thickening extends to dilute suspensions ($\phi<0.1$), although only at very high strain rates~\cite{bergenholtz2002}. Figure~\ref{figFlow}(d) summarizes these different uniform flow regimes. Shear thickening has also been proposed as the mechanism that links the rheology of Brownian hard spheres ($\sigma\lesssim1\;\mu$m) to the physics of non-Brownian (granular) hard spheres ($\sigma\gtrsim50\;\mu$m)~\cite{guy2015}.

While the shear thickening behavior is associated with an abrupt increase in the hydrodynamic stresses between the colloidal spheres, which locks them together giving rise to the large hydroclusters resistant to flow, hydrodynamic interactions between smooth surfaces predict a continuous shear-thickening transition~\cite{ball1995,foss2000jfluid,bender1996,melrose2004continuous}. The discontinuous appearance of shear thickening instead is found to depend on the small scale surface asperities of the colloidal particles. These asperities can break the lubrication layer, and can give rise to a frictional contribution. The relative motion of the particles is then akin to a stick-slip scenario triggered by the breakage of the lubrication layer~\cite{mari2015,hermes2016,morris2018,kawasaki2018,jamali2019}. Simulations have explicitly shown a transition from continuous to discontinuous shear thickening by progressively increasing the surface roughness of the particles~\cite{wang2020jrel}. The importance of surface characteristics has also been confirmed in a large variety of experimental works~\cite{fernandez2013,hsu2018,schroyen2019stress,hsiao2017}. In the absence of inertia, two types of discontinuous shear thickening are predicted to occur depending on whether the frictional particles are below or above their jamming point, where the thickened phase either flows smoothly or  is completely jammed~\cite{wyart2014}. Curiously, in experiments an additional high-frequency and low-amplitude shear orthogonal to the primary shearing flow has been shown to suppress shear thickening~\cite{lin2016pnas}.

The very high $\phi$ and high strain rate regime leads to even more surprising observations. Experiments first revealed~\cite{ackerson1988,ackerson1990}, and BD simulations later confirmed~\cite{strating1999}, that particles then form layers perpendicular to the shear gradient that slide over each other. The system therefore keeps flowing with a viscosity that suddenly drops at $\Pe_{\dot\gamma}\approx10$. Hard spheres under shear can also exhibit non-uniform flow profiles, so-called shear banding. One mechanism is the formation of an arrested band due to small variations of the local packing fraction that trigger the arrest of a much bigger region of the flow~\cite{besseling2010}. Other exotic phenomena include the formation of twinned fcc crystals and sliding layers \cite{haw1998}, strings of particles in hard-sphere fluids~\cite{cheng2012} and novel configurations which optimized packing~\cite{cohen2004prl}. The non-equilibrium phase behavior of a fluid of colloidal hard spheres under oscillatory shear was investigated in real space with experiments on PMMA colloidal suspensions and BD simulations as a function of the frequency of the oscillations and $0 \leq  \Pe_{\dot\gamma} \leq 15$, displaying a shear-induced  oscillating twinned fcc  phase, a  sliding layer phase, a string phase and a tilted layer phase \cite{besseling2012}.

Shearing hard-sphere crystals opens yet another range of phenomena. These include melting~\cite{wu2009}, shear banding~\cite{cohen2004prl,cohen2006,dhont2008}, and in the case of confined crystals buckling phenomena~\cite{schall2004}.
It is even possible to infer information about stresses between defects in hard-sphere crystals through careful analysis of particle trajectories~\cite{lin2016nmat}.

\subsection{Microrheology}
\label{sectionMicrorheology}

Trapping a colloidal probe with optical tweezers and imaging its stochastic motion provides insights into the mechanical properties of the host material~\cite{wilson2009,puertas2014}. This microrheological technique is particularly important for (biological) materials that are difficult to prepare in amounts sufficiently large for conventional rheological studies~\cite{wilhelm2008}. The approach can also be used to resolve local mechanical properties in inhomogeneous (soft) materials. 

\begin{figure}
    \centering
    \includegraphics[width=85 mm]{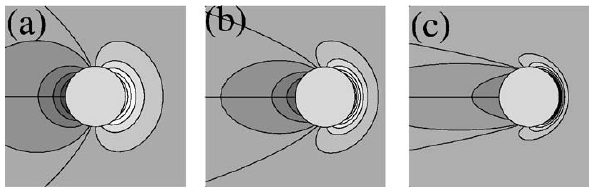}
    \caption{The microstructure $g(\vec r)$ deforms around a driven probe with (a)~$\Pe_v=0.15$, (b)~$\Pe_v=0.5$, and (c)~$\Pe_v=1.5$. Reprinted from Squires and Brady, Phys. Fluids \textbf{17}, 073101 (2005) \cite{squires2005}, with the permission of AIP Publishing.}
    \label{figMicro}
\end{figure}

Depending on whether the probe is forced or not, one distinguishes active from passive microrheology. The latter exploits the fluctuation-dissipation theorem, which greatly simplifies the analysis, but analyzing the data from driven probes requires a pre-existing model of the host material. In this context, hard spheres have emerged as a particularly useful reference. \citet{squires2005} provide a comprehensive analysis of the deformation of the microstructure (Fig.~\ref{figMicro}) around a hard probe with diameter $\sigma_\text{probe}$ forced through a bath of hard spheres with drift P\'eclet number $\Pe_v=v(\sigma+\sigma_\text{probe})/(2D_0)$. Here, $v$ is the probe speed. Analytical expressions for the viscosity are derived in the limits of small and large $\Pe$ yielding an accurate extrapolation to intermediate P\'{e}clet numbers. Other than fluid media, forcing a probe through a crystal~\cite{vossenthesis,dullens2011} and a colloidal glass~\cite{habdas2004,gazuz2009,gruber2016} have been studied.

\subsection{Other out-of-equilibrium phenomena}

Time-resolved confocal microscopy has been used to study other out-of-equilibrium phenomena. One such example is the dynamics of colloidal particles in externally created energy landscapes, such as optical potentials generated by interfering laser beams\footnote{Due to the external forcing in these systems, precise control over interparticle interactions to be hard is often unnecessary, but interactions nevertheless need to be sufficiently short-ranged.}. Other significant examples include ordering of polystyrene spheres in quasiperiodic patterned potentials~\cite{mikhael2008}, driving colloidal monolayers through time-dependent fields~\cite{bohlein2012,brazda2018}, diffusion in random landscapes~\cite{evers2013}, and the transmission of forces through dense colloidal aggregates~\cite{williams2016}.

Following the trajectories of single colloidal particles has also been instrumental in experimentally verifying fluctuation theorems~\cite{wang2002,carberry2004,blickle2006}, and in verifying a class of exact relations from nonequilibrium statistical physics starting with Jarzynski's seminal work relation~\cite{jarzynski1997}. (See~\citet{seifert2012} for a comprehensive theoretical review.) In a nutshell, thermodynamic (current-like) quantities like work and heat can be extended to single stochastic trajectories, which are therefore described by probability distributions constrained by the (near-universal) fluctuation theorems. These relations have also been tested in colloidal suspensions~\cite{solano2015}, and  recent experimental results, including with colloidal particles, have been reviewed by~\citet{ciliberto2017}.

More recently, ``active'' colloidal particles have moved into focus. A wealth of interfacial phoretic mechanisms between molecular and colloidal solutes can be exploited to generate self-sustained gradients that move with the colloidal particles and lead to directed motion (revealed as a correlation in displacements absent in passively diffusing particles). One phenomenon where simulations of active hard disks (and spheres) have been, and still are, instrumental is the coexistence of dense and dilute regions following \emph{motility-induced phase separation}~\cite{cates2015}.
Given the absence of cohesive forces, these processes are genuine far-from-equilibrium phase transitions. In light of the many reviews that have been devoted to this rich and fast moving field~\cite{elgeti2015,bechinger2016,marchetti2013,janssen2019}, we will not, however, delve further here.

%% file: GlassTransition/glassTransition.tex
\section{Hard-sphere glasses and their formation}
\label{sectionGlass}

At its core, the glass problem consists of understanding how the equilibrium dynamics of a liquid that exhibits no obvious structural change grows to be so sluggish as to freeze particles in place~\cite{berthier2011}. In a materials context, while glasses are traditionally associated with cohesive supercooled liquids, dense hard spheres have also long played a key conceptual role. The geometrical simplicity of the constituent particles, in particular, enables a variety of material and theoretical approaches to explore the underlying physics. Therefore, since the 1976 numerical simulations of ~\citet{gordon1976} and ~\citet{woodcock1976}, hard spheres have 
been fueling the debate about the nature and location of the glass transition, initially with computer simulations~\cite{frenkel1980,woodcock1981pnas,woodcock1981prl,speedy1987}, and later with experimental advances, such as dynamic light scattering (Sec.~\ref{sectionLightScattering})~\cite{pusey1987prl,vanmegen1991prl,vanmegen1991pra,vanmegen1991fara}, real-space analysis (Sec.~\ref{sectionOptical})~\cite{vanblaaderen1995,weeks2000,kegel2000science,leocmach2012,hallett2018}, and rheology~\cite{mason1995,bonn2017}.

In this section, we review the contribution to our understanding of the glass transition made by work on hard spheres. We begin by providing some context (Sec.~\ref{sectionHistoricalGlass}), before discussing the early work which mainly used light scattering (Sec.~\ref{sectionReciprocalEarlyGlass}), developments made possible by real-space imaging (Sec.~\ref{sectionRealSpaceGlass}) and more recent work which has approached rather longer time scales than some of the earlier studies (Sec.~\ref{sectionDeeper}). We also consider the effects of confinement (Sec.~\ref{sectionGlassConfined}) as well as its 2d limit  (Sec.~\ref{sectionGlass2d}). We then move on to the related phenomenon of jamming (Sec. \ref{sectionJamming}, and to specific properties of glasses, such as their vibrational behavior (Sec. \ref{sectionVibrational}), aging (Sec. \ref{sectionAging}) and rheology (Sec. \ref{sectionFarGlass}).

Inevitably, this section---like many others in this review---is limited in scope. We here exclusively discuss contributions to our understanding of glasses (which has certain universal qualities ~\cite{berthier2011}) that result from hard-sphere studies. We therefore do not include important work that used other systems. For a more complete picture, we refer the reader to reviews on the glass transition~\cite{berthier2011} and on specific aspects such as mean field/high dimensional work~\cite{charbonneau2017}, dynamical heterogeneity~\cite{berthier}, dynamical and structural length scales~\cite{karmakar2014}, local structure ~\cite{royall2015physrep}, jamming~\cite{torquato2010rmp,liu2010,vanhecke2010,charbonneau2017,arceri2022}, and aging~\cite{arceri2020} and rheology ~\cite{bonn2017}.

In this section, we also make a departure from the nomenclature used in the rest of this review. Hard spheres, of course, have no liquid phase (Fig.~\ref{figHSPhaseDiagram}, Sec.~\ref{sectionEquilibriumPhaseBehavior}). However, because the glass transition in molecular systems is typically driven by cooling a liquid below its melting point, such that it is \emph{supercooled}, the analogous behavior in hard spheres is to compress the system beyond its  freezing point $P_f$ (or $Z_f$, for the reduced pressure $Z=\beta P/\rho$) as the control parameter~\cite{berthier2009pre}. Therefore, in keeping with much of the hard-sphere glass literature, and to emphasize the analogy with molecular systems, we refer to fluid state points thus compressed as supercooled liquids (instead of supercompressed fluids). Similarly,  we shall refer to state points far beyond $Z_f$ as being ``deeply supercooled''. While the (reduced) pressure $Z$ emphasizes the link with molecular systems~\cite{berthier2009pre}, some of the literature expresses state points in terms of the volume fraction. Here we prefer $Z$, but when a particular reference uses $\phi$, we will also often use it.

\subsection{Historical theoretical developments and persistent challenges}
\label{sectionHistoricalGlass}

\begin{figure}[t]
\includegraphics[width=90mm]{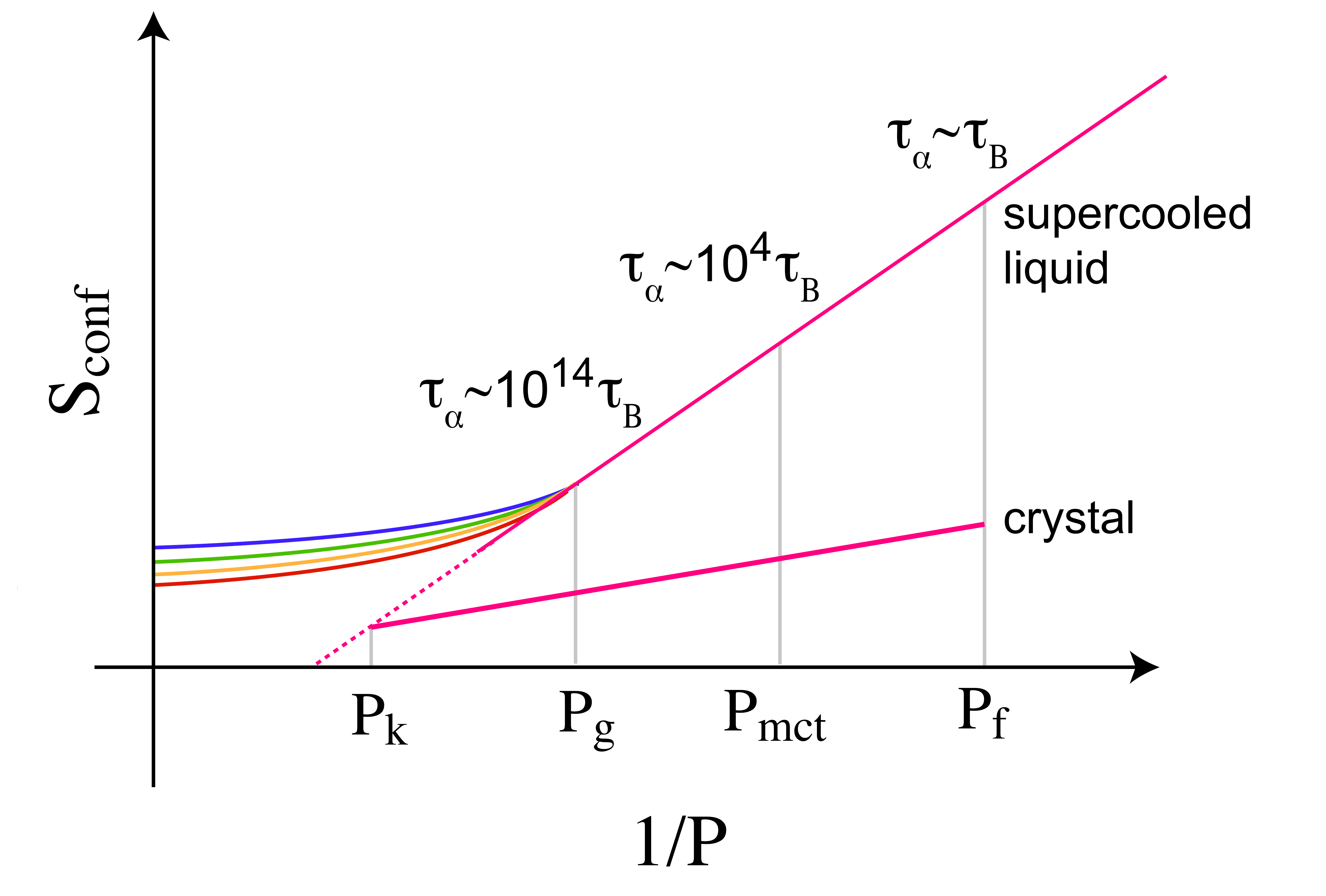}
\caption{\textbf{Roadmap to the glass transition in hard spheres.}
Configurational entropy $S_\mathrm{conf}$ as a function of the inverse pressure (osmotic pressure in the case of colloidal systems). Typically, $S_\mathrm{conf}$ of molecular liquids decreases faster than that of crystals as a function of temperature. The configurational entropy of the crystal is non-zero due to defects. At some pressure $P_k$, the (supercooled) liquid configurational entropy would fall below that of the crystal. $P_g$ is the operational glass transition, which is mapped from that of molecules where the structural relaxation time $\tau_\alpha$ typically exceeds that of the liquid by $10^{14}$ (corresponding to 100s for molecules). Further compression (fast relative to $\tau_\alpha$) leads to non-equilibrium states (colored lines). $P_\mathrm{mct}$ is the MCT crossover; $P_\text{f}$ is the freezing point. We plot this figure in terms of pressure to emphasise the connection with molecular systems~\cite{cavagna2009,kauzmann1948}.
}
\label{figCavagnaHS}
\end{figure}

Understanding the glass transition has been an active area of research for over a century~\cite{berthier2011}, but as it pertains to hard spheres, the 1980s saw the independent emergence of two major microscopic theories of the glass transition: a density-functional description of amorphous solids~\cite{singh1985,wolynes1985,baus1986,lowen1990}, and kinetic-theory based~\cite{dorfman2021} mode-coupling theory (MCT) of glasses~\cite{bengtzelius1984,leutheusser1984, kirkpatrick1985,barrat1989,fuchs1992,goetze}. Using spin glass models as inspiration, the random first-order theory (RFOT) of glasses was then proposed to unify the two descriptions~\cite{kirkpatrick1987,lubchenko2007}. In short, this framework posits the existence of a dynamical (or mode-coupling) transition at which ergodicity is lost--at reduced pressure $Z_d$ (or $Z_\mathrm{mct}$)--due to phase space then subdividing into a number of metastable states so large that it is accompanied by a thermodynamic entropy contribution, i.e., a configurational entropy (or complexity).

While this description is now understood to hold in high dimensions ~\cite{charbonneau2017,parisi2020}, in 2d and 3d, the situation is more subtle. This largely reflects the fact that this ideal scenario neglects \emph{activated} processes that can restore ergodicity beyond the MCT transition (thus turning it into a crossover) perhaps even up to the point at which the configurational entropy $S_\mathrm{conf}=S-S_\mathrm{vib}$ of the supercooled liquid is extrapolated to become equal to that of the crystal~\cite{kauzmann1948}, where $S_\mathrm{vib}$ is the vibrational contribution to the entropy~\cite{berthier2019configentropy}. In the case of molecular liquids, this would happen at a finite (Kauzmann) temperature. The analogous scenario for hard spheres is sketched in Fig.~\ref{figCavagnaHS} with the configurational entropy of the supercooled liquid becoming equal to that of the crystal at a certain pressure $P_k$. (Here we use pressure to connect with the crystal in the spirit of work with molecular systems~\cite{cavagna2009,kauzmann1948}.) In Sec.~\ref{sectionDeeper}, we see that measurements of the configurational entropy using advanced computer simulation methods confirm this picture~\cite{berthier2017}.

Now the drop in configurational entropy is understood to be accompanied by an increase in a structural length scale ~\cite{lubchenko2007}, which would 
diverge at $P_k$ (or $Z_k$) (although the transition may be avoided ~\cite{stillinger2001,royall2018jpcm}). Such a length scale may be accessed by \emph{point-to-set} correlations, which measure the impact of a frozen, disordered boundary (often spherical) on the fluid behavior away from it (often the center of the sphere)~\cite{biroli2008}. Cooperative relaxation at deep supercooling (for $P>P_\mathrm{mct}$ or $Z>Z_\mathrm{mct}$) leads to \emph{dynamical heterogeneity} with some regions relaxing slower or faster than one another in supercooled liquids~\cite{berthier}. The size of these dynamically heterogeneous regions is then characterized by a dynamical length scale.

Despite its very considerable success and broad epistemic reach, RFOT is far from being the only theory of the glass transition to have been formulated and used to interpret results from colloidal hard spheres. A well-studied structural mechanism for dynamical arrest has its roots in the work of ~\citet{frank1952}, who hypothesized that local five-fold symmetric arrangements of particles would inhibit crystallization. This line of thought was later extended into a dynamical theory built on geometric frustration which imagines domains of particles in five-fold symmetric local environments which grow with supercooling~\cite{tarjus2005}. A related proposal identifies growing structural length scales including regions of local crystalline order~\cite{tanaka2022,tanaka2005i,tanaka2005ii,tanaka2005iii}.

Yet another view revolves around facilitated dynamics which places much emphasis on dynamic correlations and the dynamical facilitation (DF) of particle mobility ~\cite{chandler2010}. While based on simplified kinetically constrained models, it has been applied to particulate systems. DF invokes 
a \emph{dynamical} phase transition between a phase in which the system relaxes quickly (the active phase) and a glassy phase with very low mobility (the inactive phase). In systems with non-trivial thermodynamics, the inactive phase has a lower configurational entropy than the active phase (somewhat akin to the crystal in the Kauzmann scenario, Fig.~\ref{figCavagnaHS})~\cite{royall2020}. Given this background and context for the glass transition for the purposes of this review, we now consider the research carried out using (or at least inspired by) colloidal hard spheres.

From a practical standpoint, a colloidal hard-sphere glass former must not (obviously) crystallize. One approach--inspired by atomistic systems such as metallic glasses--is to use binary systems as a means to suppress crystallisation. As Sec.~\ref{sectionBinary} makes clear, binary systems present a rather rich phase behavior, but these complex assemblies can be slow to form. Certain binary mixtures have therefore long been used as glass formers, e.g., the model metallic glass former Cu-Zr ~\cite{royall2015physrep}. A systematic study which used binary hard spheres as a model for these systems has identified suitable compositions and size ratios to suppress crystallization~\cite{zhang2014}.

For very asymmetric size ratios ($q\lesssim0.1$), smaller particles deplete larger ones, thus leading to effective attractions, as discussed in Sec.~\ref{sectionBinaryFluid}. The result is a more complex dynamical arrest scenario, reminiscent of  colloid--polymer mixtures~\cite{poon2002,royall2018jcp}, in which gelation competes with an attractive glass in addition to the usual hard-sphere glass~\cite{hendricks2015}. Another unexpected behavior includes a critical size asymmetry, at which anomalous collective transport of the small particles appears in a matrix of dynamically arrested large particles~\cite{sentjabrskaja2016}. The rest of this section, however, considers only relatively small size asymmetries.

\subsection{Reciprocal-space picture: early studies of the hard-sphere glass transition}
\label{sectionReciprocalEarlyGlass}

When hard-sphere experimental work took off in the 1980s, the relaxation time~\footnote{The relaxation time is typically defined as the time needed for structural correlations to decay to $1/e$ of their initial value.} window available spanned at most four decades (with respect to a simple colloidal fluid at a volume fraction $\phi\approx0.5$). Thermodynamic aspects, such as marked changes in configurational entropy, were therefore largely beyond both numerical and experimental reach~\cite{berthier2011}. As a result, early investigations mostly focused on the MCT description of the liquid dynamics, which predicts a dynamical divergence at a volume fraction $\phi_\mathrm{mct}\approx0.58$. For instance, Van Megen and co-workers undertook a series of studies of colloidal hard spheres from the MCT perspective ~\cite{vanmegen1991prl,vanmegen1991pra,vanmegen1993prl,vanmegen1993pre,vanmegen1998}, as was extensively reviewed by~\citet{sciortino2005}. Other particular highlights from this period include rheological studies of dense colloidal suspensions ~\cite{mason1995} (see Sec.~\ref{sectionFarGlass}). Given the relatively short relaxation time scales probed, \emph{activated} processes could then also be neglected, thus resulting in a reasonably good agreement with MCT predictions. The characteristic power-law growth of the relaxation time as well as the stretched exponential (or Kohlrauch) form of the correlators, in particular, were found to closely match expectations from MCT~\cite{vanmegen1998}.

\subsection{Real-space picture: Local structure and dynamic heterogeneity}
\label{sectionRealSpaceGlass}

\header{Local structure in real space}
With the use of particle-resolved studies information about certain features that are otherwise difficult to discern in molecular liquids started to emerge. For example, Van Blaaderen and Wiltzius ~\cite{vanblaaderen1995} identified the presence of local five-fold symmetric structures in colloidal glasses, as had been predicted by Frank's conjecture ~\cite{frank1952} and Tarjus' dynamical theory based on geometric frustration ~\cite{tarjus2005}. Subsequent efforts to probe specific predictions of that theory, such as growing frustration-limited domains of particles in five-fold symmetric motifs upon supercooling, however, uncovered very little evidence of the phenomenon at weak to moderate supercooling, $Z \lesssim Z_\mathrm{mct}$~\cite{royall2015jnonxtalsol,royall2018jcp,charbonneau2012,charbonneau2013jcp,dunleavy2015}. Although some growth of such domains was later found at deeper supercooling ~\cite{hallett2018,hallett2020} (see Sec.~\ref{sectionDeeper}), these findings nevertheless suggest that hard-sphere supercooled liquids are rather strongly geometrically frustrated in the dynamical regime up to the MCT crossover ($Z\lesssim Z_\mathrm{mct}$). Dynamical sluggishness can at best be only partially attributed to geometric frustration.

\begin{figure}[h!]
\includegraphics[width=70 mm]{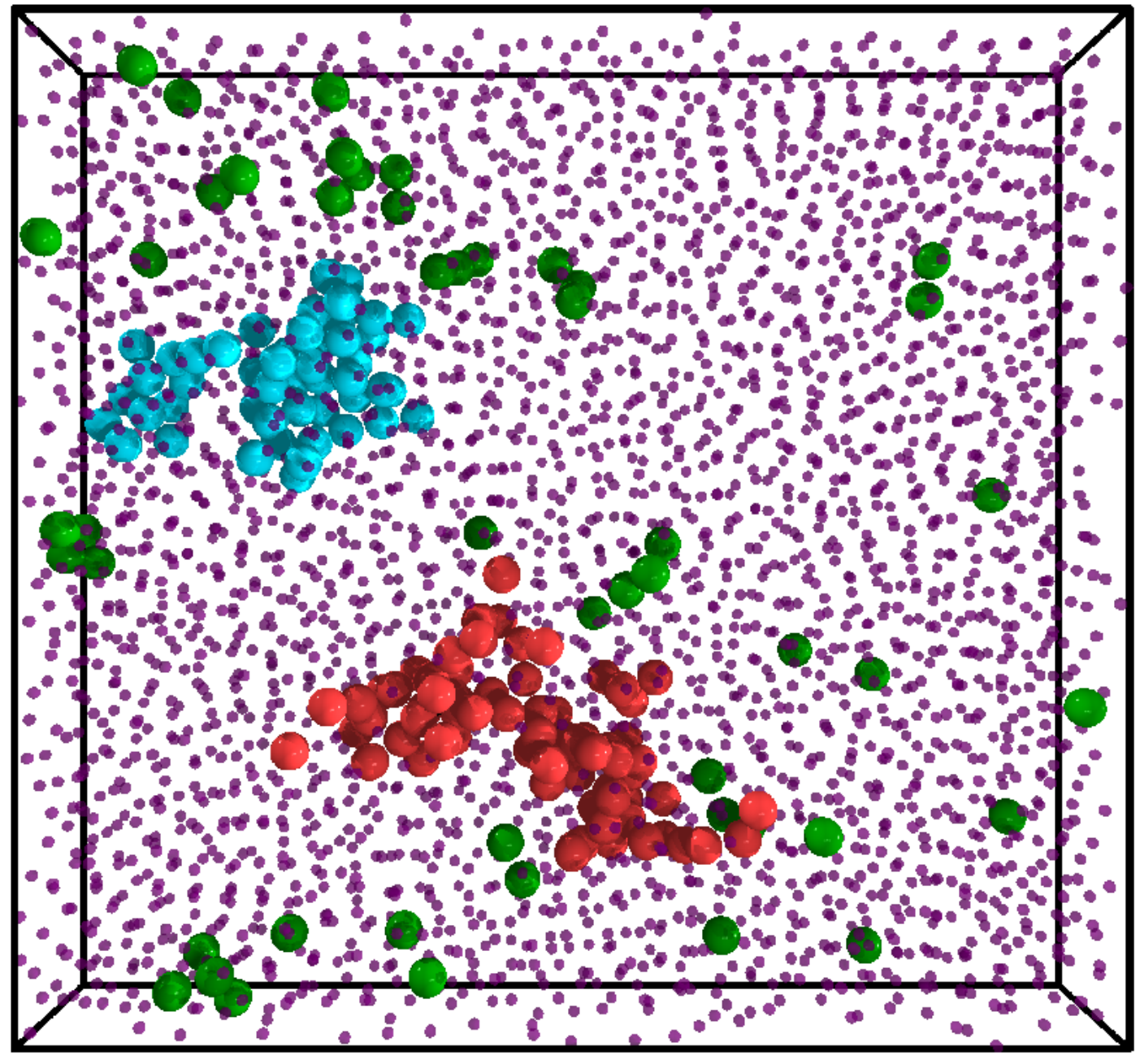}
\caption{Dynamical heterogeneity in real space is identified by highlighting the locations of the fastest (larger spheres) compared to the other (smaller spheres) particles, which are drawn smaller for clarity; the particles all have the same physical size, which is the size of the large spheres shown in this figure. Supercooled sample with $\phi\approx 0.56$, measurement time $\Delta t^*=1000$ s. The fastest particles displaced by about 0.32 $\sigma$. The red cluster contained 69 particles; the light blue cluster contained 50 particles. 
From E. R. Weeks, J. C. Crocker, A. C. Levitt, A. Schofield, and
D. A. Weitz, 2000, Science 287, 627~\cite{weeks2000}. Reprinted with permission from AAAS.}
\label{figEricWeeks}
\end{figure}

\header{Tracking colloids in supercooled liquids in time: dynamical heterogeneity}
An important contribution to our understanding of the glass transition has come through time-resolved particle-resolved studies. In the context of the glass transition, this feat was first performed in quasi-2d systems by Rice and coworkers~\cite{marcus1996,marcus1999,cui2001} and was soon after extended to 3d systems by Weeks {\em et al.} and by Kegel and Van Blaaderen ~\cite{weeks2000,kegel2000science}. Later work investigated spatially correlated clusters of slow particles ~\cite{weeks2002} that percolate across the system~\cite{conrad2006}. This work brought forward clear evidence of spatially heterogeneous dynamics as the liquid grows increasingly viscous (thus validating earlier computer simulation ~\cite{perera1996}). Colloidal observations are arguably the most explicit experimental evidence for dynamical heterogeneity in glass-forming systems, a key discovery in the field in recent decades (see Fig.~\ref{figEricWeeks}) ~\cite{ediger2000}.

This feature has since been variously interpreted, including as the fluctuations associated with an avoided dynamical transition predicted by mean-field (RFOT) theory~\cite{berthier2011} and as constituting an integral part of dynamical facilitation (DF). More recently, real-space experiments using hard spheres have also been leveraged to gain insight into DF  and to observe the dynamical and structural-dynamical phase transitions in experiment~\cite{pinchaipat2017,abou2018} and simulation ~\cite{campo2020}. Other tests of that theory include the prediction of the structural relaxation time in 2d~\cite{isobe2016} and 3d~\cite{ortlieb2023}  and the identification of the elementary units of relaxation, so-called excitations, which are localized, short-term relaxation events, seen in 2d experiments using optical tweezers ~\cite{gokhale2014,gokhale2016jsm,gokhale2016advphys}. By contrast, in other studies correlations have been found between short- and long-term relaxation associated with RFOT (see Sec.~\ref{sectionDeeper})~\cite{mishra2019}.

The various interpretations leveraged to make sense of these studies reflect an incomplete first-principle understanding. For instance, even though microscopic proposals for the origin of \emph{collective} fluctuations--traditionally captured by the dynamical susceptibility, $\chi_4(t)$--were advanced soon after dynamical heterogeneity was reported~\cite{berthier}, only recently has a first-principle explanation (based on displacements being correlated  along different dimensions of space) emerged for their \emph{single-particle} counterpart--traditionally encoded by the non-Gaussian parameter, $\alpha_2(t)$~\cite{biroli2021,folena2022}. Put simply, the consideration of dynamical heterogeneities remains an active area of research.

\header{Correlation between structure and dynamics in supercooled hard spheres}
The work which considered five-fold symmetry above notwithstanding, a number of studies have demonstrated that a significant amount of information about the dynamics is encoded in the local structure of supercooled hard spheres. ~\citet{marin2020} demonstrated a strong link between the number of local tetrahedral clusters in mixtures of hard spheres and their (local and global) dynamics. Similarly, methods based on information theory ~\cite{dunleavy2012,jack2014,dunleavy2015} and machine learning ~\cite{boattini2020, boattini2021, alkemade2022} are capable of predicting the local dynamics of glassy hard-sphere systems based on structural data alone, analogous to similar observations in other model systems (see e.g. ~\cite{bapst2020}).

In some hard-sphere systems, the growth of a static length scale has been related to a dynamical length ~\cite{kawasaki2007,tanaka2010,kawasaki2010jpcm,leocmach2012}, and to the extent to which the relaxation time increases in response to the control parameter, i.e., the fragility of the supercooled liquid~\cite{tanaka2005i,tanaka2005ii,tanaka2005iii,tanaka2022}\footnote{In molecular systems, fragility quantifies the rate of increase of the relaxation time upon cooling, with more fragile systems exhibiting a faster rate. For hard-sphere systems, the equivalent is the rate of increase of relaxation time with either $P$ (or $Z$) or $\phi$.}. In 2d systems, this length scale can be particularly significant ~\cite{tanaka2010,kawasaki2007,russo2015} (see Sec.~\ref{sectionGlass2d}). Interestingly, that length  scale often corresponds to ``medium range crystalline order'' (MRCO) ~\cite{kawasaki2007,tanaka2010}, and is therefore distinct from the five-fold symmetric local order noted above. A more general setup for detecting this length has also been considered~\cite{tong2018}. However, because most of this analysis has been carried out in the regime up to the MCT crossover $P\lesssim P_\mathrm{mct}$ (or $Z \lesssim Z_\mathrm{mct}$) range, where small length scales are typically encountered, it would be most interesting to see what happens at deeper supercooling. It is important to obtain a sharper understanding of why certain systems and analyses provide longer length scales than others ~\cite{kawasaki2007,tanaka2010,kawasaki2010jpcm,leocmach2012,dunleavy2012,dunleavy2015,marin2020,hallett2018,hallett2020} --and sometimes even longer than in molecular systems supercooled to $T_g$ ~\cite{dauchot2023} or computer simulations using swap MC (see Sec.~\ref{sectionDeeper}) ~\cite{berthier2017}. In the case of the more weakly polydisperse samples with the largest length scales, a comparison with the work of Han and coworkers ~\cite{zhang2018}, which showed a means to distinguish polycrystals and glasses could be particularly useful (see Sec.~\ref{sectionGlass2d}).

\subsection{Deeper supercooling: Beyond the mode-coupling (dynamical) crossover}
\label{sectionDeeper}

\begin{figure}
\includegraphics[width=\linewidth]{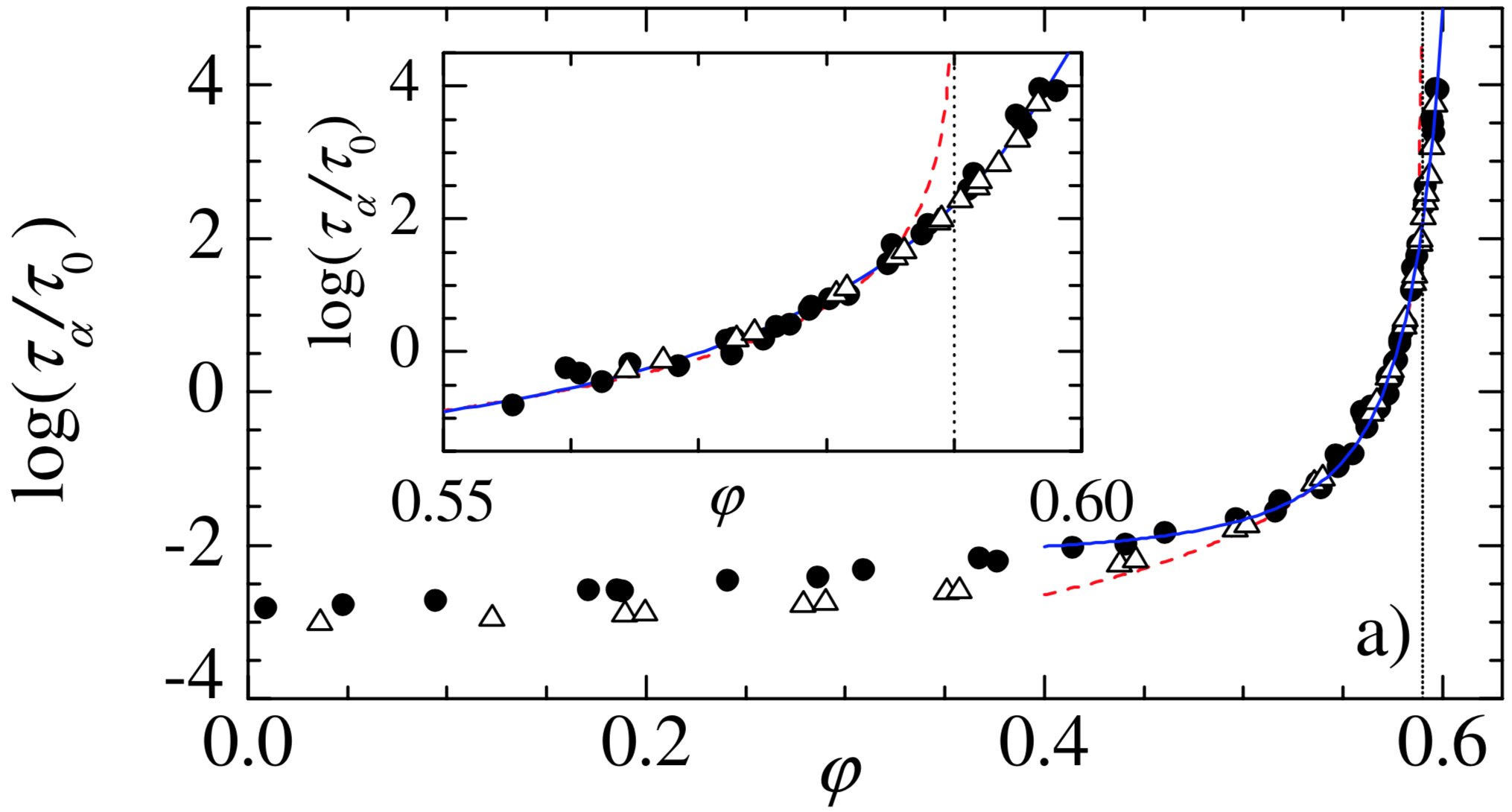}
\caption{Relaxation beyond the MCT crossover.
Relaxation time scale (for dynamic light scattering experiments (black circles) and MC simulations (open triangles), respectively, in units of ($\tau_0=1$ s and $\tau_0=7 \times 10^4$ MC steps). The red dashed line is a power-law fit to the MCT critical scaling with a transition (or, rather, a crossover) around $\phi_\mathrm{mct}=0.590$ (vertical dotted line) and exponent $\gamma=2.5$. The continuous blue line is a fit to DLS data using a modified Vogel-Fulcher-Tamman form with divergence at $\phi_\mathrm{vft}=0.637$. The inset emphasizes that the MCT singularity is absent. Reprinted figure with permission from Brambilla, G., D. El Masri, M. Pierno, L. Berthier, L. Cipelletti, G. Petekidis, and A. B. Schofield, 2009, Phys. Rev.
Lett. 102(8), 085703~\cite{brambilla2009}. Copyright (2009) by the American Physical Society.}
\label{figLucaCipeletti}
\end{figure}

Since 2010, both experiments~\cite{brambilla2009,elmasri2010,hallett2018,hallett2020} and simulations~\cite{brambilla2009} have been able to equilibrate hard-sphere liquids beyond the MCT crossover, and therefore to probe in real space the \emph{activated} processes that restore ergodicity in that regime (Fig.~\ref{figLucaCipeletti}).

\begin{figure}[h!]
\centering
\includegraphics[width=90 mm]{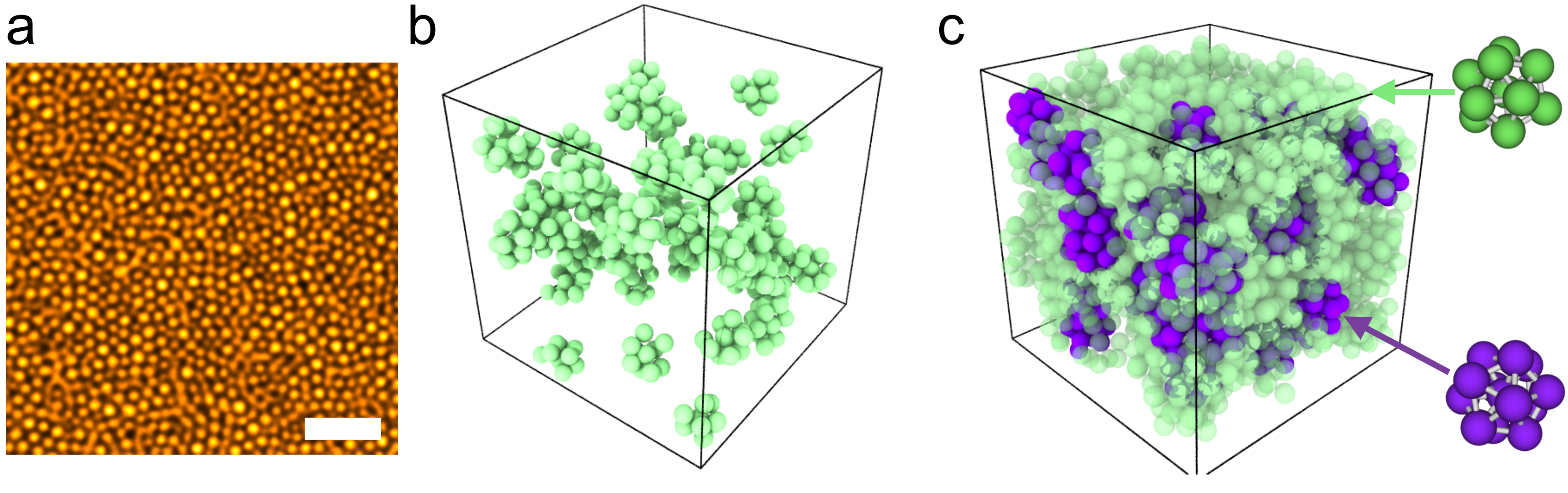}
\caption{Resolving small colloids in real space. \textbf{a}, STED nanoscopy image for $\phi=0.598$. Scale bar corresponds to 3 $\mu$m.  \textbf{b},\textbf{c}, Rendered coordinates of defective icosahedra (green, top right structure) and full icosahedra (purple, bottom right structure) for (\textbf{b}) $\phi=0.523$ and (\textbf{c}) 0.598,  respectively. Reprinted from ~\citet{hallett2018} under Creative Commons Attribution 4.0 International License  \url{http://creativecommons.org/licenses/by/4.0/}.}
\label{figJames}
\end{figure}

Using smaller colloids and concurrently improving imaging capabilities (Sec.~\ref{sectionOptical})~\cite{hallett2018,hallett2020} have markedly enlarged the range of experimentally accessible relaxation time scales (see Fig.~\ref{figJames} and Sec.~\ref{sectionSize}). Among the insights afforded--in addition to the growth of local structures with five-fold symmetry--is the increase of a structural length scale. A length scale may also be extracted from the dynamical heterogeneity ~\cite{lacevic2002}, which was found to grow concurrently with the structural length scale. Although both exhibited scaling compatible with RFOT-based arguments 
\begin{equation}\xi(Z)=\xi_0\left(\frac{1}{Z_\mathrm{vft}-Z}\right)^\frac{1}{3-\theta},
\label{eqJamesLength}
\end{equation}
where $\xi_0$ is a length at higher pressure and $P_\mathrm{vft}$ (or $Z_\mathrm{vft}$) is the compressibility corresponding to the dynamical divergence of the Vogel-Fulcher-Tamman fit and $\theta\approx2.05\pm0.1$ ~\cite{hallett2018}, this work is still well short of the 14 orders of magnitude increase in relaxation time corresponding to the operational glass transition. An intriguing means to use still smaller particles (which could access longer effective time scales) could be to sacrifice particle-resolved imaging and instead measure fluorescence recovery from photobleaching~\cite{vanblaaderen1992jcp,simeonova2004}.

An interesting new direction for experiments with small particles benefits from recent developments in synchotron intensity with X-rays~\cite{wochner2009,lehmkuhler2020,liu2022,striker2023}. This has enabled the development of methods such as X-ray photon correlation spectroscopy XPCS and microbeam X--ray scattering, X--ray cross--correlation analysis (XCCA). XPCS is the X-ray equivalent of DLS and is thus sensitive to dynamic information, though at much smaller wavelengths than DLS. Remarkably, techniques based on the latter two can reveal higher--order structure, and have been used to identify local five-fold symmetric order in systems of 100 nm diameter silica particles~\cite{wochner2009,lehmkuhler2020,liu2022}, i.e. smaller than what has been achieved even with super--resolution microscopy. Furthermore, XCCA can even be coupled with XPCS to combine higher--order structure and dynamical measurements~\cite{striker2023}. The ability to access in principle very much smaller particles shows that these methods may enable access to much longer timescales (in terms of $\tau_B$), enabling equilibration at higher reduced pressure $Z$ than has yet been achieved.

These advances, however, pale in comparison with the scale of the numerical developments based on the swap MC algorithm ~\cite{ninarello2017}. Prior studies of hard spheres had mostly considered minimally size polydisperse mixtures, which benefit from a certain proximity to theoretical frameworks and experiments, while suppressing crystallization (see, e.g.,~\cite{bernu1985,mountain1987,eldridge1995}), but Berthier and coworkers have shown that astronomical sampling gains can be obtained by enabling diameter exchanges in broadly polydisperse mixtures of hard spheres~\cite{ninarello2017}. Although these ergodicity restoring processes correspond to an extraordinarily unphysical dynamics--that at best indirectly informs our understanding of actual liquid dynamics~\cite{wyart2017,ikeda2017,berthier2019lengthscale}--the reach beyond the MCT scaling regime of resulting equilibrium configurations has no equal. It can match and even surpass the experimental glass transition of molecular liquids, where the relaxation time is some $10^{14}$ longer than in the liquid. Interestingly, the relatively high polydispersity required for swap MC to suppress crystallization seems also to reduce the degree of five-fold symmetric structures, thus suggesting a lack of universality for the kind of structural approach employed in geometric frustration~\cite{coslovich2018}.

\begin{figure}[h!]
\centering
\includegraphics[width=\linewidth]{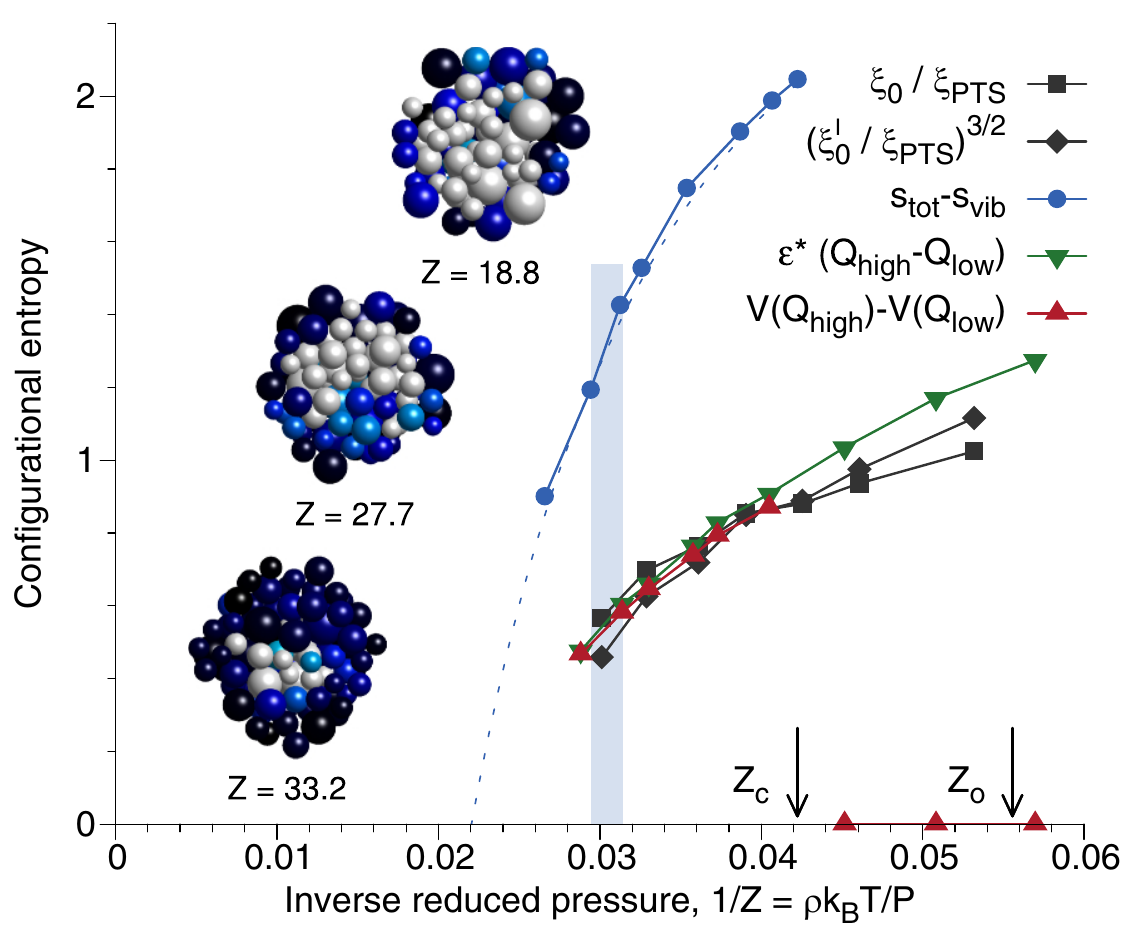}
\caption{Different estimators of the configurational entropy consistently extrapolate to their vanishing at a finite reduced pressure, congruous with the existence of a Kauzmann-type entropy crisis. The inset encodes the growing extent of amorphous order (dark colors) as pressure increases. Reproduced from Berthier, L., P. Charbonneau, D. Coslovich, A. Ninarello,
M. Ozawa, and S. Yaida, 2017, Proceedings of the National
Academy of Sciences 114, 11356~\cite{berthier2017}.}
\label{figConfEntropy}
\end{figure}

As shown in Fig.~\ref{figConfEntropy}, these configurations have notably enabled crisp complexity measurements in hard sphere~\cite{berthier2017} and hard disk~\cite[SI]{berthier2019ncomms} models. These results provide an unprecedented test of the thermodynamic complexity vanishing at a Kauzmann-like entropy crisis at $Z_k$.

These same configurations have also been used to study certain features of a proposed ergodicity restoring process. Given that equilibrium liquid configurations are metastable (in a mean-field sense) beyond the MCT crossover, it has long been proposed that a nucleation-like process should dominate dynamical relaxation. The growth of the point-to-set metastability length associated with the growing amorphous order that underlies that mechanism has even been detected~\cite{berthier2017,berthier2019lengthscale} significantly beyond what was previously possible~\cite{biroli2008}. The relationship of these observables to actual activated dynamics, however, remains far from controlled. Recent studies have isolated the contribution of an altogether different, hopping-like, relaxation mechanism~\cite{biroli2021}. The seeming robustness of the coupled (and cooperative or facilitated) relaxation of localized features~\cite{chacko2021,kapteijns2021,guiselin2022,ortlieb2023} suggest that a great deal of conceptual tension remains to be resolved.

Other approaches to assess the validity of the RFOT description have also been devised. Evidence for a drop in configurational entropy has been inferred by comparing  different regions in deeply supercooled colloid experiments ~\cite{hallett2018}. Alternatively, in deeply supercooled liquids, \emph{pinning} a fraction of the particles in place also reduces the configurational entropy, thus bringing the system closer to the Kauzmann transition without further equilibration  ~\cite{cammarota2012pnas}. In colloidal experiments, a similar setup has been achieved with optical tweezers  ~\cite{gokhale2014,gokhale2016jsm,gokhale2016advphys} and by adhering colloids to a substrate ~\cite{williams2018}. Spherical confinement (see Sec.~\ref{sectionSphericalConfinement}) has been used to investigate amorphous order in supercooled liquids through measurements akin to cavity point-to-set correlations (albeit without an equilibrated boundary). Experiments confining particles within an emulsion droplet have similarly revealed growing structural length scales~\cite{zhang2016}.

Identifying  transitions between metabasins in the free-energy landscape ~\cite{rodriguezfris2011,rodriguezfris2018} through measuring the fractal dimension of so--called cooperatively rearranging regions (CRRs) has also been investigated experimentally. Early results were found to be consistent with more compact CRRs at deep supercooling~\cite{nagamanasa2015}, but equilibrating these conventional colloidal systems past the MCT crossover, where such compaction is expected ~\cite{berthier2011}, is difficult. Smaller colloids have since confirmed the compaction of CRRs more convincingly ~\cite{ortlieb2023}. The particular scaling properties of CRR surfaces ~\cite{biroli2017} have also been measured in colloidal systems for $P \lesssim P_\mathrm{mct}$ (or $Z \lesssim Z_\mathrm{mct}$) and found to be consistent with predictions ~\cite{ganapathi2018}.

Obtaining a clear understanding of the actual relaxation  dynamics in this regime nevertheless remains fraught with theoretical and experimental challenges. Whichever way this problem moves forward, hard-sphere models and experiments will no doubt be involved in moving our comprehension along.

\subsection{The hard-sphere glass transition under confinement}
\label{sectionGlassConfined}

In molecular glass-forming systems, the effect of confinement has long been a challenge to understand, because contradictory effects on the relaxation time have been reported ~\cite{richert2011}. Unfortunately, corresponding experimental studies are few and far between, and lie in the weakly supercooled $P \lesssim P_\mathrm{mct}$ (or $Z \lesssim Z_\mathrm{mct}$) regime. They therefore cannot claim to resolve all associated difficulties. Confinement has nevertheless been shown to robustly induce layering (see Sec.~\ref{sectionConfinement}), which has a profound effect upon dynamical heterogeneity, and markedly increases the overall relaxation time~\cite{nugent2007,edmond2012}.

Using walls at which the mobility can be controlled, boundary mobility has been shown to play an important role in the relaxation of confined hard spheres  ~\cite{hunter2014}. Gradients in dynamics with respect to the boundary appear for more mobile boundaries, whereas for less mobile boundaries gradients are almost entirely suppressed. One quasi-2d system using \emph{adaptive} confinement (see Sec.~\ref{sectionConfinement}) revealed the emergence of a faster relaxation mechanism at high area fraction, leading to ``negative fragility'', that is to say, the relaxation time increasing in an ``sub--Arrhenius like'' manner ~\cite{williams2015jcp}. For a moderately polydisperse, densely packed hard-sphere fluid confined between two smooth hard walls, EDMD simulations showed the emergence of reentrant glass transitions depending on the wall separation, in agreement with MCT predictions~\cite{mandal2014}.

\subsection{The glass transition in 2d hard spheres}
\label{sectionGlass2d}

Two-dimensional glass-forming systems differ significantly from  their 3d counterpart. First, the traditional geometrical frustration argument is turned on its head. For hard disks, the local liquid structure is hexagonal, as is the crystal, and therefore no geometrical frustration is expected. As a result, simulations have revealed~\cite{kawasaki2007,tanaka2010} and experiments confirmed~\cite{tamborini2015} that structural correlations are longer-ranged than in the 3d systems mentioned in Sec.~\ref{sectionHistoricalGlass}. 
Second, Mermin--Wagner---like fluctuations result in dynamical correlations that are profoundly different from those in 3d ~\cite{flenner2015}, as confirmed in experiment ~\cite{vivek2017}.

\subsection{Vibrational properties of hard-sphere glasses} 
\label{sectionVibrational}

Molecular (rather than colloidal) glasses exhibit unusual vibrational properties. Among these is the boson peak, an excess density of states with respect to the Debye scaling of specific heat found in crystalline materials ~\cite{berthier2011}. Colloids of course exhibit overdamped dynamics and therefore do not have a proper vibrational spectrum. It is nevertheless possible to imagine a \emph{shadow system} with Newtonian interactions that features the same set of configurations ~\cite{chen2010prl}. A number of studies have thereby been able to deduce an effective density of states of soft vibrational modes ~\cite{ghosh2010prl,ghosh2010sm,kaya2010}. This approach has led to experimental evidence for such a boson peak in hard-sphere colloidal systems~\cite{ghosh2010prl,ghosh2010sm}.

The vibrational properties of a hard-sphere glass are also related to the free volume available for each particle. This free volume can be interpreted using a cell--theory----like  analysis (Sec.~\ref{sectionCellTheory}) on the real-space Voronoi volumes obtained in hard-sphere colloidal glasses. This analysis has demonstrated a  decrease in the effective (vibrational) free energy during aging ~\cite{zargar2013}. The local free energy has further been shown to display  strong spatial and temporal heterogeneity, and changes in free energy between consecutive snapshots have been correlated algebraically with particle rearrangements. The vibrational properties of the glass have also been shown to correlate with its local free energy, displaying a large excess of low-frequency modes~\cite{zargar2014,dang2022}, despite the limitations of cell theory in the context of disordered materials (see Sec.~\ref{sectionCellTheory}).

\subsection{The jamming transition and its influence} 
\label{sectionJamming}

In parallel to the study of (equilibrium) glass-forming liquids, various efforts have considered the properties of (out-of-equilibrium) glasses. Given that hard spheres when crunched sufficiently rapidly form disordered \emph{jammed} solids, these systems have played a key role in this context as well. As many thorough and complementary reviews are available on this topic~\cite{torquato2010rmp,liu2010,vanhecke2010,charbonneau2017,arceri2022}, we here focus on aspects and questions that have been neglected, especially in the context of recent advances.

\begin{figure}[h!]
\centering
\includegraphics[width=0.95\columnwidth]{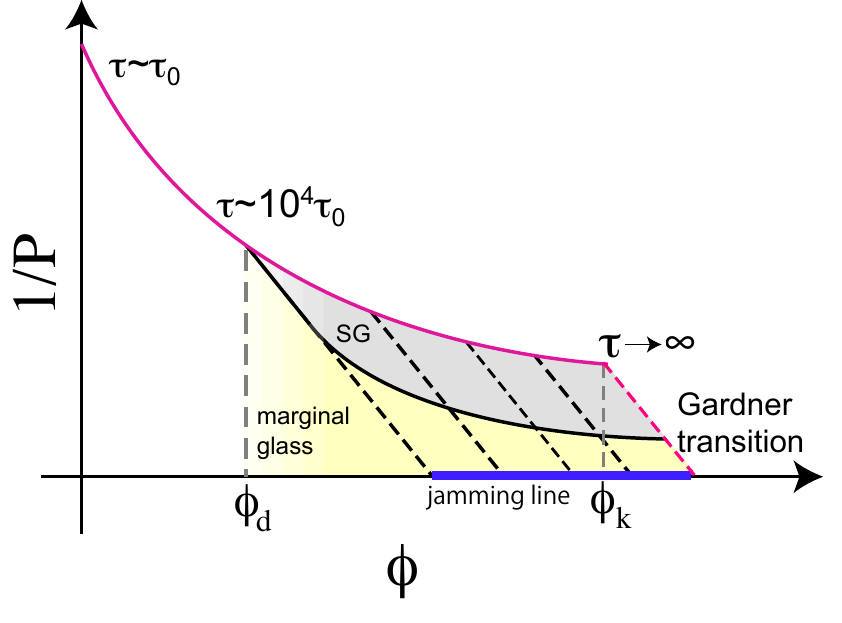}
\caption{Roadmap to the Gardner transition. The inverse pressure $1/P$ is shown as a function of the volume fraction $\phi$, with the pink line corresponding to the equilibrium equation of state.  At low volume fraction, the system is a fluid but for $\phi>\phi_d (\equiv\phi_\mathrm{mct})$ the system supports many metastable states with a range of pressures, which may be either stable glasses (dark shaded region marked SG) or marginal glasses (pale shaded region). These states are distinguished by the anomalous elastic response of the marginal glass (see main text).  The particles are hard, so jammed states correspond to $1/P=0$ (on the horizontal axis), which are all marginal glasses. Dashed lines describe compressions fast relative to $\tau_\alpha$ on the equilibrium line.
}
\label{figPatrick}
\end{figure}

In the 1960s, as the equilibrium phase diagram of hard spheres was still being debated,  Bernal and others were constructing out-of-equilibrium disordered (or glass-like) solids out of compressed ball bearings. While attempting to capture structural properties of the liquid state, they instead obtained a first controlled model of hard-sphere jamming~\cite{bernal1960,finney2013}, and of the so-called random close packing volume fraction, $\approx64\%$~\cite{scott1969}\footnote{Bernal also observed near perfect isostaticity of the interparticle contacts, anticipating a physical hallmark of the phenomenon by decades~\cite{vanhecke2010}, but did not continue.}. As computational capabilities grew and became more broadly accessible, the experimental challenges involved in these crunching studies quickly became more onerous than simulating the same process on computers. In the 1970s, Finney and coworkers therefore initiated the numerical study of amorphous close-packed  hard-sphere binary mixtures~\cite{finney1970,finney1977,boudreaux1977}.

A second approach to jamming built on the observation that crunching hard spheres results in their pressure diverging, Angell {\em et al.} proposed that singularities of the (resummed) hard-sphere virial expansion might be related to the jamming singularity~\cite{woodcock1981prl,song1988}. The simplicity of this thermodynamics-based approach was appealing, which probably explains why similar schemes were still considered decades later~\cite{kamien2007prl,woodcock2013}. Dedicated numerical efforts, however, clearly distinguished the equilibrium liquid branch from the continuum of out-of-equilibrium glass branch(es)~\cite{speedy1994,rintoul1996,robles1998}. A particularly telling (and model-free) evidence against the virial-series scheme is the non-uniqueness of the amorphous close-packed density. Depending on the preparation protocol, a wide range of possible densities can indeed be reached~\cite{brambilla2009,hermes2010jpcm,hermes2010epl,ozawa2012,charbonneau2021prl}.Although an initial explanation ascribed this range to partial crystallization of the system~\cite{torquato2000}, the effect has now been shown to exist independently of this ordering~\cite{charbonneau2017}.

~\citet{edwards1989} formulated a third influential proposal for dealing with jamming by using an equilibrium statistical mechanics-like description. Despite the approximate nature of this description for any actual jamming protocol~\cite{charbonneau2017}, it turned out to be a versatile and influential framework for granular experiments over the following decades~\cite{baule2018}. Analysis of jamming based on this scheme, however, also suffered from identifying a unique terminal density.

A fourth proposal for unifying glass formation and jamming is less prescriptive but has been more conceptually productive. In the late 1990s, ~\citet{liu1998} presented a framework for relating glass formation and jamming, commonly known as the jamming phase diagram. Although the study of hard-sphere glasses and jamming would remain fairly distinct for at least another decade, this proposal seeded a substantial effort dedicated to understanding the criticality of jamming~\cite{charbonneau2017}. 
Perhaps one of the most physically stunning features is that its criticality is largely independent of physical dimension~\cite{goodrich2012,charbonneau2021finite}. As a result, certain critical features of jamming remain robustly invariant in going from the exactly solvable limit of $d\rightarrow\infty$ -- obtained through a full replica symmetry breaking calculation -- down to amorphous packings of 2d disks~\cite{charbonneau2015,charbonneau2021finite}. The corrections that do appear are largely localized, such as rattlers (floaters), and bucklers ~\cite{charbonneau2017}, and can be geometrically identified through simple criteria. Low-energy excitations around jammed configurations, however, remain largely universal ~\cite{charbonneau2016,shimada2020,kapteijns2018}.

A key set of remaining theoretical challenges concerns the relationship between (equilibrium) hard-sphere liquids and (out-of-equilibrium) jammed configurations. A rich variety of crunching schemes have been designed for hard-sphere configurations to reach jamming~\cite{lubachevsky1990,torquato2010pre,lerner2013,charbonneau2021prl,artiaco2022}, but despite marked theoretical advances~\cite{parisi2020}, guidance even in simplified limits remains largely incomplete.

For equilibrium hard-sphere configurations well beyond the MCT crossover, a quasi-static compression analysis that neglects activated processes predicts the existence of an intermediate Gardner transition at which marginal stability emerges and then persists. By marginal stability it is meant that the system can be perturbed by the smallest external force, that is to say, its susceptibility diverges (see Fig.~\ref{figPatrick}). 
Various features of this transition have been reported in hard-sphere simulations, including growing structural and dynamical correlations~\cite{charbonneau2014,charbonneau2017}. A similar phenomenology has been predicted and reported under shear~\cite{urbani2017,jin2017,jin2018,jin2021}. In all cases, however, the thermodynamic character of this transition remains an open area of research~\cite{berthier2019configentropy,li2021}. Experimental validation of many of these findings using colloids is also an open challenge. Although clever detection schemes have been devised~\cite{hammond2020}, identifying signatures of the Gardner transition using particle-resolved studies requires a high precision of coordinate tracking with respect to the particle diameter which can be achieved in vibrated granular systems~\cite{seguin2016,xiao2022,kool2022} but is very challenging with colloids (see Sec.~\ref{sectionOptical}). 
Given the 2d nature of these granular systems, however, whatever unusual behavior reported for these systems is expected to wash out in the thermodynamic limit.

For lower density configurations, dynamical mean-field theory (DMFT) equations, which also neglect activation, have been formulated~\cite{agoritsas2019}. This DMFT would be expected to provide similar theoretical guidance to simulations and experiments, but only in very limited circumstances have they been solved thus far~\cite{manacorda2022}. In the absence of an explicit solution to the DMFT equations, evidence obtained by theoretical analogy and from numerical schemes provides some physical guidance. These indirect approaches have uncovered certain robust physical features. First, an onset liquid density $\phi_\mathrm{on}<\phi_\mathrm{MCT}$ seems to exist for various crunching algorithms~\cite{morse2022}. Below this density the jamming density is invariant; beyond it it increases. The numerical value of this low jamming density, or the identification of a crunching scheme to access it from hard-sphere configurations, however, remains the object of debate. In particular, a recent theoretical proposal suggests that a significantly lower jamming might be achievable for a particular (non-hard sphere-based) algorithm~\cite{manacorda2022}. Whether a Gardner-like behavior might also emerge from crunching low-density configurations also remains an open question~\cite{charbonneau2021prl}. In any case, little experimental consideration of these matters has yet been undertaken.

\subsection{Aging in hard-sphere glasses}
\label{sectionAging}

\begin{figure}[!t]
\begin{center}
\includegraphics[width=80 mm]{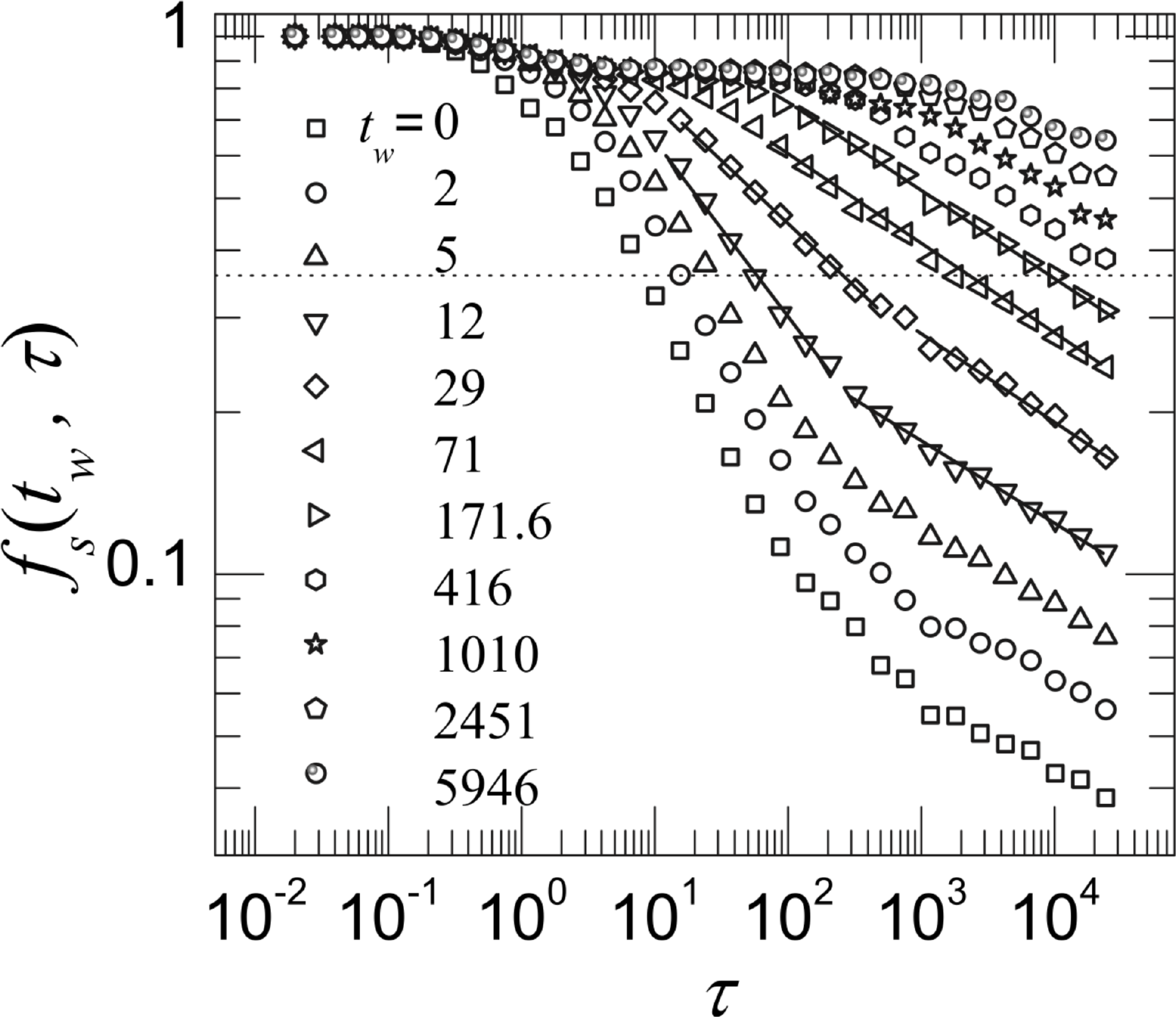}
\caption{Aging behavior in structural relaxation. Upon increasing the waiting time (different symbols correspond to different waiting times), the intermediate scattering function decays more and more slowly as a function of time (here $\tau$ in microscopic units for hard spheres). This corresponds to the system ``sinking lower and lower in its energy landscape''. Reprinted figure with permission from El Masri, D., L. Berthier, and L. Cipelletti, 2010, Phys. Rev.
E 82, 031503~\cite{elmasri2010}. Copyright (2010) by the American Physical Society.}
\label{figLucaCipelettiAging}
\end{center}
\end{figure}

Another important out-of-equilibrium phenomenon in glasses is their aging. The term then refers to the  physical properties of a material displaying a slow evolution in time after a sudden quench into the glass regime. Put differently, aging is observed when the relaxation time of the system exceeds the experimental observation time, thus breaking time translation invariance. In order to describe aging an explicit dependence on the \emph{waiting time} (the time passed since the initial quench) $t_w$ is introduced to all structural and dynamical properties.

A large body of theoretical work on aging has focused on mean-field models~\cite{cugliandolo1993,cugliandolo1994}, which describe a system that is trapped in an energy landscape in which barriers of all possible heights are present, and time correlation functions never completely decay. \emph{Trap models} provide a physically intuitive picture~\cite{bouchaud1992,denny2003} by describing phase space as a large collection of metastable states, thus resulting in a broad distribution of trapping times. These theoretical approaches make interesting predictions of \emph{universal} behavior, such as the existence of a long-time stationary regime where time-correlation functions decay as power-laws of $t/t_w$, and the possibility of defining an effective temperature, $T_\mathrm{eff}$ (or for hard spheres, $P_\mathrm{eff}$ or reduced pressure $Z_\mathrm{eff}$), to describe the downhill motion of the system in the free-energy landscape~\cite{crisanti2003}. Recent advances, however, have revealed that these broad universality claims are unwarranted. Even simple mean-field models can exhibit aging processes that are much richer~\cite{folena2022,folena2023}. As a result, the theoretical framework for describing aging remains somewhat fragile. Various studies have nevertheless explored this regime using the canonical framework of ~\citet{cugliandolo1993}.

Although the time window over which colloidal hard-sphere relaxation can be observed is limited, key insight can be obtained by accounting for the non-ergodic state of the glass phase. In this context, optical techniques are key~\cite{pusey1989physica}. A single DLS experiment measures the time-averaged time correlation function of the intensity of a single speckle. While for an ergodic system this function is equal to the ensemble-averaged one, in the glass state the sample explores only a limited region of phase space, even over very long times, and it is therefore non-ergodic.  Early studies have resolved this issue by repeating the measurements over a large number of observations~\cite{pusey1987prl}, but 
~\citet{pusey1989physica} have developed a procedure, based on approximating the fluctuating component of the density fluctuations with a Gaussian field of zero mean, to extract the intermediate scattering function by using a single measurement of the time-averaged correlation function, and a measurement of the ensemble-averaged intensity (which can be obtained quickly by scanning  rapidly the system through the laser beam). To measure slow decays, other methods have been introduced, such as the method of echoes~\cite{pham2004} (for which the sample is continuously rotated during the measurement) as well as the multispeckle~\cite{bartsch1997,cipelletti1999,elmasri2005}, and the time-resolved correlation~\cite{cipelletti2002,elmasri2005} techniques.

The first observation of aging in a hard-sphere colloidal glass was reported by ~\citet{vanmegen1998}, where the self-intermediate scattering function was measured via DLS by tuning the refractive index of a mixture of optically different, but equally sized, PMMA particles. These experiments showed the waiting-time dependence of the long-time decay of the relaxation functions. These studies were followed by multispeckle and time-resolved correlation studies~\cite{elmasri2005} that confirmed the observation of aging in measurements of the intensity of the correlation function, with relaxation times showing aging also at early times, with a possible plateau at later times. The decay of the intermediate scattering function was shown~\cite{martinez2008} to change from a simple exponential dependence at short waiting times, to an algebraic dependence on time at long waiting times, thus agreeing with the \emph{aging time superposition principle} that was deduced from mean-field models~\cite{bouchaud1992}.

In real-space experiments, dynamics was found to slow upon aging, consistent with expectations and with the reciprocal space work noted above~\cite{courtland2003}. Studies on binary hard-sphere glass formers have revealed that relaxation can be dominated by the smaller species and that these can facilitate the relaxation of larger particles~\cite{lynch2008}.

Computational studies of aging in hard-sphere systems, which can access up to six decades of relaxation times, have managed to access the decay of time correlation functions for longer waiting times. In particular, ~\citet{elmasri2010} investigated nearly hard spheres with a polydispersity of $s\sim 11.5\%$ at packing fractions in the range $\phi\in[0.553,0.662]$ and for different waiting times. Aging of structural quantities was shown to be compatible with either a power law or a slow logarithmic decay. The self-intermediate scattering function, as shown in Fig.~\ref{figLucaCipelettiAging}, displayed at least two distinct decay regimes depending on the waiting time, with the decay at long waiting times following a power law with time $f_s\sim\tau^{-a}$, with $a \sim 0.15$ suggesting that relaxations occur over a broad time window. The time $\tau_a$ after which the asymptotic stationary state is observed increases with packing fraction, and  at the higher volume fractions is often outside experimental observation. The relaxation time is found to change from a superaging exponential behavior in the waiting time before $\tau_a$, to the linear dependence in waiting time after $\tau_a$. The asymptotic regime is also characterized by a subdiffusive behavior of the single-particle dynamics, with the Van Hove functions displaying fat tails at large and small particle displacements. Particles with fast displacements are found to be around $4-5\%$ of the system and to have correlated motions limited in space, with the overall displacement of the particles during the aging regime being almost negligible. Despite the subdiffusive dynamics during the aging regime, simulations of weakly polydisperse hard spheres have shown that crystallization can still occur in the bulk~\cite{zaccarelli2009}. Such crystallization without diffusion~\cite{sanz2011,russo2012,sanz2014,yanagishima2017,yanagishima2021} is connected to avalanche-like processes, as discussed in Sec.~\ref{sectionBeyondCNT}.

\subsection{Hard-sphere glasses far from equilibrium}
\label{sectionFarGlass}

\header{Steady--state shear}
Bulk rheology experiments (Sec.~\ref{sectionRheology}) have also been used to study the far-from-equilibrium response of hard-sphere glasses ~\cite{ballesta2008,koumakis2008,koumakis2012,koumakis2013,ballesta2016}. (Phenomena specifically associated with hard-sphere fluids, including shear thinning and thickening, are covered in Sec.~\ref{sectionRheology}.) In steady state these experiments provide flow curves $\tau(\dot\gamma)$ similar to Fig.~\ref{figRheology}(b). Because relating microscopic structure and dynamics often requires theoretical and computational insights, this approach has provided a fertile ground to test theoretical predictions including, \emph{inter alia}, the extension of MCT to treat shear in the ``integration through transients'' approach~\cite{fuchs2005jpcm,brader2008,brader2012} and the effective model of soft glassy rheology~\cite{sollich1997,fielding2014}. MCT predictions of flow curves have been found to be consistent with rheology experiments on emulsions~\cite{mason1995}, microgels, save for some hydrodynamic effects that can be taken into account with rescaling~\cite{fuchs2005jcp}. Steady-state shear is amenable to confocal microscopy, provided the frame rate is fast enough ~\cite{besseling2007,besseling2009}. Particle--resolved data reveals local displacements and the effect of shear can reveal details of shear bands such as shear-concentration coupling ~\cite{besseling2010,chikkadi2014} and Bingham-like slip behavior~\cite{ballesta2008,ballesta2012}.

Strongly confined, quasi-2d systems also offer insights into flow behavior. By animating optical traps confining a circular assembly of particles, the flow field which defines the viscosity has been resolved at the single-particle level. Notably, hard-disk systems have been shown \emph{not} to exhibit any massive increase in viscosity under confinement down to a few diameters (unlike many molecular systems)~\cite{williams2022}. This effect was attributed to the absence of Van der Waals interactions. In the same system, particle-resolved data has identified the mechanism of slip between layers~\cite{williams2016}.

\begin{figure}
\centering
\includegraphics[width=\linewidth]{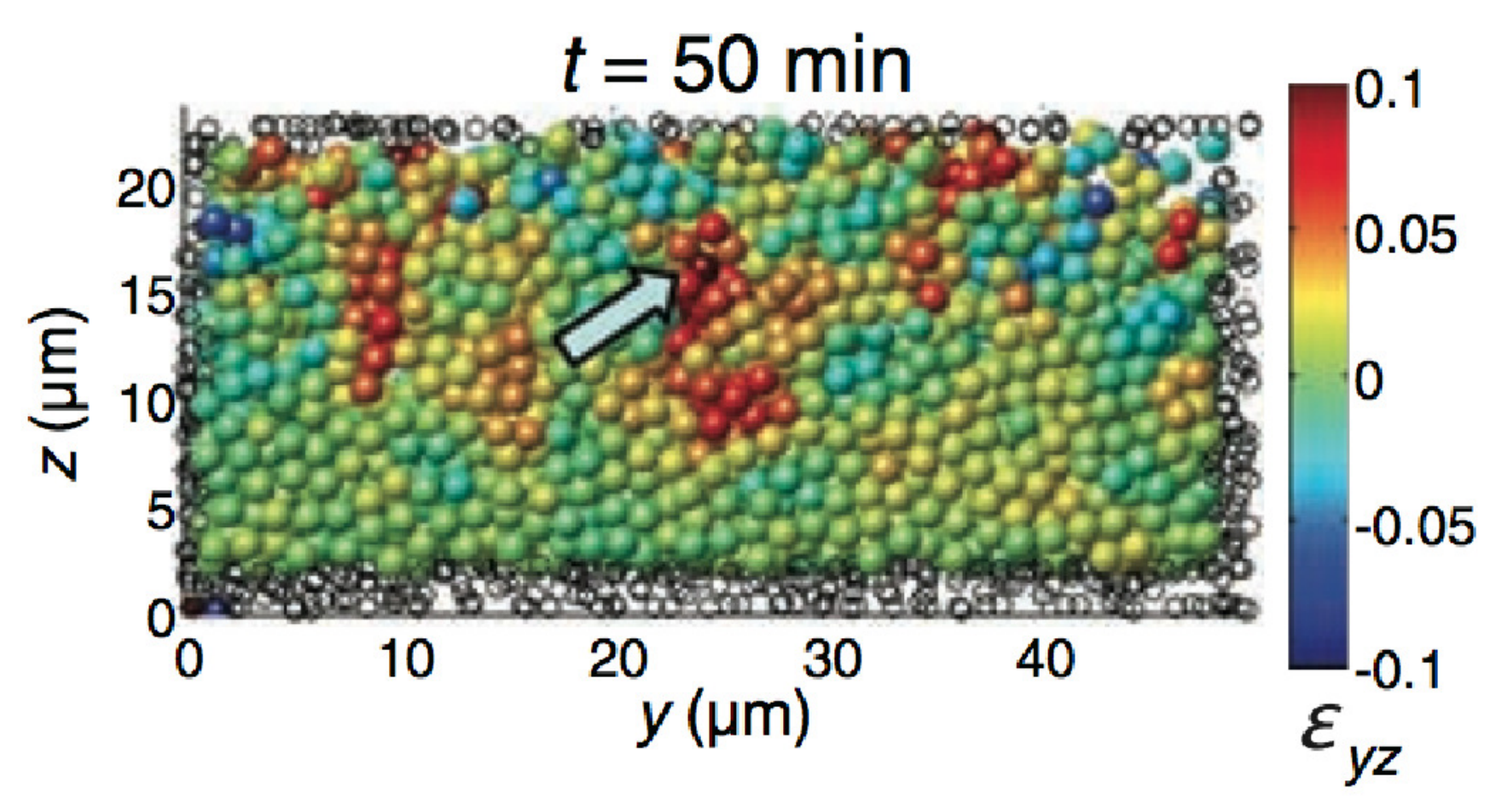}
\caption{Response of a 3d colloidal glass to linear shear strain. 
Cumulative strain $\gamma$ after 50 minutes with the particle color denoting the degree of strain. Arrow points to a shear transformation zone, which subsequently relaxed. From Schall, P., D. A. Weitz, and F. Spaepen, 2007, Science
318(5858), 1895~\cite{schall2007}. Reprinted with permission from AAAS.}
\label{figPeterSchall}
\end{figure}

\header{Yielding}
Upon applying a load to an amorphous solid, yielding occurs at a (reasonably) characteristic yield stress $\tau_y$ ~\cite{bonn2017}. Prior to yielding, the otherwise elastic response is punctuated by stress release in discrete plastic events known as shear transformation zones (STZ)~\cite{falk1998}. Under certain conditions (typically a slow shear rate, as characterized by the Weissenberg number 
(see Table~\ref{tablePeclet}), a stress overshoot is found, where the steady stress is less than the yield stress $\tau_\mathrm{steady}<\tau_y$. With confocal microscopy STZs may be directly visualized. Figure~\ref{figPeterSchall} demonstrates how tracking single particles in a dense suspension of hard spheres allows to build local strain maps and to identify STZs~\cite{schall2007}. Long-range strain correlations may also be investigated~\cite{chikkadi2011,chikkadi2012,mandal2013}. Further work using confocal microscopy has revealed a microstructural anisotropy in the extension axis where the maximum of the pair distribution function exhibits a minimum at the stress overshoot ~\cite{koumakis2012,koumakis2016}. Related work by some of the same authors found super-diffusive dynamics approaching yielding which is akin to (albeit underestimated by) that predicted by MCT. The stress overshoot was also rather weak, likely due to the measurements having been made in the supercooled liquid ($\phi\approx0.56$)~\cite{zausch2008,laurati2012,sentjabrskaja2014}. MCT and BD simulations of 2d hard disks qualitatively capture yielding predictions from the theory~\cite{henrich2009}. Binary hard sphere glasses have also been studied, and for size ratio $q=0.2$, some weakening of the system upon inclusion of the smaller species has been reported~\cite{sentjabrskaja2013}, a phenomenon likely related to the fluidization of quiescent systems with depletion attraction~\cite{pham2002,royall2018jcp}.

Creep describes the very slow flow that is observed at stresses below the yield stress $\tau_y$. The phenomenon is commonly rationalized using a bulk-rheology--based generalized Maxwell model~\cite{siebenburger2012}, but particle-based descriptions are also rich. For instance, avalanche stress relaxation behavior has been detected in creep~\cite{jacob2019}. 
Using rheology and light scattering echo, particle trajectories were further found to be partly reversible under strains which significantly exceed the yield strain ~\cite{petekidis2003}. Confocal microscopy has additionally revealed dynamically heterogeneous regions as a means to link creep and steady-state flow behavior~\cite{sentjabrskaja2015}.

\begin{figure}
\centering
\includegraphics[width=85 mm]{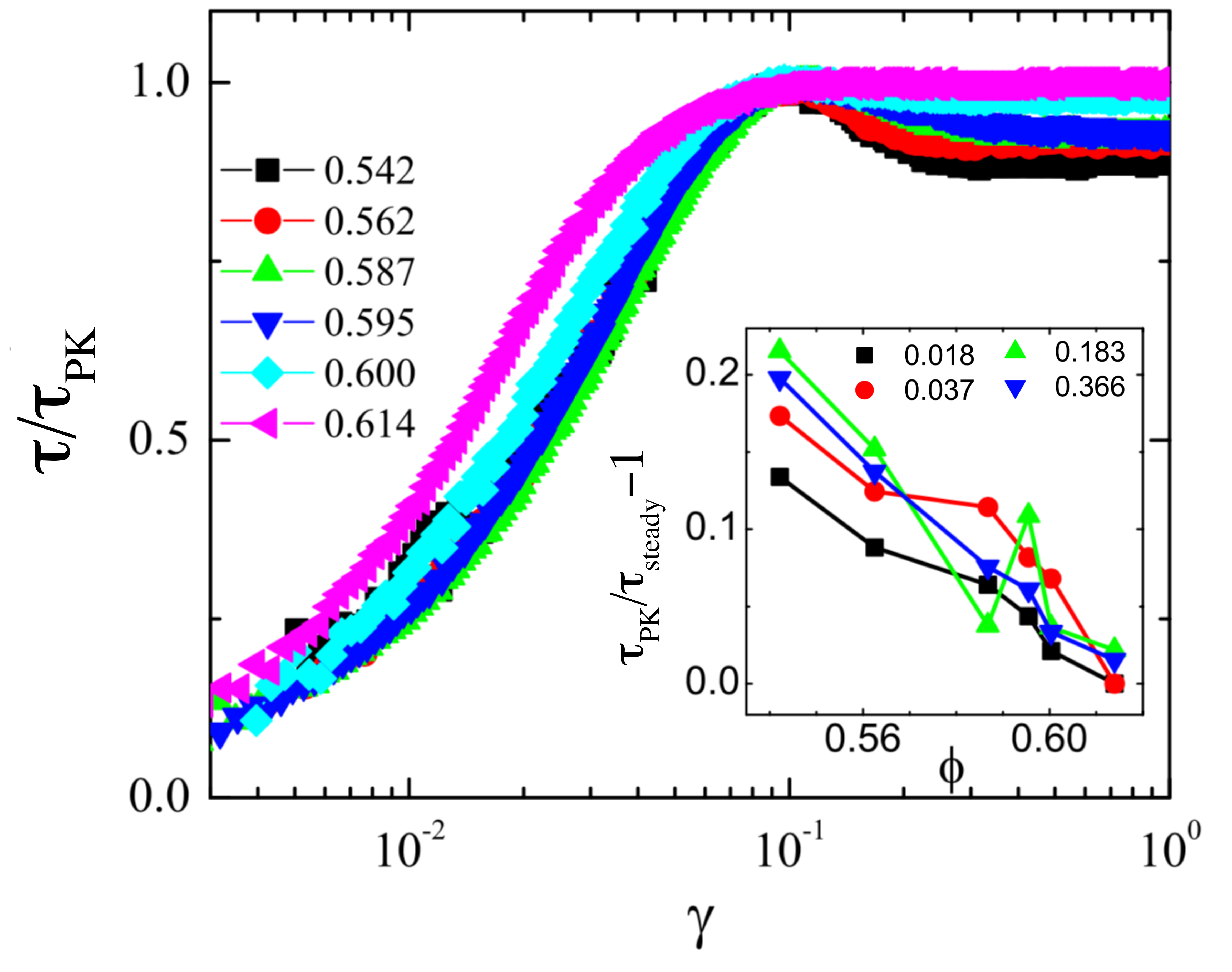}
\caption{Stress overshoot in colloidal hard spheres. Stress normalized by the peak stress $\tau_\mathrm{PK}$ at different volume fractions $\phi$ as indicated.
Inset: the stress peak height scaled by its steady state value $\tau_\mathrm{PK}/\tau_\mathrm{steady}-1$. Lines correspond to the Weissenburg numbers $\pecl_w=\dot\gamma \tau_\alpha$ 
indicated. Reprinted (abstract/excerpt/figure) with permission from Koumakis, N., M. Laurati, S. U. Egelhaaf, J. F. Brady, and G. Petekidis, 2012, Phys. Rev. Lett. 108(9), 098303~\cite{koumakis2012}. Copyright (2012) by the American Physical Society.}
\label{figStressOvershoot}
\end{figure}

Another means to probe the far-from-equilibrium behavior of a hard-sphere glass is to drag a particle through it using optical tweezers, ie microrheology (Sec.~\ref{sectionMicrorheology}). 
There it was found that the threshold force for movement of the dragged particle varies strongly with volume fraction, and it velocity fluctuations do not change near the glass transition ~\cite{habdas2004,gazuz2009,gruber2016}. Unlike particle-resolved studies where all imaged particles are tracked through its use of a (tweezed) probe particle, this latter method can in principle be applied to smaller particles, thus opening the door to deeper supercooling (see Secs.~\ref{sectionSize} and \ref{sectionDeeper}). Finally, flow in channels of colloidal glasses has received relatively little attention, but work has been done to reveal surprising oscillatory flow behavior~\cite{isa2007,isa2009}.

%% file: NucleationCrystallisation/nucleationCrystallisation.tex
\section{Nucleation and growth}
\label{sectionNucleation}

Although the thermodynamic  phase transition between the fluid and crystal phase of hard spheres is fairly well controlled (see Sec.~\ref{sectionBulk}), the kinetics of the transformation from one to the other remains an active area of research. Many fundamental questions about the process are still largely open, such as the regime of validity of classical nucleation theory (CNT), the discrepancy between numerically computed homogeneous nucleation rates and experimentally measured values, and the glass forming ability of (relatively and absolutely) monodisperse hard spheres. This section describes our current understanding of the phase transformation process and further details of some of the associated challenges.

\subsection{A primer on classical nucleation theory}
\label{sectionCNT}

One hundred and fifty years ago~\citet{gibbs1878} suggested that a first-order phase transition should proceed through the formation of a nucleus of the thermodynamically stable phase embedded in the metastable phase. Some of the spatial and temporal crystal-like fluctuations that spontaneously form in the fluid can then give rise to the macroscopic phase transformation. Gibbs proposed that this process may be viewed as two coupled homogeneous systems with a sharp interface, whereby the reversible work required to grow the nucleus determines its probability. The thermodynamic aspects of nucleus (cluster) formation were later expanded through modeling the dynamics of cluster growth by~\citet{volmer1926} and~\citet{becker1935}, who formulated CNT~\cite{kelton1991,debenedetti1996}. In particular, Becker and D\"oring proposed an infinite set of coupled equations that describe in general terms the coagulation and fragmentation of clusters of different size. From this perspective, the basic assumptions of CNT are: (\emph{i}) a single order parameter describes the size evolution of the different clusters, and (\emph{ii}) the nucleation process is Markovian, i.e., the time evolution does not depend on the state of the system at previous times and clusters can change their size only by gaining or losing a single free component.

For hard-sphere crystallization, in particular, CNT translates as follows. The fluid becomes thermodynamically metastable with respect to the stable solid phase by over-compressing it beyond the fluid-crystal coexistence pressure. Assuming that the nucleus is not strained and occupies a well-defined volume $V_\text{s}$ with area $A$, and working isothermally in the grand-canonical ensemble with grand potentials $\Omega_\alpha(T,V)=-P_\alpha V$, where $\alpha=\text{f,s}$ for the fluid and solid phases, respectively, 
the presence of a nucleus results in a change of grand potential
\begin{equation}
\label{eqCNTDOmega}
    \Delta\Omega = \Omega_\text{f}(V_\text{f}) + \Omega_\text{s}(V_\text{s}) + \bar\gamma A - \Omega_\text{f}(V) = -\Delta P V_\text{s} + \bar\gamma A
\end{equation}
with $V_\text{s}+V_\text{f}=V$ and $\Delta P=P_\text{s}-P_\text{f}$. The penalty for bringing two systems in contact is governed by the interfacial free energy $\gamma$ 
discussed in Sec.~\ref{sectionXtalFluidInterfaces}. While $\gamma$ depends on the interface's crystal orientation and lattice spacing, here we employ a scalar effective 
$\bar\gamma$, 
which  amounts to  assuming that the stochastic nucleus 
dynamics averages over orientation-dependent features.
Consequently, we consider spherical droplets with volume $V_\text{s}=\frac{4\pi}{3}R^3$ and area $A=4\pi R^2$, thus leaving the radius $R$ as the sole order parameter. The thermodynamic driving force behind nucleation is then the pressure difference $\Delta P=P_\text{s}-P_\text{f}>0$ between the pressure $P_\text{s}$ inside the nucleus compared to the pressure $P_\text{f}$ of the surrounding metastable fluid.

In order for the nucleus to grow into the equilibrium phase, it has to overcome a nucleation free-energy barrier at the critical radius $R^\ast$, as determined through 
$\frac{\partial \Delta \Omega}{\partial R}|_{R^\ast} = 0$.
We then find the well-known Laplace equation $\Delta P=2\bar\gamma^\ast/R^\ast$ provided 
$\frac{\partial \gamma (R)}{\partial R}|_{R^\ast} = 0$, 
which defines the \emph{surface of tension},  being $\bar\gamma^\ast=\bar\gamma(R^\ast)$. Eliminating the critical radius $R^\ast$ yields the barrier height
\begin{equation}
    \Delta\Omega^\ast = \frac{16\pi(\bar\gamma^\ast)^3}{3(\Delta P)^2} = \frac{1}{2}\Delta P V_\text{s}^\ast.
\end{equation}
The \emph{capillary approximation} substitutes for $\bar\gamma^\ast$ the interfacial free energy $\gamma$ of an infinite planar interface (often taken to be under bulk fluid-crystal coexistence conditions), averaged over all orientations.

In practice, computer simulations are performed at constant volume or constant (fluid) pressure counting the number of solid particles $n$ in the nucleus. We can relate the above framework to these simulations through the isothermal Gibbs-Duhem equation
\begin{equation}
    \frac{\partial\mu_\text{s}}{\partial p} = \frac{1}{\rho_\text{s}} = \frac{V_\text{s}}{n}
\end{equation}
with $\mu_\text{s}$ the chemical potential of the solid. Assuming the solid density to be independent of pressure (i.e., an incompressible solid), integration of both sides gives
\begin{equation}
    \Delta\mu = \mu_\text{s}(P_\text{s}) - \mu_\text{s}(P_\text{f}) = \frac{V_\text{s}}{n}\Delta P,
\end{equation}
and hence the nucleation barrier $\Delta G^\ast=\frac{1}{2}\Delta\mu n^\ast$ with $n^\ast$ the number of solid particles in the critical nucleus. Using the fact that the chemical potential $\mu_\text{s}(P_\text{s})=\mu_\text{f}(P_\text{f})$ is uniform, $\Delta\mu=\mu_\text{f}(P_\text{f})-\mu_\text{s}(P_\text{f})>0$ compares the chemical potentials of the fluid and solid at the same pressure $P_\text{f}$. Note that in reality the presence of a curved fluid-crystal interface leads to strain inside the crystal nucleus~\cite{mullins1984}. As a result,~\citet{montero2020} have shown that the mechanical pressure inside hard-sphere crystal nuclei can in fact be lower than that of the surrounding fluid, and have argued that $\Delta P$ in the Laplace equation should be interpreted as the pressure difference between the two (bulk) phases at equal chemical potential $\mu_\text{f}$ as we have done in here.

\begin{figure}
\centering
\includegraphics[width=0.9\linewidth]{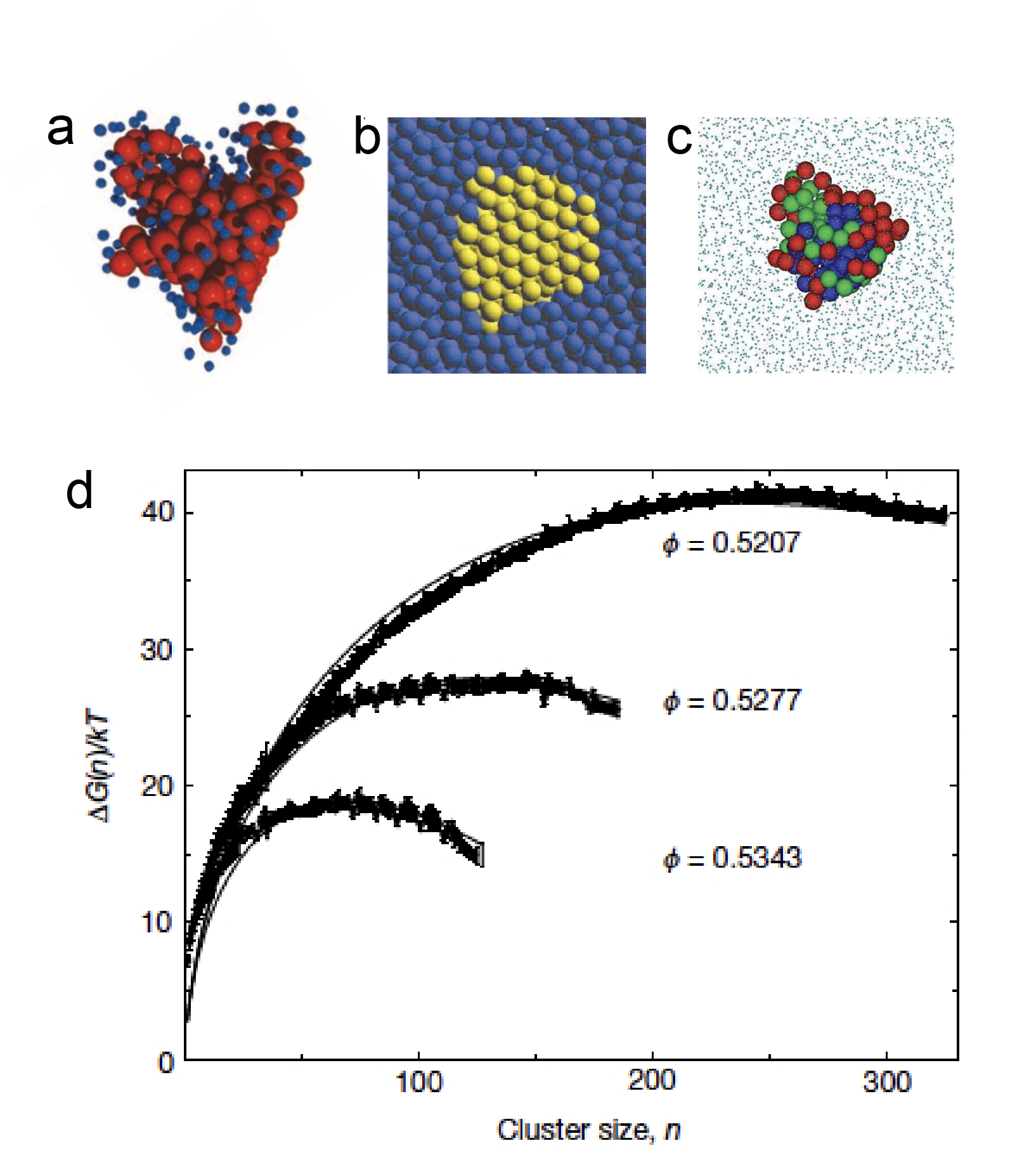}
\caption{Nuclei in metastable hard-sphere fluids: detection and free energy measurement.
(a) Nucleus detected in nearly hard spheres by confocal microscopy. Here the red particles are identified as crystal and blue particles have at least one crystal-like bond according to BOO parameters. Reproduced from~\citet{gasser2001}. Experimental detection of a \emph{critical} nucleus is particularly challenging (see text). Reprinted with permission from AAAS.
(b-c) Critical nuclei obtained from umbrella sampling simulations. (b)  A  critical crystal nucleus  embedded in a metastable hard-sphere fluid  at $\phi=0.5207$~\cite{auer2004jcp}. Reprinted from~\cite{auer2004jcp}, with the permission of AIP Publishing.
(c) Critical cluster  in a metastable fluid (smaller particles) at $\phi=0.5355$~\cite{filion2010}. Different colors are  for different criteria to detect crystalline particles. Reprinted from~\cite{filion2010}, with the permission of AIP Publishing.
(d) Free-energy barriers  (symbols) as a function of the largest crystalline cluster size $n$  from umbrella sampling simulations at different $\phi$~\cite{auer2004jcp}. 
Continuous lines are fits to the CNT functional form. Reprinted from~\cite{auer2004jcp}, with the permission of AIP Publishing.
}
\label{figNucleus}
\end{figure}

Figure \ref{figNucleus} (d) depicts  free-energy barriers as a function of the nucleus size  $n$, obtained from numerical simulations of a hard-sphere fluid at different supersaturated pressures (or, equivalently, metastable fluid volume fractions $\phi$). Typical snapshots of the critical nucleus  from numerical simulation results are shown in Fig.~\ref{figNucleus} (a, b and c).

As the nucleation barrier $\Delta G^\ast$ increases, spontaneous fluctuations that might give rise to a cluster of size $n^\ast$ grow rarer. These rare events are activated processes, for which the average waiting time between events is orders of magnitude longer than the time needed for the event itself to take place.  Rare events are therefore intrinsically difficult to investigate.  The most valuable observable to study nucleation, accessible both experimentally and  numerically, is therefore the nucleation rate $J$, i.e., the number of independent critical nuclei (that give rise to crystal formation) formed per unit time and volume. In the rare-event regime~\cite{turnbull1949},  the CNT steady-state  nucleation rate can be estimated from the product of two terms: the probability to form the critical nucleus, $\exp({-\beta \Delta G^\ast})$, and the kinetic prefactor $\kappa$, which describes the frequency at which the Gibbs free-energy barrier is crossed. By accounting for the number of critical nuclei whose size fluctuates at the top of the free-energy barrier, we obtain~\cite{kelton1991}
\begin{equation}\label{eqRate}
 J = \kappa e^{-\beta \Delta G^\ast} = \left( \rho_\mathrm{fluid} \frac{24 D_L n^{*{2/3}}}{\lambda^2} Z_\mathrm{corr}  \right) e^{-\beta \Delta G^\ast},
 \end{equation}
where $\kappa$ expresses the attempt frequency of attaching/detaching a  particle from the critical nucleus  (or jump frequency) per unit volume, and depends on: (\emph{i}) the  metastable fluid density $\rho_\mathrm{fluid}$, (\emph{ii}) the long-time diffusivity $D_L$ and jump distance $\lambda$, $6 D_L/\lambda^2$, and (\emph{iii}) the number of available attachment sites on a spherical critical nucleus of size  $n^\ast$, $ 4(n^\ast)^{2/3}$. The Zeldovich correction factor  $Z_\mathrm{corr}$ accounts for the fact that during the steady-state nucleation process, the concentration of critical nuclei is not truly an equilibrium concentration~\cite{zeldovich1942}.

\begin{figure*}[!t]
\centering
\includegraphics[width=140mm]{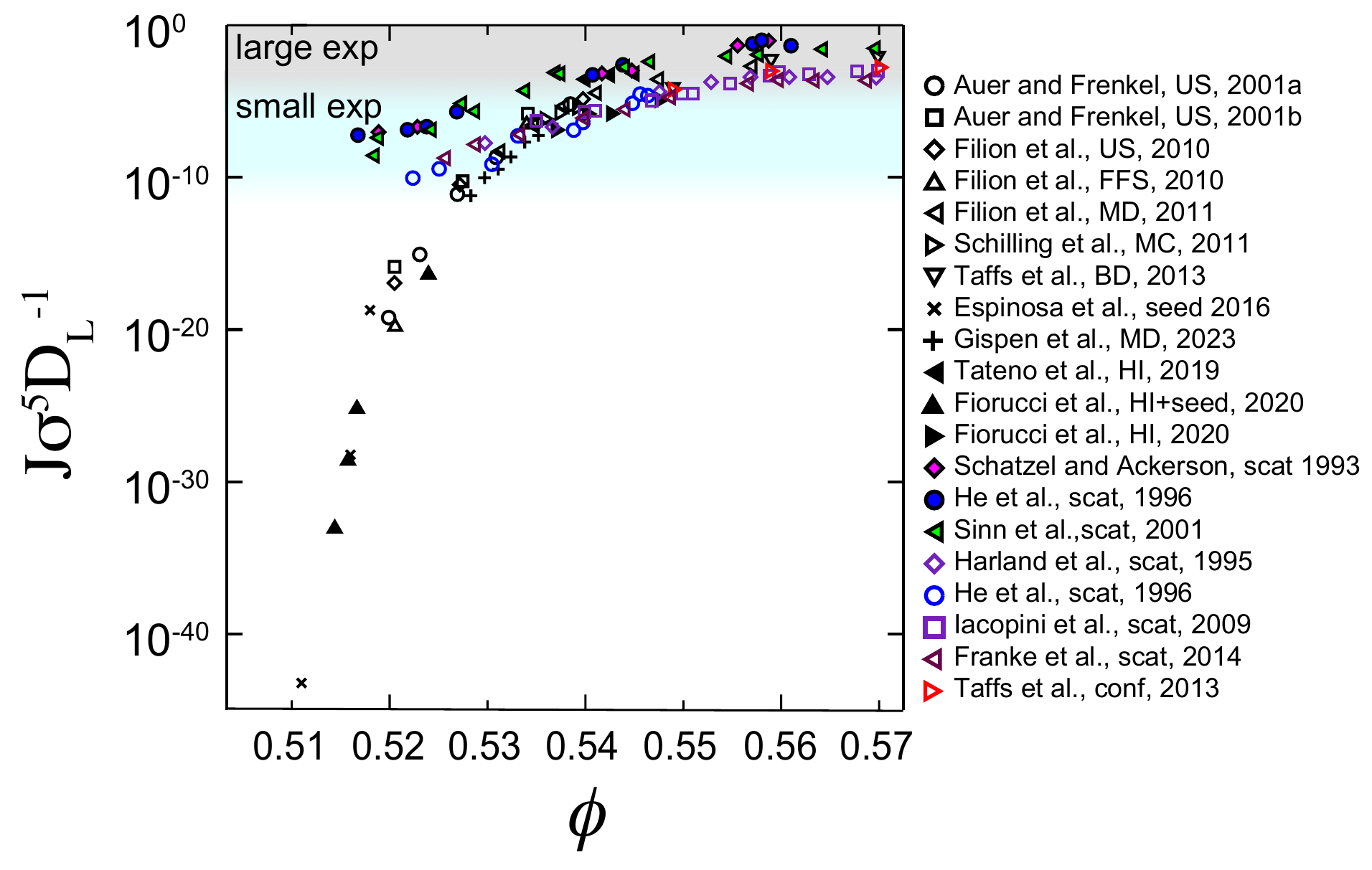}
\caption{Reduced nucleation rates $J\sigma^5/D_L$ for hard spheres as a function of the supersaturated fluid volume fraction $\phi$, comparing experiments (colored symbols) with simulations (black symbols). Simulations are further divided in simulations with (full triangles) and without (open symbols) hydrodynamic interactions. 
Experiments are (approximately) divided into (nearly) density matched (open symbols) and non-density matched (full symbols). The method used in each case is: umbrella sampling (US), forward flux sampling (FFS), molecular dynamics (MD), Monte Carlo (MC), seeding (seed), Brownian dynamics (BD), and  hydrodynanic interactions (HI) for the simulations; light scattering (scat) and confocal microscopy (conf) for the experiments. The fluid volume fraction is taken as that quoted in the original paper. Grey shading pertains to the dynamical regime accessible to experiments with relatively large colloids (see Sec.~\ref{sectionSize}). The regime accessible to experiments using smaller colloids is shaded blue.
}
\label{figOneToRuleThemAll}
\end{figure*}

As one can immediately deduce from Eq.~\eqref{eqRate}, the larger the driving force $\Delta \mu$ for nucleation to occur, the lower the free-energy barrier height $\Delta G^\ast$, the smaller the  critical nucleus size $n^\ast$, and the higher the nucleation rate $J$. Figure \ref{figOneToRuleThemAll} reports several nucleation rates $J$ as a function of $\phi$ taken from both experiments and numerical simulations. As shown in Fig.~\ref{figOneToRuleThemAll}, while the experimental and numerical nucleation rates nicely agree  at high $\phi$ (or large $\Delta \mu$), marked discrepancies are observed when $\phi \lesssim 0.525$. The possible origins of this discrepancy have been extensively discussed in the literature and are here reviewed in Sec.~\ref{sectionDiscrepancy}. Before doing so, however, we first revisit the experimental context for these measurements.

\subsection{Early nucleation experiments}
\label{sectionEarly}

Light scattering techniques have long been a mainstay of nucleation rate measurements in colloidal suspensions (see Sec.~\ref{sectionLightScattering})~\cite{schatzel1992,schatzel1993,harland1995}. The approach, which tracks the time evolution of the static structure factor of the crystallizing suspension,  $S(k,t)$, detects contributions from both the crystal, $S(k)_\mathrm{xtal}$, and the fluid, $S(k)_\mathrm{fluid}$ 
\begin{equation}
S(k,t)=X(t)S_\mathrm{xtal}(k)+(1-X(t))S_\mathrm{fluid}(k),
\label{eqSqNucleation}
\end{equation}
where $X(t)$ is the crystalline fraction. Given that the small size of the crystallites results in broadening of the signal, the average linear size can be estimated according to  
\begin{equation}
L_\mathrm{nucl}(t)=\frac{2\pi K}{w_q(t)},
\label{eqLNucleation}
\end{equation}
where $w_q(t)$ is the width of the peak at half maximum and $K$ is the Scherrer constant (for a crystal of cubic shape, $K = 1.155$~\cite{langford1978}). The number density of (average sized) crystals is therefore
\begin{equation}n(t)=\frac{X(t)}{L_\mathrm{nucl}^3(t)}.
\label{eqNucleusDensity}
\end{equation}
~\citet{harland1995} carried out experiments before density matching of colloids was widely used, using particles with  $\sigma=800$ nm.  By determining the colloidal number density of crystals as a function of time $n(t)$, a nucleation rate could then be extracted (see Fig.~\ref{figOneToRuleThemAll}). Later experiments performed in the same lab~\cite{harland1997} with smaller particles ($\sigma=400$ nm) showed similar trends. The physical interpretation of the results, however, was not altogether robust. On the one hand, the results successfully agreed  with classical nucleation theory, in that they could be fitted to Eq.~\eqref{eqn:rate} (using the theoretical equations of state of the fluid and crystal to determine $\Delta\mu$) to extract a reasonable value of the fluid--crystal interfacial free energy $\bar \gamma$. On the other hand, compared to the earlier results of Ackerson and coworkers~\cite{schatzel1992,schatzel1993}, the adimensional nucleation rate, $J \sigma^5/D_L$, seemed to depend on particle size, with larger particles nucleating faster than smaller ones~\cite{he1996}, which it clearly should not.

These early experiments sought to probe the physics of nucleation using colloidal systems as test beds, given the experimentally tractable time-- and length--scales then at hand. While at low supersaturation, most of the CNT assumptions appear reasonable, beyond the rare-event (or activated) regime the situation is clearly more complex. In particular, the fluctuations in crystallizing systems at $\phi\gtrsim0.54$ appear to reach a kinetic spinodal limit~\cite{schatzel1992,schatzel1993}, in which the nucleation barrier for the 
fluid to crystallize becomes so small that nucleation is no longer rare, but takes place on the same time scale as that of the structural relaxation of the fluid (see Sec.~\ref{sectionBeyondCNT}).

\subsection{Light scattering  versus  real-space experiments}

Light scattering measurements remain state of the art in the weak supersaturation (or rare-event nucleation) regime~\cite{palberg2014}. The technique is perfectly suited for the relatively small colloids, with $\sigma=200$--$500$ nm, then used to keep the long time scale associated with nucleation experimentally tractable (see Secs.~\ref{sectionSize} and~\ref{sectionStructureDynamics}). It provides, however, only limited microscopic information about the structure of the nucleus.

Particle--resolved studies 
allow measurements of the critical nucleus shape and size distribution, along with the detailed structure and dynamics of the surrounding colloidal fluid as shown in Fig.~\ref{figNucleus}(a)~\cite{ivlev,elliot2001,gasser2009,gasser2001,wood2018,taffs2013}. It can therefore be used to test some of the CNT assumptions, such as whether the solid phase has a density close to its bulk value, and whether crystalline defects play a role during the formation of the critical nucleus. Alas, real-space information comes at a high experimental price. As reflected in Fig.~\ref{figOneToRuleThemAll}, conventional real-space analysis requires larger colloids, 2-3 $\mu$m in size~\cite{ivlev}, which diffuse  much slower than those commonly used in light-scattering experiments (see Fig.~\ref{figOneToRuleThemAll} and Sec.~\ref{sectionSize}). 
Even at fairly high supersaturation, $\phi \approx 0.53$, crystallisation can take days~\cite{taffs2013}. Furthermore, the analysis is laborious, so that obataining a statistically meaningful number of nuclei of different sizes and shapes is a major challenge.

\subsection{The nucleation rate discrepancy}
\label{sectionDiscrepancy}

As mentioned in Sec.~\ref{sectionEarly}, early nucleation experiments followed the expected physical trend, namely a low nucleation rate at weak supersaturation that increases with volume fraction and eventually decreases with increasing viscosity  (upon approaching the glass transition, see Sec.~\ref{sectionGlass})~\cite{palberg1999}. Note that this decrease is not visible in Fig.~\ref{figOneToRuleThemAll} due to the rescaling of the nucleation rate with the diffusion constant $D_L$, which decreases sharply  with increasing volume fraction.
However, upon comparing experimental rates~\cite{palberg2014}  with numerical predictions obtained using  rare-event sampling methods  developed since the mid--1990s, marked discrepancies were observed.

The  computations by~\citet{auer2001prediction}, in particular, uncovered a pronounced difference in the \emph{slope} of the nucleation rate curve (see Fig.~\ref{figOneToRuleThemAll}). Since then, different numerical approaches have been used to assess these predictions. From the middle of the fluid-crystal coexistence region and as long as the nucleation barrier is at least a few times larger than the thermal energy, i.e., ($0.52\lesssim\phi\lesssim 0.54$), biasing methods such as seeding~\cite{espinosa2016}, umbrella sampling~\cite{auer2001prediction, filion2010,filion2011jcp} and forward flux sampling~\cite{filion2010,richard2018jcpkinetics,richard2018jcpthermodynamic} can be used. For $\phi\gtrsim 0.528$, nucleation is even accessible through direct numerical simulations~\cite{kawasaki2010pnas,filion2011jcp,gispen2023arxiv} and first-passage time methods~\cite{richard2018jcpkinetics}. In the density regimes where these methods overlap, they provide consistent estimates, thus confirming a 15 order of magnitude increase of the nucleation rate in going from $\phi=0.52$ to $\phi=0.55$. By contrast, experiments at best display a variation of five orders of magnitude of the nucleation rate over the same volume fraction range (see Fig.~\ref{figOneToRuleThemAll}).

This staggering gap between simulations and experiments of well over ten orders of magnitude has been dubbed ``the second--worst discrepancy in physics''~\cite{russo2013}--the first being the 120 orders of magnitude discrepancy between the value of the cosmological constant and the quantum energy of the vacuum. Accounting for this mismatch largely remains an open problem, but given the strong consistency between different simulation approaches it is certainly tempting to conclude that some  experimental effects are not properly taken into consideration. Potential candidates are plentiful: from experimental effects that are not accounted for in simulations, to experimental errors in determining the true homogeneous nucleation rate or volume fraction. Claims of a more fundamental lack of understanding of the nucleation process are dubious, as the discrepancy is more pronounced in the low supersaturation regime where the CNT description is most justified. Notably, while the discrepancy in the \emph{nucleation rate} is significant, that in the volume fraction is rather less. Indeed a shift of $\delta \phi = 0.01$, which is akin to the uncertainty on $\phi$ (see Sec.~\ref{sectionAccuracy})  would largely alleviate the discrepancy. What it would not explain is why the error in volume fraction is apparently \emph{consistently underestimated.} Here we first review efforts made to address the discrepancy before returning to the accuracy of the phase boundaries.

In order to understand this discrepancy, the dependence of the nucleation rate on polydispersity~\cite{auer2001suppression,pusey2009}, electrostatic interactions~\cite{auer2002}, hydrodynamics~\cite{radu2014,tateno2019,fiorucci2020}, sedimentation~\cite{russo2013,wood2018,ketzetzi2018}, and external walls~\cite{espinosa2019} have been considered. We discuss these different effects below.

\header{Polydispersity} 
Experimental colloidal systems inevitably exhibit size polydispersity $s$ (see Secs.~\ref{sectionBulk} and \ref{sectionBinary}), which can considerably alter the crystallization transition~\cite{fasolo2003}. For moderate polydispersity ($s\lesssim 5--6\%$), even though diffusivity is only negligibly affected~\cite{zaccarelli2009}, the coexistence curves shift to significantly higher $\phi$~\cite{sollich2010}. The impact of this shift on nucleation rates was considered already in the early papers of Auer and Frenkel~\cite{auer2001suppression}. For $s=5\%$, the shift of the nucleation curve makes $\phi$ about 0.015 higher than for the equivalent nucleation rate in the monodisperse case. However, this shift essentially disappears when rescaling the volume fraction based on the freezing density of the system~\cite{filion2010}.  Crucially, polydispersity does not alter the \emph{slope} of the nucleation rate curve. In other words, the rate curves for different $s$ are roughly parallel. Although the shape of the particle size distribution (and not only $s$) affects both the nucleation and growth kinetics, as shown experimentally by Sch\"{o}pe and Van Megen~\cite{schope2006pre,schope2007}, colloid syntheses that deliver significantly different polydispersity distributions (due to, for example,  secondary nucleation) give relatively similar sets of experimental nucleation rates (Fig.~\ref{figOneToRuleThemAll}). It therefore seems quite reasonable to conclude that polydispersity does not contribute significantly to the nucleation rate discrepancy.

\header{Electrostatic Charge}
The possibility of residual charges on the colloids was also considered early on~\cite{auer2002}, and modeled through a hard-core Yukawa or Debye-Hückel potential as in Eq.~\eqref{eqYuk}.
Because systems of interest typically use non--aqueous solvents with low dielectric constants (see Sec.~\ref{sectionInteractions}), the degree of electrostatic charging is very low (especially compared to aqueous systems), and hence linear Poisson--Boltzmann theory should hold~\cite{royall2006}. Some work where no salt was added may exhibit density-dependent interactions due to counterion condensation as a function of volume fraction~\cite{royall2006}. The effective charge then drops as a function of volume fraction, thus affecting the  mapping to an effective hard-sphere volume fraction in a qualitatively similar direction as the observed discrepancy. Evidence for some deviation from centro--symmetric interactions between the colloids implicit in linear PB theory in these non--aqueous systems has been detected in crystals~\cite{reinke2007}. Since such effects are highly parameter specific~\cite{royall2006}, it seems rather unlikely that all experiments in this regime, which have used fairly different conditions (solvent dielectric constant,  chemical composition of colloids and solvent), would exhibit a quantitatively consistent behavior. In any case, these experiments concern confocal microscopy studies using larger colloids in a density-matched solvent. As Fig.~\ref{figOneToRuleThemAll} shows, it is the experiements that used smaller colloids that feature in the regime of the discrepancy (blue shaded region in Fig.~\ref{figOneToRuleThemAll}). Our analysis in Sec.~\ref{sectionInteractions} and the dashed blue line in Fig.~\ref{figU}(a), in particular, suggest that interactions in these systems are very close to hard spheres and exhibit little density dependence compared to those in~\citet{royall2006}.

The dominant contribution is therefore expected to be the softening of the interaction potential due to residual charges. At the hard-core Yukawa or Debye-H\"{u}ckel, the effects of residual charges on the nucleation rate were considered  by Auer \emph{et al.}~\cite{auer2002}. They showed that, compared to hard spheres, introducing a small charge increases the nucleation rate via two mechanisms: (\emph{i}) at constant density, because the supersaturation increases (given the shift in fluid-crystal phase boundaries to lower packing fractions), and (\emph{ii}) at constant supersaturation (i.e., the chemical potential difference $\mu-\mu_\text{coex}$), because of a considerable reduction of the fluid-crystal interfacial free energy. However, it should be noted that~\citet{auer2002} did not consider the weak, longer-ranged electrostatic interactions pertinent to the studies using smaller PMMA particles in low dielectric constant solvents (Sec.~\ref{sectionInteractions} and Fig.~\ref{figU}(a) blue dashed line).

A recent numerical study~\cite{dejager2022} has shown that for highly-screened electrostatic interactions the phase behavior depends nearly completely on the screening length, with a nucleation barrier that increases with increasing screening length at fixed supersaturation (i.e., measuring the barriers at constant $\phi/\phi_f(\kappa\sigma)$, where $\kappa\sigma$ is the inverse screening length and $\phi_f$ is the freezing volume fraction). This trend does not explain the nucleation rate discrepancy between experiments and simulations, as it suggests that charge effects would speed up nucleation rates rather than slow them down~\cite{royall2013myth}. Moreover, as the dashed blue line in Fig.~\ref{figU}(a) and discussion in Sec.~\ref{sectionInteractions}  indicate, the smaller particles used in the experiments that reached the discrepancy regime had very large Debye lengths, ie a different interaction to those considered here.

In short, electrostatic interactions shift but do not change the density dependence of the nucleation rate. It is therefore hard to see how the observed discrepancy between simulations and experiments could  
be explained through residual charges alone.

\header{Hydrodynamics}
The nucleation rate discrepancy between experiments and simulations hints at possible unexpected nucleation pathways that lead to the efficient formation of large nuclei at small supersaturation in experiment. A particularly careful consideration of long-range hydrodynamic interactions, which can alter the aggregation kinetics of colloidal particles, is therefore in order.

Properly accounting for hydrodynamics in numerical simulations requires specialized techniques that couple the dynamics of hard spheres with  either a continuum or a coarse-grained representation of the surrounding fluid solvent (see Sec.~\ref{sectionSimHydro}). The high numerical cost of these methods considerably shortens the observable time scales compared to standard BD (or similar) simulations, thus restricting the consideration of nucleation for an even smaller range of (fairly high) volume fractions. Because this regime is typically where numerical and experimental nucleation rates agree, these simulations offer but limited physical insight.

Several groups have nevertheless tried to quantify the relevance of hydrodynamics. ~\citet{radu2014} first considered the effect of hydrodynamics in simulations via the \emph{stochastic rotational dynamics} approach. Surprisingly, an increase of the nucleation rate with increasing solvent viscosity  at high volume fractions was then observed. However, these results are caused by the way the hydrodynamic interactions were implemented~\cite{fiorucci2020}. In particular, their  implementation ignored the excluded-volume interactions and the mass difference between the solvent and the colloidal particles, both of which impact the resulting hydrodynamic interactions. 
To examine the effect of hydrodynamics on hard-sphere nucleation,~\citet{fiorucci2020} computed  nucleation rates at  high volume fractions using brute-force MD simulations with hydrodynamics implemented via the stochastic rotational dynamics method. The results, however, agreed well with studies that neglect hydrodynamic interactions.

To evaluate the importance of hydrodynamics on nuclei dynamics at low volume fractions, seeding has been considered.
~\citet{tateno2019} directly solved the Navier-Stokes equations with the fluid particle dynamics method, while~\citet{fiorucci2020} used stochastic rotational dynamics. Both approaches convincingly demonstrated  that mass transport was considerably slowed down by hydrodynamic lubrication effects, and that if the nucleation rate was rescaled by the long-time diffusion coefficient, it would be  practically unaffected by hydrodynamics (see full color symbols in Fig.~\ref{figOneToRuleThemAll}).

~\citet{fiorucci2020} also examined the coupling of hydrodynamics  with different solvent viscosities, different sedimentation rates, and different softness of the interaction potential, finding that these effects did not significantly alter the local fluid structure. Existing results therefore suggest that hydrodynamics is not likely to be the main origin for the discrepancy.

\header{Sedimentation}
Another effect that is frequently neglected by simulations is gravity. As introduced in Sec.~\ref{sectionFarFromEq}, the strength of gravity is measured by the gravitational P\'eclet number $\Pe_g$, Eq.~\eqref{eqSedPecl}. The nucleation rates reported in Fig.~\ref{figOneToRuleThemAll} can be grouped according to the sedimentation strength~\cite{russo2013,palberg2014,wood2018}: those that pertain to \emph{weak} ($\Pe_g \lesssim 0.01$) and \emph{strong} sedimentation ($0.1\lesssim \Pe_g \lesssim 1$) differ, with the latter nucleating  much faster. 
~\citet{schatzel1993} and~\citet{sinn2001}, which studied suspensions of colloidal particles in a non-density matched solvent, i.e., with $10^{-1}\lesssim \Pe_g\lesssim 1$, both obtained comparably high nucleation rates. By contrast, work which studied density matched colloidal suspensions with $\xi/\sigma\sim 100$ and $\Pe_g\sim 10^{-2}$ (either by employing small particles, or by using swelling microgels) reported nucleation rates that cluster at lower values, and are therefore a few decades closer to the simulation results~\cite{harland1997,iacopini2009,franke2011}.

An assessment of the relative importance of the sedimentation time scale with respect to that of crystallization was carried out with BD simulations~\cite{russo2013}. Relative to experiments, this work considered small system sizes and timescales (compared to experiments) for  $\Pe_g \lesssim 1$ and found that the nucleation events are the same as in a gravity-free environment for $\phi\gtrsim 0.525$, whereas for $\phi\lesssim 0.525$ sedimentation occurs on shorter time scales than nucleation. Significant deviations therefore have to be expected with respect to the zero gravity case. While gravitational effects can induce density fluctuations that significantly enhance the nucleation rate, the precise mechanism by which this occurs is still unknown. In addition, for experimental system sizes (where the sample height can be $\sim10^4$ or $10^5\sigma$, compared to $\sim10^2\sigma$ in simulation), nucleation typically proceeds faster than sedimentation (see Sec.~\ref{sectionEquilibriumPhaseBehavior})~\cite{paulin1990}. One possibility could be that hydrodynamic interactions affect the structure of hard-sphere systems undergoing sedimentation, as has already been shown in the case of strong confinement (see Sec.~\ref{sectionNonEquilibriumSedimentation})~\cite{wysocki2009}. It is therefore tempting to imagine that the higher--order fluid structure might somehow be affected by the hydrodynamic interactions associated with out-of-equilibrium sedimentation, thus influencing the nucleation rate.

\begin{figure}[h!]
\centering
\includegraphics[width=75mm]{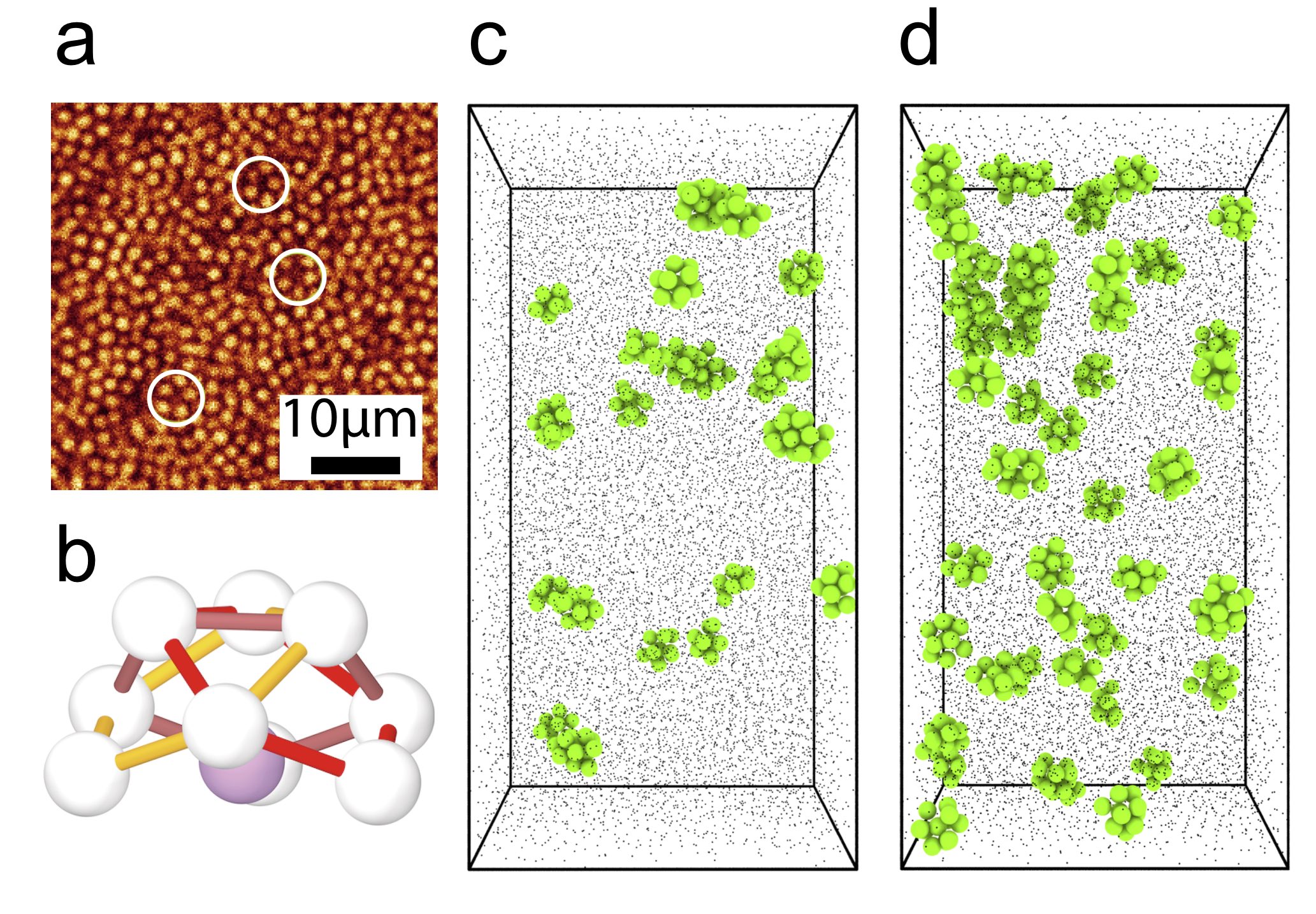}
\caption{
(a) A single slice through a 3d confocal image stack. Three five-membered rings are circled. 
(b) Diagram of the defective icosahedron (10B). Five-membered rings are indicated in yellow, red, and dark red. (c,d). Visualization of the effect of sedimentation upon the structure of the hard-sphere fluid at an average $\phi=0.45$. 
(c) shows a sedimenting system ($\Pe_g \simeq 1.5$), 
(d) shows a density matched system.  
This \emph{experimental} data is  rendered after all particle centers have been located with particle tracking. Green particles are those  in defective icosahedra, particles not in defective icosahedra are rendered as smaller grey points. 
Reprinted from~\cite{wood2018}, with the permission of AIP Publishing.}
\label{figNick}
\end{figure}

~\citet{wood2018} used particle-resolved experiments to investigate the population of five-fold symmetric structures that, as originally postulated by~\citet{frank1952}, have the effect of suppressing nucleation due to an increase in interfacial free energy
between the crystal and fluid~\cite{taffs2016}. They measured the population of five-fold symmetric structures in sedimenting hard-sphere systems with $\Pe_g\approx1$, and found that the population of such structures drops by a factor of two with respect to the non-sedimenting density matched case (Fig.~\ref{figNick}). However, simulations that considered hydrodynamic interactions showed insignificant differences in the quantity of clusters exhibiting five-fold symmetry in  hard-sphere fluids when exposed to a gravitational field with $\Pe_g \leq 2$ ~\cite{fiorucci2020}. It is possible that this difference may be explained by the way the boundary conditions are implemented (periodic in the case of the simulations, and hard walls for the experiments). Yet for the only direct comparison of higher--order structure in sedimenting hard spheres that we are aware of to show such a significant discrepancy between experiment and simulation surely merits further investigation. After all, crystal nuclei themselves originate from higher-order fluid structures.

The authors then carried out computer simulations with umbrella sampling. To mimic the impact of sedimentation, a system in which the population of five-fold symmetric structures was reduced so as to match these experiments was prepared using biased MC simulations. Evaluating the nucleation barrier using umbrella sampling found that a drop in five-fold symmetry indeed reduced the nucleation barrier considerably (by $11\,\kB T$ at $\phi=0.52$, corresponding to a nucleation rate increase of approximately five orders of magnitude), but still insufficiently to account for the discrepancy (see Fig.~\ref{figOneToRuleThemAll}). Furthermore, Wood \emph{et al.}~\cite{wood2018} used $\Pe_g=$ 1.5, which  at the top end of those experiments that sediment strongly.

Further investigation is required to determine whether other aspects of the sedimentation process, such as the impact of shear flows and other hydrodynamic effects that couple to gravity, could explain the discrepancy.

\header{Non-homogeneous nucleation rates}
Several authors have also challenged the assumption that the nucleation rate measured in experiments is that of a homogeneous nucleation  process~\cite{gasser2001,espinosa2019,wohler2022}, pointing to the fact that 
other nucleation and growth channels could be responsible for the discrepancy. For one thing, colloidal fluids are unavoidably in contact with the walls of the sample cell, which can be a source of nucleation sites. Heterogeneous nucleation will be presented in a more general context in Sec.~\ref{sectionHeterogeneous}, but for now let us  consider its putative contribution to the discrepancy.

To elucidate the effect of different possible surfaces which could lead to heterogeneous nucleation,~\citet{espinosa2019} numerically studied the competition between homogeneous and heterogeneous crystallization as a function of wall type, fluid density and system size. For flat walls and surfaces randomly coated with non-overlapping spheres of a diameter three times larger than those in the fluid (as used in some experiments~\cite{ziese2013}) heterogeneous nucleation overwhelmed homogeneous nucleation  for $\phi<$0.535. By contrast, when the coating is done with non-overlapping spheres with the same diameter as those of the fluid (as done in other experiments~\cite{taffs2013}), nucleation was more likely to occur in the bulk, given how suppressed heterogeneous nucleation was. However, to the best of our knowledge, the older experiments, which used light scattering 
~\cite{harland1995,harland1997,he1996,schatzel1992,schatzel1993} to obtain nucleation rates in the discrepancy regime, did not coat surfaces  (``sintering'' of polydisperse particles onto the sample cell walls is typically done for more recent, real-space experiments~\cite{gasser2001,ziese2013,taffs2013}). In any case, in the earlier light scattering experiments, the data was then taken in the center of the sample cell, far from the wall, and it is hard to believe that macroscopic iridescent crystals could have migrated from the walls to the center of the cell without those carrying out the experiments noticing. Furthermore, how such a large crystal of hard spheres could sustain the gravitational stresses in these systems which were not density matched (see Sec.~\ref{sectionDiscrepancy}) is entirely unclear. Any heterogeneous nucleation effects would then likely arise from impurities or from colloid clusters that were not completely dispersed in the sample preparation. Although the resulting nucleation rate estimates that account for crystallite formation both in the bulk and at the walls have been argued to coincide with the experimental results the above discussion suggests this coincidence is fortuitous.

Recently, it has also been proposed that Bragg scattering signals measure poly-crystalline growth from different nuclei, whose size is described by the Avrami law, and therefore cannot access the true homogeneous nucleation rate~\cite{wohler2022}. Correcting the rate by adding secondary nucleation events occurring at the interface of the already-formed crystals, was found to putatively account for the difference with the simulated rates. However, how nuclei can have on average $\sim 10^{10}$ domains (the amount presumably required to resolve the discrepancy) is unclear. If we assume that there are $10^3$ domains in each direction, i.e., $10^9$ in total, and that each had a typical size of 10 particle diameters, for 500 nm diameter particles this amounts to nuclei 5 mm in each direction, a macroscopic object! Such a macroscopic object would, a the very least, rapidly sediment so that it seems unclear how it could form and survive long enough t be analyzed.

Despite these putative leads, a definitive understanding of the origin of the discrepancy between experimental and simulation nucleation rates remains an outstanding challenge, possibly explained by one or more of the mechanisms described above, or by effects that have not yet been considered. This brings us back to the question of the accuracy with which phase boundaries have been determined. Our discussion in Secs.~\ref{sectionAccuracy} and ~\ref{sectionEquilibriumPhaseBehavior} suggest that an error of $\delta \phi = 0.01$ is reasonable. In light of this observation, and the failure to find a clear-cut physical explanation, one may perhaps enquire if there really is a discrepancy at all, given that one could largely make it disappear by shifting the experimental data by adding $\delta \phi = 0.01$ to it.

One argument that has been made is that the \emph{slope} of the experimental data is different to that of the simulations. With the current (perhaps erroneous) determination of $\phi$, then the statement that the slope is different need not hold. Moreover, the lowest nucleation rates measured in experiment of around $J_\mathrm{exp}^\mathrm{min}\sim 10^{-5} \sigma^5 D_L^{-1}$ are \emph{over 25 orders of magnitude} higher than the lowest rates determined in simulation (Fig.~\ref{figOneToRuleThemAll}). So while it could be possible to reconcile the experimental and simulated rates, what is clear is that massively slower rates have been determined in simulation.

Clearly, experimental measurements at lower nucleation rates are desirable and it would seem that using smaller colloids or nanoparticles could be a way forward here, see Sec. \ref{sectionSize}. It may still be possible to achieve some reduction in size using light scattering, but perhaps neutron or X-ray scattering (in particular the recent developments noted in Sec.~\ref{sectionRealSpaceGlass})~\cite{wochner2009,liu2022,lehmkuhler2020}) could be experimental methods of choice in this case. Before leaving this topic, we emphasize one clear mystery: While the accuracy of phase boundary measurement could account for the magnitude of the change in volume fraction required to address the discrepancy, why do experiments systematically \emph{underestimate} the effective volume fraction?

\subsection{Homogeneous nucleation in binary hard-sphere mixtures}
\label{sectionHomegeneousBinary}

In principle, when dealing with a multicomponent suspension, such as a binary mixture,  CNT cannot be straightforwardly generalized, see e.g.~\citet{ni2011} and references therein, where several attempts are discussed. When  crystal nucleation takes place, fractionation indeed comes into play. The composition of the metastable fluid phase and that of the nucleating crystal phase then differ from the compositions of the coexisting bulk phases which has repercussions on how to apply the capillary approximation and how to define the interfacial free-energy~\cite{oxtoby1994,Laaksonen1999,ni2011}.

\begin{figure}[!t]
\centering
\includegraphics[width=80mm]{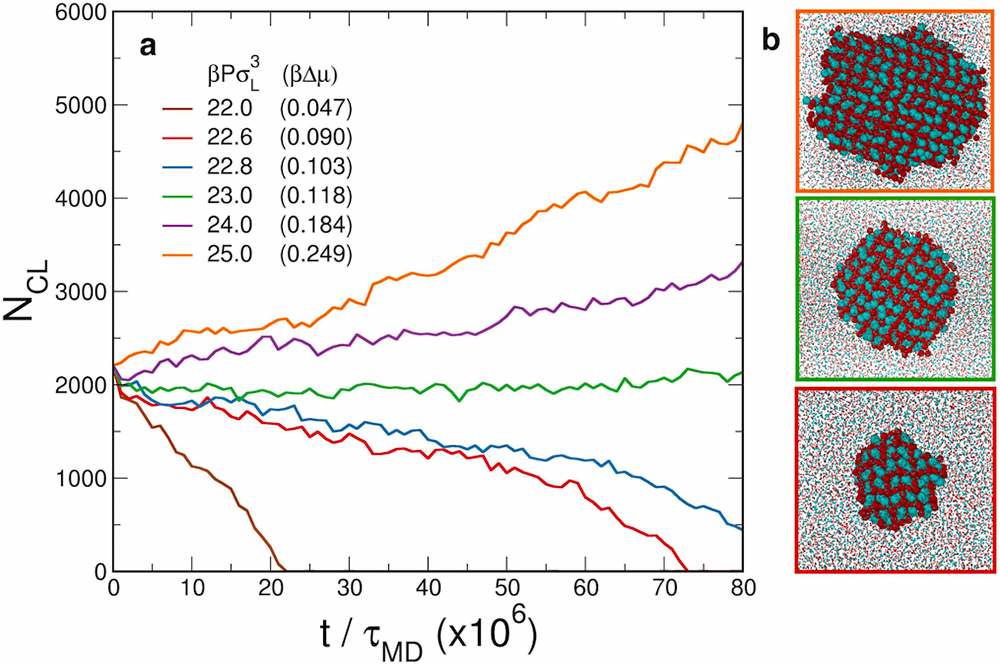}
\caption{Largest cluster size $N_\mathrm{CL}$ with Laves phase symmetry as a function of time $t$ using the seeding approach in MD simulations of a binary mixture of nearly hard spheres in the $NPT$ ensemble at composition $x_L=N_L/(N_S+N_L) = 1/3$ and a diameter ratio $q = 0.78$ for varying pressures. The initial seed size is 2205 particles of the MgZn$_2$ Laves phase. The snapshots in (b) show the melting of the seed at pressure $\beta P \sigma_L^3=  22.6$ (red box), growth of the seed at $\beta P \sigma_L^3 = 25$ (orange box), and a stable seed at the critical pressure $\beta P \sigma_L^3= 23$ (green box). The large (small) spheres are colored blue (red). Fluid particles (particles with a disordered neighborhood) are reduced in size for visual clarity.
Reprinted from~\cite{dasgupta2020} under CC-BY-NC-ND license.
\label{figBinaryNuclei}}
\end{figure}

~\citet{ni2011} numerically studied crystal nucleation in varying binary mixtures of hard spheres. They first investigated the effect of the order parameter on the cluster composition for nucleation of a substitutional solid solution of identical hard spheres but tagged with different colors, and concluded: (\emph{i}) the composition of noncritical clusters depends on the order parameter choice, but can be explained by the predictions from CNT nevertheless; and (\emph{ii})  the properties of the critical cluster do not depend on the order parameter choice. In addition, these authors studied the nucleation of an interstitial solid solution in a binary hard-sphere mixture with $q=0.3$. It was found that, for a suitable choice of order parameter, the composition of noncritical clusters is determined by the chemical equilibrium condition of the small spheres in the crystal nucleus and the fluid phase. One may expect to observe such a chemical equilibrium of the small species in the case of a highly asymmetric binary hard-sphere mixture, where the small spheres can diffuse throughout the whole system including the nucleated crystal.  For less asymmetric binary hard-sphere mixtures, where the small spheres cannot diffuse freely in the solid cluster, chemical equilibrium of the smaller species is harder to maintain, especially when the nucleated crystal phase has long-range crystalline order for both species as in the case of a superlattice structure~\cite{filion2011prl,riosdeanda2017}. It would be interesting to investigate what other mechanisms the system resorts to in order to maintain  equal chemical potentials of the two species between the fluid and the crystal nucleus.

A particularly enticing application of  binary hard-sphere mixtures is their putative role in synthesizing colloidal crystals with diamond and pyrochlore structures, which are characterised by wide photonic band gaps at low refractive index contrasts~\cite{hynninen2007}. Direct assembly is not deemed possible, but given a self-assembled binary mixture of colloidal spheres into a close-packed MgCu$_2$ Laves phase, one could selectively remove one of the sublattices to obtain  these low-coordinated crystalline structures. Excitingly, Laves phases are proven to be thermodynamically stable in binary hard-sphere systems with $q\approx0.8$, and clusters of Laves phases have been reported to form spontaneously in simulations~\cite{bommineni2020, marin2020}. Additionally, it was shown using the seeding approach (see Fig.~\ref{figBinaryNuclei}) that the barrier heights coincide for all three Laves phases, which is to be expected, as the free-energy differences between the three bulk phases are extremely small~\cite{dasgupta2020}.   Softened spheres  have further been shown to enhance crystallization of Laves phases, by suppressing the degree of five-fold symmetry in the binary fluid phase~\cite{dasgupta2020}. "These structures have nevertheless only been observed to spontaneously crystallize in hard-sphere experiments with sub-micron-sized colloids~\citet{schaertl2018} . We posit that the slow assembly dynamics may be the culprit, which could be addressed using smaller colloids, or even nanoparticles."

For more asymmetric hard-sphere mixtures, the AB$_{13}$ crystal  structure, analogous to the NaZn$_{13}$ phase, has been predicted to be stable in binary hard-sphere mixtures  30 years ago~\cite{eldridge1993nature}. The kinetic pathway for homogeneous nucleation of the icosahedral AB$_{13}$  crystal from a binary fluid phase of nearly hard spheres  has been numerically studied by~\citet{coli2021acsnano}, making use of an artificial neural network to identify the AB$_{13}$  phase from the binary fluid phase and the competing fcc crystal phase. Interestingly,  AB$_{13}$  crystal nucleation proceeds via a co-assembly process with large spheres and icosahedral small-sphere clusters simultaneously attaching to the nucleus. Even though the binary fluid phase is highly structured and exhibits local regions of high BOO parameter\footnote{Three-dimensional BOO parameters detect regions of local crystal-like order. They generalize the 2d BOO introduced in Sec.~\ref{sectionBulk} to quantify hexagonal ordering.}, the kinetic pathway follows CNT.

\subsection{Heterogeneous and seeded nucleation}
\label{sectionHeterogeneous}

\begin{figure}
\centering
\includegraphics[width=70 mm]{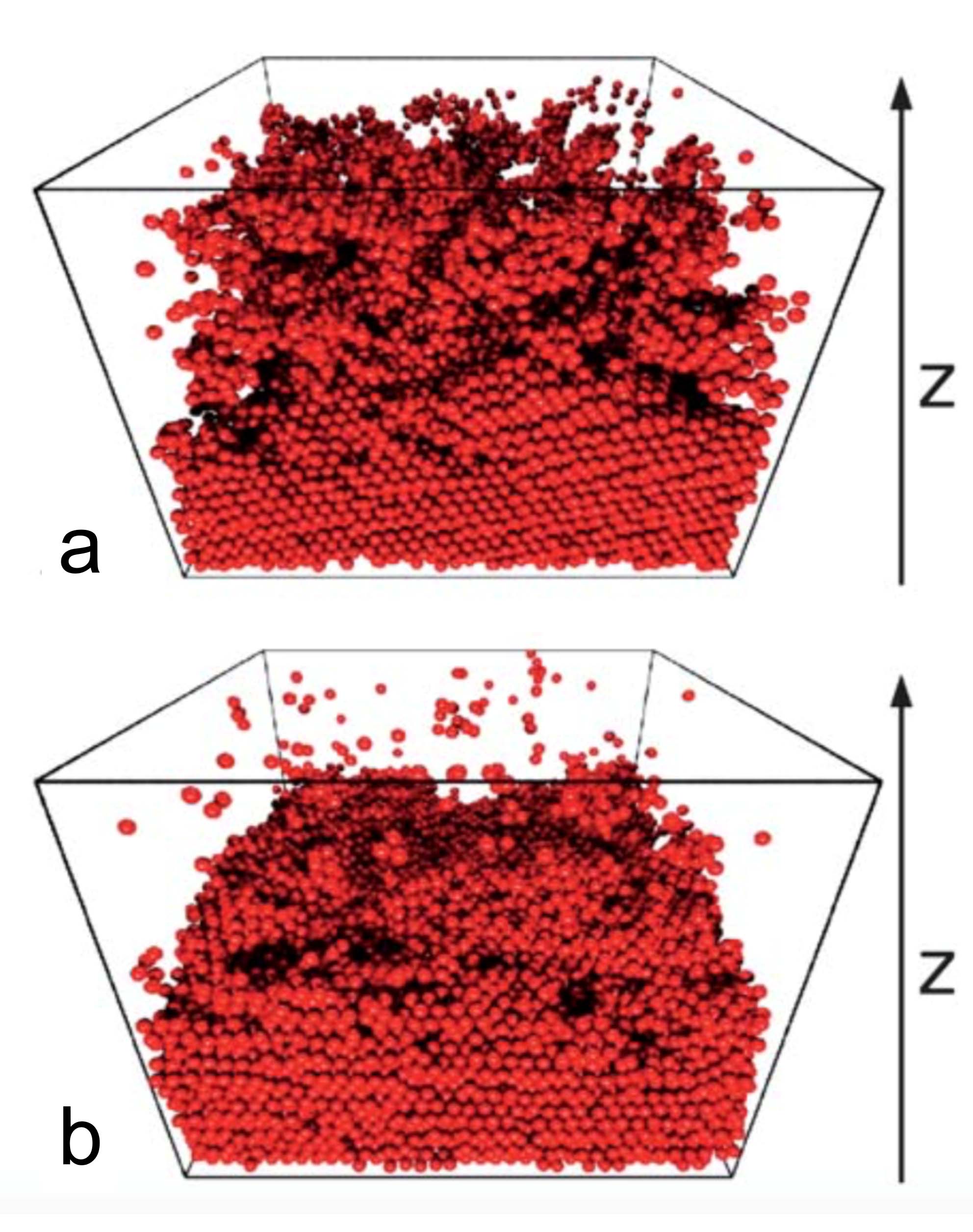}
\caption{Heterogeneous crystallization induced by a flat wall. Crystal domains found after sedimentation on a flat substrate, obtained from (a) confocal microscopy and (b) BD simulations. For both snapshots, the overall packing fraction is  $\phi=0.52$, the snapshots are taken after a waiting time of 
$t=62 \tau_B$, and only the crystalline particles are shown.
Reproduced from~\cite{sandomirski2011} with permission from the Royal Society of Chemistry. }
\label{figStefanEgelhaaf}
\end{figure}

Homogeneous nucleation of a metastable suspension of hard spheres has been extensively studied using numerical methods. Because such nucleation events are rare, and because the critical nucleus could form anywhere in the system and at any time, experimental detection is particularly challenging. Heterogeneous nucleation, by contrast, happens whenever the phase transition is assisted by inhomogeneities, such as walls (either flat~\cite{wette2009,cacciuto2005} or structured~\cite{hermes2011,heni2000,xu2000,teeffelen2008}), or  impurities~\cite{cacciuto2004,devilleneuve2005jpcm,devilleneuve2005science,sandomirski2011}. Nuclei of the stable phase then form at the surface of these exogenous bodies, thus facilitating their detection. As a result, for hard-sphere colloids the process has been studied extensively by means of both simulations and experiments.

In general, heterogeneous crystallization is controlled by the size, structure and composition of the seed as well as its inhomogeneities. For simplicity, however, standard seed models mostly consider flat and curved walls. Flat, unstructured walls, in particular, maximize the nucleation and crystal growth rates~\cite{kose1976}. Because flat walls do not strain the growing crystal, they also minimize the concentration of defects within the structure. Figure~\ref{figStefanEgelhaaf} shows two snapshots of crystal domains formed after sedimentation on a flat wall, both for (a) experiments with confocal microscopy and (b) BD simulations~\cite{sandomirski2011}.  For heterogeneous nucleation at a flat wall, the free-energy cost of forming a critical nucleus can be orders of magnitude smaller than for homogeneous nucleation.~\citet{auer2003} showed that at $\beta P\sigma^3=12.1$ ($\phi \simeq 0.497$) the  free energy barrier to nucleation goes from $\Delta G^*_\text{hom}=1334\,\kB T$ in the homogeneous case to $\Delta G^*_\text{het}=17\,\kB T$ for heterogeneous nucleation of a crystal growing with the (111) plane parallel to the flat wall, thus increasing the nucleation rate by roughly $570$ orders of magnitude over that of homogeneous nucleation under the same thermodynamic conditions. The reduction of the free-energy barrier is predicted from CNT to be
\begin{equation}
    \Delta G^*_\text{het}=\Delta G^*_\text{hom}\frac{(2+\cos\theta)(1-\cos\theta)^2}{4},
\end{equation}
where $\theta$ is the contact angle between the fluid and solid phases with the wall, and is given by $\cos\theta=\gamma_\text{wf}-\gamma_\text{ws}/\gamma_\text{fs}$, where $\gamma$ is the interfacial free energy, and the subscripts $w$, $s$, and $f$ refer to the wall, solid and fluid phases. The condition for which $\gamma_\text{ws}+\gamma_\text{fs}-\gamma_\text{wf}\leq 0$ corresponds to complete wetting of the crystalline surface on the wall. Using the values $\gamma_\text{wf}=1.99 \kB T/\sigma^2$~\cite{heni1999}, and, for the (111) plane, $\gamma_\text{ws}=1.42 \kB T/\sigma^2$~\cite{heni1999}, and $\gamma_\text{fs}=0.55 \kB T/\sigma^2$~\cite{benjamin2015}, we obtain $\gamma_\text{ws}+\gamma_\text{fs}-\gamma_\text{wf}=-0.02 \kB T/\sigma^2$, i.e., complete wetting as discussed in Sec.~\ref{sectionInterfaces}. The presence of a barrier $\Delta G^*_\text{het}=17\,\kB T$~\cite{auer2003} significantly higher than the thermal energy is thus at odds with CNT predictions. These differences can be somehow adjusted if the contribution from the wall-crystal-melt line tension~\cite{auer2003} (with a further correction to include its curvature dependence) is taken into account, and if the value of $\gamma_\text{wf}$ is allowed to deviate from its bulk value. Analysis of the simulation trajectories~\cite{auer2003} show that the crystal grows first laterally on the wall rather than extending into the bulk, indicating that heterogeneous nucleation happens close to the wetting threshold.

The effectiveness of a templated wall to induce epitaxial growth has been considered in experiments~\cite{hoogenboom2003prl}, simulations~\cite{cacciuto2005} and theory~\cite{heni2000}. In experiments, the structure and size of the seed cluster can be controlled at will by fixing colloidal particles with laser tweezers or/and putting a prescribed structure to the under-cooled or over-compressed colloidal fluid. In the epitaxial growth of colloidal hard-sphere crystals~\cite{hoogenboom2003prl}, a structured wall (favoring hcp crystallization) was offered as template to the colloidal solution. Perfect hcp-crystal growth was achieved for template unit cells that are isotropically stretched compared to the bulk unit cell dimensions. By contrast, isotropically compressed templates give rise to the growth of a perfect fcc. Large mismatches, however, suppressed crystallization. The computational study of~\citet{cacciuto2005} showed that disorder in the template can also suppress nucleation if the displacement of the template particles from their lattice sites is comparable to that specified by the Lindemann criterion ($10\%$ of the nearest-neighbor distance), while the template retains its full effectiveness for smaller displacements.

\begin{figure}[h!]
\centering
\includegraphics[width=85mm]{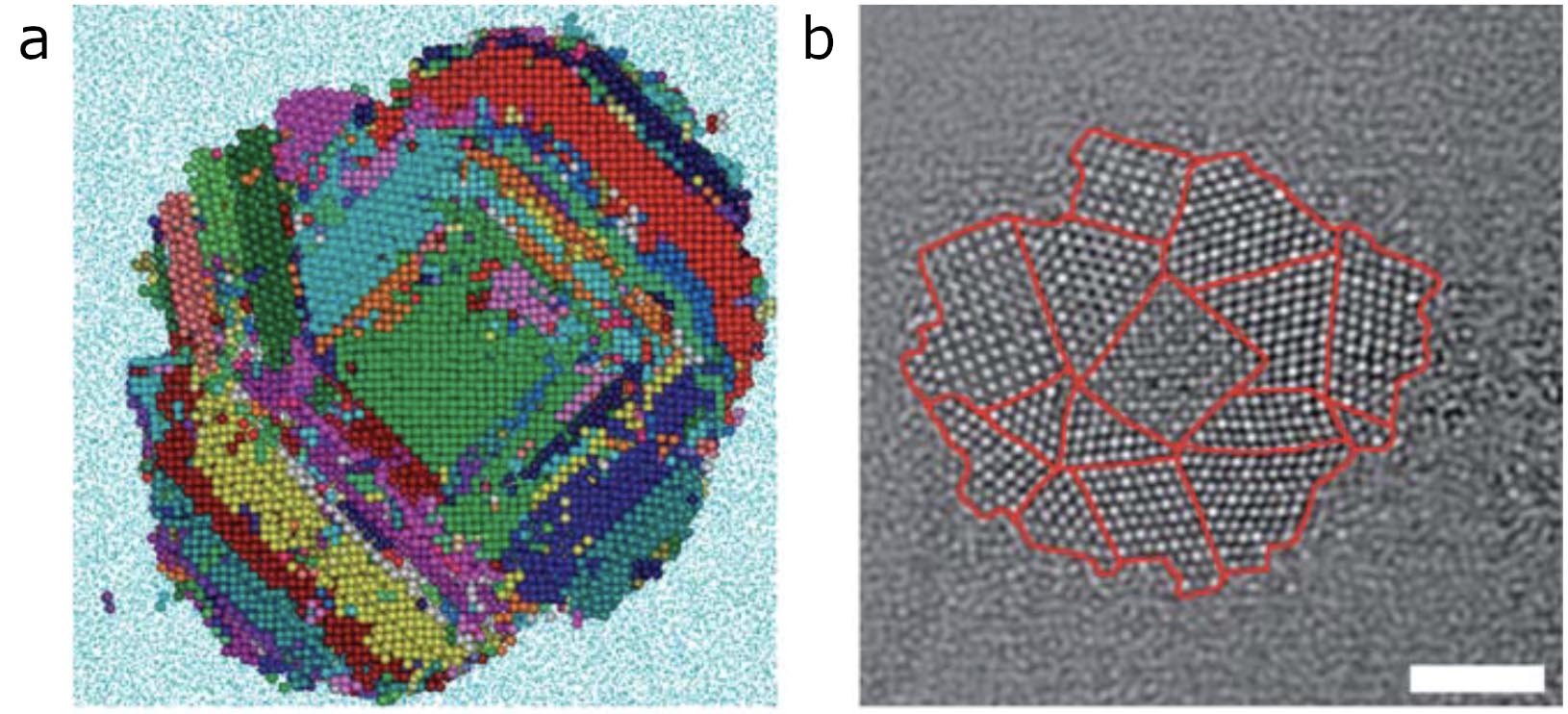}
\caption{Seeding a hard-sphere crystal from a square seed in (a) simulations at $\phi=0.51$, and (b) experiments with optical tweezers.
In (a), the colors correspond to different crystal grains, crystalline particles are drawn at their normal size, while fluid particles are drawn as dots. In (b), the scale bar is $10\,\mu m$. Reproduced from~\cite{hermes2011} with permission from the Royal Society of Chemistry.}
\label{figMichiel}
\end{figure}

~\citet{hermes2011} have experimentally studied crystal nucleation while initiating the phenomenon by means of a seed structure using optical tweezers. They showed that the nucleation barrier can be lowered by introducing a 2d seed structure into the bulk of a supersaturated fluid, resulting into large crystallites like that shown in Fig.~\ref{figMichiel}. Unlike~\citet{cacciuto2004}, they did not find  hexagonal seeds to be good nucleating agents. However, the square seed worked remarkably well, inducing nucleation at low supersaturation, although the resulting fcc crystals had significantly more defects than crystals obtained using hexagonal seeds.

As mentioned in Sec.~\ref{sectionDiscrepancy}, fully disordered templates, in which spheres are absorbed on a flat wall at random positions, completely suppress nucleation if the spheres of the template have sizes comparable to the size of the hard spheres in the fluid, but as the size of the template spheres increases (at least larger than $3\sigma$) heterogeneous nucleation from the walls is still favored over homogeneous nucleation~\cite{espinosa2019}.

\begin{figure}[h!]
\centering
\includegraphics[width=75mm]{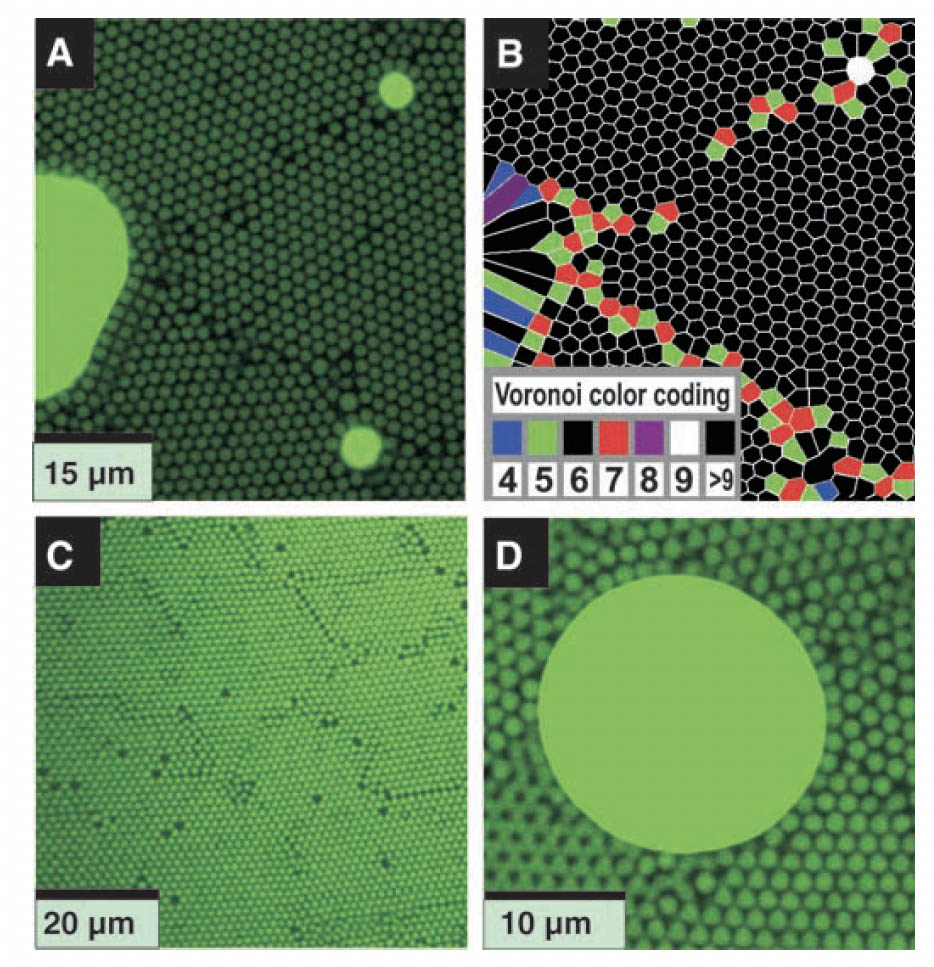}
\caption{Experimental work showing crystallization in the presence of curved walls.
(A) Grain boundaries connect impurities. 
(B) Voronoi representation of (A) shows that the defects nicely capture the grain boundary.
(C) In a sample without impurities, grain boundaries have annealed. 
(D) The mobile layer of single-particle thickness around an impurity with diameter ratio $\alpha=0.13$ appears fluid-like. 
Reprinted from~\cite{devilleneuve2005} with permission from AAAS.}
\label{figVisualisation}
\end{figure}

Spherical seeds are but a specific case of heterogeneous nucleation being promoted by curved walls. And just like the crystal growth rate increases with the radius of spherical impurities~\cite{devilleneuve2005}, strongly curved walls suppress nucleation by straining the growing crystal~\cite{cacciuto2004,devilleneuve2005}. Crystals which form on small spherical seeds further have an hexagonal structure with the (111) plane bending around the surface of the seed. Seed curvature is accommodated through grain boundaries and defects, and the crystal accumulates elastic energy. Small impurities  accumulate at grain boundaries of nuclei that are nucleated homogeneously in the sample, thus stabilizing polycrystalline samples~\cite{dullens2008,devilleneuve2009}. As a function of the seed diameter, one therefore observes a transition from a regime where heterogeneous nucleation is suppressed compared to homogeneous nucleation to one where it dominates. Simulations suggest that the barrier for heterogeneous nucleation is lower than that for homogeneous nucleation for seeds of radius larger than five particle diameters~\cite{cacciuto2004}. This value of the threshold radius is not far from the value found for a flat wall with randomly absorbed spheres discussed earlier, i.e., $3\sigma$~\cite{espinosa2019}.

Both simulations~\cite{cacciuto2004} and experiments~\cite{sandomirski2014,allahyarov2015} have reported that an interesting phenomenon occurs once the elastic stress balances the interfacial energy gained through the heterogeneous nucleation on the seed.  The crystal nucleus then detaches from the seed before reaching its critical size, but continues to grow, eventually reaching the seed, which then acts as an impurity that hinders further growth. This process also prevents the spherical seed from acting as a crystallization catalyst--the catalyst becomes poisoned--as heterogeneous nucleation is inhibited by the detached nucleus nearby. For hard spheres, the phenomenon takes place for seeds of size of about $30\sigma$~\cite{allahyarov2015}, but this multi-step nucleation scenario is expected to occur in all situations in which the seed induces a structural mismatch compared to the equilibrium crystal lattice. Because a perfect match is impossible to achieve, this mechanism is expected to be quite general and important for the control of crystal morphology. All these phenomena are also observed for hard spheres assembling in spherical confinement, i.e., a negative curvature or concave surface, as demonstrated by simulations, and discussed in Sec.~\ref{sectionConfinement}.D. In this case as well, heterogeneous nucleation is favored over homogeneous nucleation, and crystallization begins at the boundary of the spherical confinement, forming a polycrystalline system consisting of 20 tetrahedral fcc domains. When the inward-growing fcc domains cannot accommodate the strain induced by the curvature of the spherical confinement, the crystalline layers at the spherical boundary melt and later recrystallize, forming an ``anti-Mackay" cluster~\cite{denijs2015}. Cylindrical seeds have also been considered~\cite{auer2004,cacciuto2004,sandomirski2014}. Qualitatively, the situation is similar to what is observed with spherical seeds, although with larger nucleation rates~\cite{cacciuto2004}. The presence of only one principal curvature direction--compared to two for spheres--indeed reduces the strain on the growing crystal.

\subsection{Crystal growth}

Following nucleation, crystallization proceeds through crystal growth, which is well dynamically separated from the formation of the initial activated process. The growth process is characterized by a crystal-fluid interface with a thickness that ranges from 8 to 16 layers, as confirmed in experiments~\cite{dullens2006prl}, simulations~\cite{auer2003,zykova2010} and theory~\cite{hartel2012,oettel2012}. At this stage, the number of crystalline layers grows linearly with time~\cite{ackerson1995,derber1997,sandomirski2011}, at a rate that is maximal around $\phi=0.52$ in simulations and $\phi=0.53$ in experiments~\cite{sandomirski2011}. The non-monotonicity is ascribed to the competition between the increasing driving force to crystallize and the decreasing diffusivity $D_L$ of hard spheres as $\phi$ increases~\cite{sandomirski2011}. Crystal growth is indeed so rapid compared to particle diffusion that a depletion front in the fluid in contact with the crystal has been observed in experiments and simulations~\cite{ackerson1995,derber1997,sandomirski2011}. At later times, crystal growth slows down, crystal layers expand slightly, and the depletion zone vanishes.

Light scattering experiments by Sch\"{o}pe and collaborators have studied in detail the process of crystal growth in polydisperse hard spheres~\cite{schope2007,iacopini2009}. At volume fractions slightly above melting, it was found that crystal growth is ripening dominated, with the average crystal growth scaling as a power law in time. Upon increasing $\phi$, growth gets increasingly hindered, and the initiation of ripening-like growth is further delayed. Therefore, while samples close to coexistence conditions tend to achieve the highest possible
crystal structure quality, samples above melting form crystals with many defects that are later annealed over the whole crystallization process, and with particle rearrangements mostly occurring at grain boundaries.

Experiments of crystal growth in microgravity~\cite{cheng2001prl} have observed dentritic growth, which is absent in normal gravity. Here the growth is observed to proceed with a dramatic increase of crystallinity (as measured from the intensity of the scattered light), but without large changes in the average linear dimension of the crystals. The root of this difference, however, remains largely unexplained.

\subsection{Challenges to the fundamental assumptions of CNT}
\label{sectionChallengesCNT}

As discussed in Sec.~\ref{sectionCNT}, CNT provides a simple yet powerful framework to understand and analyze nucleation data for a large variety of processes. The theory nevertheless depends on several assumptions and approximations, and deviations from these have led to non-classical versions of the theory.

A fundamental assumption of CNT is its reliance on a single order parameter for  describing first-order phase transitions~\cite{prestipino2014,jungblut2016,russo2016jcp,lutsko2019}. This simplification is often appropriate for transitions between phases with the same symmetry, such as the gas-to-liquid phase transition, but the liquid-to-crystal phase transition is different in the sense that both the translational and orientational symmetries of the liquid are then broken. In 2d hard disks at equilibrium, this distinction results in orientational symmetry being broken separately from the translational symmetry, as discussed in Sec.~\ref{sectionBulk}. Even though both translational and orientational orders share a same equilibrium phase transition in 3d hard spheres, the two features can evolve differently during the out-of-equilibrium nucleation process
~\cite{russo2016mrs,duran2018,lutsko2019,rogal2022}. Experiments~\cite{schatzel1993,schope2006pre,schope2007,schope2006prl,franke2014,tan2014,iacopini2009} and simulations~\cite{schilling2010prl,schilling2011jpcm,russo2012,leocmach2013,russo2016jcp,berryman2016} on precursors, i.e., in the regions where small pre-critical nuclei are formed, evince that more than one reaction coordinate is involved in hard-sphere nucleation. Another central assumption of CNT is that the order parameter is taken to be a slow variable with Markovian dynamics~\cite{richard2018jcpkinetics},
an assumption that has been challenged recently~\cite{kuhnhold2019}. These two effects, however, could be argued to be fairly mild deviations from CNT. The rest of this subsection describes two more significant effects.

\header{Polymorph selection}
During the crystal nucleation process, hard spheres can assemble in one of the infinitely-many Barlow packings. As in Sec.~\ref{sectionBulk}, we here distinguish three different  \emph{polymorphs} (or rather \emph{polytypes}): the face-centered cubic (fcc), the hexagonal close packed (hcp), and the random hexagonal close packed (rhcp) crystals.
Because these polymorphs have nearly equal free energies, with only $\sim 10^{-3}\,\kB T$ per particle favoring  fcc~\cite{bolhuis1997} (see Sec.~\ref{sectionBulk}), one might expect that the early stages of nucleation should produce an equal amount of fcc and hcp (often in the form of rhcp), with the fcc becoming the stable crystalline structure only for large crystallites.
Some computational studies have reported large deviations from these predictions; the proportion between fcc and hcp within the nucleus during the later stage of nucleation varies from 2 to 1, to 3 to 1~\cite{filion2010,russo2012}. However, experimental work using light scattering powder crystallography found the initial crystal to be highly random (rhcp) with no preference for one of another polytype, although a preference for fcc was detected at long times (see Sec.~\ref{sectionBulk} ~\cite{kegel2000jcp,martelozzo2002}.

\begin{figure}[h!]
\centering
\includegraphics[width=75mm]{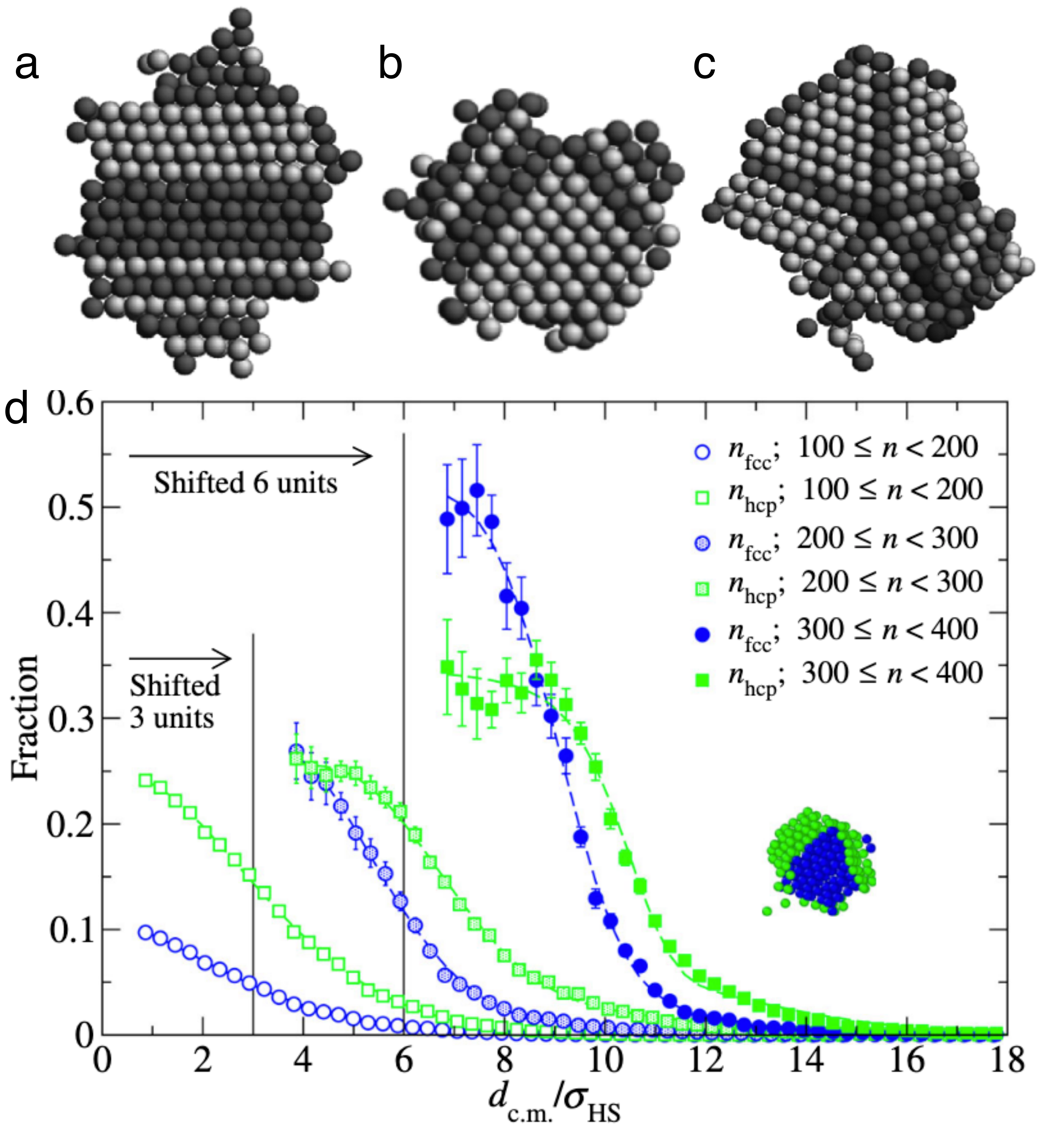}
\caption{Typical grain boundaries formed during  hard-sphere nucleation (light spheres are fcc, dark spheres are hcp): 
(a) stacking faults, 
(b) tetrahedrally shaped fcc domains bounded by stacking faults, 
(c) five-fold coherent grain boundary. Reproduced from~\cite{omalley2003}.
(d) The average radial fractional composition of nuclei  calculated with respect to the center of mass for different cluster sizes. The snapshot presents the section of a typical nucleus with 400 particles. fcc and hcp particles are in blue and green, respectively. Reproduced from~\cite{leoni2021} under CC BY 4.0 License.}
\label{figGrain}
\end{figure}

A recent simulation work~\cite{leoni2021} on hard-sphere nucleation has considered the sensitivity of the fcc to hcp ratio on the choice of order parameter. While low-dimensional order parameters built on simple BOO parameters found an excess of the fcc phase, high-dimensional order parameters built from a complete set of atomic descriptors for the local environment surrounding each sphere  found that the total ratio between fcc and hcp is close to unity for sufficiently large nuclei. The radial composition of the nuclei was also found to be inhomogeneous, with a relative preference for fcc in the core compared to the shell. This preference for fcc is attributed to its small entropic gain  in that it allows for stacking disorder to appear along four different directions, compared to a single direction of both hcp and rhcp crystals (the direction perpendicular to the basal plane). Stacking in multiple directions is accompanied by the formation of a coherent five-fold grain boundary at the intersection of the different stacking directions, as shown in Fig.~\ref{figGrain}c. The formation of this five-fold coherent grain boundary was found to be very abundant in the early stages of nucleation~\cite{omalley2003,leoni2021}, as shown in Fig.~\ref{figGrain}, thus resulting in more compact nuclei with an fcc core that emerges from the more diffuse pre-critical nuclei. This mechanism provides a microscopic explanation for the abundance of fcc in the cores. It is also an example of two-step nucleation governed by finite-size effects, in that the relative abundance of the different phases changes with nucleus size. Two-step nucleation mechanisms had previously often been invoked to explain onion-like 
structures~\cite{santra2013,tan2014,kratzer2015,guo2016,eslami2017,adorf2019}.

\begin{figure}[h!]
\centering
\includegraphics[width=85mm]{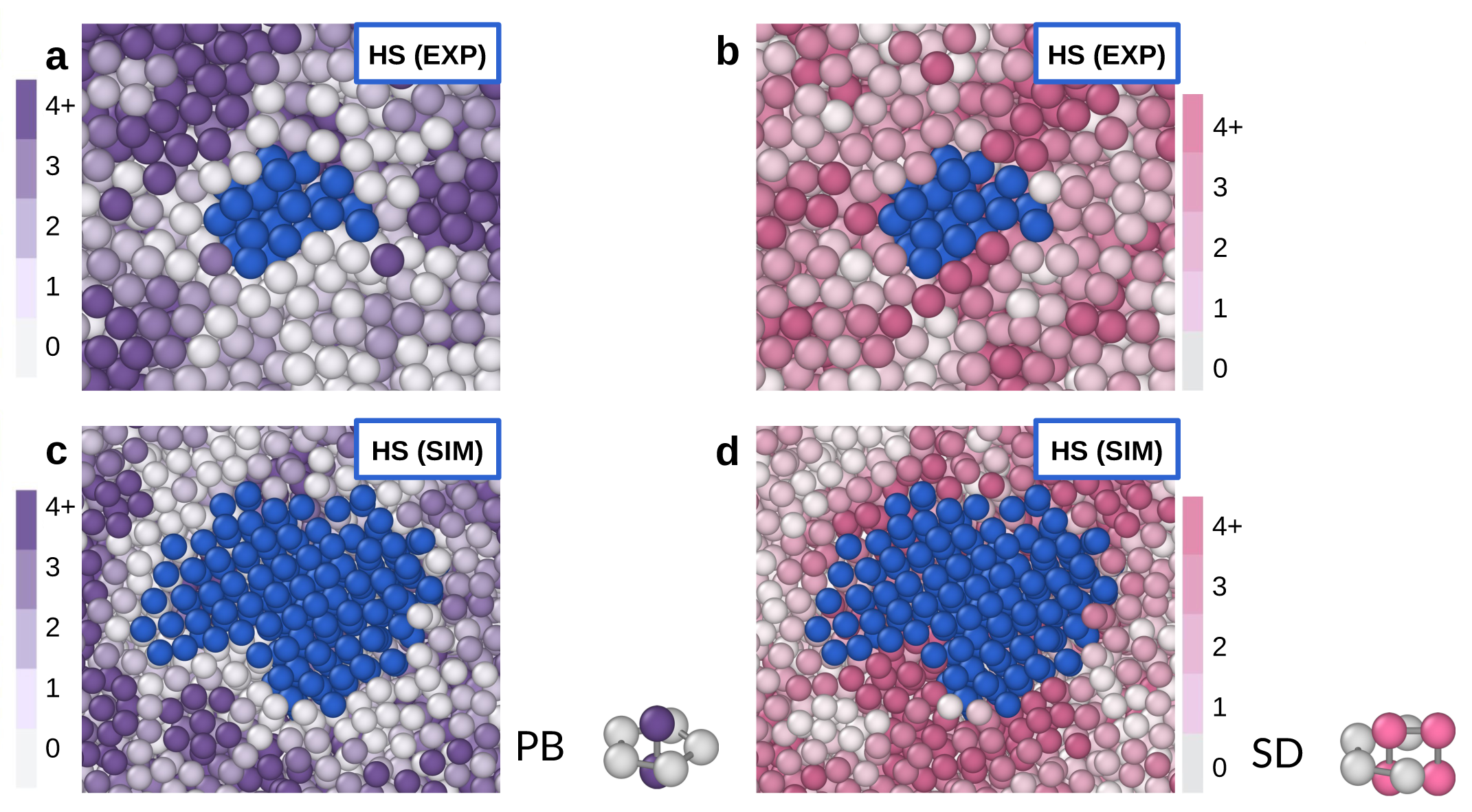}
\caption{
The mechanism of polymorph selection in hard sphere nucleation.
PB (a,c) and SD (b,d) clusters during crystal nucleation in hard spheres around $\phi=0.54$. 
(a,b) PMMA spheres imaged with confocal microscopy. 
(c,d) Nearly hard spheres with WCA potential interactions. 
Shown are cut-through images of crystal nuclei. The core of the crystal nucleus is colored dark blue, while the rest of the particles are colored following the scale bar on the left (a,c) or right (b,d) depending on the number of PB (a,c) or SD (b,d) clusters each particle belongs to. Adapted  from~\cite{gispen2023acsnano} under CC BY 4.0 License.}
\label{figPolymorph}
\end{figure}

It has also recently been shown  that the polymorph selection leading to an  abundance of fcc at the core of hard-sphere nuclei is already hidden in the metastable fluid phase as shown in Fig.~\ref{figPolymorph}~\cite{gispen2023acsnano}. Applying the topological cluster classification method of analyzing higher-order fluid structure (Sec.~\ref{sectionStructureDynamics})~\cite{malins2013tcc} to both simulation and experimental data revealed that two geometric motifs, or clusters, siamese dodecahedra (SD) and pentagonal bipyramids (PB), play particularly important roles. 
PB are known to suppress nucleation~\cite{taffs2016}, while SD form a crucial link between fcc crystals and the fivefold symmetric fluid: they feature elements of both in their structure. Thus SD is able to ``fit'' an fcc nucleus into the hard-sphere fluid. This finding presents a geometric mechanism for polymorph selection of fcc over hcp~\cite{gispen2023acsnano}.

\header{Capillarity approximation}
Another central tenet of CNT is that crystalline nuclei are governed by the same bulk values as the flat fluid-crystal interface. Although this assumption is clearly violated in the small nucleus regime, CNT can still be considered an effective theory that provides the correct scaling for different thermodynamic quantities, albeit with renormalized constants. For example, an artificially high value of the interfacial free energy  needs to be introduced to match computed nucleation rates with CNT predictions. Compared to the flat-interface (equilibrium) value of $\beta\gamma_\infty\sigma^2\simeq 0.56$~\cite{espinosa2016,bultmann2020} (see Sec.~\ref{sectionAdvancedSimulation}), a fit of the nucleation data typically requires values of $\beta\gamma\sigma^2\simeq 0.76$~\cite{richard2018jcpkinetics} for the quasi-equilibrium nucleation process.

\begin{figure}[t]
\centering
\includegraphics[width=\linewidth]{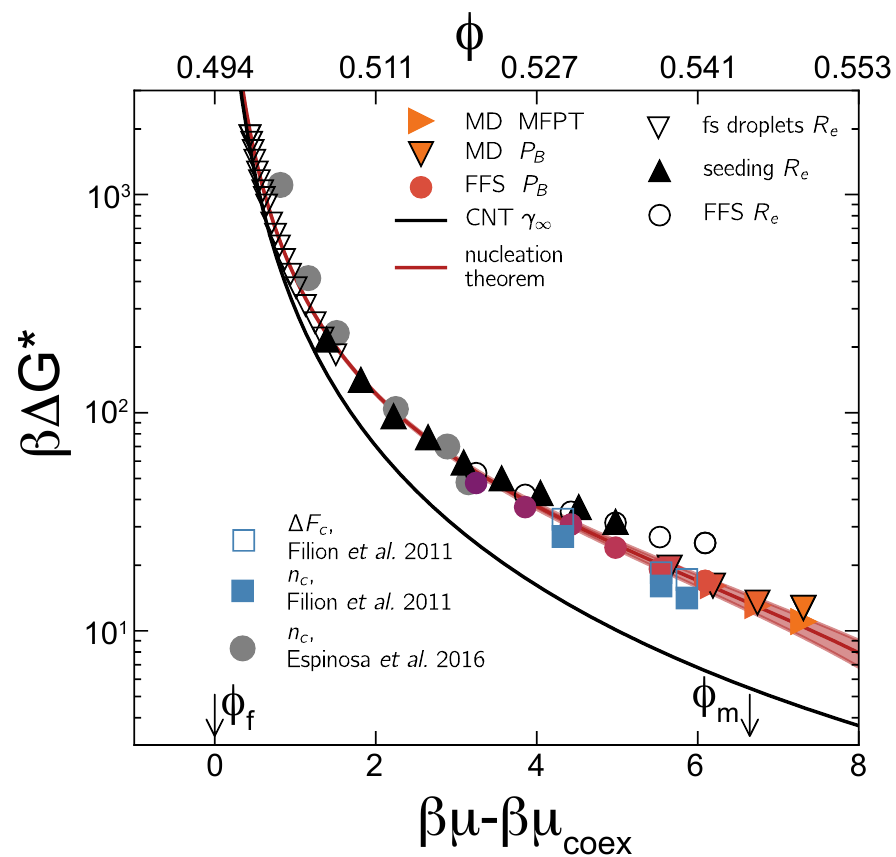}
 \caption{Nucleation work $\Delta G^\ast$ as a function of the supersaturation
obtained from unbiased MD simulations (down and right triangles), forward flux sampling (circles), and umbrella sampling (squares). The black solid line is the CNT prediction based on the bulk values of the pressure difference and surface tension. The thick red solid line is obtained from the nucleation theorem. The data are plotted as a function of chemical potential with respect to phase coexistence (lower axis) from which volume fraction is determined, using the Carnahan-Starling relation  in Eq.~\eqref{eqCS}. Plotted simulation data has been obtained in the $NPT$~\cite{filion2011jcp,espinosa2016} and $NVT$ ~\cite{richard2018jcpthermodynamic} ensembles. Reproduced from~\citet{richard2018jcpthermodynamic}, with the permission of AIP Publishing.}
\label{figRichardWork}
\end{figure}

Note that the nucleation theorem~\cite{hill1962} 
\begin{equation}
\Delta G^\ast(\mu_\text{f}) = \Delta G^\ast(\mu_0) - \int_{\mu_0}^{\mu_\text{f}} d\mu \Delta N^\ast(\mu)
\label{eqNucleationTheorem}
\end{equation}
provides a route for calculating the nucleation work from density profiles of critical nuclei, which can be obtained very precisely (e.g. through seeding, cf. Sec.~\ref{sectionSimulation}). Here, $\mu_0$ denotes a reference state point and $\Delta N^\ast$ is the average number of additional particles in critical nuclei compared to the bulk fluid with chemical potential $\mu_\text{f}$. Advantageously, this quantity does not suffer from the ambiguities associated with order parameters used to identify solid-like particles. Figure~\ref{figRichardWork} shows that the prediction from Eq.~\eqref{eqNucleationTheorem} agrees very well with other estimates of the nucleation work. More importantly, it shows that the CNT prediction employing the bulk values $\gamma_\infty$ for the interfacial tension (and pressure difference) severely underestimates the actual nucleation work.

\subsection{Beyond the CNT regime: Crystallization at high volume fractions}
\label{sectionBeyondCNT}

\begin{figure*}
\centering
\includegraphics[width=170mm]{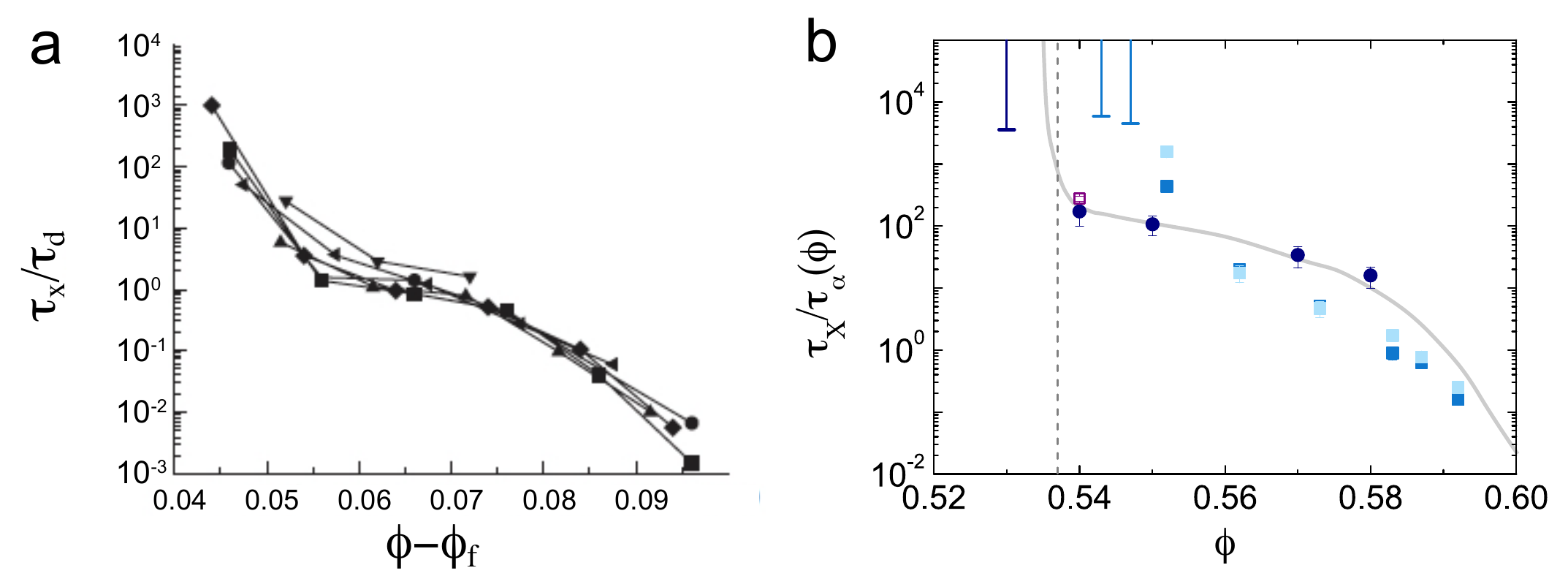}
\caption{ Spinodal nucleation.
(a) Crystallization times $\tau_X$  scaled
by the time $\tau_D$ for a particle to diffuse one diameter as a function of $\phi - \phi_f$. (Recall that monodisperse system freeze at $\phi_f=0.492$.). The different points represent the polydispersities ranging from $s=0$ up to $6\%$. The monodisperse system is reported with filled circles. Reproduced from~\cite{pusey2009} with permission from the Royal Society.
(b) Crystal nucleation times scaled by the relaxation time $\tau_\alpha$. 
Circles are experimental data obtained via particle-resolved studies, light and dark squares are simulation data for $s=4\%$ polydisperse systems with $N=2048$ and $N=10,976$, respectively. The unfilled square is for a monodisperse system with $N=10,976$. Dashed lines are melting estimated as described in Sec.~\ref{sectionEquilibriumPhaseBehavior}. The solid line is to guide the eye. Error bars extending upwards are lower bounds for crystallization times determined from experiments (light lines) and simulations (dark lines) which did not crystallize.
Reproduced from~\cite{taffs2013} with permission from the Royal Society of Chemistry.
}
\label{figXtalTimes}
\end{figure*}

Up to $\phi\approx0.55$, the CNT description qualitatively captures the homogeneous nucleation of a monodisperse suspension of hard spheres (see Sec.~\ref{sectionCNT}).~\citet{auer2001prediction} and~\citet{filion2010} have calculated the nucleation free-energy barrier height for $\phi\in[0.521,0.534]$ and found it to drop rapidly upon further increasing $\phi$. An approximate extrapolation~\cite{pusey2009} suggests that the nucleation barrier height becomes of the order of $\kB T$ around  $\phi=0.55-0.56$. In other words, the barrier can then be easily crossed and crystal nucleation is no longer a rare event (in the thermodynamic sense). The lifetime of the metastable liquid is then so short that nucleation can only be studied following rapid density quenches. Depending on the supersaturation of this quench, crystallization proceeds via two distinct mechanisms. For $0.55 \le \phi \le 0.58 $, the metastable fluid crystallises via a \emph{spinodal}-like mechanism. For $\phi \ge 0.58$, the fluid forms an out-of-equilibrium \emph{glass-like} state and crystallization proceeds via an  \emph{avalanche}-mediated mechanism.

\header{{\it Spinodal}-like crystallization}
General phase transformations in the binodal region proceed via a nucleation mechanism,  a process initiated by finite-amplitude,  localized fluctuations in the metastable fluid. By contrast, phase transformations in the unstable region occur via a spinodal decomposition mechanism~\cite{gunton1983}, identified by non-local fluctuations with an infinitesimal amplitude. At the (mean-field) spinodal point~\cite{gunton1983}, the nucleation free-energy barrier vanishes, and the stable phase spontaneously grows. The deterministic nonlinear amplification of order parameter fluctuations~\cite{bray1994} is then  governed by the Cahn-Hilliard equation~\cite{cahn1957}.

An analogous mechanism has  been proposed to explain crystallization for deeply supersaturated ($\phi\gtrsim0.55$) hard-sphere fluids, where the nucleation free-energy barrier becomes negligible~\cite{pusey2009}. Crystallization then proceeds via spatially diffuse collective motion, analogous to  spinodal decomposition~\cite{trudu2006,cavagna2009,yang1988}.

\begin{figure}[h!]
\includegraphics[width=85mm]{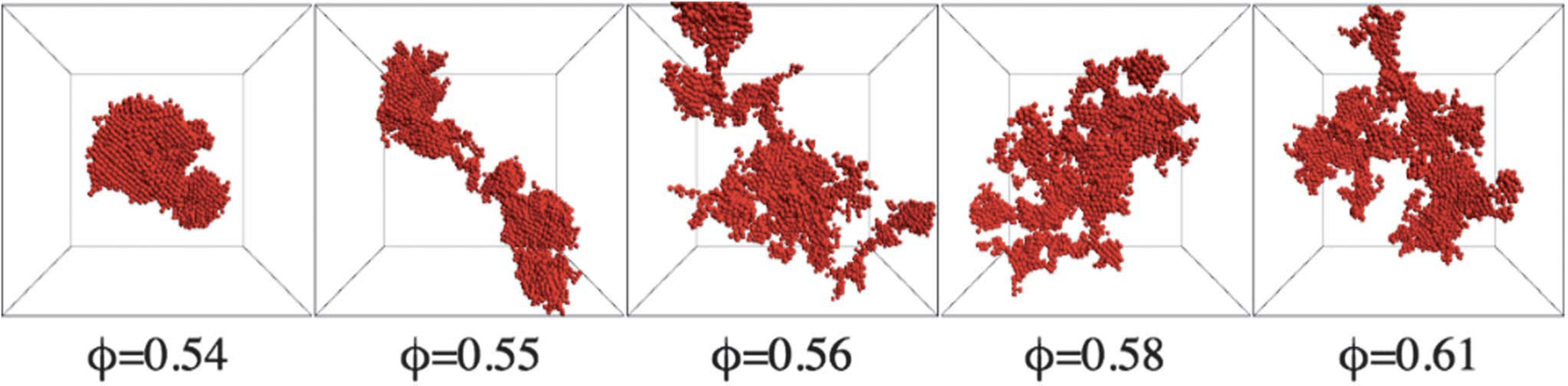}
\caption{ Snapshots of typical largest crystalline clusters as a function of $\phi$ at fixed cluster size (s= 5000). This size is achieved when ($\phi$, X) are respectively
(0.54, 0.08), (0.55, 0.15), (0.56, 0.15), (0.58, 0.14), (0.61, 0.12). Periodic boundary conditions are taken into account and clusters are centered in the
simulation box. Reproduced from~\citet{valeriani2012} with permission from the Royal Society.}
\label{figSpinodalClusters}
\end{figure}

When comparing the crystallization time (the time at which crystallinity reaches 20\% of the sample, for example) to the time needed for a particle to diffuse one diameter in the fluid, different behaviors have indeed been detected depending on the degree of metastability~\cite{pusey2009}. (Numerical results~\cite{pusey2009,taffs2013} have been shown to agree with experiments~\cite{harland1997,taffs2013}.) As shown in Fig.~\ref{figXtalTimes}, at low metastability, the system crystallizes via  a nucleation and growth process. Particles freely diffuse before crystal nucleation takes place; a crystalline nucleus has to grow large enough to overcome the nucleation free-energy barrier,  before macroscopic crystallization takes place. 
By contrast, for $\phi \gtrsim 0.55$, nucleation proceeds without particles  moving beyond their own diameter. A local re-arrangement of particle positions suffices for crystallization to proceed ~\cite{kelton1991,zanotto}.  This \emph{spinodal nucleation} regime is characterised by a large driving force for crystallization and a vanishing free-energy barrier~\cite{cavagna2009,trudu2006,pusey2009,klein1986}, thus resulting in a large  density of small crystal nuclei~\cite{pusey1986,vanmegen1993nature,schatzel1993}.  At even higher supersaturations, $\phi \gtrsim 0.56$, 
clusters of crystalline particles (Fig.~\ref{figSpinodalClusters}) heterogeneously percolate, following many small nuclei coming into contact. Clusters do not grow completely at random, but particles become solid-like in the vicinity of solid regions~\cite{sanz2011,valeriani2012}.

A similar behavior has been reported in  the  experimental work of~\citet{vanmegen1993nature} and~\citet{schatzel1993}. In this work, the crystallite size was found to markedly decrease  with increasing concentration. Their picture is coherent with the idea of a spinodal nucleation regime, in which increasing supersaturation  leads to  nucleation on an ever decreasing spatial scale.

\header{Crystallization in the glassy regime}
At higher volume fractions still, hard spheres exhibit glassy dynamics (see Sec.~\ref{sectionGlass}). Particle size polydispersity is then key to controlling crystallization. As described in Sec.~\ref{sectionPolyPhase}, when $s>5--6\%$, crystallization is suppressed because the system cannot form a stable crystal with the same composition as the fluid. Crystal formation then requires either size fractionation~\cite{fasolo2003,martin2003}, or the assembly of Laves phases~\cite{bommineni2019}, both necessitating transport over larger distances (see Sec.~\ref{sectionEquilibriumPhaseBehavior}). For $s<5\%$, however, neither the structural relaxation times of a dense colloidal suspension~\cite{henderson1996,foffi2003,sear2000} nor its dynamics~\cite{zaccarelli2009} are significantly affected by $s$.

For $\phi\approx0.57-0.58$, the crystallization time becomes smaller than the structural relaxation time of the fluid (Fig.~\ref{figXtalTimes} (a) and (b))~\cite{pusey2009,taffs2013}. For $\phi > 0.58$, even though particles move on average less than one particle diameter crystallization can still proceed~\cite{pusey2009}. Numerical results have shown that a suspension of monodisperse hard spheres ~\cite{williams2001,rintoul1996}  could crystallise even for supersaturations of out-of-equilibrium systems approaching random close packing $\phi_\mathrm{rcp}\approx0.64$ (see Sec.~\ref{sectionJamming}). It  even proceeds  on  numerical time scales short compared to the aging time of the amorphous hard-sphere system (see Sec. \ref{sectionAging})~\cite{zaccarelli2009}. In this glassy regime, crystalline clusters percolate as they do in the spinodal regime, even as the relaxation dynamics slows down by several orders of magnitude. In other words, the structure  of the growing crystals appears to be unaffected by particle dynamics~\cite{valeriani2012}.

As in computer simulations,  the experimental findings of Van Megen and Underwood in normal gravity~\cite{vanmegen1993nature} demonstrated the presence of a change in the crystallization mechanism at packing fraction  $\phi=0.58$. They detected homogeneous nucleation of compact nuclei for $\phi < 0.58$ and  asymmetric nuclei for $\phi > 0.58$. For high volume fraction, crystallization happened without particle diffusion. Quite possibly,  nucleation was heterogeneously induced  by preformed  nuclei, which may not have been fully shear--melted prior to the experiment.
By contrast, experiments in microgravity  showed that slightly polydisperse colloidal suspensions of PMMA hard spheres  rapidly crystallize in  bulk when $\phi>0.58$~\cite{zhu1997,cheng2001}.

\header{Avalanche-mediated devitrification}
Crystallization from a deeply over-compressed suspension occurs without the need for macroscopic diffusion~\cite{sanz2011}, but instead through the gradual formation of crystalline patches~\cite{sanz2011,sanz2014}.  The  crystallization  mechanism has been related to a series of discrete avalanche-like dynamical events~\cite{kwasniewski2014}, characterized by a spatio-temporal burst of particle displacements on a sub-diameter scale~\cite{sanz2014,montero2017,kwasniewski2014}.

\begin{figure}[h!]
\centering
\includegraphics[width=85mm]{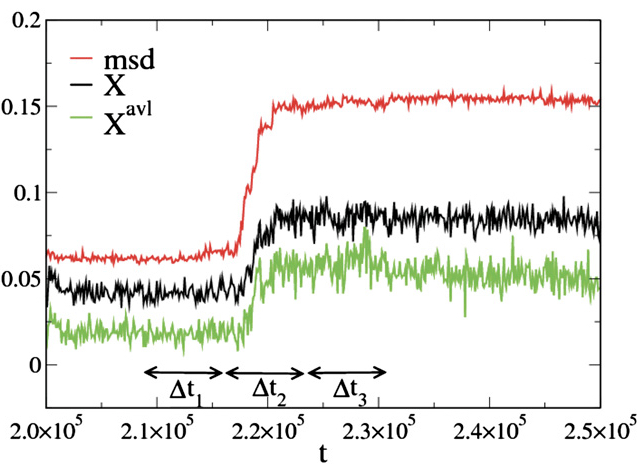}
\caption{Development of intermittent, heterogeneous dynamics in concentrated hard-sphere suspensions. Crystallinity $X$ (in black) and mean squared displacement (MSD, in red) versus time around the step-like crystallization event shown at $t = 2.2 \times 10^{5}$. The green curve, $X^{\mathrm{avl}}$, is the fraction of avalanche particles that are solid-like. Reproduced from~\citet{sanz2014} under PNAS License to Publish.}
\label{figAvalanche}
\end{figure}

As shown in Fig.~\ref{figAvalanche}, during a quiescent interval ($\Delta t_1$) most particles rattle locally in their cages and less than 1\% undergo significant displacements. At $\Delta t_2$,  a burst of displacements is recorded, with around 25\% of all particles moving more than $3 \sigma$.
After that, the system returns to quiescence ($\Delta t_3$). 
Such a sequence of events corresponds to an avalanche, and particles that move more than $3\sigma$ during the jump are deemed ``avalanche particles”. The particle dynamics was thus shown to be intermittent: quiescent periods of motion within the cage of neighbors are punctuated by avalanches in which a correlated subset of particles undergo cage-breaking displacements~\cite{sanz2014}.
Interestingly, crystallinity (in black) and  mean squared displacement (in red)  jump simultaneously, suggesting that the avalanches responsible for particle displacement are also responsible for crystallization.

The structural propensity to locally crystallize in a   ``medium-range crystalline order'' region~\citet{kawasaki2010pnas} is converted into actual crystallinity by the small random disturbances provided by the displacement avalanche. 
Given that  the  crystal is locally denser than the glass, the avalanche-mediated crystallization process leads to an increase of the local free volume, thus facilitating a higher  mobility of particles next to the newly formed crystalline region. Therefore, crystallization  proceeds by a sequence of stochastic  
events leading to an increased number of crystalline particles, correlated in space by emergent local avalanches~\cite{valeriani2011}.

Recently, Yanagishima and coworkers~\cite{yanagishima2017} have explored the mechanism behind avalanches, arguing that aging and devitrification are both triggered by a small number of particles that are driven to rearrange in regions of low density and BOO parameter. In particular, avalanches are accompanied by a transient loss of mechanical equilibrium which facilitates a large cascade of motion. The connection between mechanical rigidity and glassy dynamics was explored by~\citet{yanagishima2021}. By artificially minimizing the heterogeneities in the force network between particles, ultra-stable glasses that are free from aging and crystallization were obtained

Avalanches~\cite{sanz2014} have further been related to dynamical heterogeneity in that clusters of particles with high/low mobility evolve similarly in space and time. (A description of dynamical heterogeneity in the context of 
overcompressed
hard-sphere fluids is given in Sec.~\ref{sectionRealSpaceGlass}.) While direct visualization of avalanches in experiment has yet to be performed, measurements of dynamical heterogeneity (Fig.~\ref{figEricWeeks}) 
indicate that this is possible. In these systems, avalanches and dynamical heterogeneity might share a more general tendency to develop in regions of the system akin to ``soft spots'' regions of high displacement in low-frequency quasi-localised phonon modes~\cite{brito2009} . A deeper analysis would, however, be needed to better understand this issue.

In a glassy hard-sphere suspension (see Sec.~\ref{sectionRealSpaceGlass}) ~\citet{weeks2000}, using 
confocal microscopy, observed  small local rearrangements, detecting avalanches although smaller than those  in  simulations~\cite{sanz2014}. Whereas ~\citet{kwasniewski2014} reported intermittent dynamics due to avalanches  by means of  X-ray photon correlation spectroscopy.   It is still an open question whether these differences are due to different experimental protocols or to some other effect.

\subsection{Nucleation in external fields}
\label{sectionNucleationExternal}

\header{Nucleation under gravity} As discussed in Sec. \ref{sectionDiscrepancy} and Sec.~\ref{sectionEquilibriumPhaseBehavior}, colloidal hard-sphere crystallisation usually occurs under gravity. (The equilibrium case of a system confined by a wall is considered in Sec.~\ref{sectionSinglePlanarWalls}.)
First we consider crystal growth for sedimenting particles on flat surfaces.~\citet{hilhorst2010} have shown that in sedimentary colloidal crystals, obtained from dispersions with high initial volume,  that persistent fcc crystals are favored by the presence of slanted stacking faults, while regions devoid of these defects tend to grow as a random hexagonal close packed (rhcp) structure. Simulations by~\citet{marechal2011} instead attributed the formation of fcc to the free-energy difference between fcc and hcp, and not to the presence of these slanted stacking faults. They also showed that the amount of fcc increases upon lowering the sedimentation rate or decreasing the initial volume fraction.

Compared to the case of crystal growth on a flat wall, crystal growth from a templated surface has shown that it is possible to obtain surprisingly large defect-free crystals.~\citet{jensen2013} controlled growth of fcc crystal by centrifugation (up to $3000~g$) on fcc (100) templates, in contrast to what  is observed for (111) and (110) faces (and flat walls) where high centrifugation rates result in defective or amorphous crystals. These results found confirmation in the simulation work of~\citet{dasgupta2017}, which also confirmed the growth of large defect-free crystals from the (100) fcc face.

Out of equilibrium, a limiting P\'{e}clet number has been identified for crystallisaton to occur, which depends upon the volume fraction~\cite{ackerson1999}. Gravity has also been found to broaden the interface been crystal and fluid with respect to the equilibrium case. Here the largest value of the gravitational Peclet number used was $\Pe_g=0.7$~\cite{dullens2006prl}.

\header{Nucleation under confinement} A more complex scenario is that of confinement between two hard walls. In equilibrium a range of structures is obtained (Sec.~\ref{sectionQuasi2dConfinement}). 
By means of single-particle resolution video microscopy of colloidal films,~\citet{peng2015} demonstrated that transitions between square and triangular lattices  occurred via a two-step diffusive nucleation pathway involving liquid nuclei (due to the low fluid-crystal interfacial energy).
Such a two-step nucleation process  has also been observed in the case of a system confined by two parallel walls separated by four diameters~\cite{qi2015,peng2015}. 
Here a solid-solid phase transition was considered in MD and MC simulations. The transition from a solid consisting of five crystalline layers with square symmetry ($\square$) to a solid consisting of four layers with
triangular symmetry ($4\triangle$) was shown to occur through a nonclassical nucleation mechanism: a precritical fluid cluster within which a cluster of the stable $4\triangle$ phase then grows with one step~\cite{qi2015}. As already discussed in Sec. \ref{sectionNucleation}.F  crystallization of nanoparticles in spherical confinement proceeds via heterogeneous nucleation~\cite{denijs2015}, and very recently, it has been shown to exhibit a bifurcation to decahedral or icosahedral structures~\cite{frumbah2023}.

\begin{figure*}[t]
    \centering
    \includegraphics[width= 130 mm]{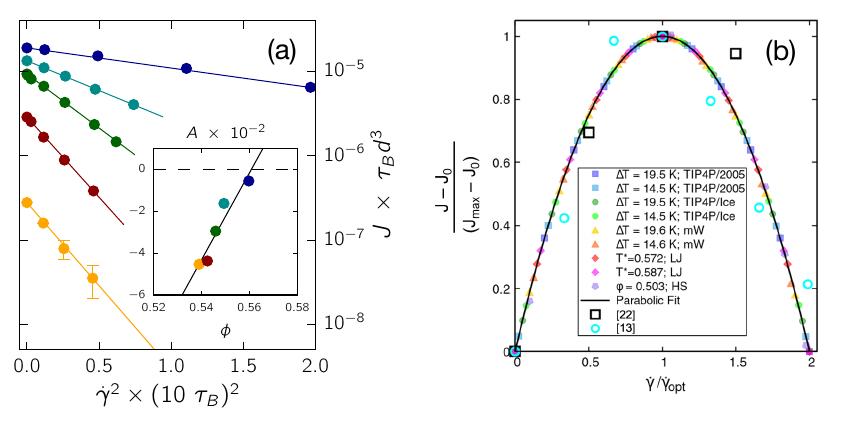}
    \caption{Nucleation under shear. (a)~Nucleation rate $J$ as a function of strain rate for several packing fractions decreasing from top $\phi=0.56$ (blue) to bottom (yellow). The inset shows the slope, which decreases as $\phi$ increases (faster nucleation). Reprinted from~\citet{richard2015}. (b)~Shifted and normalized nucleation rates $J$ as a function of strain rate $\dot\gamma/\dot\gamma_\text{opt}$ for several model liquids including hard spheres at $\phi=0.503$ very close to the binodal showing a non-monotonic behavior. Reprinted from~\citet{goswami2021}. Copyright 2021 by the American Physical Society.}
    \label{figNucShear}
\end{figure*}

\header{Nucleation under shear}
Nucleation in the presence of uniform driving, i.e., sedimentation with $\Pe_g<1$, has already been discussed as a possible cause of the nucleation rate discrepancy (Sec.~\ref{sectionDiscrepancy}). Another important geometry is simple shear flow.
One might expect that steadily shearing a hard-sphere suspension would destroy the entropic forces that favor the crystal, and thus prevent the formation of the crystal starting from the supersaturated melt. Experiments nevertheless show that for weak strain rates ($\Pe_{\dot\gamma}\ll 1$) nucleation persists~\cite{wu2009}. The effect, however, strongly depends on $\phi$ with two regimes that are delineated by the nucleation spinodal (Sec.~\ref{sectionBeyondCNT}), at which the free-energy barrier in quiescent suspensions vanishes. In the activated regime for $\phi<0.56$, nucleation is suppressed by shear and the nucleation rate drops as we decrease $\phi$ [Fig.~\ref{figNucShear}(a)]. This reduction can be rationalized by invoking an increase of the (effective) free-energy barrier\footnote{Although in the driven melt we should rather think of the reversible work to form a solid droplet instead of an equilibrium free-energy difference.} $\Delta F({\dot\gamma})-\Delta F_0\propto\dot\gamma^2$~\cite{blaak2004}, cf.~Fig.~\ref{figNucShear}(a). This increase is dominated by the additional elastic work to strain the solid nucleus~\cite{mura2016}, and a more careful numerical investigation of the effective parameters (the elastic modulus and pressure difference) revealed again a strong dependence on droplet size~\cite{richard2019}.

For packing fractions close to the binodal, it has been posited that the kinetic prefactor is linearly enhanced, leading to non-monotonic nucleation rates $J(\dot\gamma)-J_0\propto1-(\dot\gamma/\dot\gamma_\text{opt}-1)^2$ with the maximum at strain rate $\dot\gamma_\text{opt}$. Figure~\ref{figNucShear}(b) shows the normalized nucleation rate as a function of strain rate for a range of model liquids~\cite{goswami2021}. Going to the opposite limit of high packing fractions ($\phi>0.56$), the quiescent barrier $\Delta F_0$ vanishes and the shear flow again facilitates nucleation through increasing the mobility of single particles, thus helping to overcome the arrested dynamics at high packing fraction~\cite{wu2009,richard2015,koumakis2008}.

Binary systems under shear offer the possibility to model the effects of polydispersity (which appear comparable to quiescent systems)~\cite{masshoff2020}. These studies also hold the potential to explore fractionation in crystallization, which is predicted from numerical simulation and theory (see Sec.~\ref{sectionPolyPhase}).

\subsection{Melting}
\label{sectionMelting}

While hard-sphere crystallisation has received the lion's share of the attention, crystal melting is also of physical interest~\cite{lowen1994,dash1999}. Normally, crystal melting occurs at an existing fluid-crystal interface, without the need for a nucleation event. However, nucleation can play a role in the bulk melting of hard-sphere crystals. Even in the early days of hard-sphere simulations, it was noted that ``in the metastable extension of the crystalline phase [\ldots] small finite systems can resist melting indefinitely~\cite{bennett1971}.'' Such crystals are indeed expected to persist up to the Born limit of mechanical stability, at which the elastic moduli of the crystal vanish~\cite{born1940,wang1997}. Between that spinodal-like point and the thermodynamic melting transition, nucleation is expected to be the pathway  through which equilibrium is attained. Numerical work has shown the melting to be strongly algorithm dependent~\cite{isobe2015}. Controlled studies of nucleation are much more recent.~\citet{wang2018sm} have revealed that the melting spinodal is found well before the Born instability criterion, but that homogeneous nucleation efficiently melts an fcc crystal. Quantitative assessments of the CNT scenario for this process, however, remain largely unexplored.

\begin{figure}[h!]
\centering
\includegraphics[width=85mm]{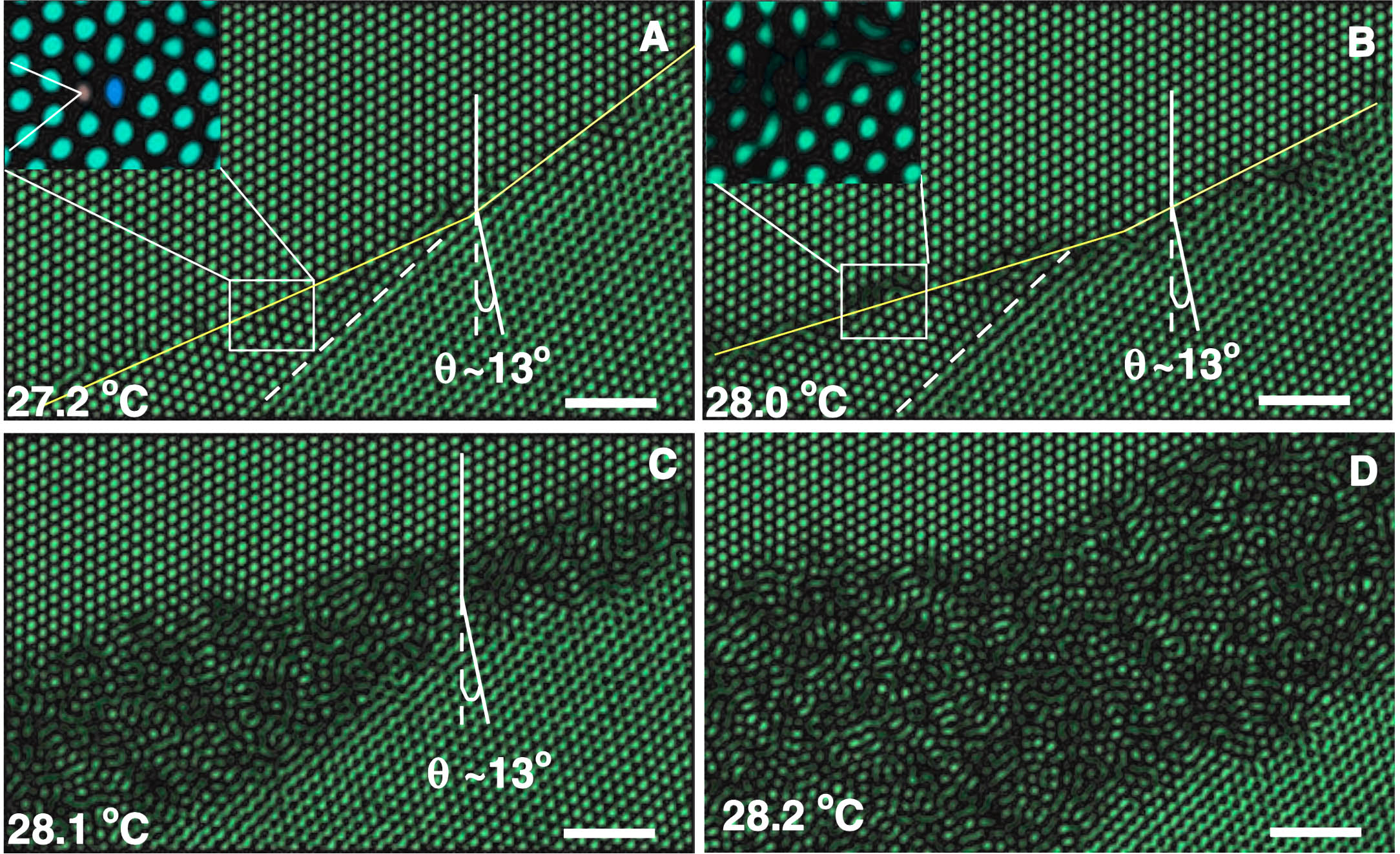}
\caption{Premelting of the colloidal crystal at a grain boundary. The figure shows bright-field images at different temperatures (i.e., particle volume fractions) of two crystallites separated by a grain boundary (crystallites tilted at an angle $\theta$ 
with respect to one another). 
(a) Sample at 27.2 $^\circ$C. The solid and dashed lines show the grain boundary and a partial dislocation, respectively. The grain boundary cuts the two crystals along two different planes (the yellow line has two slopes). It is composed of an array of dislocations; the two extra planes are indicated by lines in the inset. (c) Sample at 28.0 $^\circ$C. The grain boundary starts to premelt; nearby particles undergo liquid-like diffusion (inset). The partial dislocation, denoted by the dashed line, is not affected. (c and d) The same sample at 28.1$^\circ$C and 28.2$^\circ$C, respectively. The width of the premelt region near the grain boundary increases. Scale bars, 5 m$\mu$. Reprinted from~\citet{alsayed2005}
with permission from AAAS. 
}
\label{figArjunYodh}
\end{figure}

Experimentally, non-equibrium melting is quite challenging to explore. In order to melt a hard-sphere crystal, one needs to reduce the volume fraction \emph{in--situ}, which is itself typically hard. There are therefore rather few examples. One solution is to use microgel colloids whose effective diameter may be tuned with temperature. These have been shown to melt, providing insight into local mechanisms~\cite{alsayed2005}. As shown in Fig.~\ref{figArjunYodh}, 3d melting appears to be initiated at grain boundaries. In quasi-2d systems, consistent with later equilibrium work (Sec.~\ref{sectionEquilibrium2d}), two-step melting from the crystal to a hexatic phase and from the hexatic to the fluid phase was observed~\cite{han2008}.

 An ingenious development of this approach is to use laser-induced localized heating of a microgel system. Here, the system was (mildly) heated locally using the lens of the microscope. An attraction of this method is that the melting is initiated in the middle of the system and we can be confident that homogeneous melting is observed, far from any wall effects. Melting precursors were observed in the form of local particle-exchange loops surrounded by particles with large displacements rather than defects. The nucleus size, shape and time-evolution were found to deviate from CNT~\cite{wang2012science}. Another option is to produce an equilibrium sedimentation profile and then to invert it~\cite{turci2016jsm}. This approach of course means that melting is observed in a system with somewhat inhomogenous density, but may be achievable in most experimental hard-sphere systems without the need for a specialized set-up.

An intriguing form of re-entrant surface melting has been observed in hard spheres in experiment. Here an increase in mobility was found in the layer by the wall, when the bulk of the system was crystalline (and exhibited negligible mobility). This increase in mobility was attributed to a 2d-like behavior at the wall~\cite{dullens2004}. Other examples of the role of interfaces include studies of thin (at the colloidal scale) films. Depending on the film thickness, thicker films melted from grain boundaries  while thin solid films melt in one step
~\cite{peng2010}.

%% file: SummaryOutlook/summaryOutlook.tex
\section{Summary and Outlook}

We close this review by taking stock of what we have learned from experiments and related work on colloidal hard spheres, and consider what the future might hold. Undeniably, as one of the most fundamental model systems in colloidal and statistical physics, hard spheres have played a key role in shaping our understanding of a wide range of phenomena. This progress has been partly driven by the inherent simplicity of the model, and partly by the successful intertwining of theoretical and simulation approaches with the experimental realization of hard-sphere colloids. We begin by summarizing the general areas that are well-understood, before moving on to what we regard as outstanding challenges.

\subsection{What have we learned?}

As we discussed in Secs.~\ref{sectionRealizing} and \ref{sectionMeasuring}, a number of reliable methodologies have been developed to realize hard-sphere--like colloidal particles over a range of size scales. Similarly, simulation methods for hard spheres are by now very well-established (Sec.~\ref{sectionSimulation}). Naturally, a key challenge in comparing these is synthesizing particles with interaction potentials as close as possible to the hard-sphere ideal, and obtaining reliable quantitative estimates of discrepancies. While up to six significant figures can be achieved in molecular systems, at this point the characterization of state points in colloidal systems in experiments is accurate to at (very) best three (Sec.~\ref{sectionAccuracy}). Although improvements will no doubt continue to emerge, many studies highlighted in this review show that in many circumstances good agreement already exists between experiments and simulations. The impact of even slight softness should nevertheless continue to be explored (see, e.g.,~\cite{taffs2013, royall2018jcp, dasgupta2020, dejager2022}), and should be carefully considered when interpreting experimental results (Sec.~\ref{sectionMapping}).

In terms of methodological advances, a key recent advance has been the development of experimental systems that are truly close to hard spheres for 3d confocal microscopy work, where the larger colloids required need more care in producing hard sphere like behavior (Sec.~\ref{sectionInteractions})~\cite{kale2023,kodger2015,liu2016,royall2013myth}. Other experimental developments include coupling light scattering with rheology~\cite{tamborini2012} as well as deploying spatially resolved light scattering~\cite{golde2016}, with which one may soon probe dynamical heterogeneity at deeper experimental supercooling than achieved thus far. In parallel, the techniques used in particle-resolved studies have been extended to smaller particles which access longer timescales in terms of their Brownian time~\cite{hallett2018}. This may prove useful in probing dynamics at deeper experimental supercooling. The approach could also be used to investigate crystal nucleation in the regime of supersaturation thus far accessed only by light scattering (with small particles) but with real-space resolution, to probe the hard-sphere nucleation discrepancy. Other experimental developments which could be brought to bear to address that discrepancy include seeding with optical tweezers~\cite{hermes2011,curran2014}, with which a careful study for a range of well--controlled state points  may be expected to yield considerable insight. Experimentally inferring local forces and stresses~\cite{lin2016nmat,dong2022} may also reveal details into failure mechanisms of amorphous and crystalline solids.

Self-assembly of hard-sphere systems in confinement still often yields surprises, such as the emergence of crystal structures that are not favored in bulk systems (Sec.~\ref{sectionSphericalConfinement}), and is likely to continue to do so in the future, especially when coupled with size dispersity. Similarly, the careful exploration of various ranges of size ratios in binary hard sphere mixtures (Sec.~\ref{sectionBinary}),   has revealed  the existence of (metastable) fluid-fluid demixing as well as the formation of complex crystals. Self-assembly of yet other crystal phases at size ratios that have not been (as carefully) explored is expected, given the wealth of structures known to exist at infinite pressure \cite{hopkins2012}. More generally, the consideration of binary mixtures naturally enriches all of the other phenomena discussed in this review and plays a particularly important role in the study of glass forming systems.

How external forces and fields modify the structure and dynamics of hard spheres has been explored in Sec.~\ref{sectionFarFromEq}. The solvent plays an important role in out-of-equilibrium systems, in that it then hydrodynamically couples forces. The careful treatment of hydrodynamic interactions  (Sec.~\ref{sectionSimHydro}) has therefore significantly improved our understanding of phenomena such as sedimentation and shear thickening. Theoretical insights have further led to a rather comprehensive picture connecting deformations of the microstructure to the macroscopic material properties probed in rheology experiments.

Hard spheres have played a key role in both computer simulation  and the experimental study of glass physics (see Sec.~\ref{sectionGlass}). As experimental systems and simulation techniques evolve to probe structure and dynamics at ever deeper supercooling, hard-sphere studies will hopefully lead to an even stronger framework for understanding glass formation.  If it were possible to use still smaller particles in experiment than those used by~\citet{hallett2018}, then those could be a means to probe, for example, the additional increase of dynamical length scales and the hierarchy in dynamical behavior in experiment~\cite{scalliet2022,ortlieb2023}. Related to this would be the structural relaxation mechanism at deep supercooling. Failure in amorphous soft materials and, in the context of hard spheres, glasses is a most promising area for future work. This brings together multiple challenges of aging non--equilibrium materials. For example, few studies have addressed the Gardner transition (Sec.~\ref{sectionJamming}) with colloids and it is tempting to imagine that a combination of shear and force measurement may be a way to do so. A further exciting possibility would be to use hard-sphere experiments to investigate the ductile--brittle transition recently found in computer simulation~\cite{ozawa2018}. Developments in X-ray scattering (see Sec.~\ref{sectionRealSpaceGlass}) that reveal higher--order structure and dynamical information may be able to play a key role in equilibrating samples much closer to the glkass transition, and also in annealing samples that might exhibit brittle behavior~\cite{wochner2009,lehmkuhler2020,liu2022}.

Regarding the crystallization at high concentrations, it is still not clear whether the high-concentration crystallization in colloid experiments results from residual crystals in the samples arising from their preparation. Concerning experiments, \citet{vanmegen1993nature} suggested that the shear-melting process by which amorphous colloid samples were prepared could leave small shear-aligned nuclei on which a crystal could grow. Concerning simulations, there is  the possibility that crystals could be formed during compression of the system to the high concentration regime. These crystals could later on act as seeds for further crystallization.

Another important challenge is the observation of avalanche crystallization in experiments. Since they are rare events, one might need a huge amount of data to analyze and store, in order to find avalanches in confocal microscopy experiments (as those in~\citet{weeks2000}). Concerning the (light or X-ray) scattering experiments (as those in~\citet{kwasniewski2014}), one might vary the size of the scattering volume and use a small number of particles to be able to clearly detect intermittency with long quiescent periods. The reason for suggesting this approach is that in a larger volume containing many particles avalanches might occur simultaneously in different regions of the system, and  the dynamics would appear much more homogeneous.

\subsection{Open challenges}

We now turn to areas in which relatively little work has been done, or in which major developments are sorely needed. At the experimental level, a major bottleneck with particle--resolved studies is the quantity of data produced. Experimental papers with very few coordinate sets analysed are common~\cite{dong2022}. Often the technique seems to promise much, based on the exquisite precision of the data, but the quantity obtained presents the kind of statistically meaningful analysis that would be required to fully exploit the technique. While great strides have been made in the ability to handle large data sets, the issue often lies in the acquisition of the data, which is limited by the scanning of a single lens across a sample. It therefore is not easy to see how this can be improved massively, but if some means can be found to increase the rate of data acquisition then many challenges relating to quantity of data might be addressed. For example, imagine being able to compare the shapes of nuclei of a certain size between experiment and simulation in the context of the nucleation discrepancy. A further issue is that the system can evolve during the time taken to scan. One direction that has been explored is to ``freeze'' the solvent, by polymerizing it \emph{in-situ}. This enables the imaging of much larger regions, and images containing in excess of a million particles have been obtained (the typical values is around ten thousand)~\cite{vanderwee2019}.

In monodisperse hard spheres in equilibrium, interfaces stand out as a major challenge. As discussed in Sec.~\ref{sectionInterfaces}, numerical and theoretical methods to obtain the interfacial free energy and stiffness vary by up to 30\%. Meanwhile, experimental measurements show even greater discrepancy, often differing from simulation and theory by a factor of two or more. Given the importance of these quantities in the context of nucleation, there is a clear interest in arriving at generally agreed values for interfacial free energies and stiffness, and for well-controlled experimental validation. This discrepancy might stem from a high sensitivity of interfacial properties to ``small'' deviations from hard-sphere behavior. Put differently, if accurate values were established for the hard-sphere interfacial free energy and stiffness, these results could prove valuable benchmarks for assessing how close to a hard sphere a given (experimental) system is. Another interfacial phenomenon, grain boundaries, remains largely unexplored in 3d hard spheres. In simulation this may reflect the need for large system sizes, but in experiment, there is much to be done.

Our knowledge of hard-sphere phase behavior becomes more clouded when mixing more than two particle sizes. Due to the vastness of the resulting parameter spaces, the exploration of crystallization in ternary and higher-component systems has been relatively sparse~\cite{stucke2003, wang2016, koshoji2021}, and hence a clear overview of the resulting phase behavior is largely missing. For hard-sphere systems with relatively large polydispersity, the equilibrium phase behavior is essentially unknown. Simulations suggest the possible stability of complex crystal structures~\cite{lindquist2018, bommineni2019}, but whether these are indeed thermodynamically stable or realizable in experiments, remain open questions. Therefore, the possibility of surprising new physics in mixtures of hard spheres should not be discounted.

The nucleation process in monodisperse hard-sphere systems remains a topic of intense debate. As discussed in Sec.~\ref{sectionNucleation}, a number of reasons have been proposed for the ongoing discrepancy between experimentally measured and theoretically predicted nucleation rates of the hard-sphere crystal phase. However, resolving this issue will remain difficult as long as real-space experiments have trouble probing the time scales required to observe nucleation in the low supersaturation regime where this discrepancy is observed. One possibility would be to probe pre-critical nuclei. That experiment~\cite{wood2018} and simulation~\cite{fiorucci2020} disagree wildly in measurement of higher-order structure in sedimenting metastable fluids suggests that this might be a fruitful line of enquiry.

Alternatively, using smaller particles could enable particle-resolved studies in the discrepancy regime. Such experiments will have to rely on advances in imaging techniques (see, e.g., Refs.~\cite{hell2007,hallett2018}), as well as in (possibly machine-learning-based) algorithms for obtaining particle coordinates from lower-resolution images (Sec.~\ref{sectionOptical}). As noted above, the potential for the new X-ray scattering methods to access smalelr partcoels may also be important here~\cite{wochner2009,lehmkuhler2020,liu2022}.

Perhaps another approach may yield fruit here, and this would be to determine the volume fraction accurately or measure the nucleation rate at very much lower supersaturations. Disappointing as it may be, the possibility that the discrepancy could be resolved by a correct determination of (effective) volume fraction absolutely cannot be ruled out. However, if there is an issue with experimental determination of the volume fraction, it is \emph{systematically underestimated} and why this should be remains a profound mystery. That there are two lines in the experimental data (Fig.~\ref{figOneToRuleThemAll}) with significantly differing rates of sedimentation might be seen as a smoking gun. However investigations of the effects of sedimentation which can change the higher--order structure of the system, reducing the barrier to nucleation did not fully address the discrepancy~\cite{wood2018,fiorucci2020}.

Experimental investigations of polymorph selection pathways remain few and far between (save for~\citet{gispen2023acsnano}, see Sec.~\ref{sectionChallengesCNT}). Confocal microscopy experiments have observed the real-space formation of relatively ordered precursor structures with different symmetries, but understanding their conversion into crystalline nuclei remains a challenge. Furthermore, the observation in simulation of a preference for fcc~\cite{leoni2021,gispen2023acsnano} is not reproduced in experiment which seems to form rhcp~\cite{kegel2000jcp,martelozzo2002}. A systematic comparison of polymorph selection at well-matched state points between experiment and simulation would seem to be in order.

Overall, the main lesson learned from the rich history and broad applicability of the hard-sphere model is the impressive complexity that emerges even from a simple system. Depending on their density, size dispersity, and environment, hard spheres demonstrate that energetic interactions are not a requirement for, e.g., the formation of amazingly complex (quasi)crystal structures, for probing the complex and heterogeneous dynamics of glassy materials, or gaining insight into far-from-equilibrium phenomena. Investigating these different processes in hard spheres instead provides a framework for understanding these same phenomena for systems driven by more complex interactions, regardless of whether they consist of atoms, molecules, or more complex colloidal building blocks. As a result, hard spheres will inevitably continue to function as a key reference system in statistical physics and colloid science, and as a testing ground for newly developed theoretical and computational techniques. Moreover, as we have highlighted in this review, hard spheres continue to hold surprises, challenges, and unresolved questions that will inevitably spark new research directions in the years and decades to come.